\documentclass[aps,pra,twocolumn,groupedaddress,amsmath,amssymb,superscriptaddress]{revtex4-2}
\usepackage{graphicx}  
\usepackage{dcolumn}   
\usepackage{bm}        
\usepackage{verbatim}   
\usepackage{ mathrsfs }
\usepackage[colorlinks=true,linkcolor=blue,citecolor=blue,allcolors=blue]{hyperref}
\numberwithin{equation}{section}
\usepackage{float}

\begin{document}

\title{Quench dynamics of two interacting atoms in a one-dimensional anharmonic trap}

\author{I. S. Ishmukhamedov}
\email{i.ishmukhamedov@mail.ru}
\affiliation{Bogoliubov Laboratory of Theoretical Physics, Joint Institute for Nuclear Research, Dubna, Moscow Region 141980, Russian Federation}
\affiliation{Al-Farabi Kazakh National University, Almaty 050040, Kazakhstan}
\affiliation{Institute of Nuclear Physics, Almaty 050032, Kazakhstan}
\begin{abstract}
A temporal response of two interacting particles to a quench of the coupling strength in one-dimensional harmonic and anharmonic traps is explored. The coupling strength is changed from repulsive to attractive interactions and vice versa. The time evolution of the fidelity, wave packet, one-body reduced density matrix and momentum distributions is analyzed in details. It was found that impacts of the pre- and postquench states interchange during the dynamics. In the case of the anharmonic trap additional contribution of the center-of-mass excited states comes into play and the whole evolution becomes significantly altered. Yet the impact of the pre- and postquench states in some cases still could be identified. The quench dynamics of the ground and excited states are considered.
\end{abstract}

\maketitle

\section{Introduction}

The physics of ultracold atoms has become one of the most promising fields of research. It has gained much attention in variety of ways such as in a simulation of the Breit-Wheeler pair production for QED study \cite{titov}, a search of improved methods of sympathetic cooling \cite{melezhik}, a study of precision spectroscopy \cite{korobov}, a study of a transport of a Bose-Einstein condensate (BEC) \cite{nesterenko}, a study of Efimov physics \cite{roudnev}, a study of Josephson junctions \cite{shukrinov} and many more. A quantum computer \cite{hartke} is one of the prominent examples of the application that this research can provide. All this is made possible due to a possibility of precise controlling of the interaction between particles, the geometry in which they are localized and their state.

The external potential that traps atoms is usually approximated by a potential of a harmonic oscillator. This assumption led to many insightful models, which successfully explained the underlying physics \cite{busch,olshanii,petrov}. Yet as was found in \cite{haller,kestner,sala,peng,zephania,wang} anharmonic corrections to the trapping potential are needed to be considered, in order to explain resonances, which are not covered by the harmonic approximation of the potential. Another difficulty is connected with the interaction potential between the particles, since not every theoretical model can easily incorporate it into the model correctly. For example, tunneling rates calculated within a Wentzel-Kramers-Brillouin (WKB) theory lacks of accuracy due to its inability to include the interaction potential into the consideration \cite{gharashi}. Nevertheless, such problems can be overcomed by means of advanced numerical methods \cite{vinitsky,penkov} and then applied to tunneling problems \cite{dobrzyniecki,giacosa,dobrzyniecki2,ishmukh,ishmukh2,ishmukh3}.

Nonequilibrium dynamics is present in almost every real physical system \cite{mistakidis,mistakidis2,mistakidis3,mockel,kerin}. Nuclear reactions in such stars as the sun and neutron star, superfluid turbulence, mesoscopic electrical circuits and many more are the examples where this dynamics takes place. Setups with ultracold atoms opens up quite a reliable facility to investigate such physics with an unprecedented accuracy. It is possible to localize even two atoms subject to an external field and control their interaction strength in a wide range of its intensity \cite{zuern,zuern2}. Also, such setups allow one to reduce the geometry of the trapping potential to make it almost a one-dimensional one, which not only simplifies matters significantly, but also leads to new unique phenomena, available exclusively in one-dimensional systems \cite{cazalilla}. Thus, exploring the system with two interacting atoms in a one-dimensional trapping potential can be served as a natural step towards understanding of the basic, yet important principles of quantum physics and the quench dynamics in particular.

Following \cite{mistakidis}, where the system of two interacting atoms with a bosonic spatial symmetry in a one-dimensional harmonic trap is investigated, we also consider a similar system, but with some changes, leading to a more complete and even new understanding of the quench dynamics of the system. The quench dynamics is proceeded by changing the coupling strength of the interaction between the particles from repulsive to attractive and vice versa for an intermediate value of the coupling. The initial and final states are called pre- and postquench states, respectively. The work \cite{mistakidis} analyzes an impact of the postquench states on the quench dynamics in terms of the time-evolution of the fidelity, wave packet, one-body reduced density matrix and momentum distribution. It was found that the impact of these states is such that the system occupies postquench excited states for some time during its evolution. The current paper, on the other hand, considers also the prequench states and in the same manner analyzes their impact on the quench dynamics. It was found that the contribution of these states is quite significant and may be even leading in some cases. Moreover, the inclusion of anharmonic corrections to the trapping potential alters the quench dynamics dramatically. Due to the coupling of the relative and center-of-mass motions in the anharmonic trap the center-of-mass excited states become involved and the whole oscillation pattern is distorted considerably.

The remainder of this paper is organized as follows. Section~\ref{model} introduces the Hamiltonian of the two-particle system and the methods for solving of the corresponding Schr\"{o}inger equation. The section also provides results of the computed energy spectrum. Sections~\ref{quench} and \ref{quench2} discuss the results of the quench dynamics of the system. Impacts of both the pre- and postquench states on the dynamics are scrutinized by analyzing the evolution of the fidelty, wave packet, one-body reduced density matrix and momentum distribution. Section~\ref{summary} summarizes all the obtained results and provides an outlook.

\section{Model Hamiltonian}
\label{model}
We consider the Hamiltonian of two identical atoms, each with mass $m$,
\begin{eqnarray}\label{ham}
\nonumber
H=&-&\frac{\hbar^2}{2m}\frac{\partial^2}{\partial x_1^2}-\frac{\hbar^2}{2m}\frac{\partial^2}{\partial x_2^2}+V(x_1)+V(x_2)
\\
&+&V_{\textrm{int}}(x_1-x_2)\,\,,
\end{eqnarray}
where the potentials $V(x)$ describe an external trap potential. We assume that the atoms are subjected to an optical lattice potential, which has the form
\begin{eqnarray}\label{lat}
V_{\textrm{opt}}(x)=V_\textmd{d}\sin^2\left(\frac{2\pi}{\lambda}x\right),
\end{eqnarray}
where $V_d$ defines the depth of the trap and $\lambda$ corresponding wavelength. We parameterize \eqref{lat} as
\begin{eqnarray}
V_{\textrm{opt}}(x)=-\frac{\hbar\omega}{12\alpha}\sin^2\left(\sqrt{-6\alpha}\dfrac{x}{\ell} \right),
\end{eqnarray}
where the trap anharmonicity $\alpha$ and frequency $\omega$ are introduced as
\begin{eqnarray}
\alpha=-\dfrac{8\pi^2 \hbar}{12\lambda^2 m \omega}, \hspace{.3cm}
\omega=\dfrac{2\pi}{\lambda}\sqrt{\frac{2|V_\textmd{d}|}{m}}
\end{eqnarray}

We consider the value $\alpha=-0.03$ for the anharmonic trap. This value approximately corresponds to the trap frequency $\omega=2\pi\times14.5$ kHz and wavelength $\lambda=1.064\times10^{-4}$ cm, which are used in the Innsbruck experiment \cite{haller} for $^133$Cs atoms.

\begin{figure}
	\centering
	\includegraphics[width=6cm,clip]{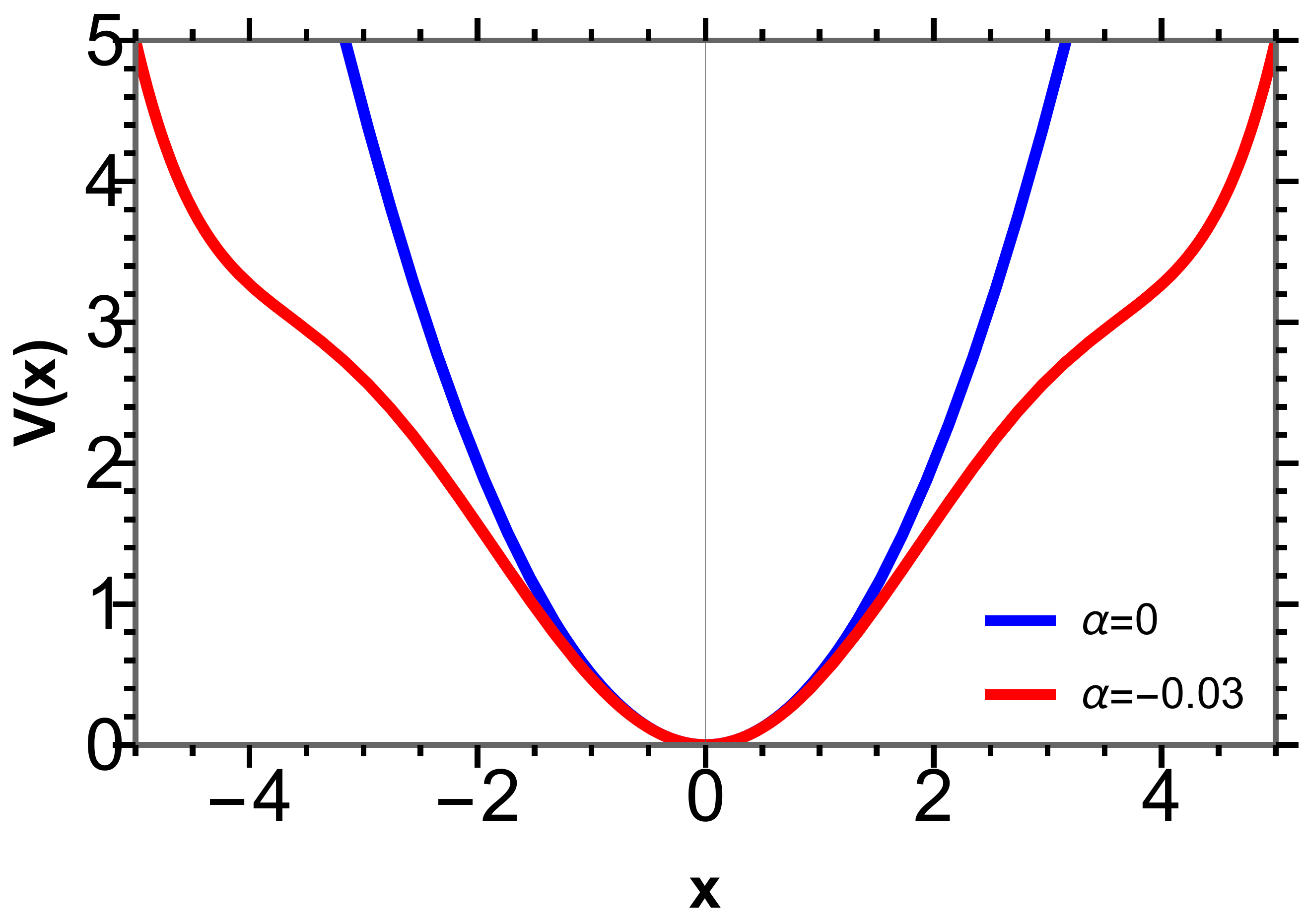}
	\caption{(Color online) Trap potential $V(x)$ \eqref{trap} for $\alpha=0$ (blue line, the harmonic trap) and $\alpha=-0.03$ (red line, anharmonic trap) in the oscillator units.}\label{trap_plot}
\end{figure}

We restrict ourselves to the case of a single-well trap by expanding and cutting off the series' terms up to the sixth-order in $x$:
\begin{eqnarray}\label{trap}
V(x)=\hbar \omega \left(
\frac{1}{2}\left(\frac{x_j}{\ell}\right)^2+\alpha  \left(\frac{x_j}{\ell}\right)^4
+\frac{4 \alpha ^2}{5}\left(\frac{x_j}{\ell}\right)^6 \right),
\end{eqnarray}
Such approximation of the trap potential is quite an accurate representation of a single site of an optical lattice \cite{grishkevich}. The plot of the potential is shown in Fig.~\ref{trap_plot}. In the following we use the oscillator units, where length is defined as $\ell=\sqrt{\frac{\hbar}{m\omega}}$ and energy as $\hbar\omega$.

The interatomic potential $V_{\textrm{int}}(x_1-x_2)$ is chosen in the Gaussian form
\begin{eqnarray}\label{gauss}
V_{\textrm{int}}(x_1-x_2)=-V_{\textrm{G}}\exp\left\{-\frac{(x_1-x_2)^2}{2r_0^2}\right\}
\end{eqnarray}
with $V_{\textrm{G}}$ and $r_0$ defining the depth and range of the interaction. As has been shown in \cite{kestner,ishmukh4,gharashi} the value $r_0=0.1$ well describes a short-range interaction, which takes place for low energies.

We also would like to switch from the parameter $V_{\textrm{G}}$ to a more commonly used parameter in the physics of ultracold atoms - the one-dimensional coupling strength $g$. This can be done by adjusting $V_{\textrm{G}}$ to yield the same bound state energy as that for the zero-range potential $\sqrt{2}g\delta(x_1-x_2)$, where $\sqrt{2}$ is used for a convenience as in \cite{busch,mistakidis}. 

\begin{figure}
	\centering
	\includegraphics[width=7cm,clip]{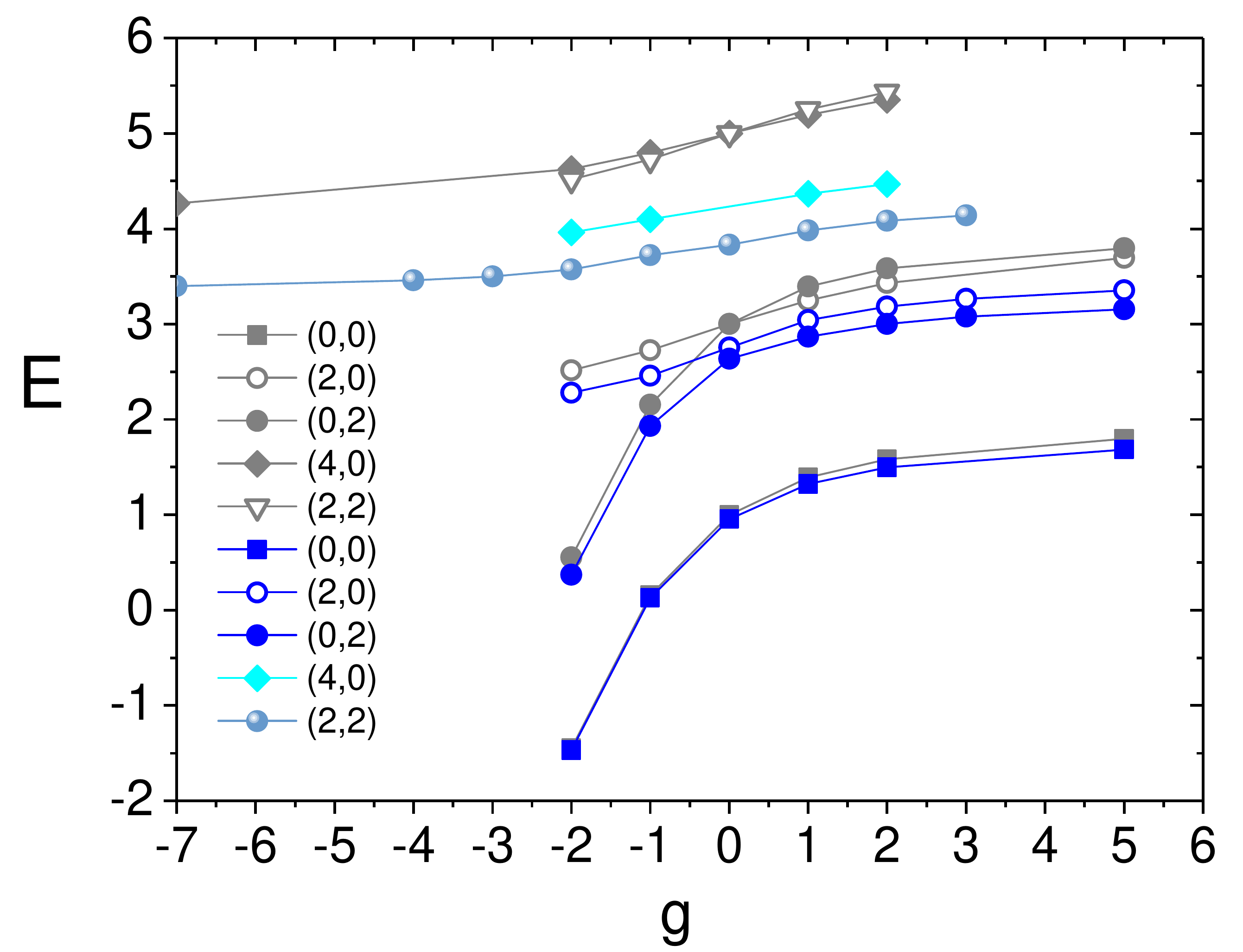}
	\caption{(Color online) Bound energy levels for the ground and excited states in the cases of harmonic (gray lines) and anharmonic (color lines) traps as functions of the coupling strength $g$. The indices $(n,N)$ refer to the quantum numbers of the relative and center-of-mass motions.}\label{energy}
\end{figure}

As a first step, we solve the stationary Schr\"{o}inger equation with the Hamiltonian \eqref{ham} and the potentials \eqref{trap} and \eqref{gauss}. This task is accomplished by using a method in \cite{ishmukh5}, where the problem is reduced to the problem of finding eigenvalues and eigenvectors by means of the shifted inverse power method. The obtained solutions are served as initial states when solving the time-dependent Schr\"{o}inger equation. In Fig.~\ref{energy} the energy levels of the Hamiltonian \eqref{ham} as functions of the coupling strength $g$ in the cases of both the harmonic $\alpha=0$ and anharmonic $\alpha=-0.03$ trap potentials are shown. The indices $(n,N)$ refer to the quantum numbers of the relative and center-of-mass motions. The avoided crossings of the energy levels of the anharmonic trap are due to the rotational symmetry breaking, which lifts the degeneracy of the energy levels of the harmonic trap.

The quench dynamics of the two-body system \eqref{ham} is governed by the time-dependent Schr\"{o}inger equation
\begin{eqnarray}
	i\frac{\partial \Psi(x_1,x_2,t)}{\partial t}=H(x_1,x_2)\Psi(x_1,x_2,t).
\end{eqnarray}
To find its solution we employ the split-operator method, which leads to
\begin{equation}\label{split}
	\begin{aligned}	
	&\Psi(x_1,x_2,t+\Delta t)=\exp\left\{-i \frac{\Delta t}{2\hbar} V_{\textrm{int}}(x_1-x_2)\right\}
	\\ 
	&\times\exp\left\{-\frac{i \Delta t H_1(x_1)}{\hbar}\right\}
	\exp\left\{-\frac{i \Delta t H_2(x_2)}{\hbar}\right\}
	\\
	&\times
	\exp\left\{-i \frac{\Delta t}{2\hbar} V_{\textrm{int}}(x_1-x_2)\right\}
	\\
	&\times \Psi(x_1,x_2,t)
	\end{aligned}
\end{equation}
where
\begin{eqnarray}
	H_j(x_j)=-\frac{\hbar^2}{2m}\frac{\partial^2}{\partial x_j^2}+V(x_j),\hspace{.5cm}j=1,2.
\end{eqnarray}
are the single-particle Hamiltonians with the trap potential \eqref{trap}. The accuracy of the equation \eqref{split} is correct up to $\mathcal{O}(\Delta t^3)$, where $\Delta t$ is a step of the time variable. The spatial derivatives are approximated by the central differences with the sixth-order accuracy. To use \eqref{split} we use the initial state, which we discussed above.

To analyze the quench dynamics we consider the overlap integral $\mathcal{Q}$
\begin{equation}
	\mathcal{Q}=\left|\int\int dx_1dx_2 \Psi_{n,N}(x_1,x_2)\Psi(x_1,x_2,t)\right|
\end{equation}
between the stationary states $\Psi_{n,N}(x_1,x_2)$ and the time-evolved state $\Psi(x_1,x_2,t)$. The overlap between the initial state and the evolved state is called the fidelity $F(t)$.

Another quantity of interest is the momentum distribution
\begin{equation}\label{momentum}
	n(k,t)=\frac{1}{2\pi}\int\int dx_1 dx_1' \rho^{(1)}(x_1,x_1',t)e^{-ik(x_1-x_1')},
\end{equation}
of the one-body reduced density matrix
\begin{eqnarray}\label{rho}
	\rho^{(1)}(x_1,x_1',t)=\int dx_2 \Psi(x_1,x_2,t)\Psi^{\ast}(x_1',x_2,t).
\end{eqnarray}
The momentum distribution is experimentally accessible via time-of-flight measurements \cite{bloch}. 

\section{Quench dynamics from repulsive to attractive interactions}\label{quench}

In this section the initial state is prepared for $g=2$ and at $t>0$ the interaction strength is switched to $g=-2$. First, we consider the case of the harmonic trap and the ground initial state. We analyze the evolution of the fidelity, wave packet, one-body reduced density matrix and momentum distribution. Next, we perform similar analysis for the anharmonic trap and consider afterwards, in the last parts of the section, the quench dynamics starting from excited states.

\subsection{Harmonic trap, $\alpha=0$, the ground state, $(0,0)$.}

\begin{figure}
\centering
\textbf{$\alpha=0$}\par\medskip
\includegraphics[width=8.5cm,clip]{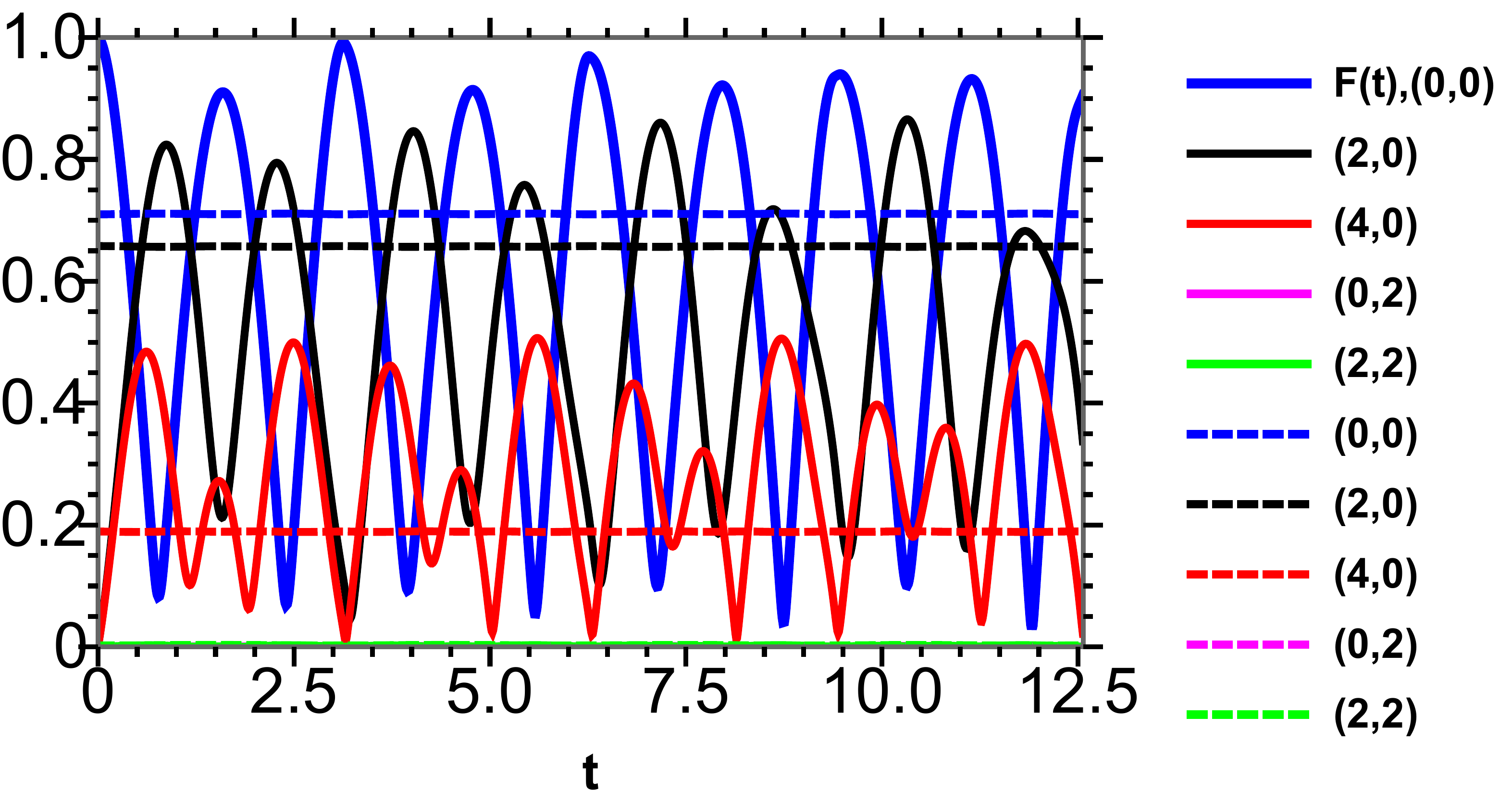}
\caption{(Color online) Fidelity $F(t)$ and the overlap integrals $\mathcal{Q}$ between the time-evolving state $\Psi(x_1,x_2,t)$ and different pre- (solid lines) and postquench (dashed lines) states in the case of the harmonic trap. The indices $(n,N)$ refer to the states with the quantum numbers of the relative and center-of-mass motions. The system of the two atoms is prepared in the ground state with $g=2$ and quenched to $g=-2$.}\label{fwd}
\end{figure}

The time evolution of the fidelity of the quench dynamics for the ground state is shown in Fig.~\ref{fwd}.

Fig.~\ref{fwd} shows the overlap integrals $\mathcal{Q}$ between the time-evolving state $\Psi(x_1,x_2,t)$  and stationary ground and even excited states $\Psi_{n,N}(x_1,x_2)$, where $(n,N)$ denotes quantum numbers defining the states. In Fig.~\ref{fwd} we see that the fidelity and the two overlaps for the prequench states $(2,0)$ (black solid line) and $(4,0)$ (red solid line), oscillate almost in antiphase: the maximum value of $F(t)$ is reached when the other ones takes their minimum values. This indicates that impacts of the prequench states are significant. The system oscillates periodically and approximately returns to its initial state at $t=\pi$. As can be seen, the oscillations alter with time. In Fig.~\ref{fwd} one can also notice the overlaps between the postquench ground and excited states, depicted as dashed lines. These overlaps are constant in time. The reason for that is that the time-dependent solution can be expanded over a sum of postquench stationary states \cite{mistakidis}, which are orthogonal to each other and hence only the corresponding overlap integral in this sum will remain. Thus, the system undergoes a state transition between the initial state and different pre- and postquench states.

\begin{figure}
	\centering
	\textbf{$\alpha=0$}\par\medskip
\begin{tabular}{ccc}
$t=0.001$ & $t=0.1$ & $t=\pi/8$ \\
\includegraphics[width=.15\textwidth]{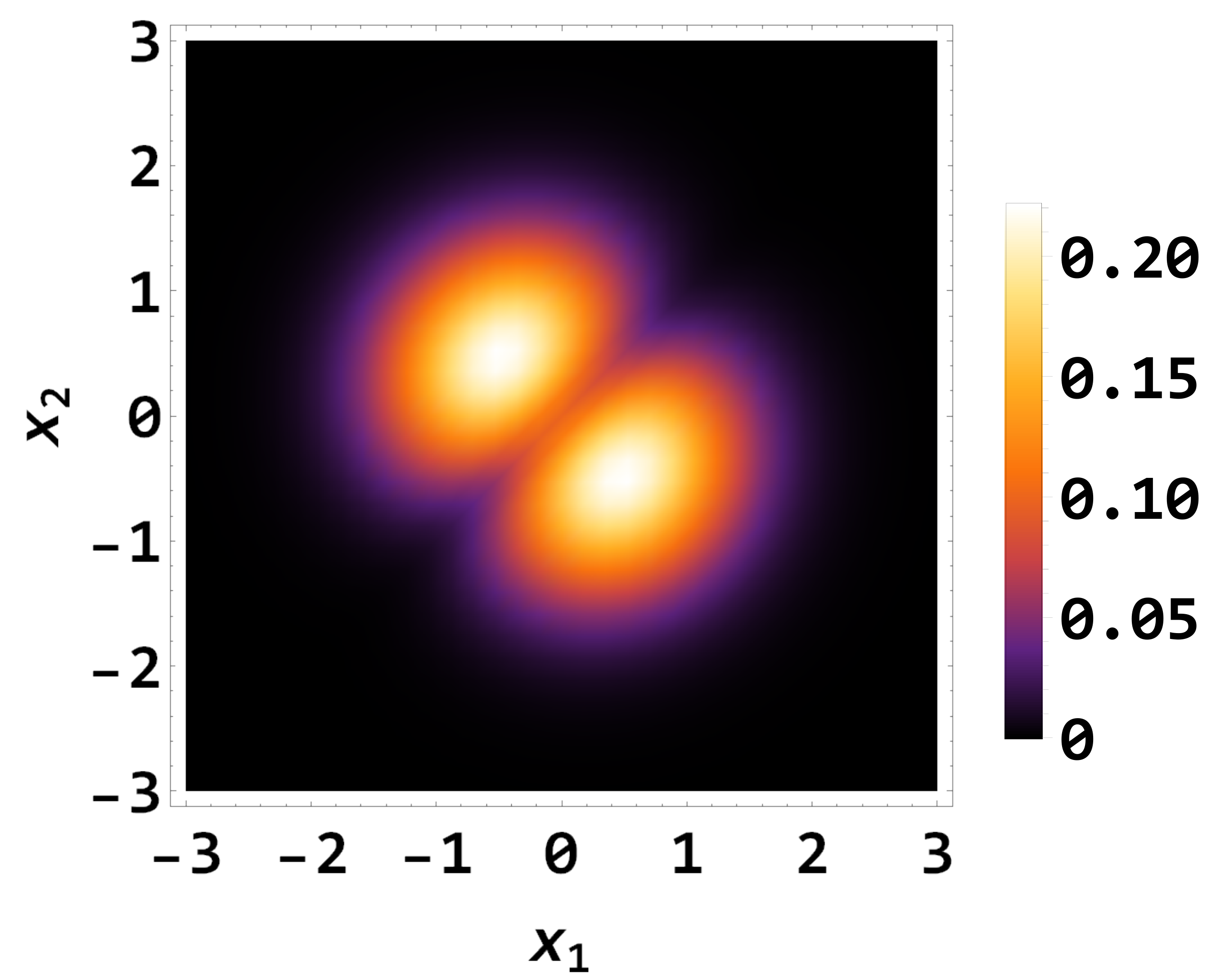} &
\includegraphics[width=.15\textwidth]{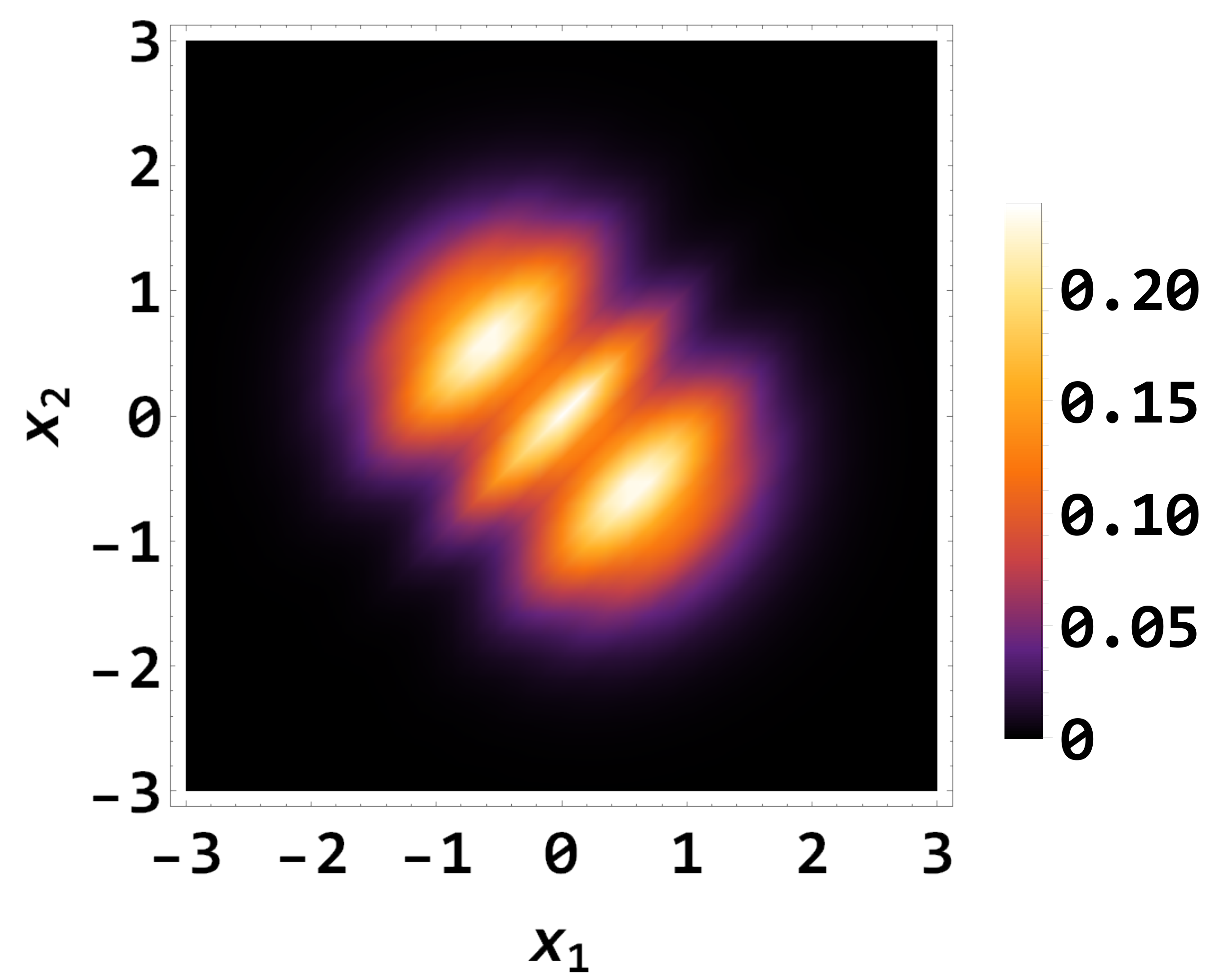} &
\includegraphics[width=.15\textwidth]{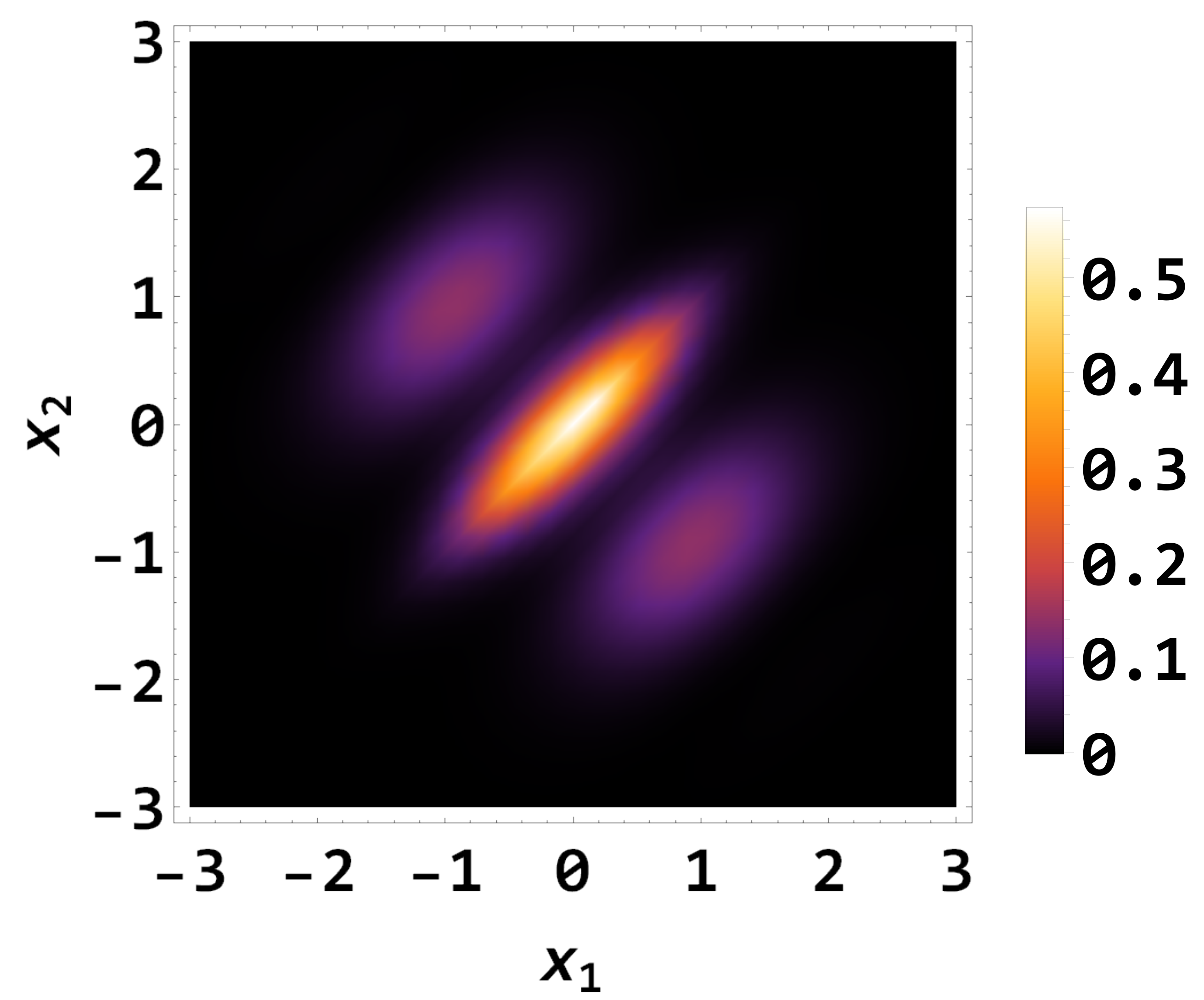} \\
$t=\pi/4$ & $t=\pi/2$ & $t=\pi$\\
\includegraphics[width=.15\textwidth]{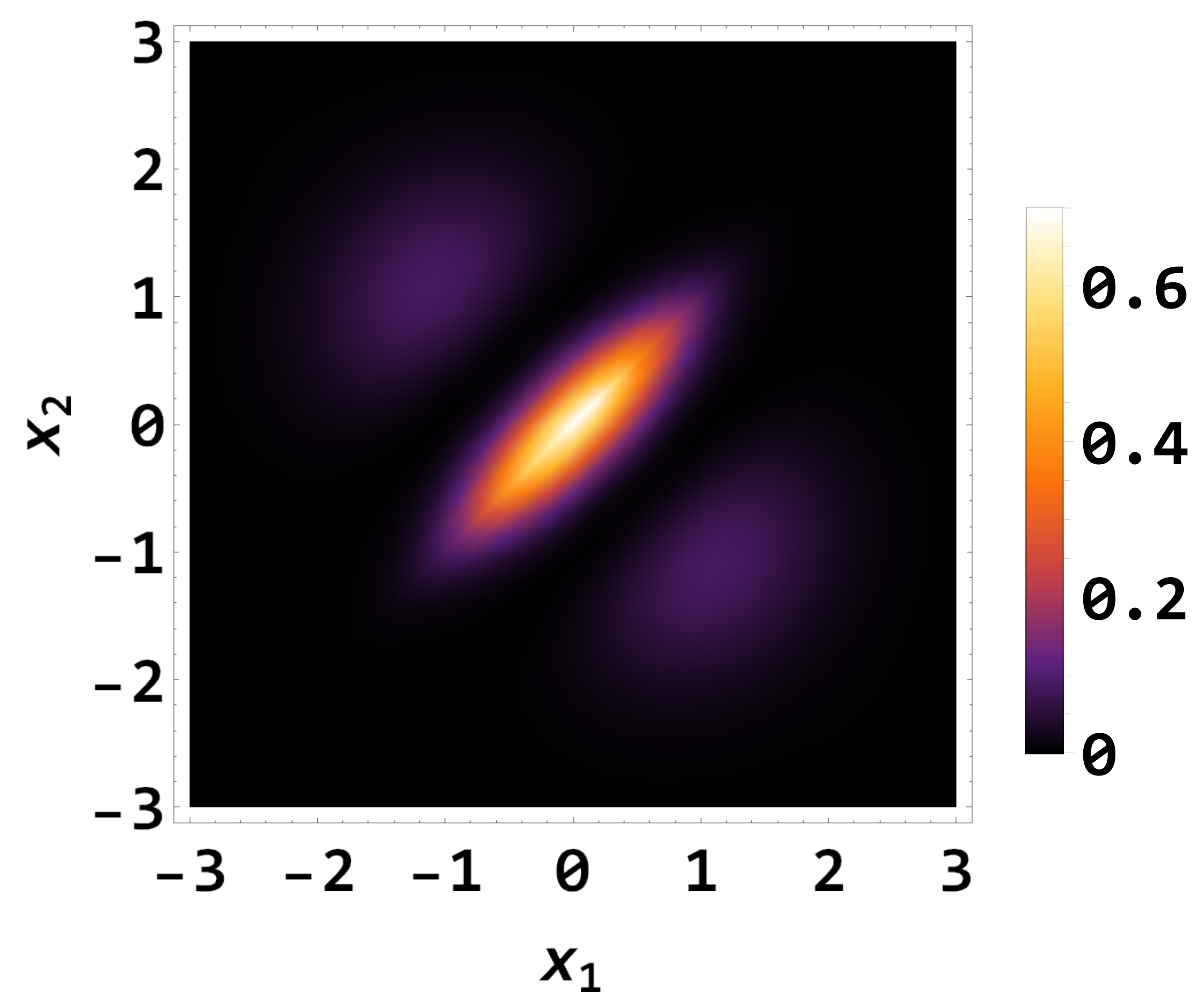} &
\includegraphics[width=.15\textwidth]{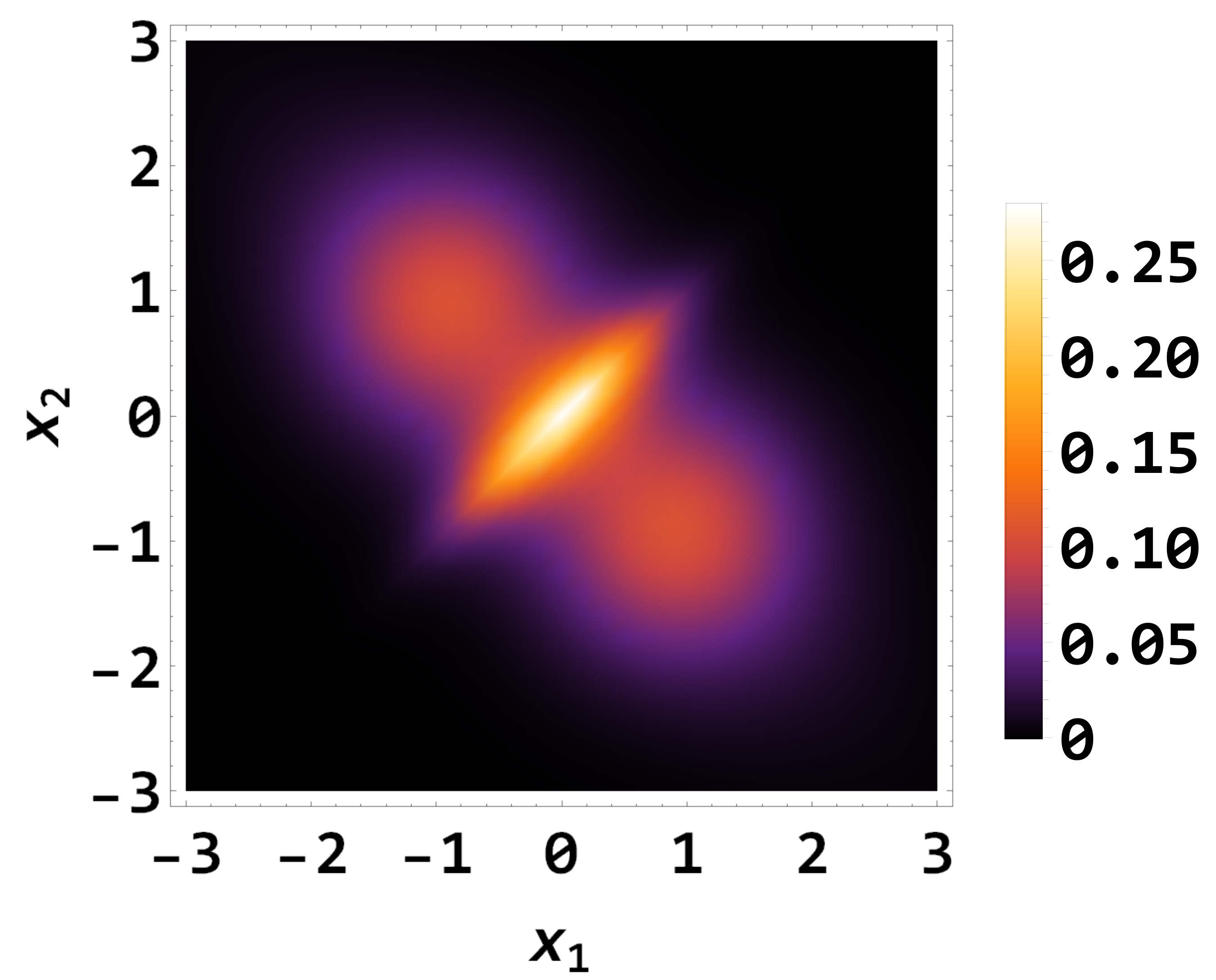} &
\includegraphics[width=.15\textwidth]{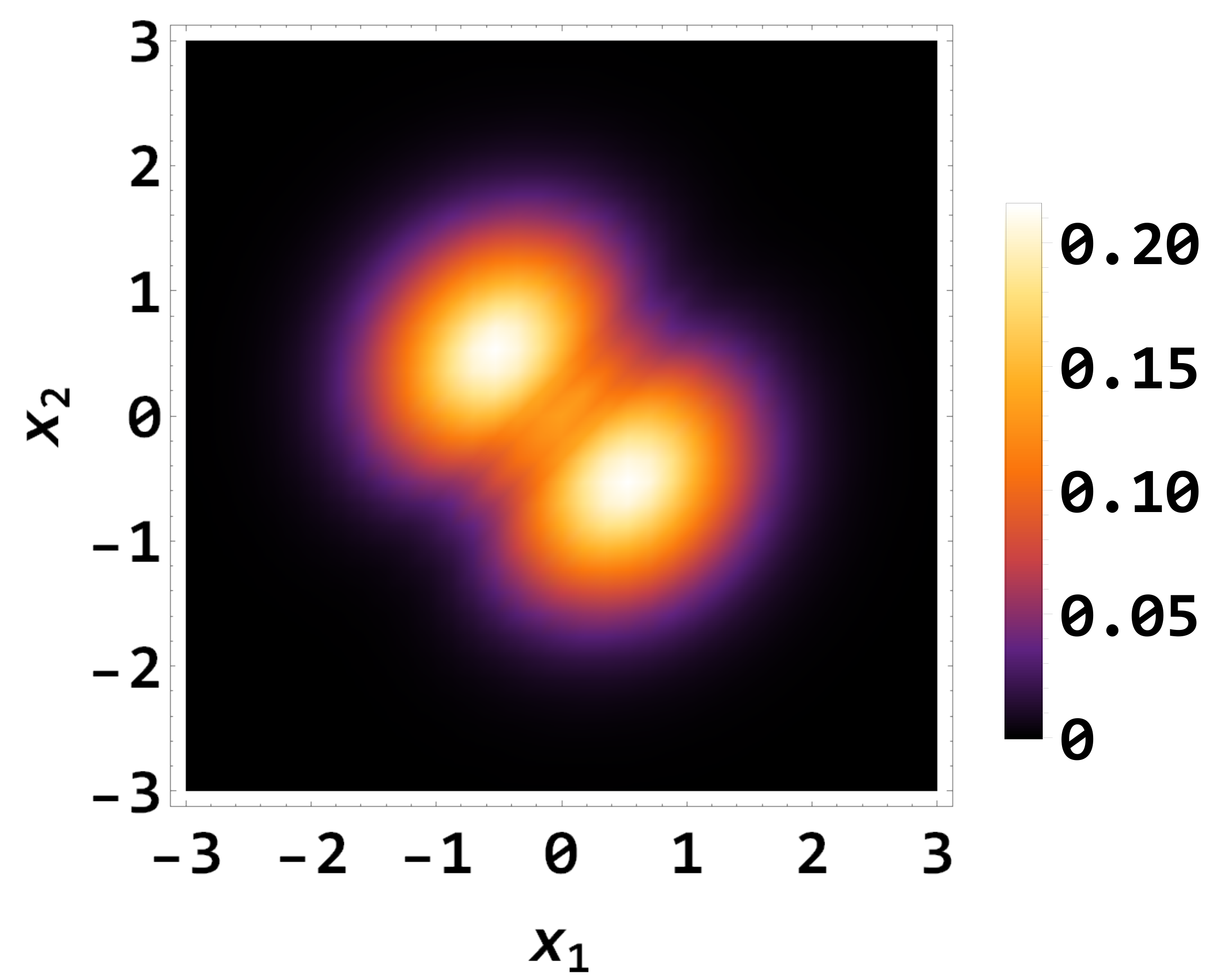} \\
\end{tabular}
\caption{(Color online) Evolution of the probability density $|\Psi(x_1,x_2,t)|^2$.}\label{fwd_wf}
\end{figure}

The evolution of the wave packet is shown in Fig.~\ref{fwd_wf}, in which we can see that, as time passes, the initial two-hump structure of the wave packet evolves into a three-hump one. The third hump appears at the center, which is due to the presence of the attractive potential well, describing the interaction between the particles. When $t=\pi$ the wave packet returns to its initial state and the third hump dissolves.

\begin{figure}
	\centering
	\textbf{$\alpha=0$}\par\medskip
	\begin{tabular}{ccc}
		$g=2,(0,0)$  & $g=2,(2,0)$ & $g=2,(4,0)$ \\
		\includegraphics[width=.15\textwidth]{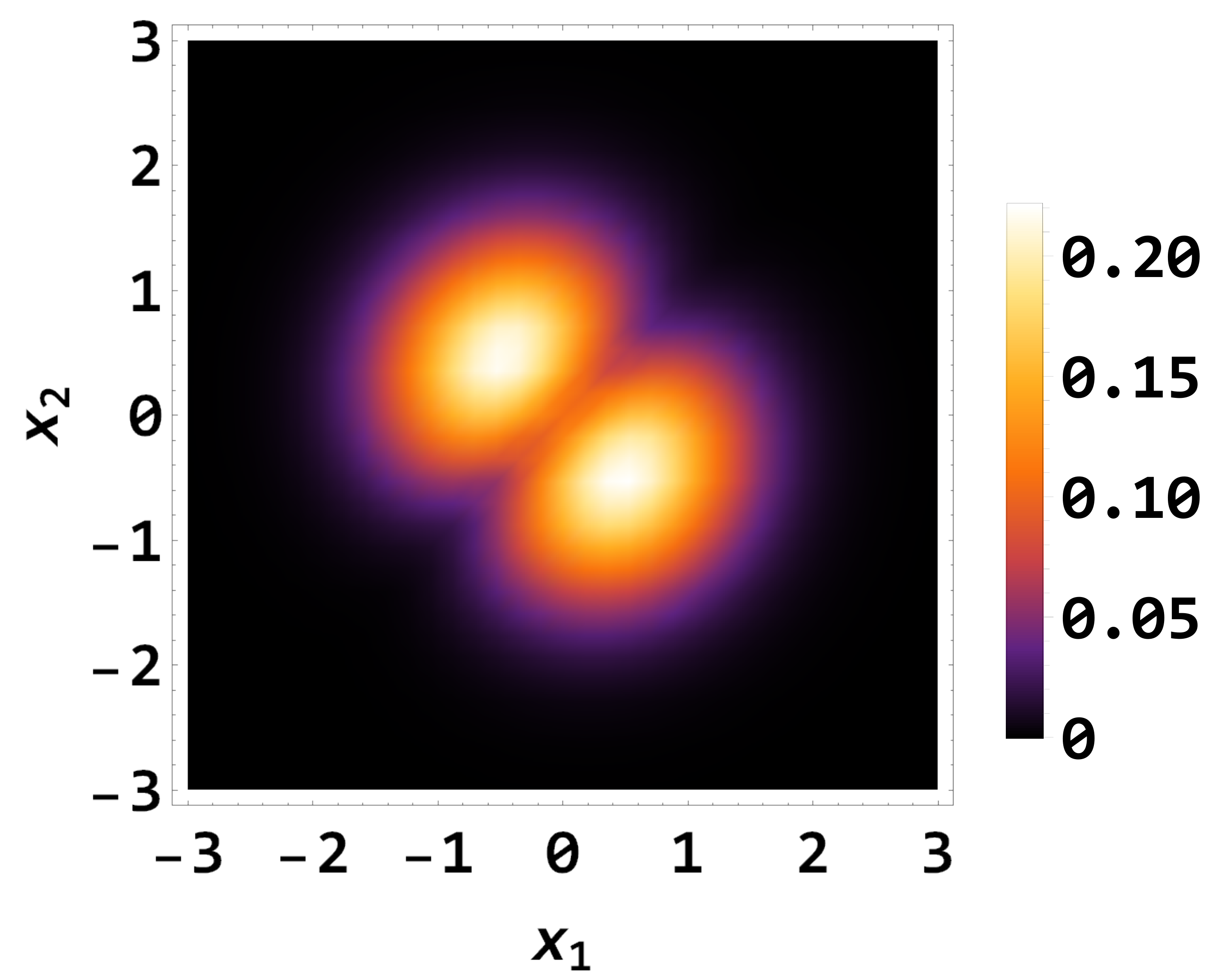} &
		\includegraphics[width=.15\textwidth]{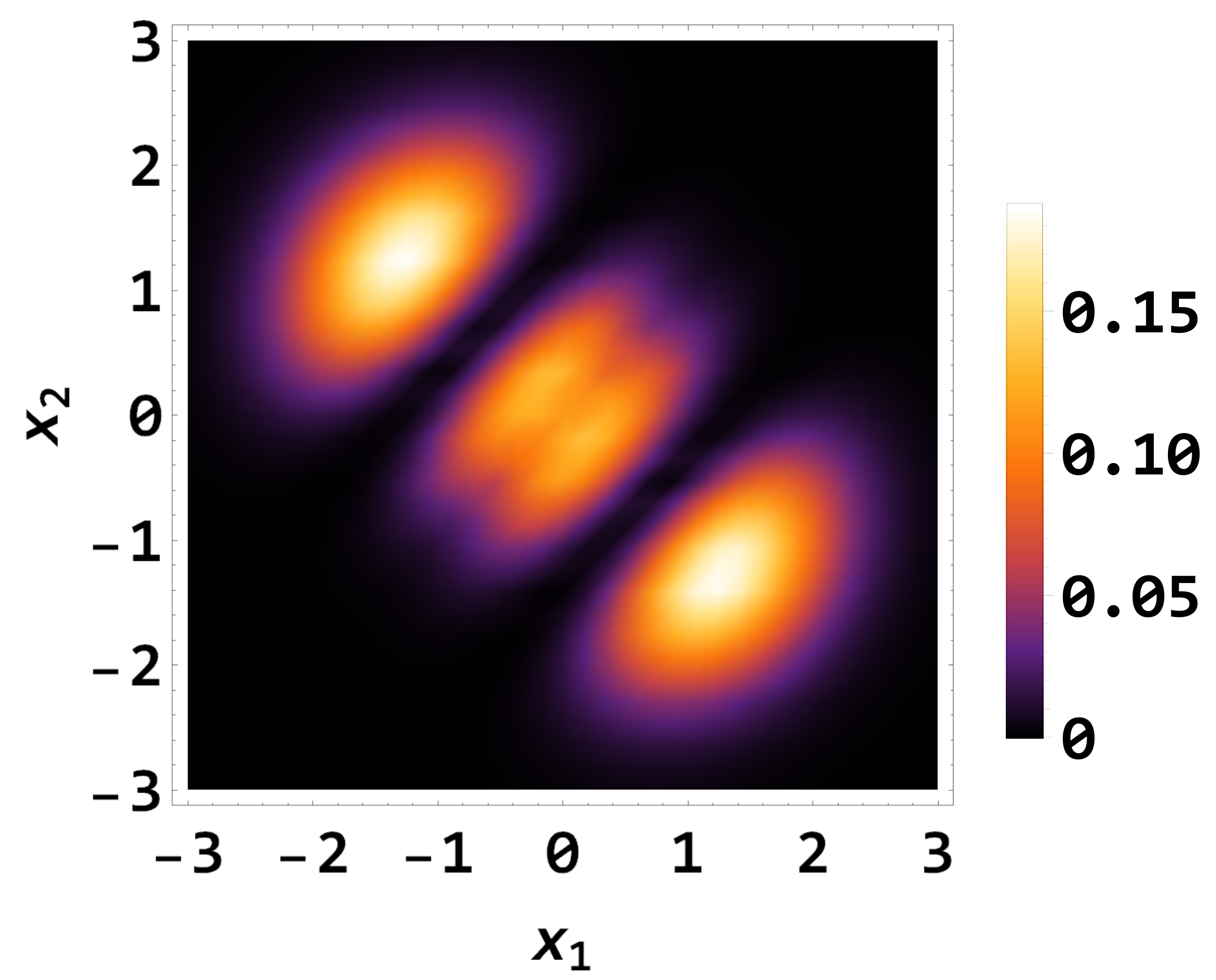} &
		\includegraphics[width=.15\textwidth]{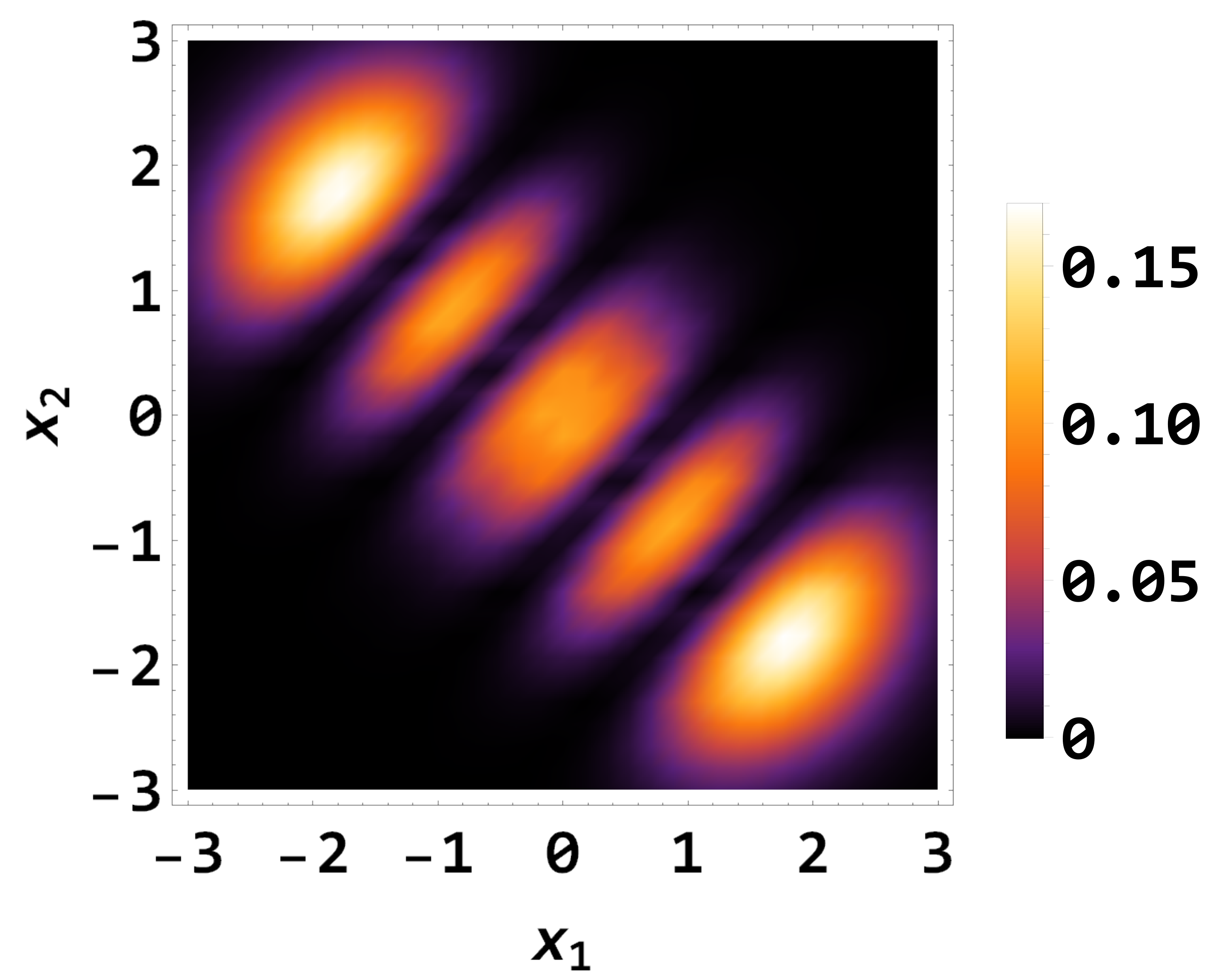} \\
		$g=-2,(0,0)$  & $g=-2,(2,0)$ & $g=-2,(4,0)$ \\
		\includegraphics[width=.15\textwidth]{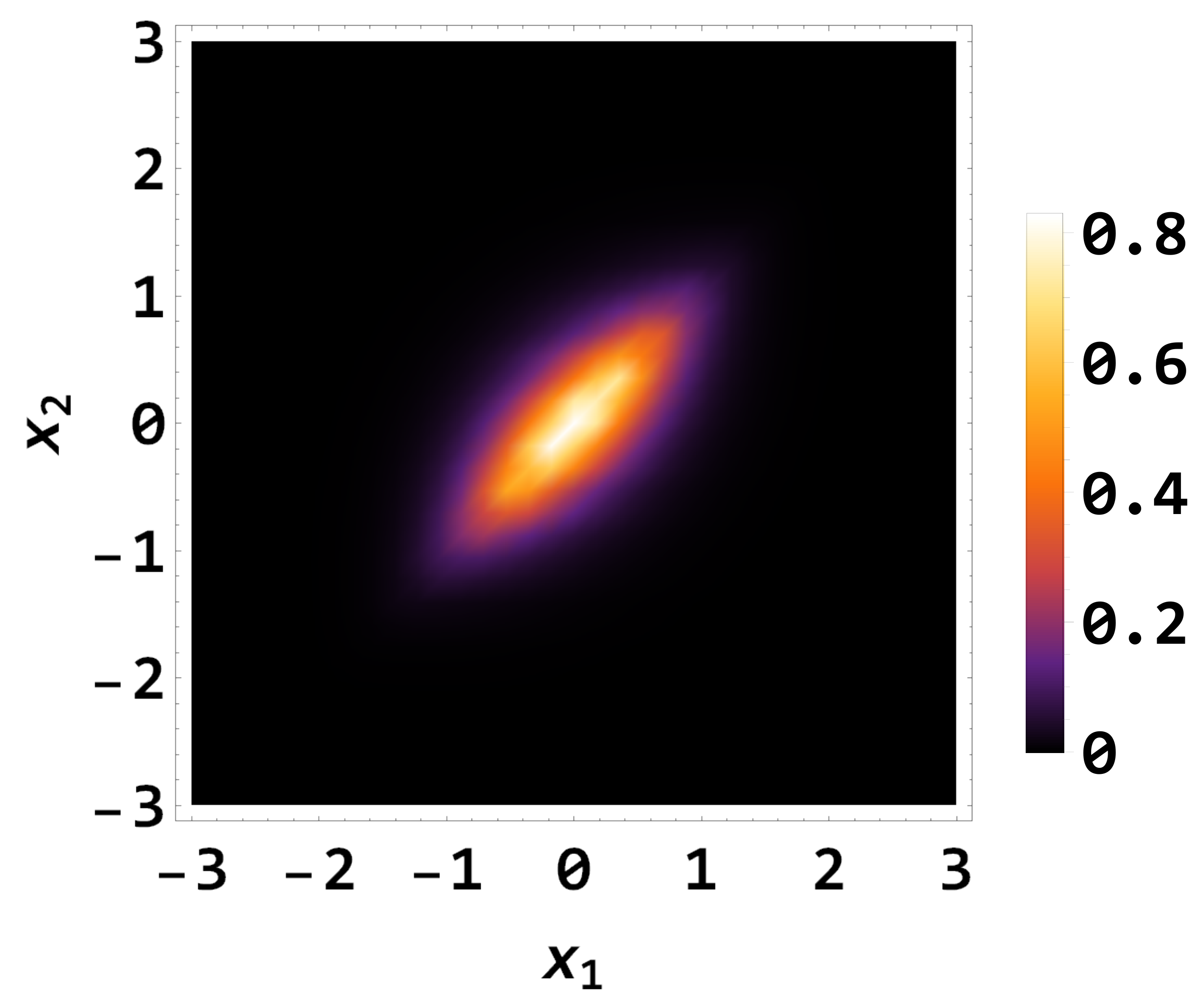} &
		\includegraphics[width=.15\textwidth]{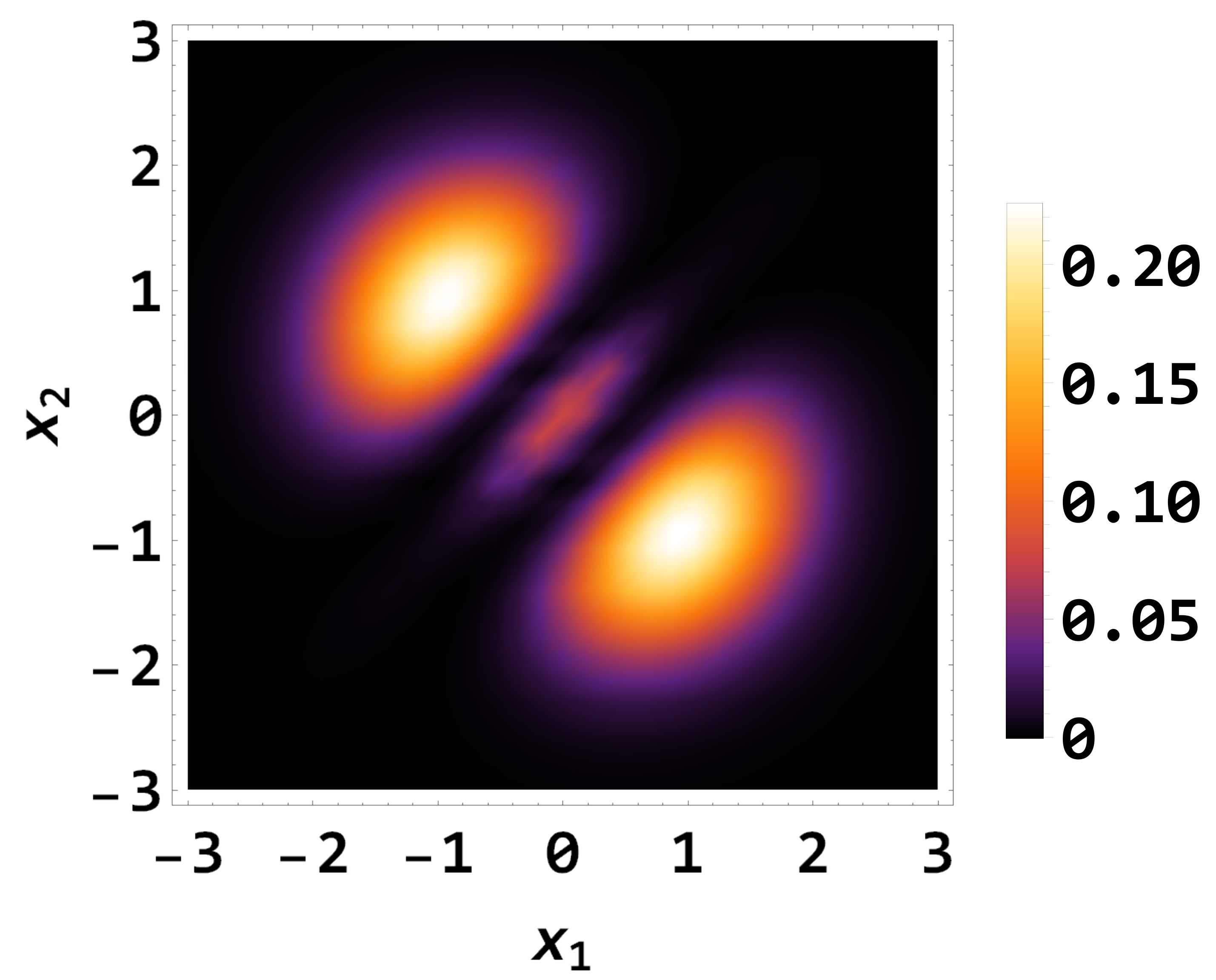} &
		\includegraphics[width=.15\textwidth]{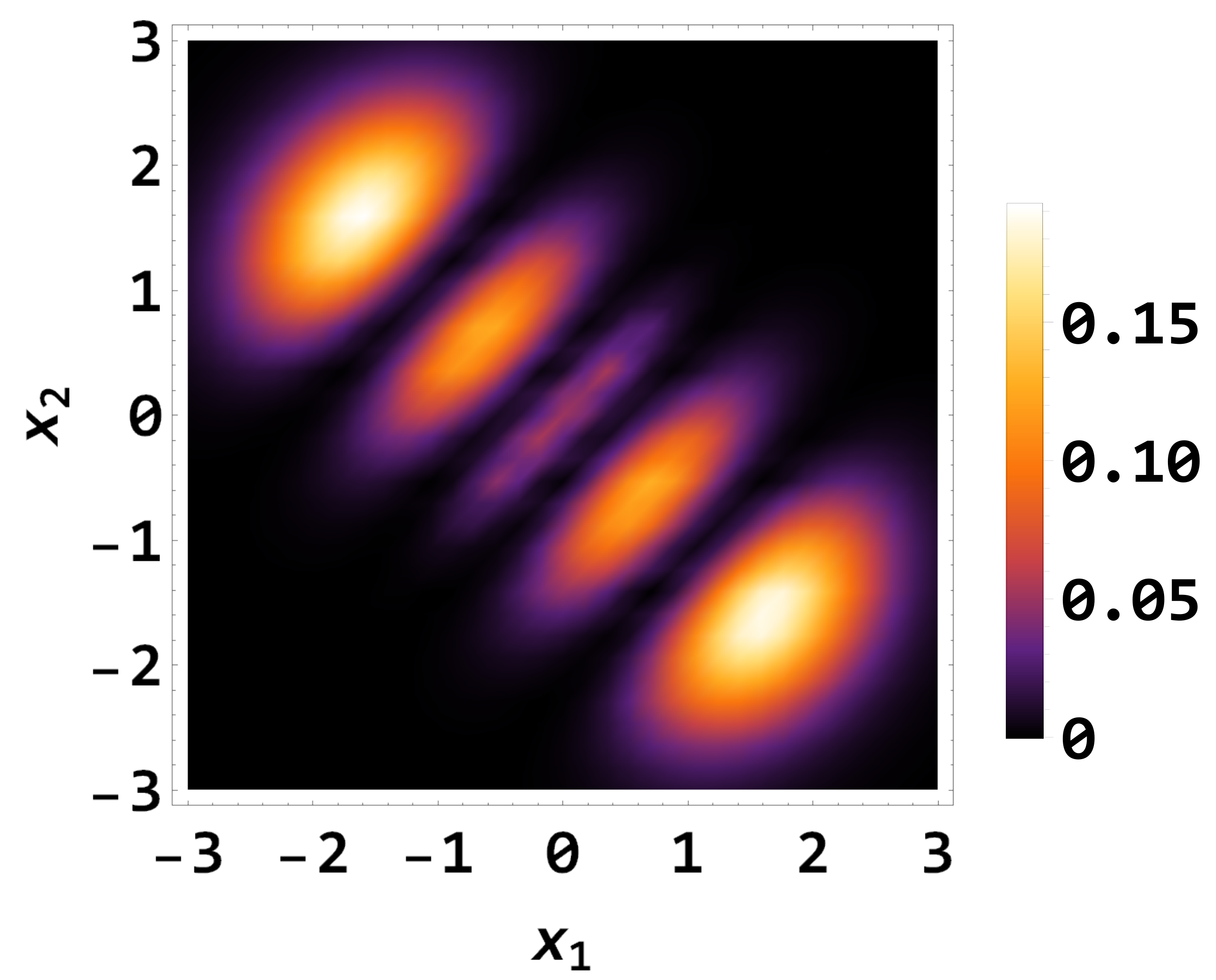} \\
	\end{tabular}
	\caption{(Color online) Probability density $|\Psi(x_1,x_2,t=0)|^2$ for different pre- and postquench stationary states.}\label{fwd_wf0}
\end{figure}

We plot probability densities of different pre- and postquench stationary states in Fig.~\ref{fwd_wf0} in order to match those with the snapshots of the probability density evolution in Fig.~\ref{fwd_wf}. The transition between different states of the system can be interpreted in terms of the hump positions of the wave packet $|\Psi(x_1,x_2,t)|^2$ and the hump positions of the probability densities $|\Psi(x_1,x_2,t=0)|^2$ of the considered ground and excited states. As can be noticed in Fig.~\ref{fwd_wf0}, most of these probability densities also have humps at the center and similar humps on sides, which is why $|\Psi(x_1,x_2,t)|^2$ overlaps with these states.

\begin{figure}
	\centering
	\textbf{$\alpha=0$}\par\medskip
	\begin{tabular}{ccc}
		$t=0.1$ & $t=\pi/8$ & $t=\pi/4$ \\
		\includegraphics[width=.15\textwidth]{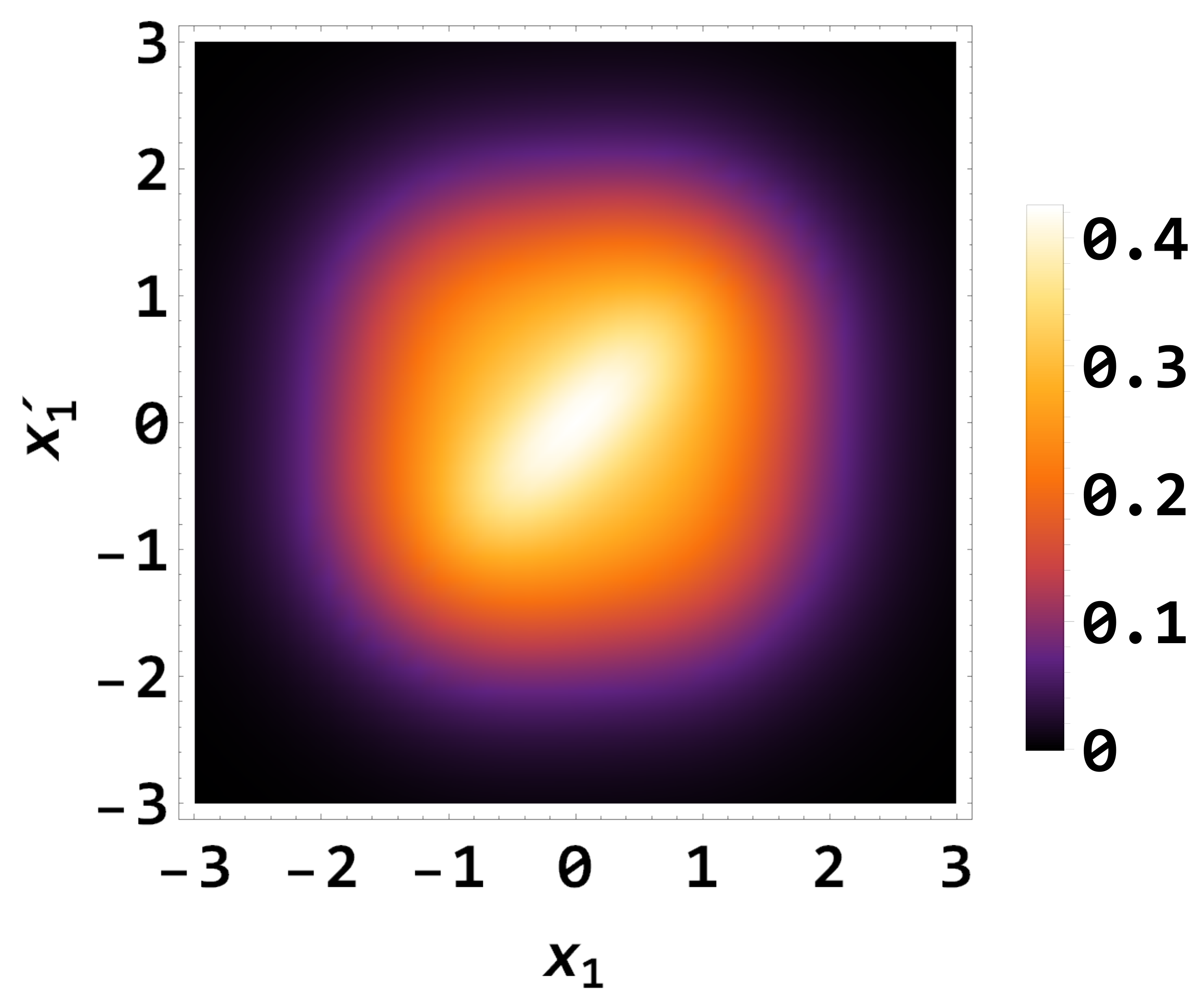} &
		\includegraphics[width=.15\textwidth]{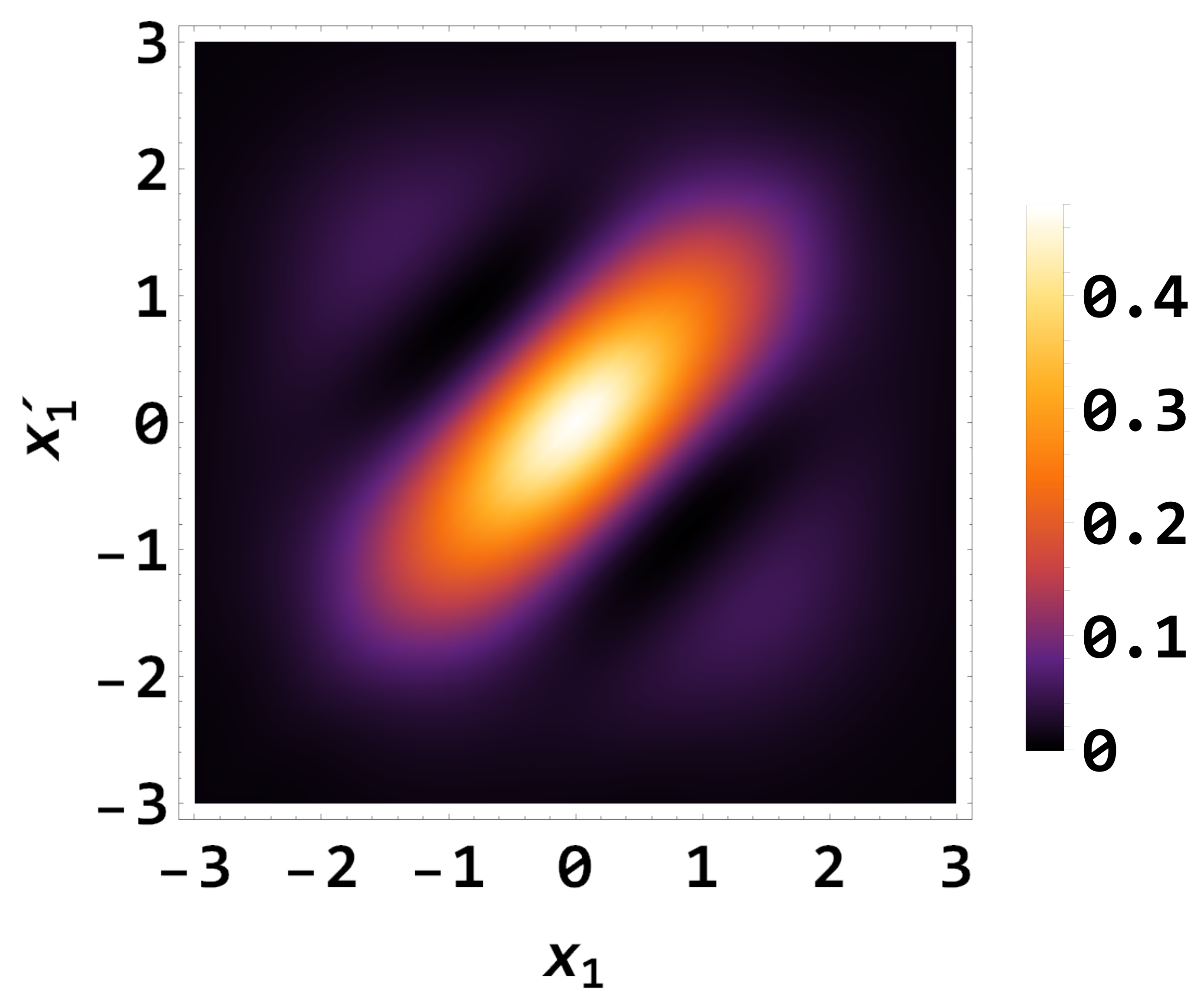} &
		\includegraphics[width=.15\textwidth]{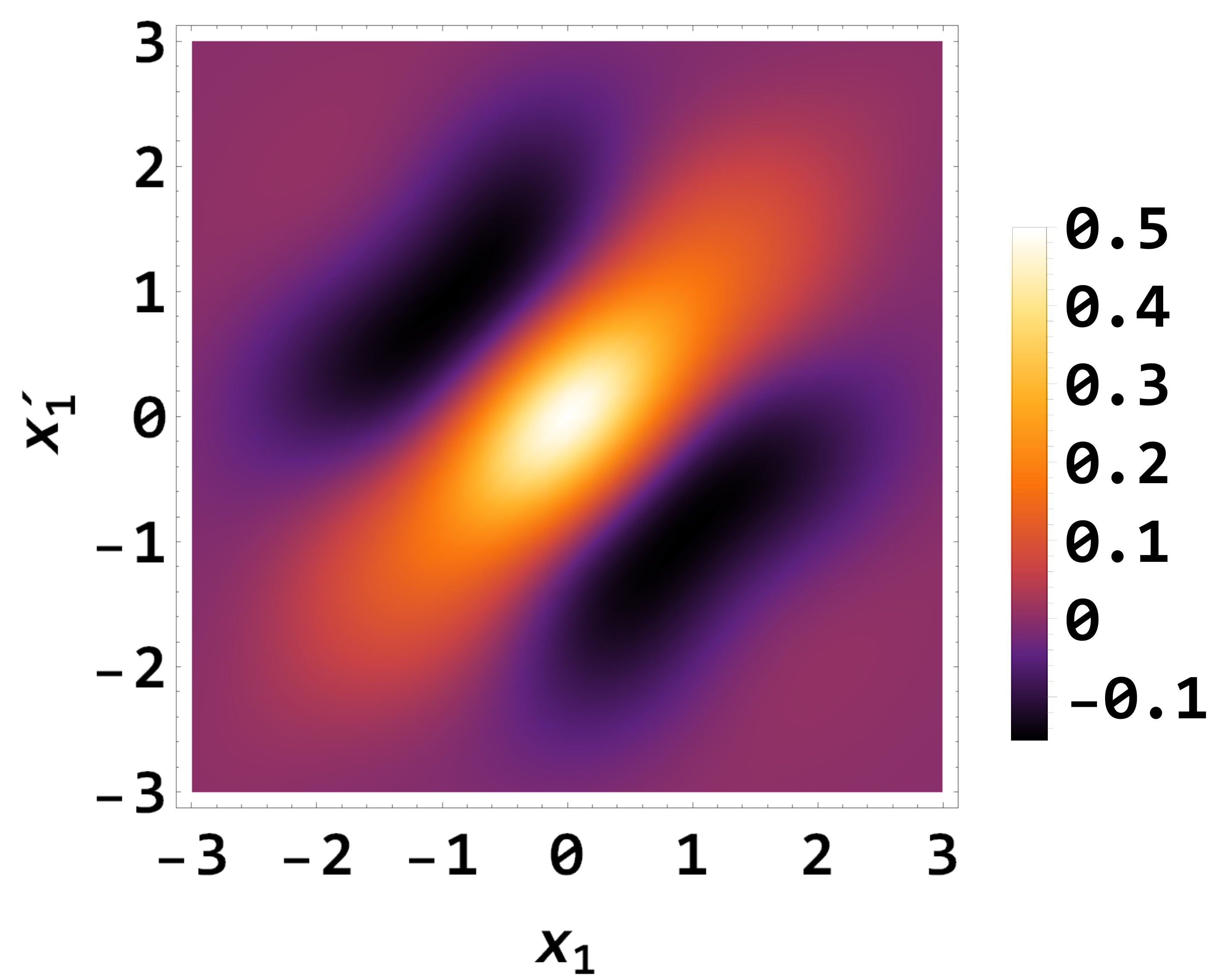} \\
		$t=\pi/2$ & $t=\pi$ \\
		\includegraphics[width=.15\textwidth]{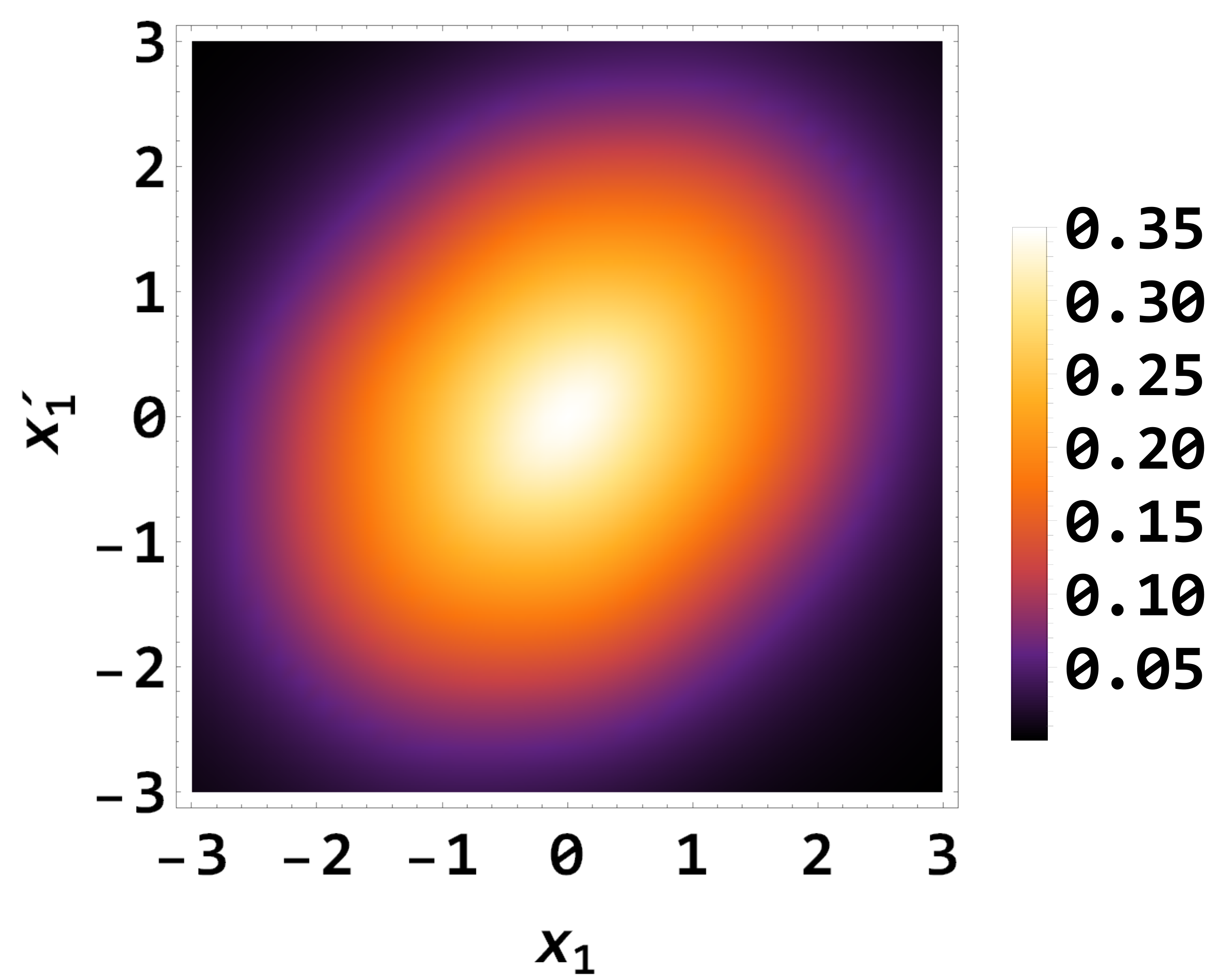} &
		\includegraphics[width=.15\textwidth]{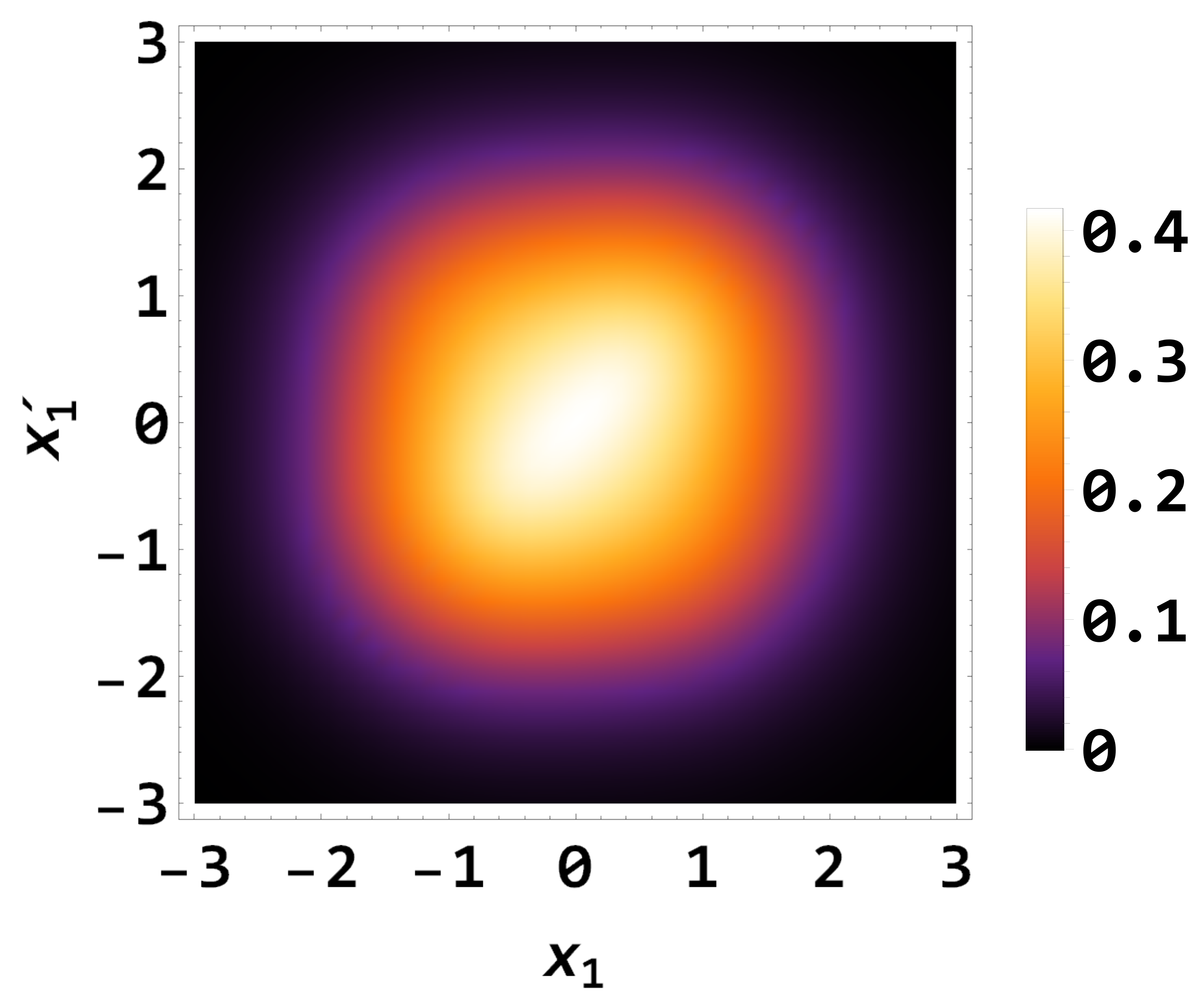} \\
	\end{tabular}
	\caption{(Color online) One-body reduced density matrix, $\rho^{(1)}(x_1,x_1';t)$, at different time-instants.}\label{fwd_rho}
\end{figure}

\begin{figure}
	\centering
	\textbf{$\alpha=0$}\par\medskip
	\begin{tabular}{ccc}
		$g=2,(0,0)$  & $g=2,(2,0)$ & $g=2,(4,0)$ \\
		\includegraphics[width=.15\textwidth]{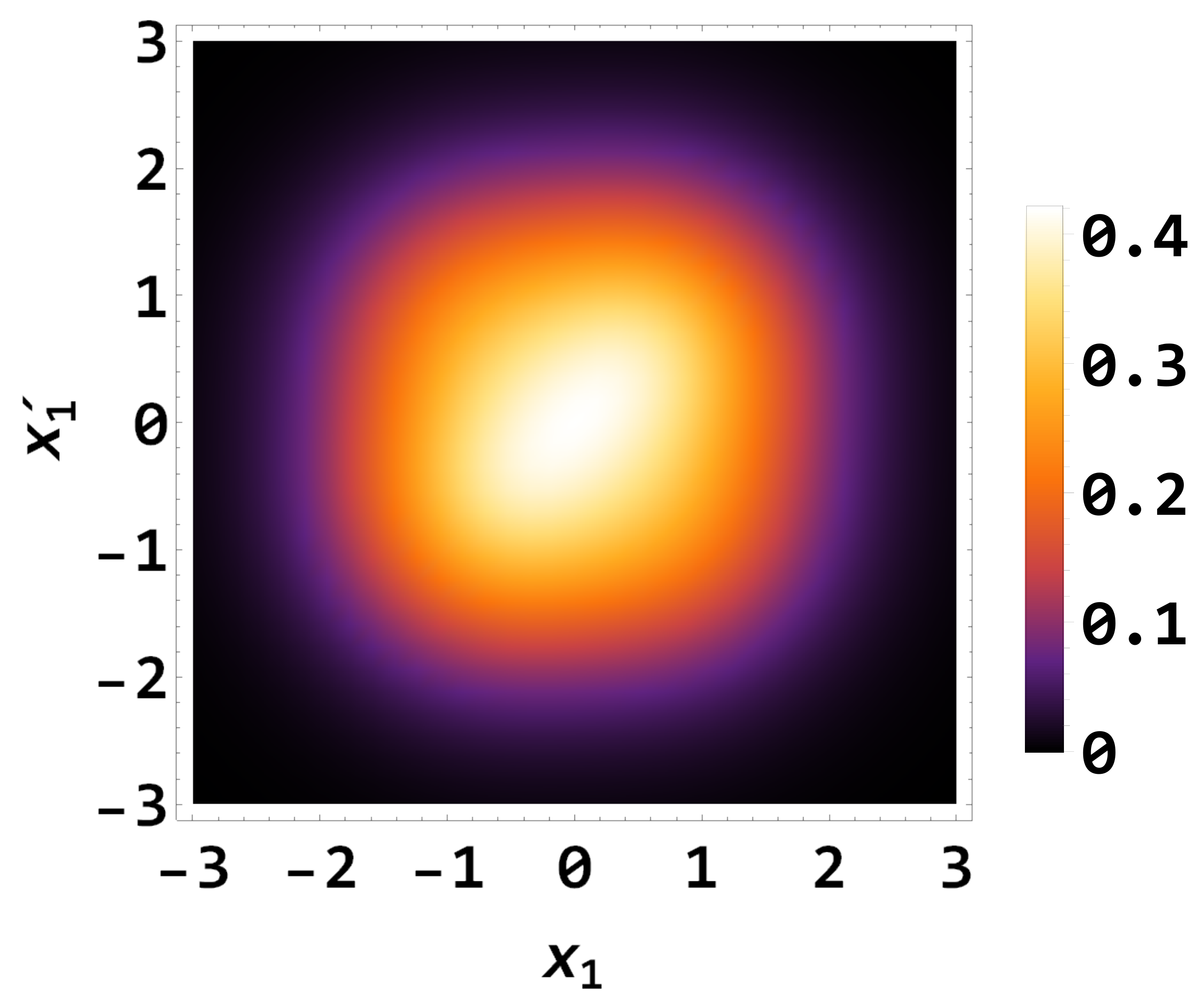} &
		\includegraphics[width=.15\textwidth]{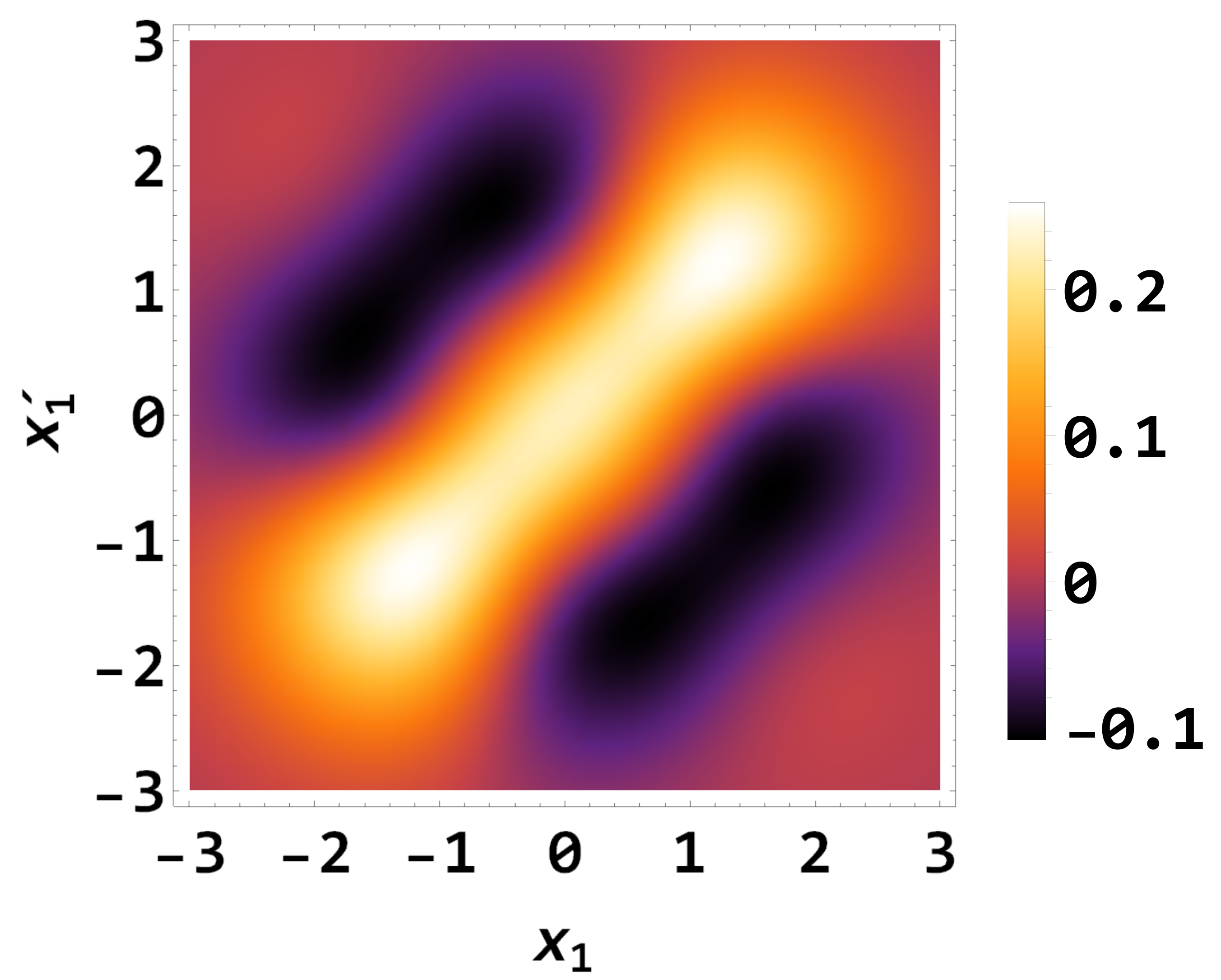} &
		\includegraphics[width=.15\textwidth]{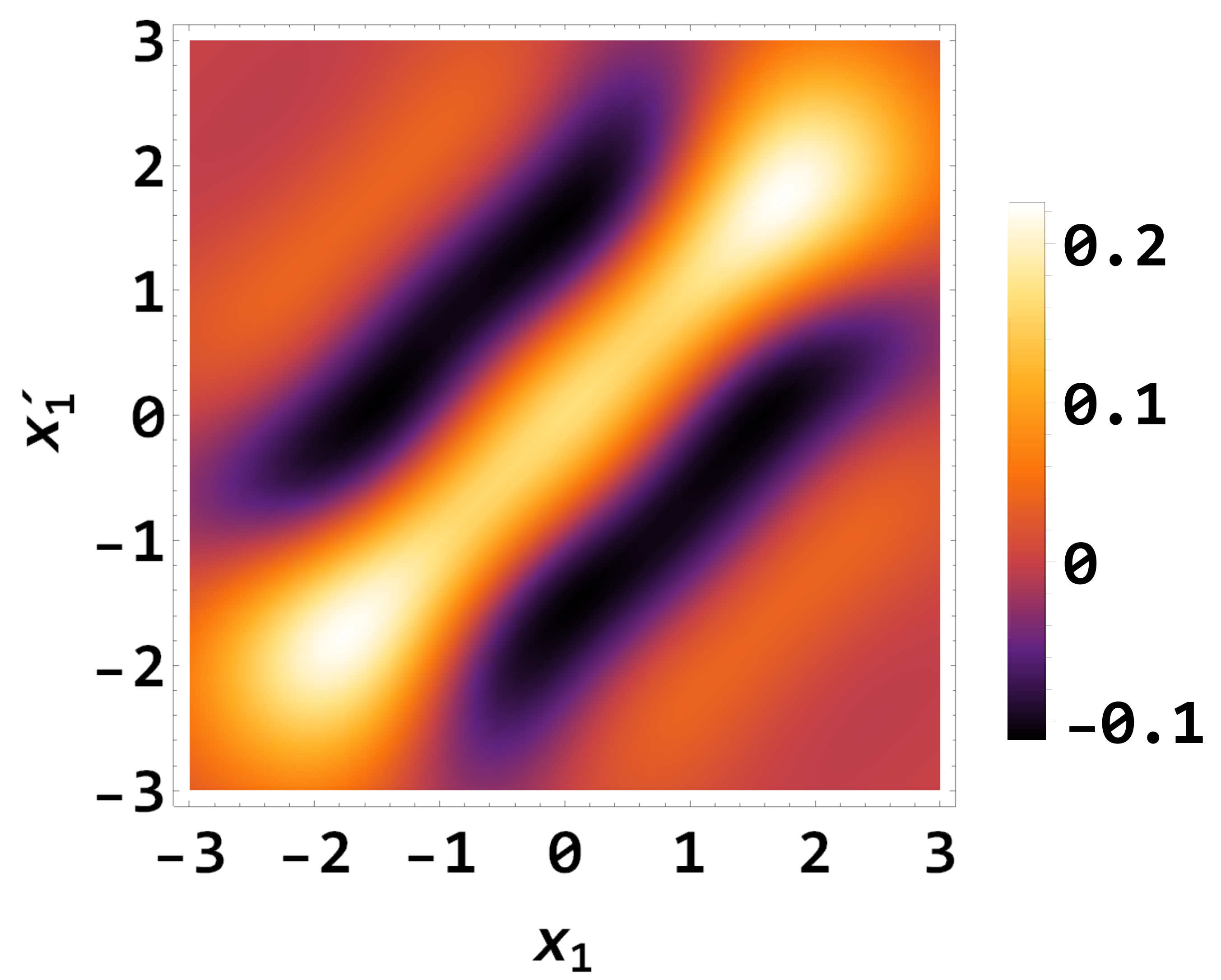} \\
		$g=-2,(0,0)$  & $g=-2,(2,0)$ & $g=-2,(4,0)$ \\
		\includegraphics[width=.15\textwidth]{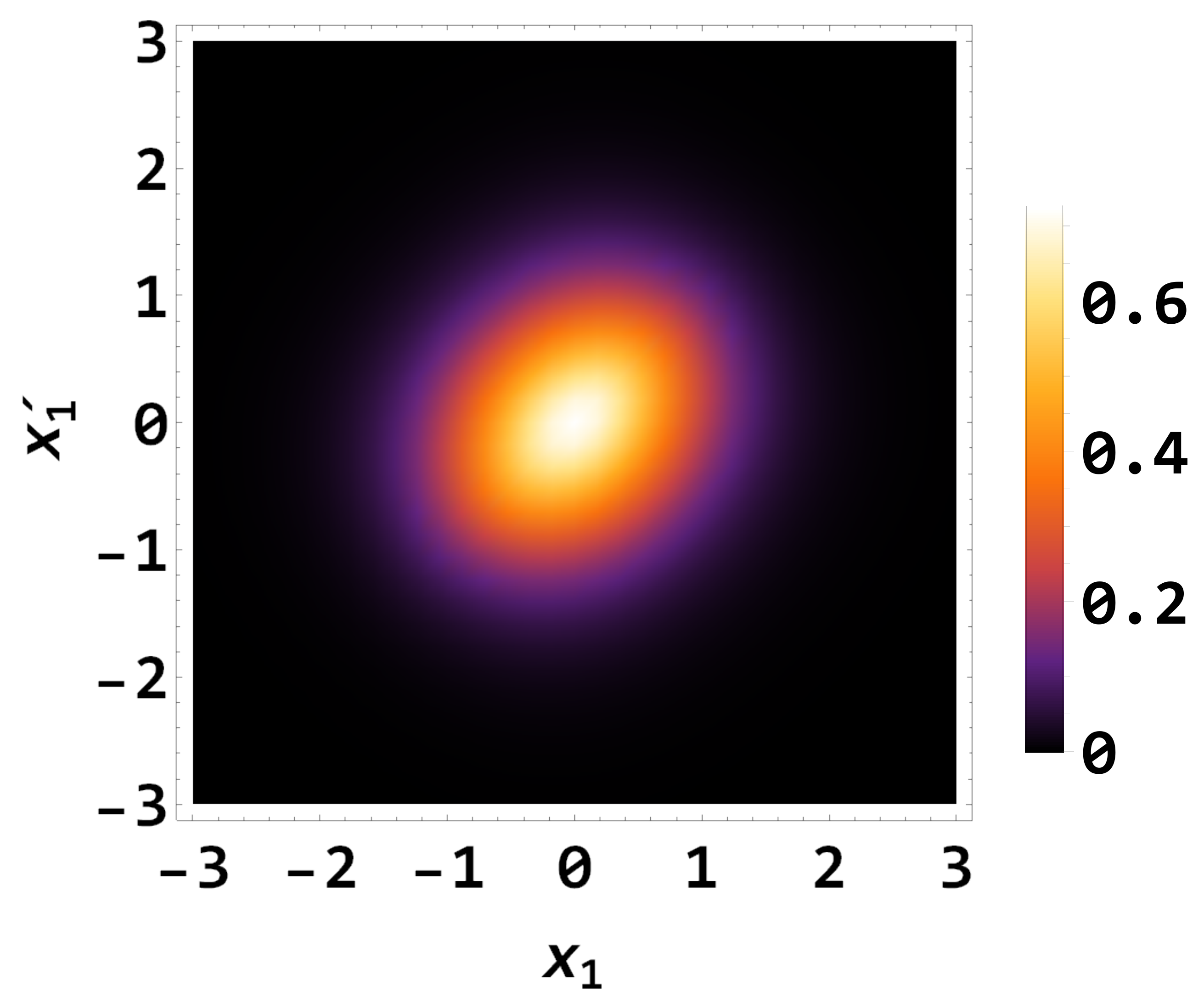} &
		\includegraphics[width=.15\textwidth]{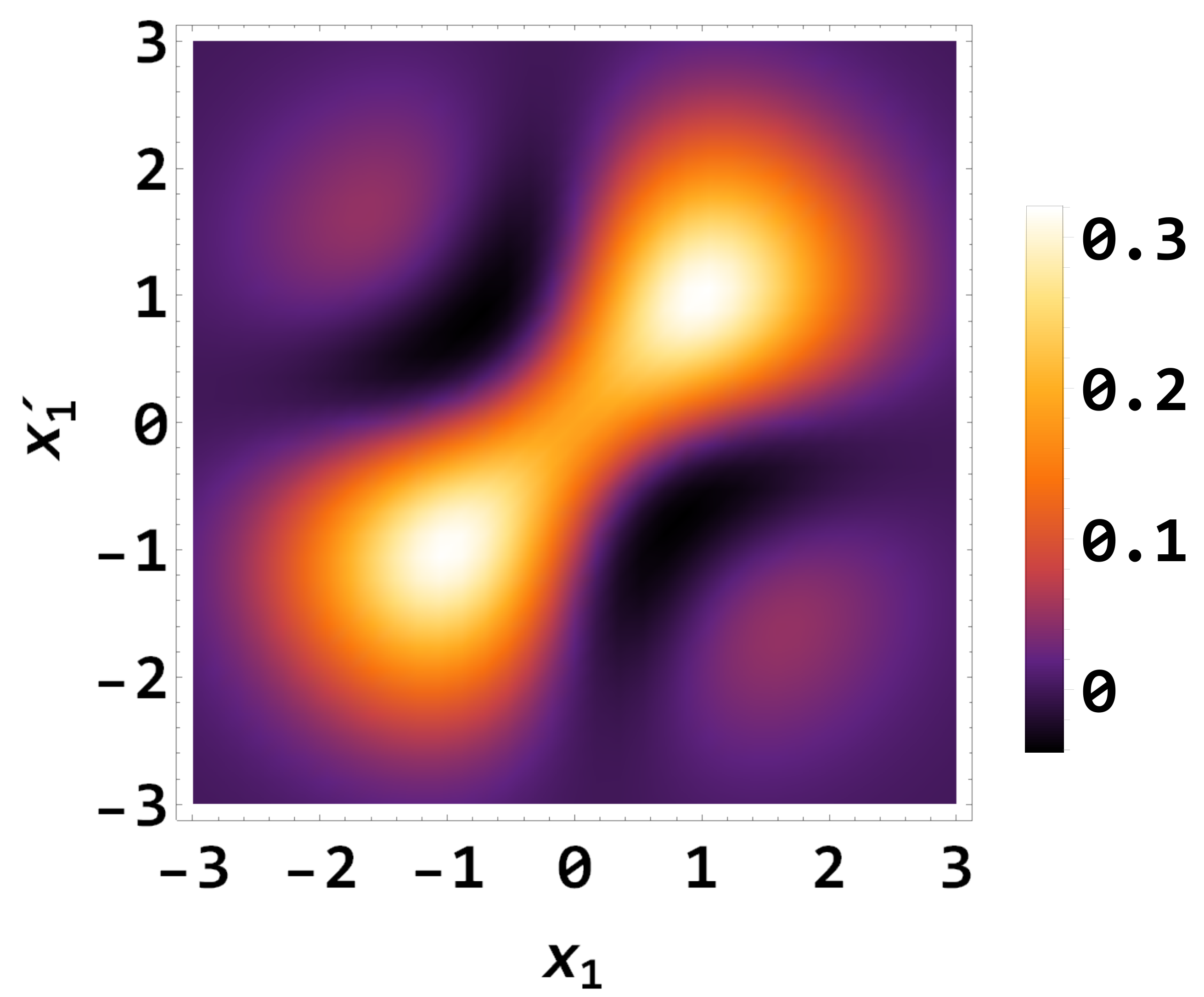} &
		\includegraphics[width=.15\textwidth]{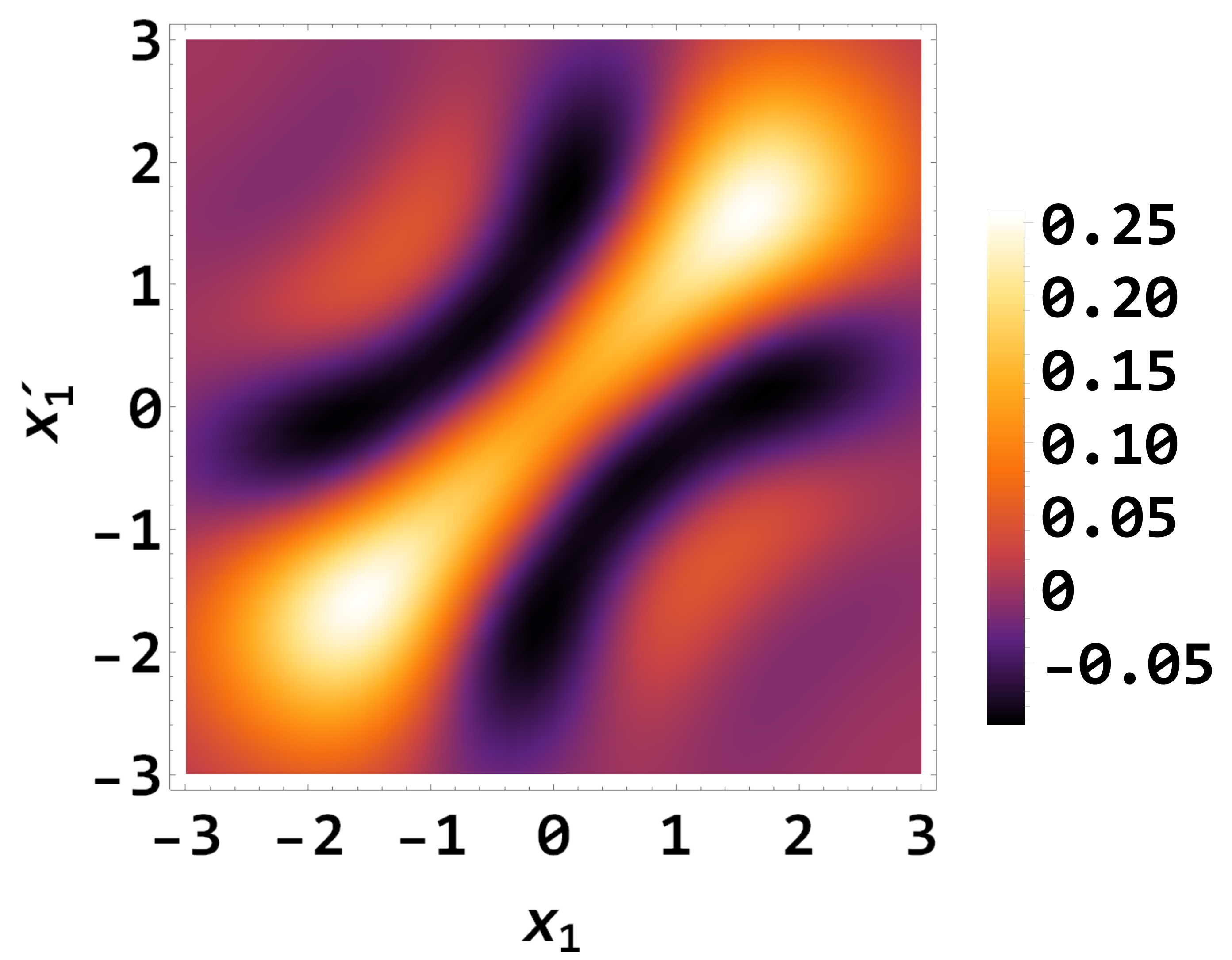} \\
	\end{tabular}
	\caption{(Color online) One-body reduced density matrix, $\rho^{(1)}(x_1,x_1';t=0)$, for different stationary states.}\label{fwd_irho}
\end{figure}

Similar analysis can be performed for the one-body reduced density matrix $\rho^{(1)}(x_1,x_1';t)$. In Fig.~\ref{fwd_rho} we plot the evolution of $\rho^{(1)}(x_1,x_1';t)$, where we can see that during its evolution the hump at the center gets narrowed and two humps on the side appears. At the end of this time period the density matrix approximately returns to its initial form. When comparing $\rho^{(1)}(x_1,x_1';t)$ with the one-body reduced density matrix $\rho^{(1)}(x_1,x_1';t=0)$ in Fig.~\ref{fwd_irho} for different stationary ground and excited states one can notice that there is a quite similar hump distribution for the excited states.

\begin{figure}
	\centering
	\textbf{$\alpha=0$}\par\medskip
	\includegraphics[width=7cm,clip]{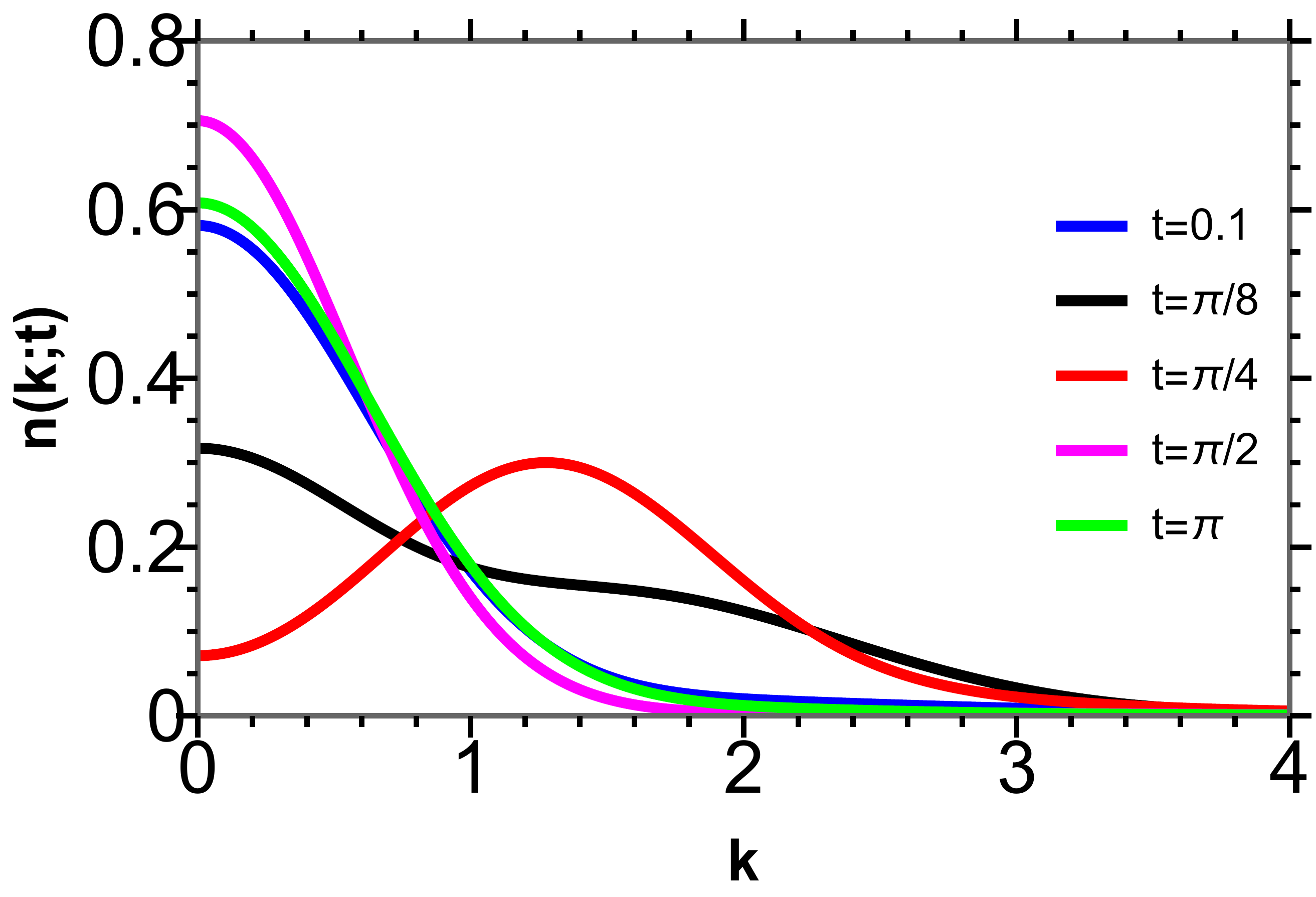}
	\caption{(Color online) Evolution of the momentum distribution $n(k;t)$ for the harmonic trap.}\label{fwd_nk_all}
\end{figure}

Let us plot the time-evolution of the momentum distribution $n(k;t)$ in the time interval $0.1\le t \le \pi$ (Fig.~\ref{fwd_nk_all}). The momentum distribution $n(k;t)$ is symmetrical with respect to the momentum $k$, therefore we consider only positive momentum. The shape of $n(k;t)$ at $t=0.1$ has a pronounced peak around $k=0$ (blue line). As time passes, the shape deforms by lowering of the zero-momentum peak and shifting it closer to higher values of $k$ (red line). Such a tendency, as we we will see below, is due to the system's transition into excited states. By the time reaches $t=\pi/2$ the shape of $n(k;t)$ reverts to the shape with the pronounced zero-momentum peak. At $t=\pi$ the shape returns almost to its initial shape, indicating the periodic oscillation of the system's dynamics.

\begin{figure}
	\centering
	\textbf{$\alpha=0$,~~~$t=\pi/8$}\par\medskip
	\begin{tabular}{cc}
		Prequench states &  Postquench states \\
		\hspace{-.5cm}\includegraphics[width=.25\textwidth]{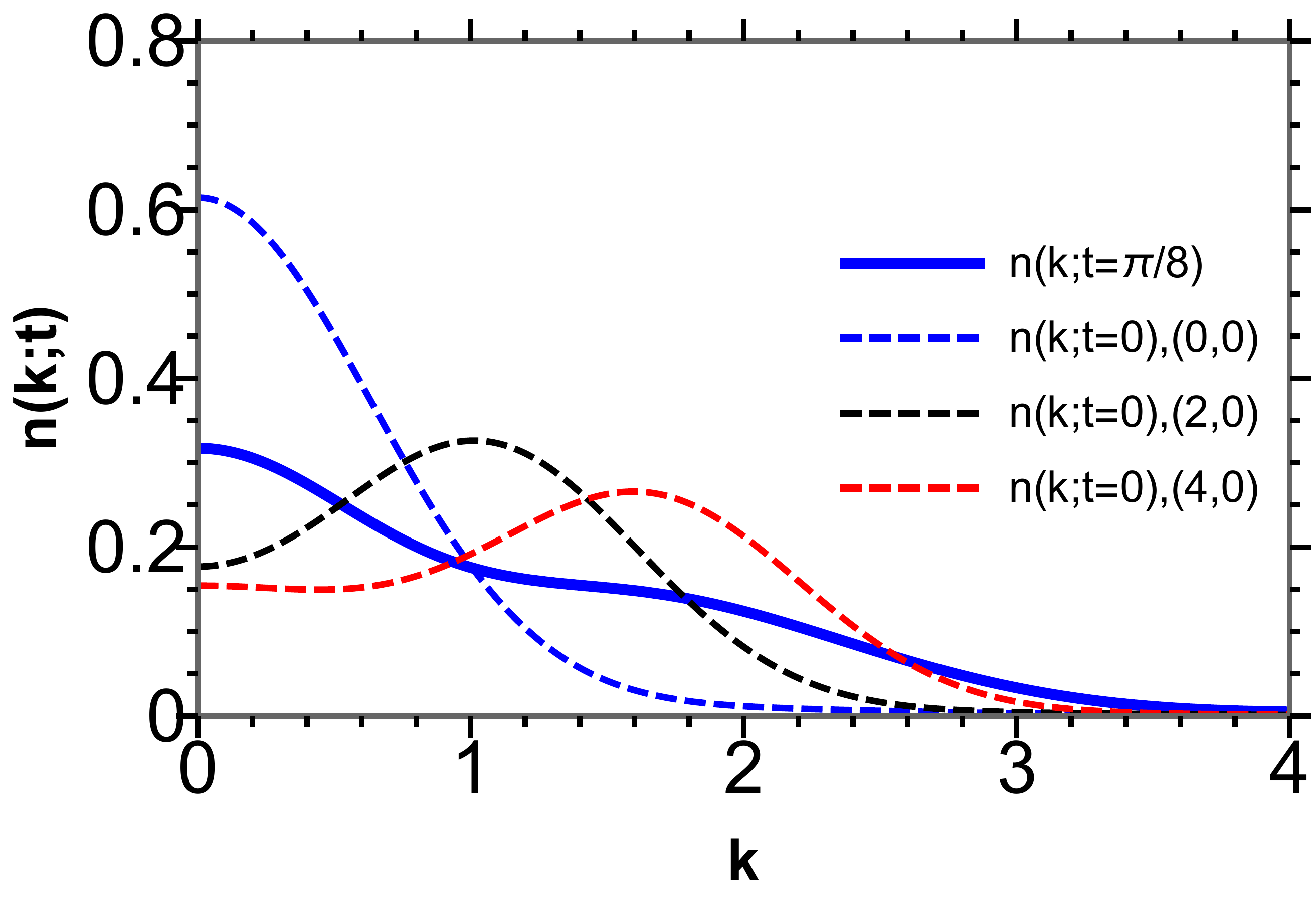} &
		\includegraphics[width=.25\textwidth]{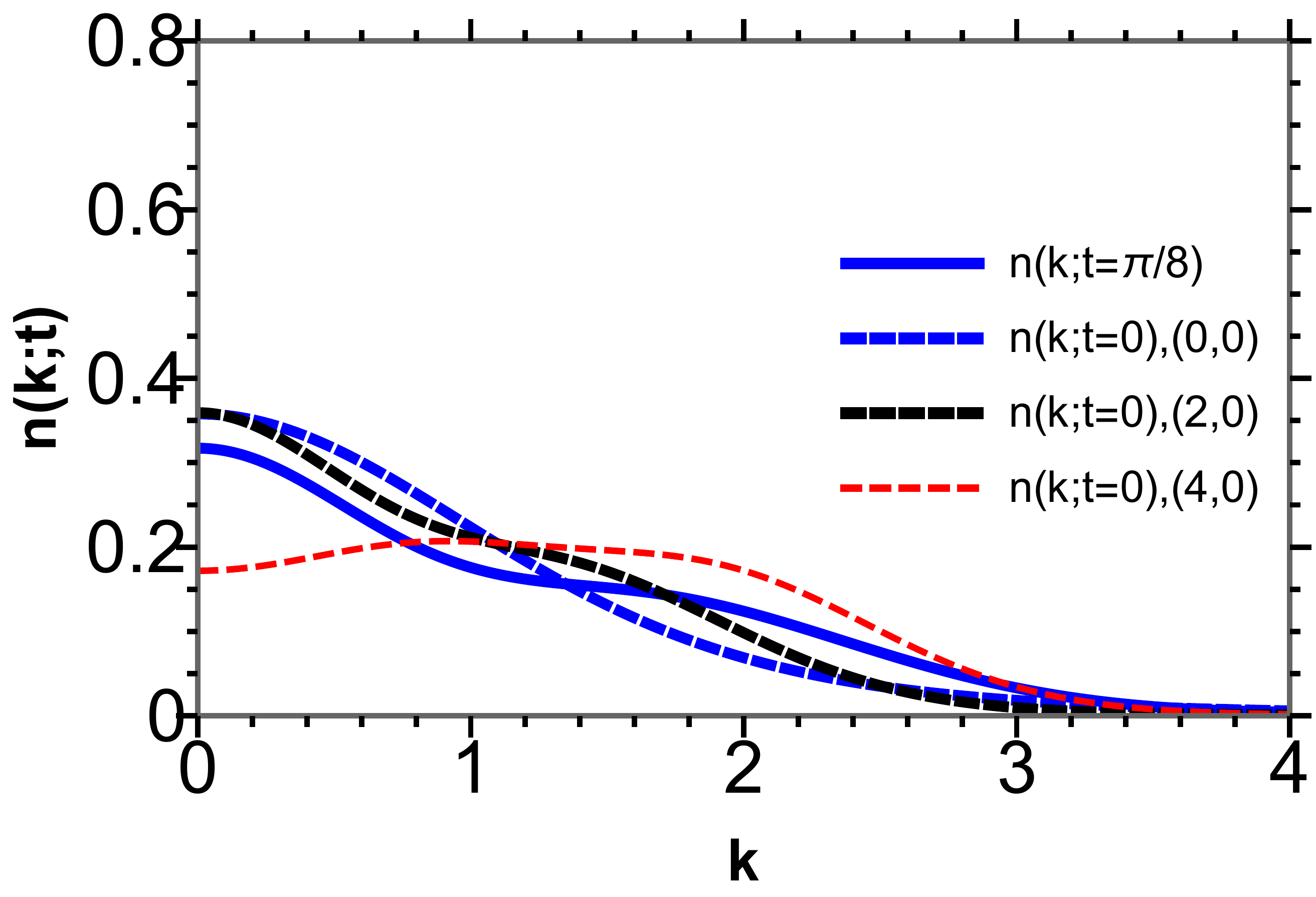} \\
	\end{tabular}
	\caption{(Color online) Comparison of the momentum distribution $n(k;t)$ (blue line) with the momentum distributions $n(k;t=0)$ for the pre- and postquench states (dashed lines).}\label{fwd_nk_12}
\end{figure}

\begin{figure}
	\centering
	\textbf{$\alpha=0$,~~~$t=\pi/4$}\par\medskip
	\begin{tabular}{cc}
		Prequench states &  Postquench states \\
		\hspace{-.5cm}\includegraphics[width=.25\textwidth]{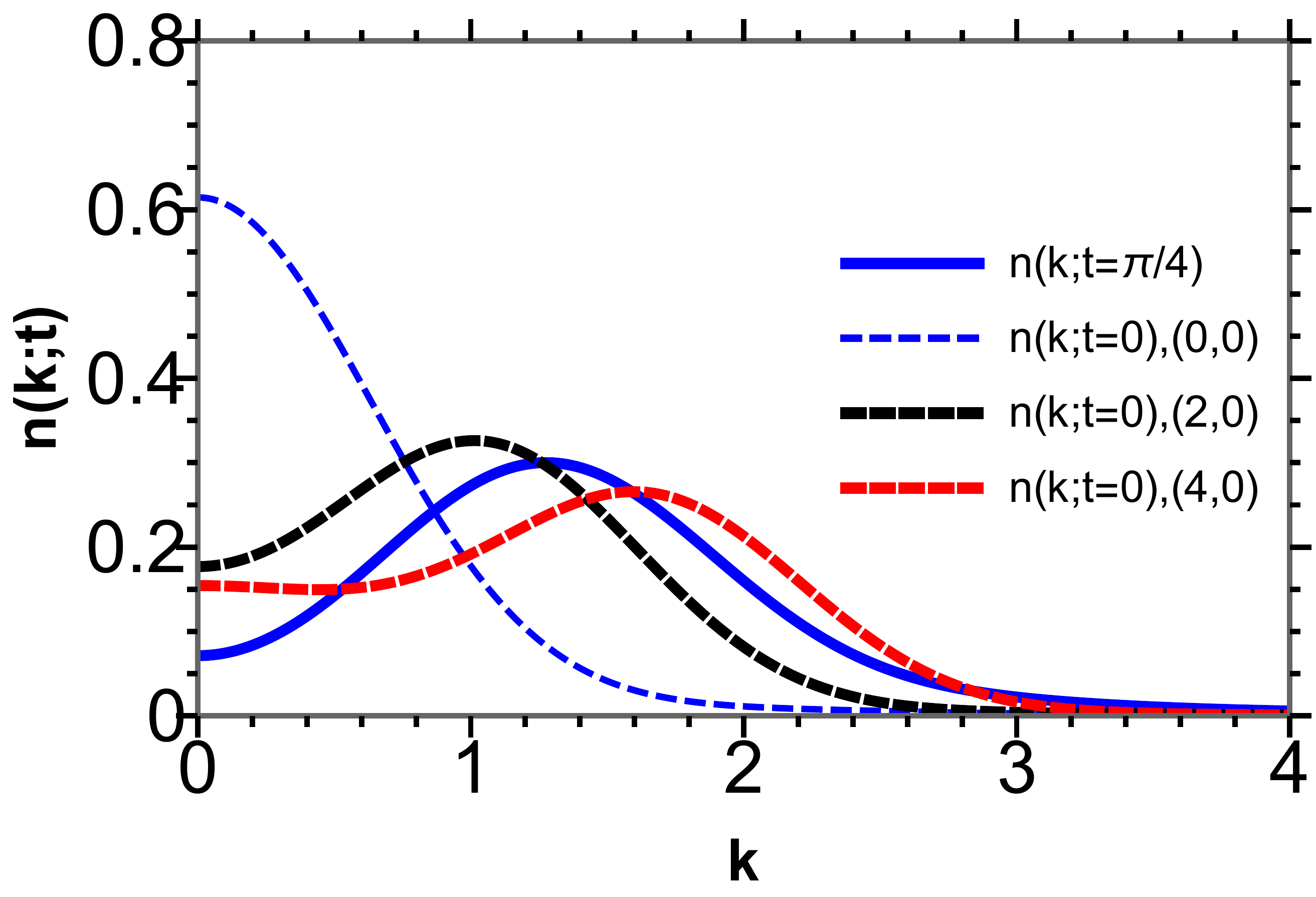} &
		\includegraphics[width=.25\textwidth]{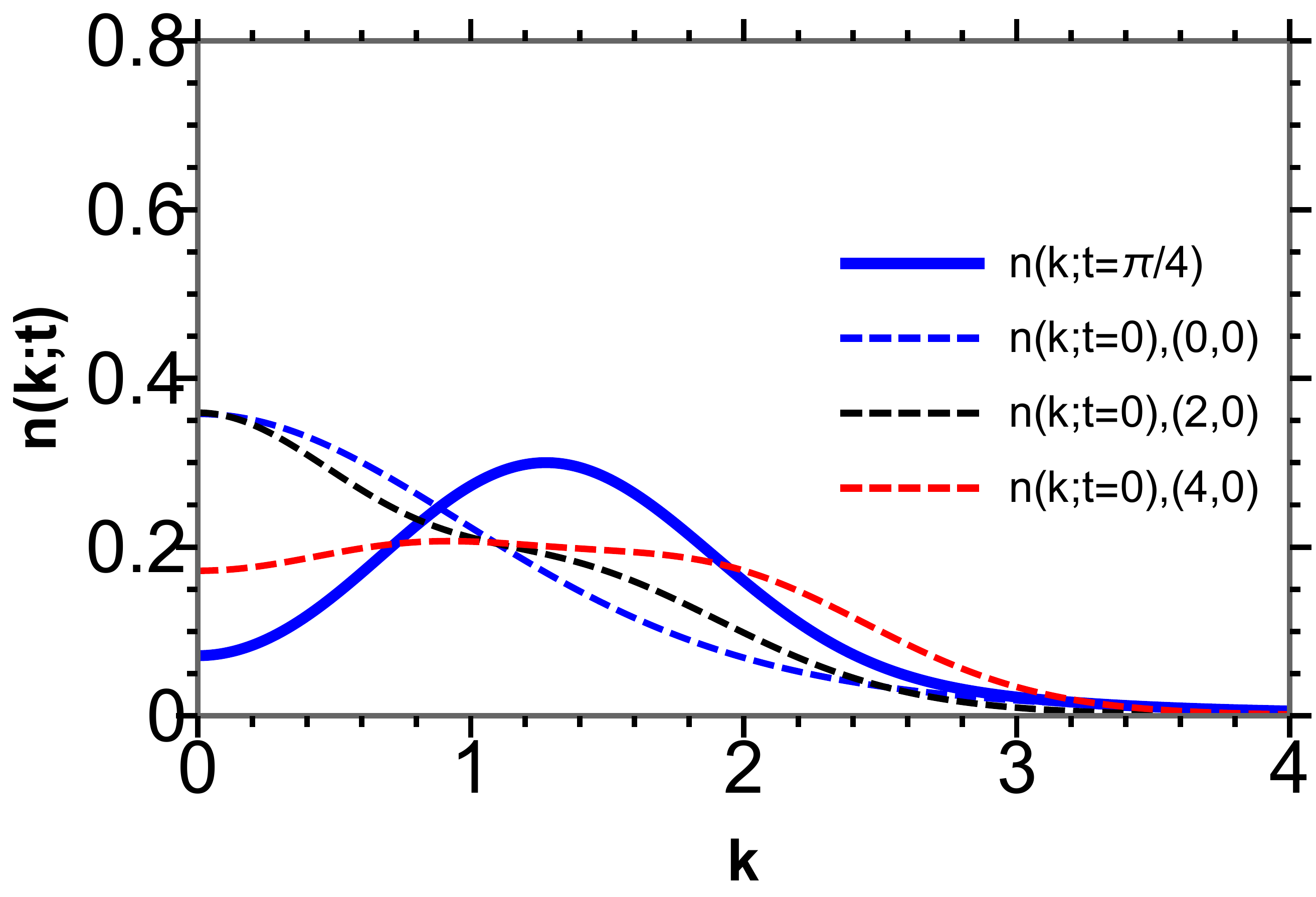} \\
	\end{tabular}
	\caption{(Color online) Comparison of the momentum distribution $n(k;t)$ (blue line) with the momentum distributions $n(k;t=0)$ for the pre- and postquench states (dashed lines).}\label{fwd_nk_34}
\end{figure}

The impact of the excited states can be noticed by analyzing the shapes of the momentum distribution $n(k;t)$. In Fig.~\ref{fwd_nk_12} and Fig.~\ref{fwd_nk_34} we compare the momentum distribution $n(k;t)$ with momentum distributions $n(k;t=0)$ for different pre- and postquench states. In Fig.~\ref{fwd_nk_12} we see that the shape of $n(k;t)$ at $t=\pi/8$ (blue line) resembles the shapes of $n(k;t=0)$ for the postquench ground state (blue dashed line) and excited state $(2,0)$ (black dashed line). This postquench states overlaps quite significantly with the wave packet, according to Fig.~\ref{fwd}, hence the similarity of the shapes. The shapes of $n(k;t=0)$ in Fig.~\ref{fwd_nk_34} for the prequench excited state $(2,0)$ (black dashed line) and $(4,0)$ (red dashed line) are quite similar to the shape of $n(k;t)$ at $t=\pi/4$. This is quite consistent with Fig.~\ref{fwd}, where approximately at $t=\pi/4$ the overlaps for these prequench excited states take their maximum values. 

\subsection{Anharmonic trap, $\alpha=-0.03$, the ground state, $(0,0)$}

\begin{figure}
\centering
\textbf{$\alpha=-0.03$}\par\medskip
\includegraphics[width=8.5cm,clip]{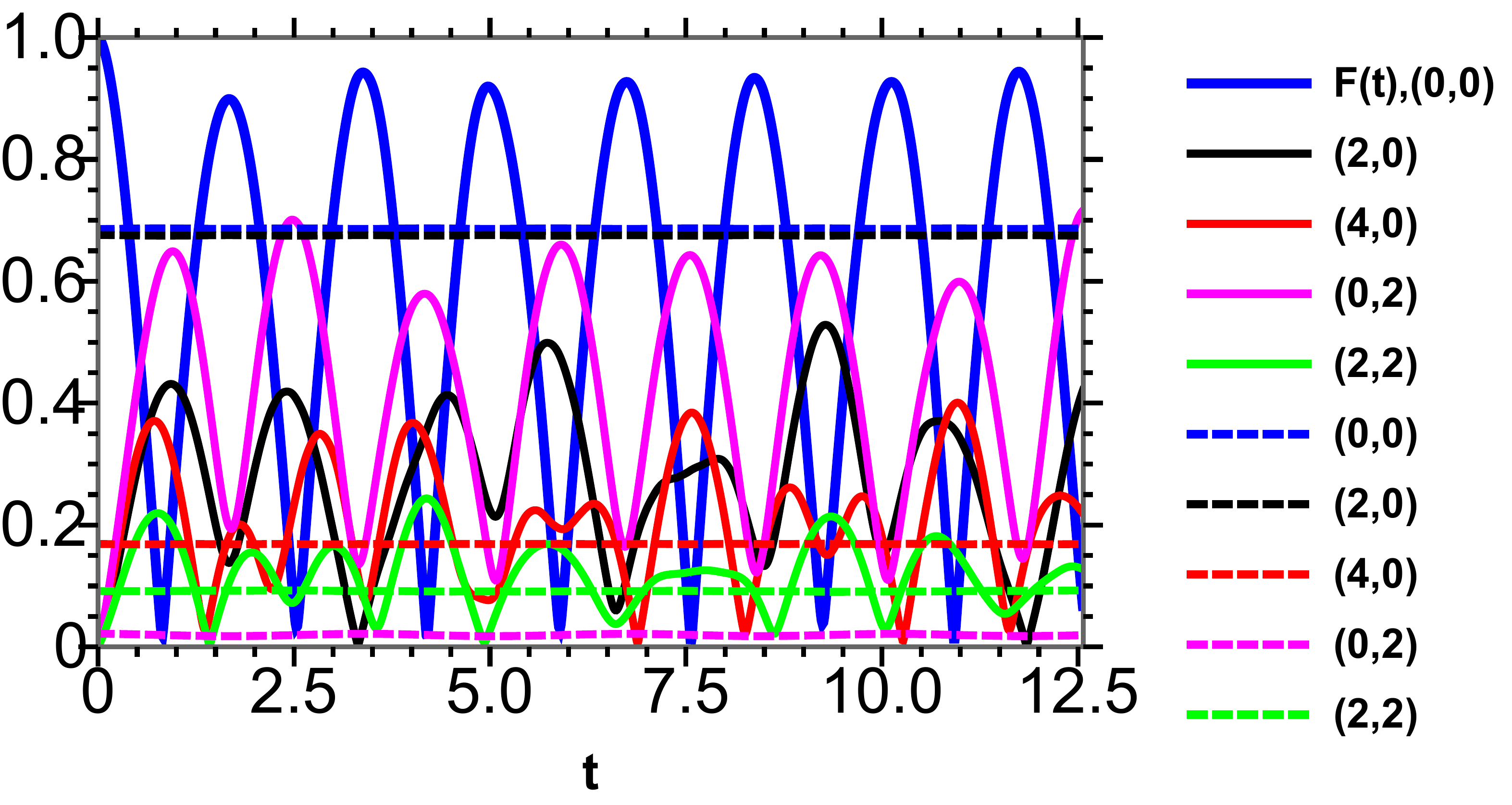}
\caption{(Color online) Fidelity $F(t)$ and the overlap integrals $\mathcal{Q}$ between the time-evolving state $\Psi(x_1,x_2,t)$ and different pre- (solid lines) and postquench (dashed lines) states in the case of the anharmonic trap. The indices $(n,N)$ refer to the states with the quantum numbers of the relative and center-of-mass motions. The system of the two atoms is prepared in the ground state with $g=2$ and quenched to $g=-2$.}\label{fwd_al}
\end{figure}

Now let us turn the anharmonicity on, $\alpha=-0.03$, and plot dynamics of the overlap integrals $\mathcal{Q}$ (Fig~\ref{fwd_al}). 
The number of contributing states, in this case, significantly increases. Due to the coupling of the relative and center-of-mass motions, additional impacts of the center-of-mass excited states are considerable. The overlaps of the prequench states $(0,2)$ and $(2,2)$ oscillate approximately in antiphase with the fidelity $F(t)$. The period of $F(t)$ becomes slightly larger and the amplitude becomes smaller in comparison with the case of the harmonic trap, $\alpha=0$.

\begin{figure}
	\centering
	\textbf{$\alpha=-0.03$}\par\medskip
\begin{tabular}{ccc}
$t=0.001$ & $t=0.1$ & $t=\pi/8$ \\
\includegraphics[width=.15\textwidth]{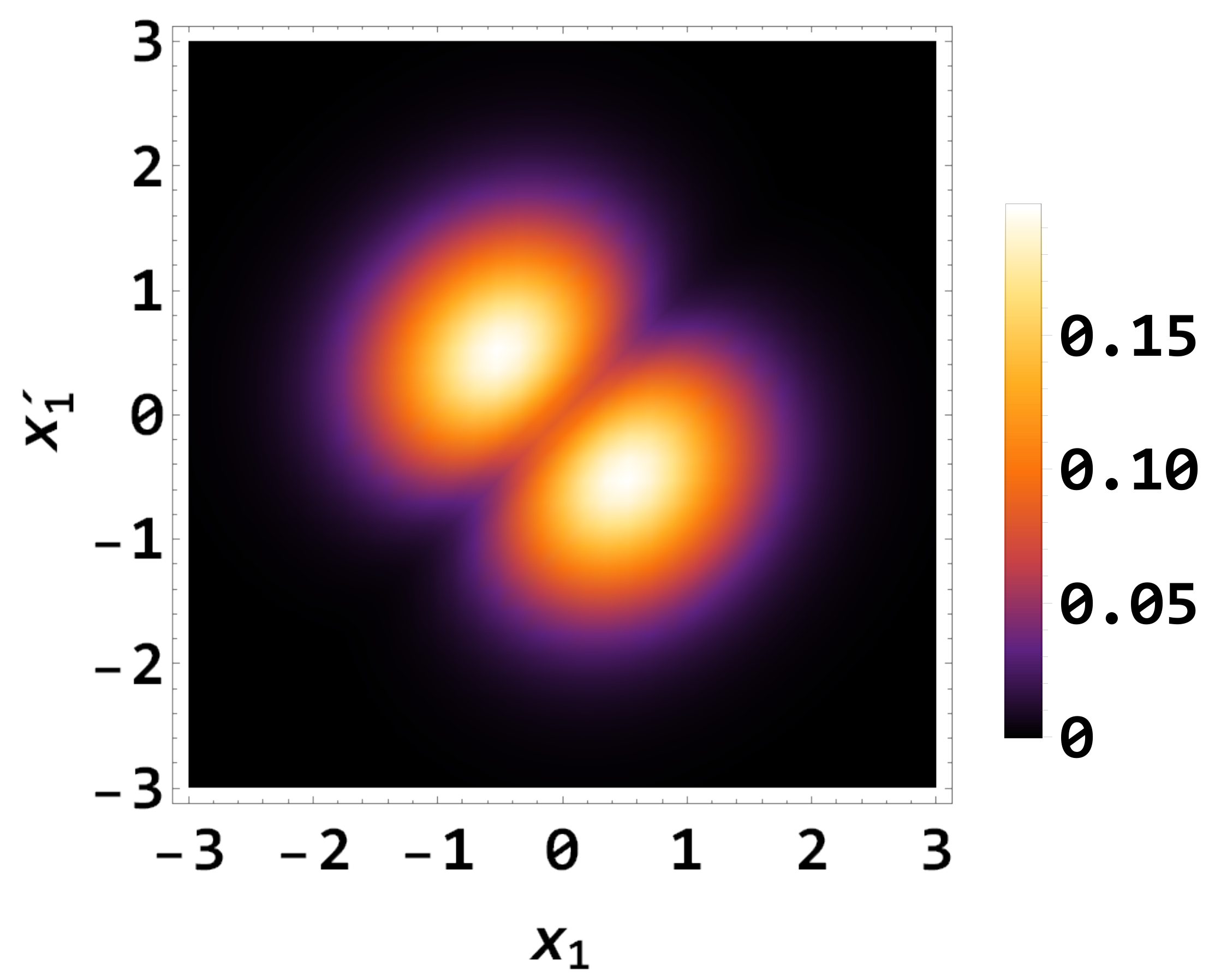} &
\includegraphics[width=.15\textwidth]{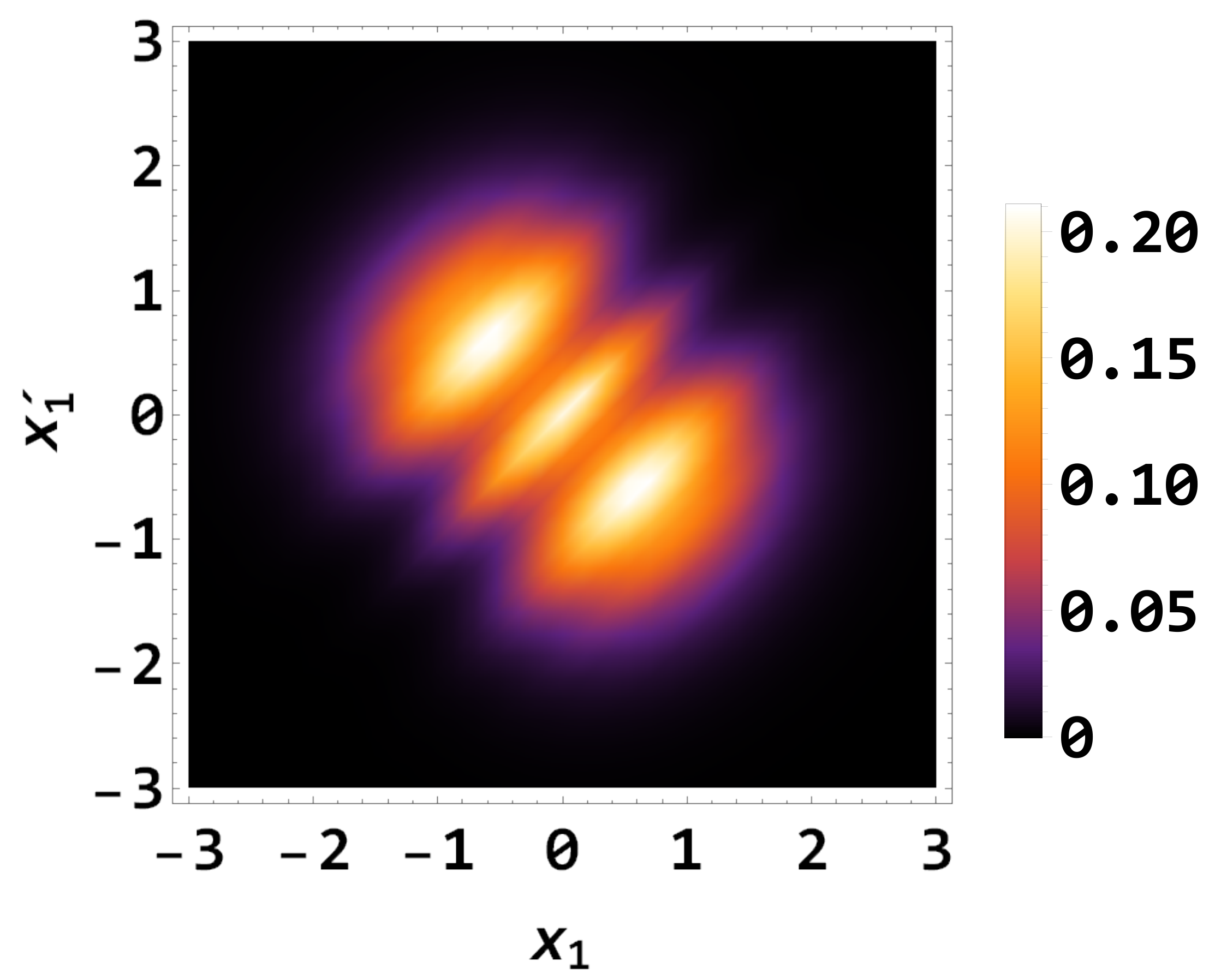} &
\includegraphics[width=.15\textwidth]{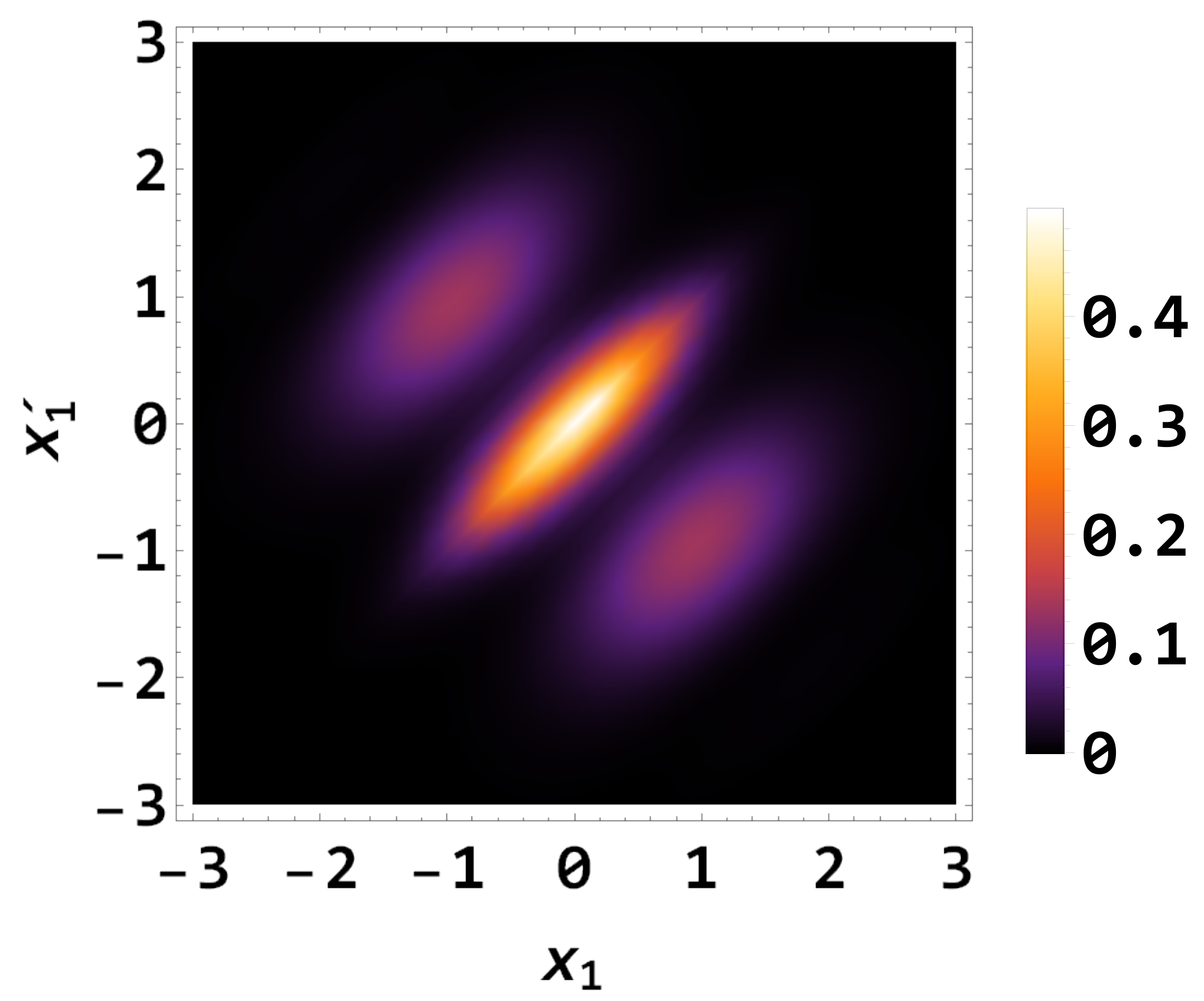} \\
$t=\pi/4$ & $t=\pi/2$ & $t=\pi$\\
\includegraphics[width=.15\textwidth]{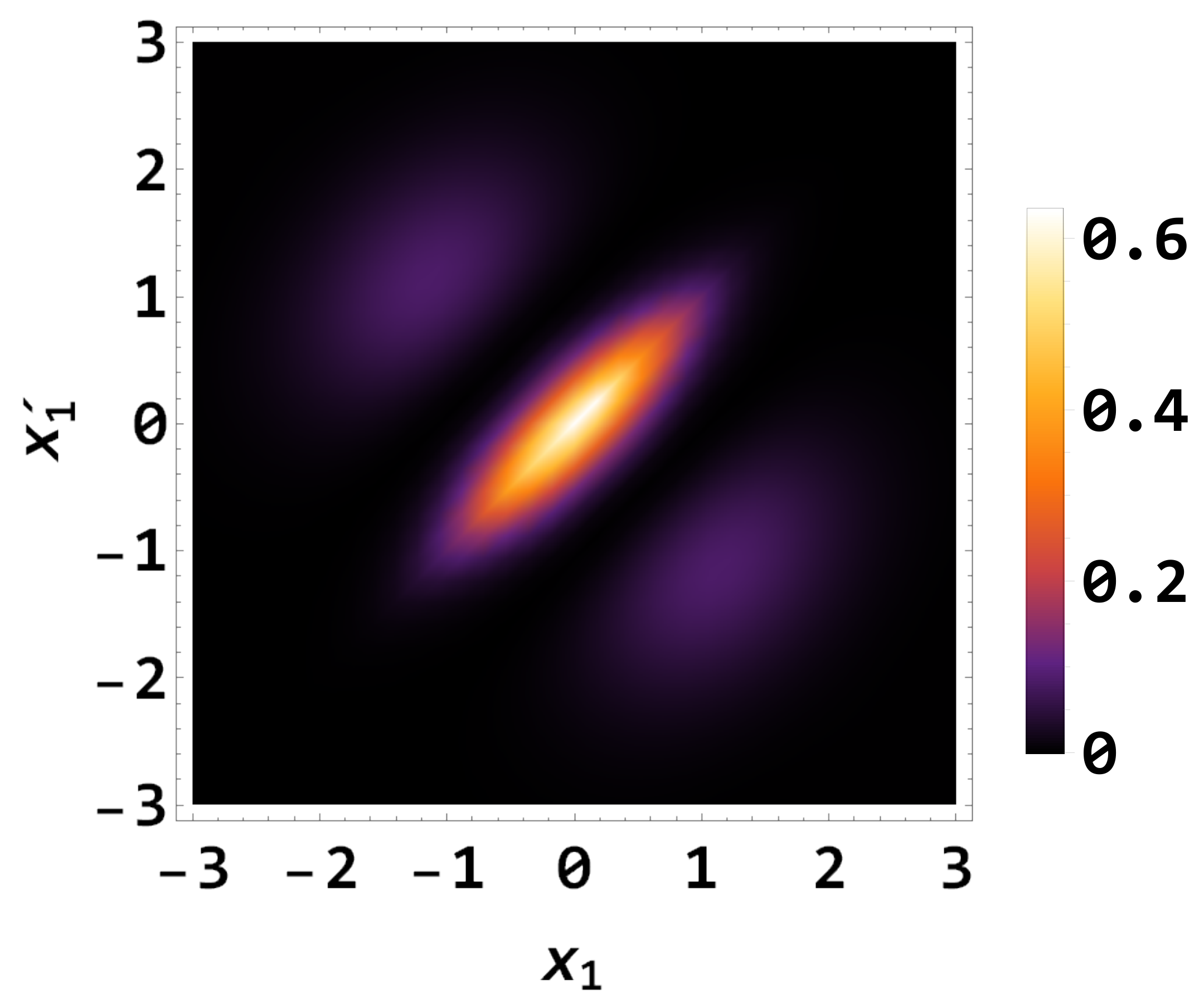} &
\includegraphics[width=.15\textwidth]{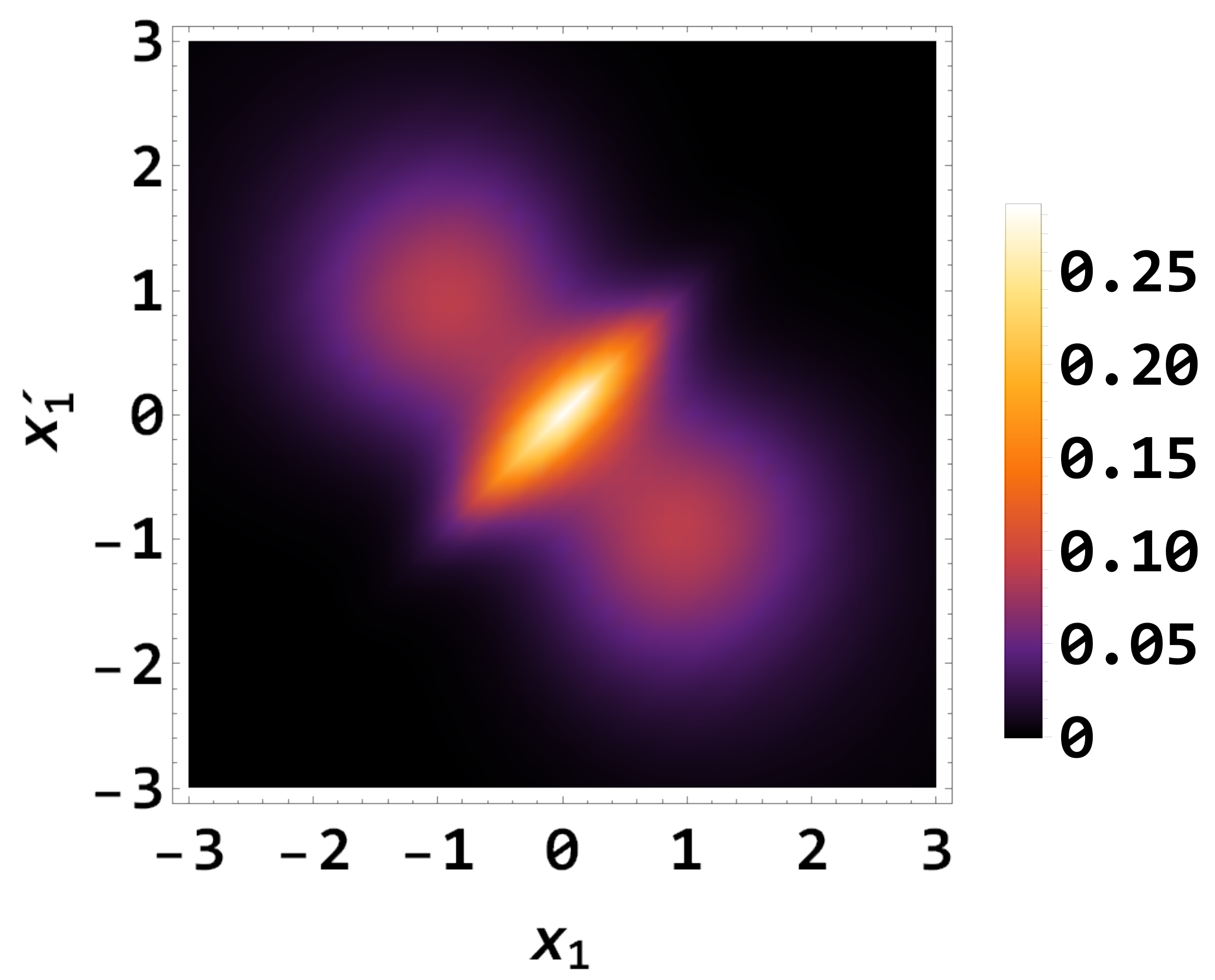} &
\includegraphics[width=.15\textwidth]{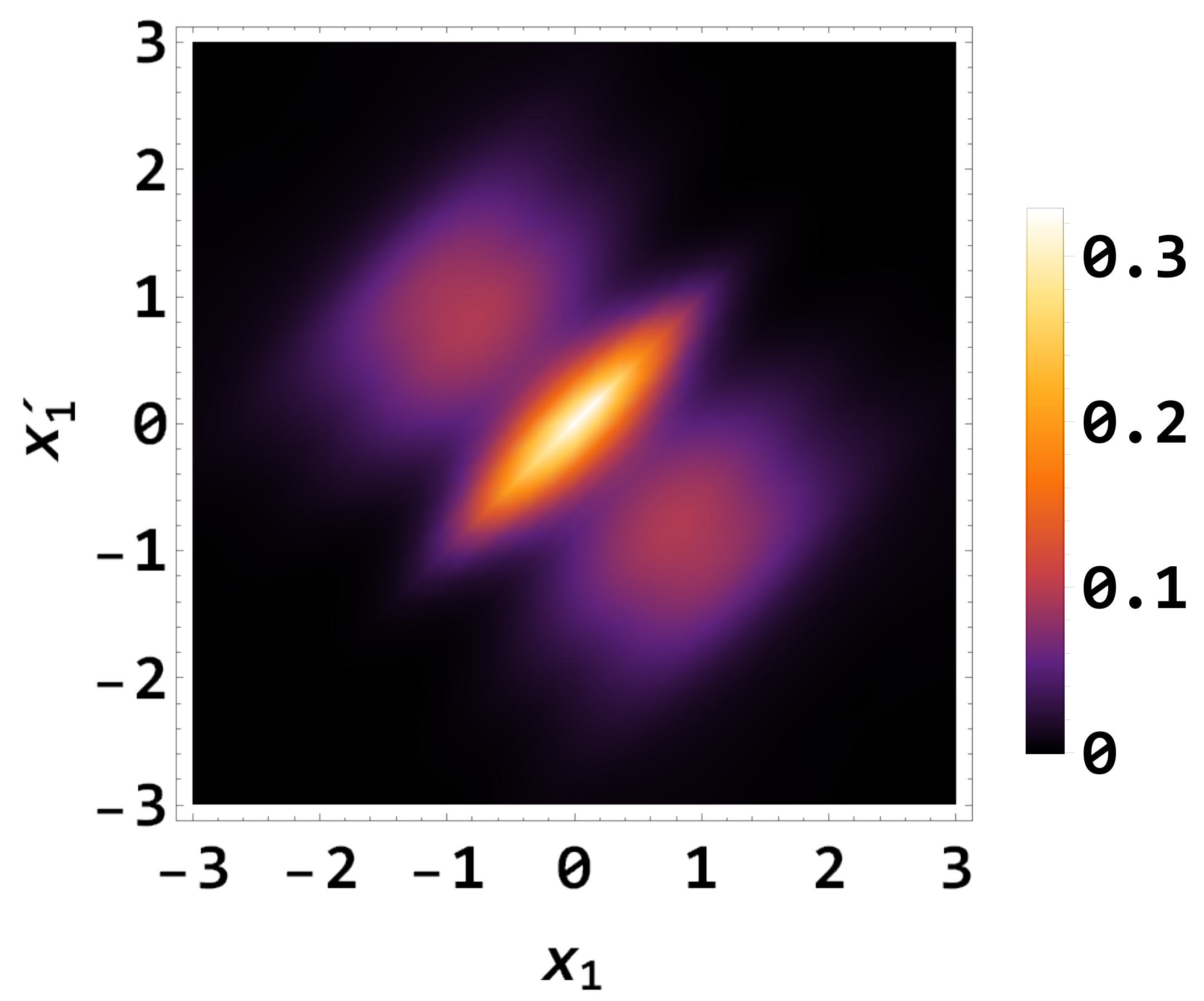} \\
\end{tabular}
\caption{(Color online) Evolution of the probability density $|\Psi(x_1,x_2,t)|^2$.}\label{fwd_al_wf}
\end{figure}

The evolution of the wave packet is plotted in Fig.~\ref{fwd_al_wf}. As time evolves, the initial two-hump structure of the wave packet starts to deform and a third hump in the center appears, just as in the case of the harmonic trap. At $t=\pi/8$ and $t=\pi/4$ the third hump gets larger values than the values of the two other humps. At $t=\pi$ the wave packet still differs from the initial one, which shows that the period of oscillation is shifted. Also, at much closer look some ripples could be noticed. These will be more pronounced in Section~\ref{section_bwd_al} in Fig.~\ref{bwd_al_wf}.

\begin{figure}
	\centering
	\textbf{$\alpha=-0.03$}\par\medskip
	\begin{tabular}{ccc}
		$g=2,(0,0)$  & $g=2,(2,0)$ & $g=2,(4,0)$ \\
		\includegraphics[width=.15\textwidth]{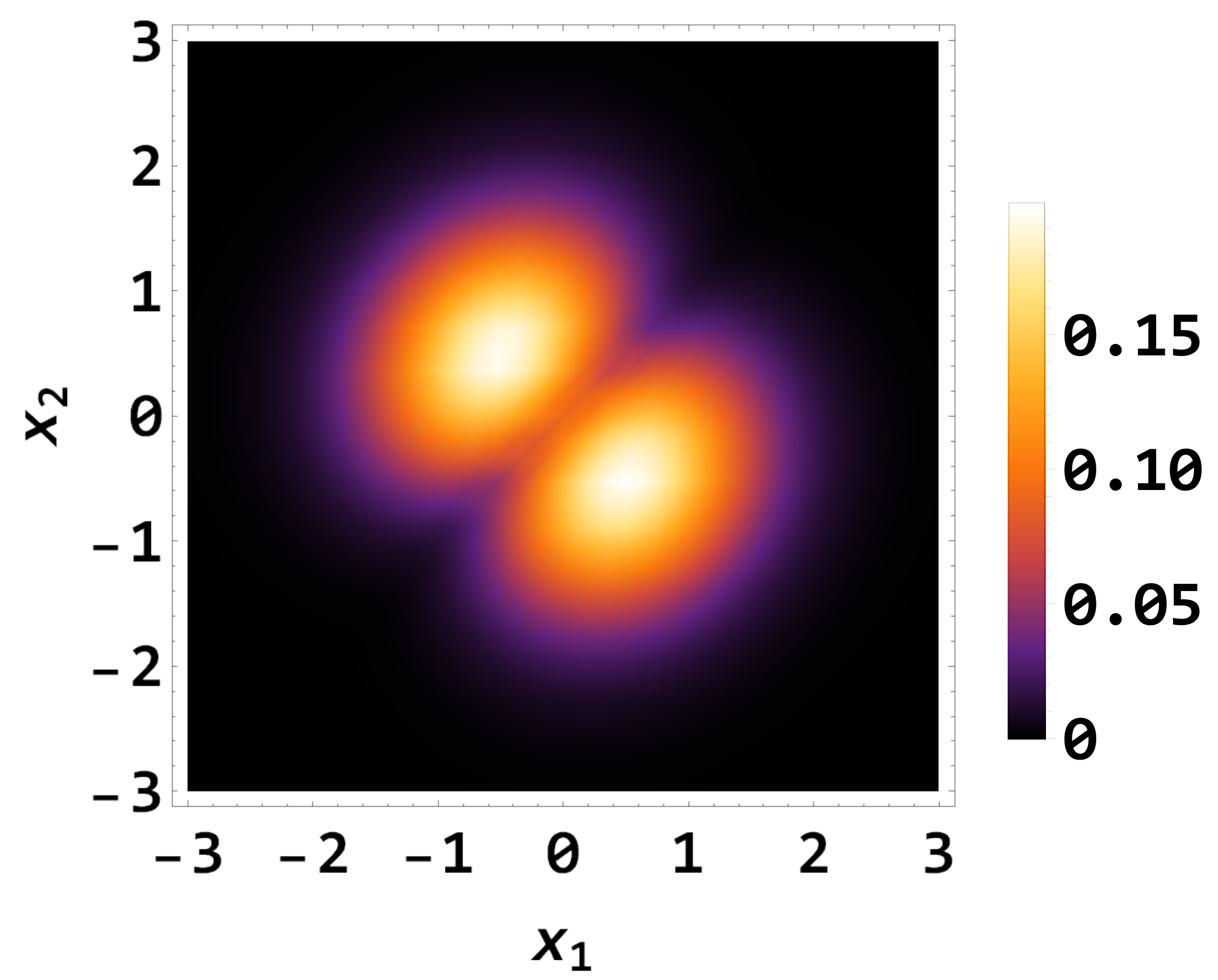} &
		\includegraphics[width=.15\textwidth]{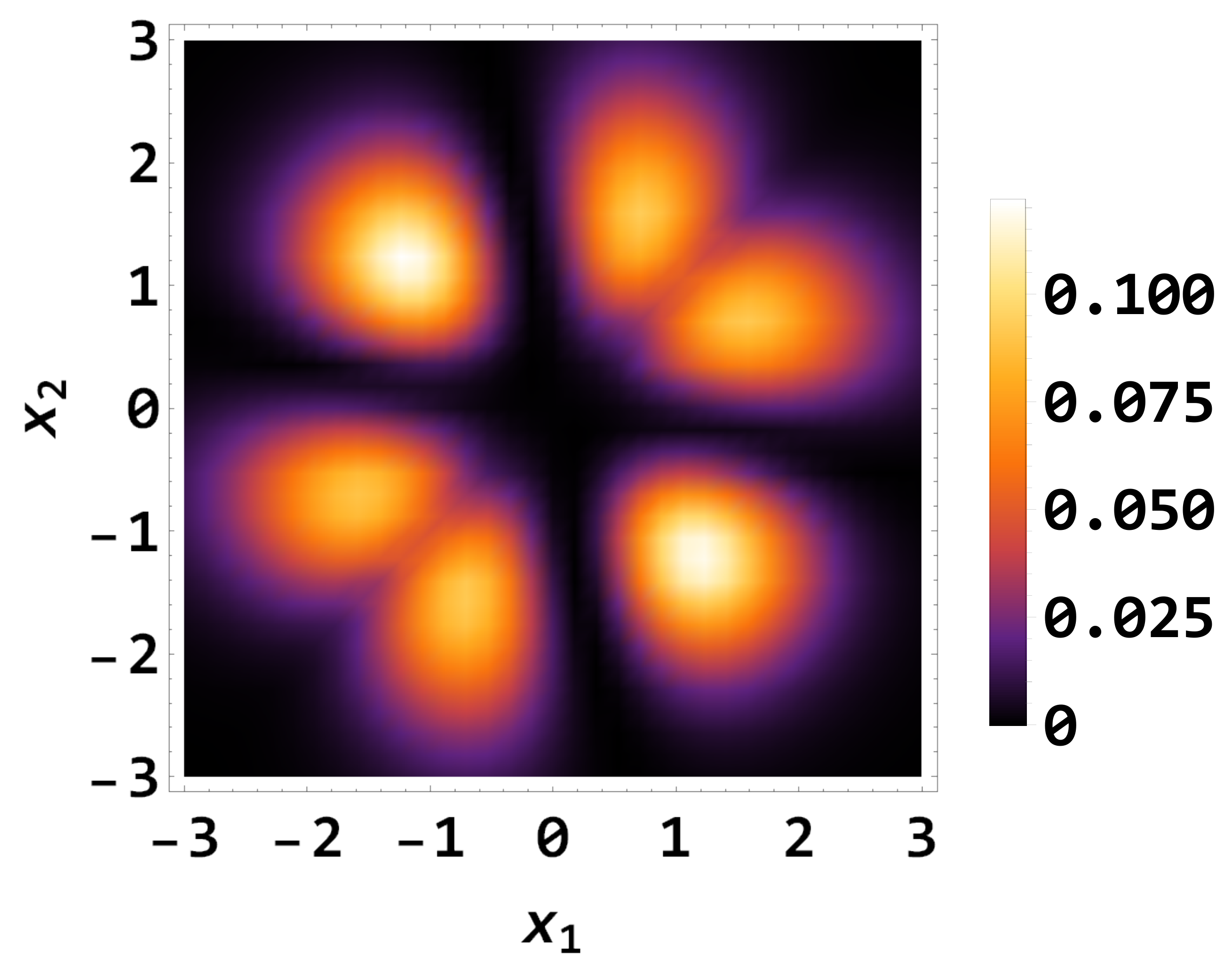} &
		\includegraphics[width=.15\textwidth]{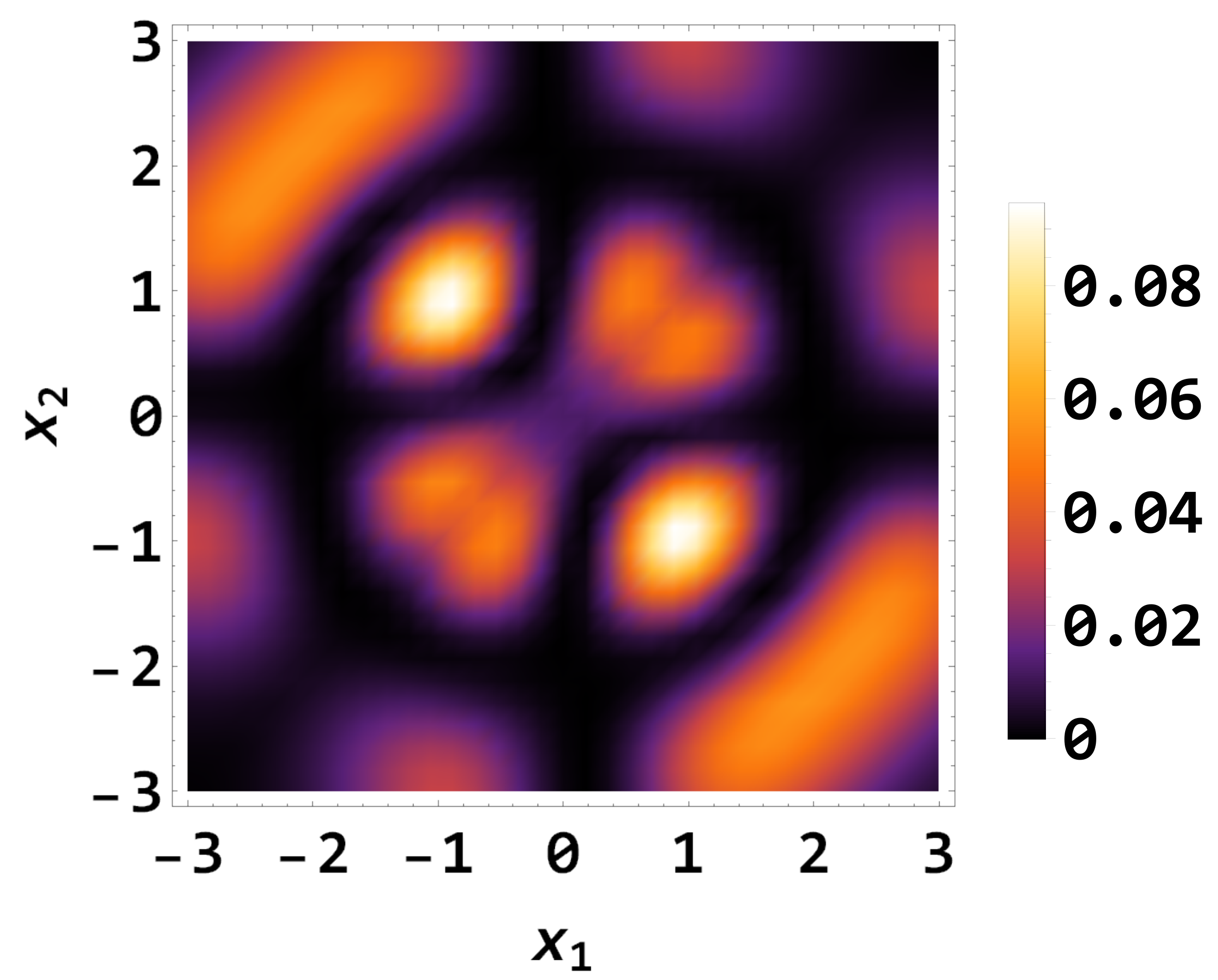} \\		
		$g=-2,(0,0)$  & $g=-2,(2,0)$ & $g=-2,(4,0)$ \\
		\includegraphics[width=.15\textwidth]{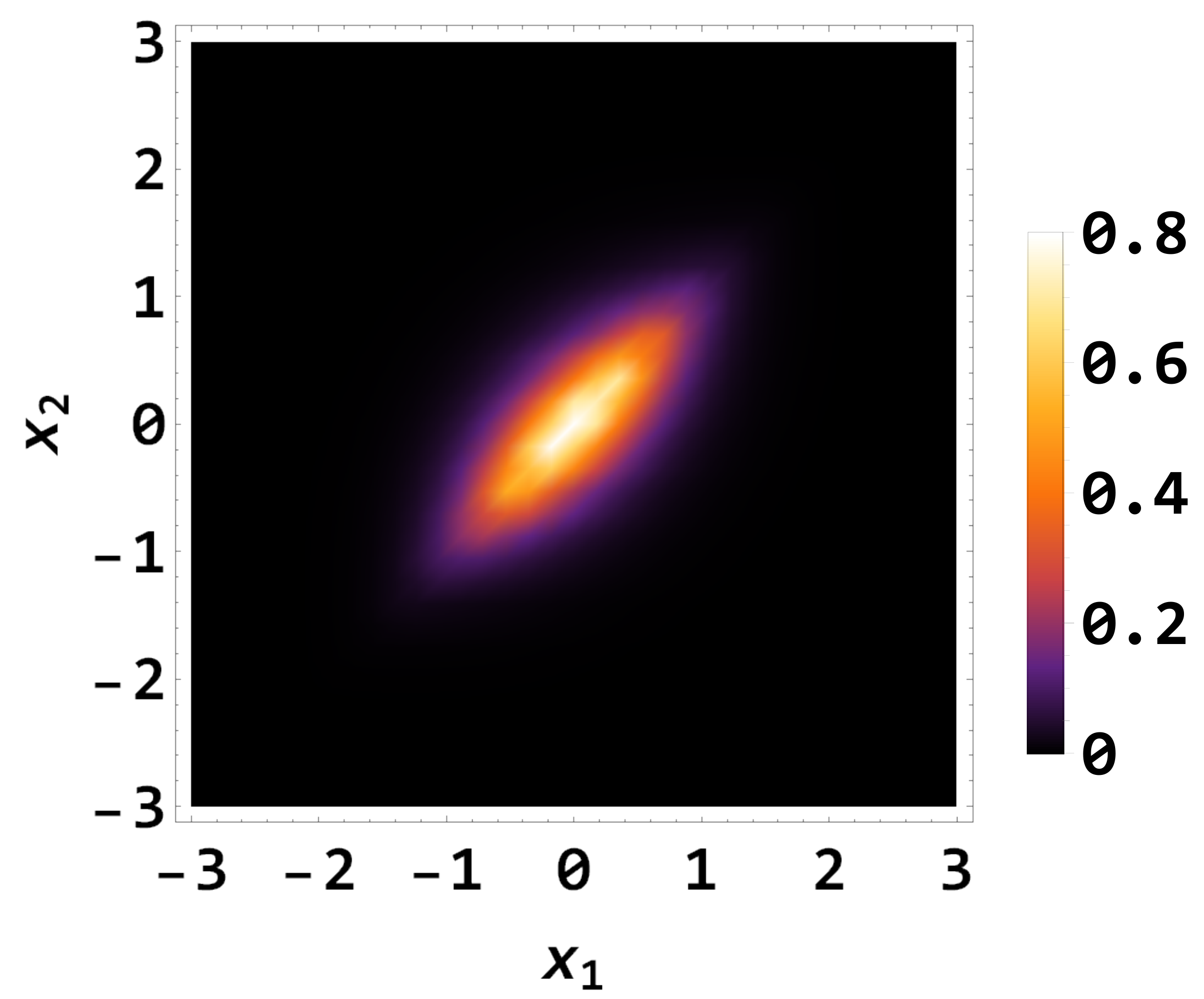} &
		\includegraphics[width=.15\textwidth]{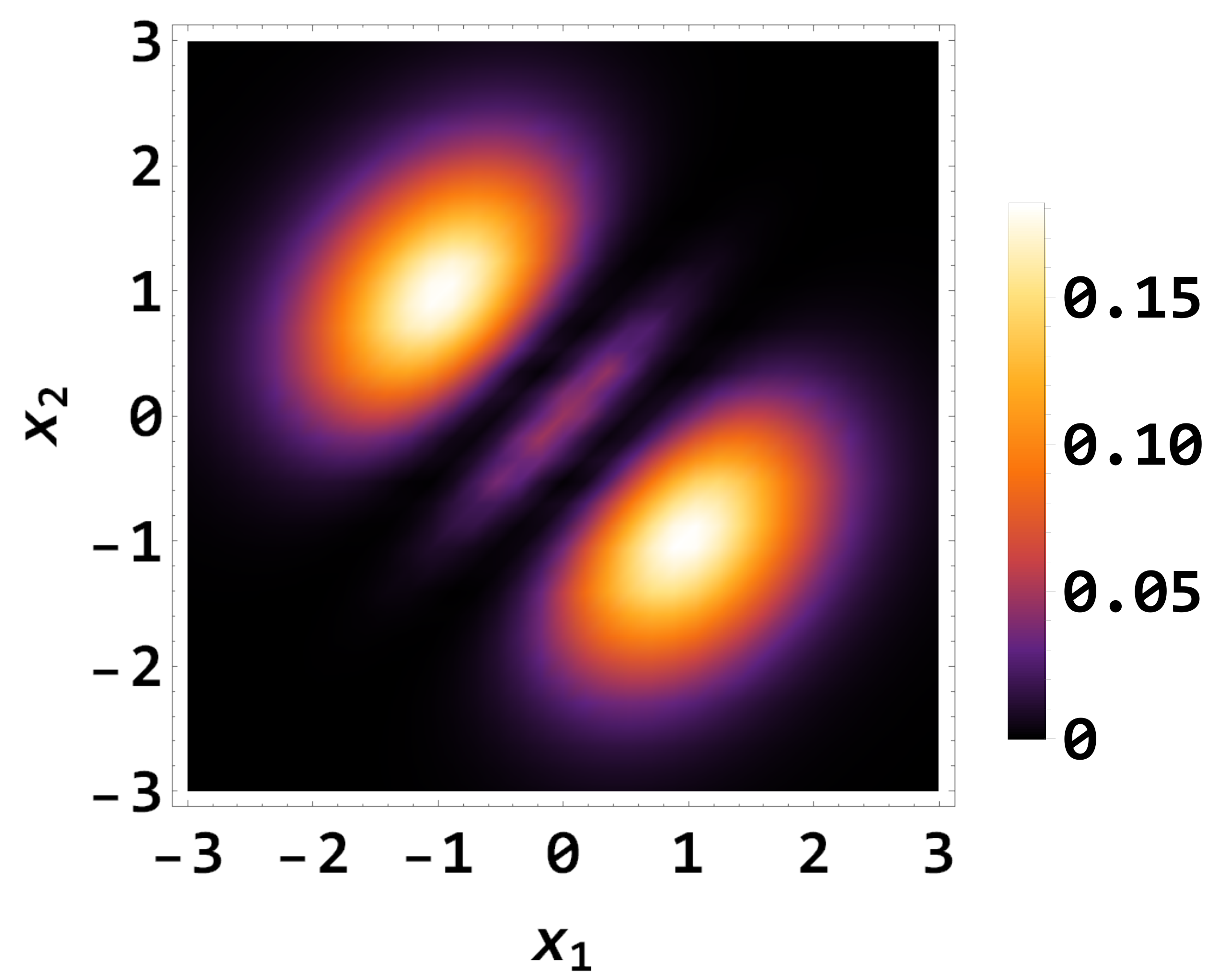} &
		\includegraphics[width=.15\textwidth]{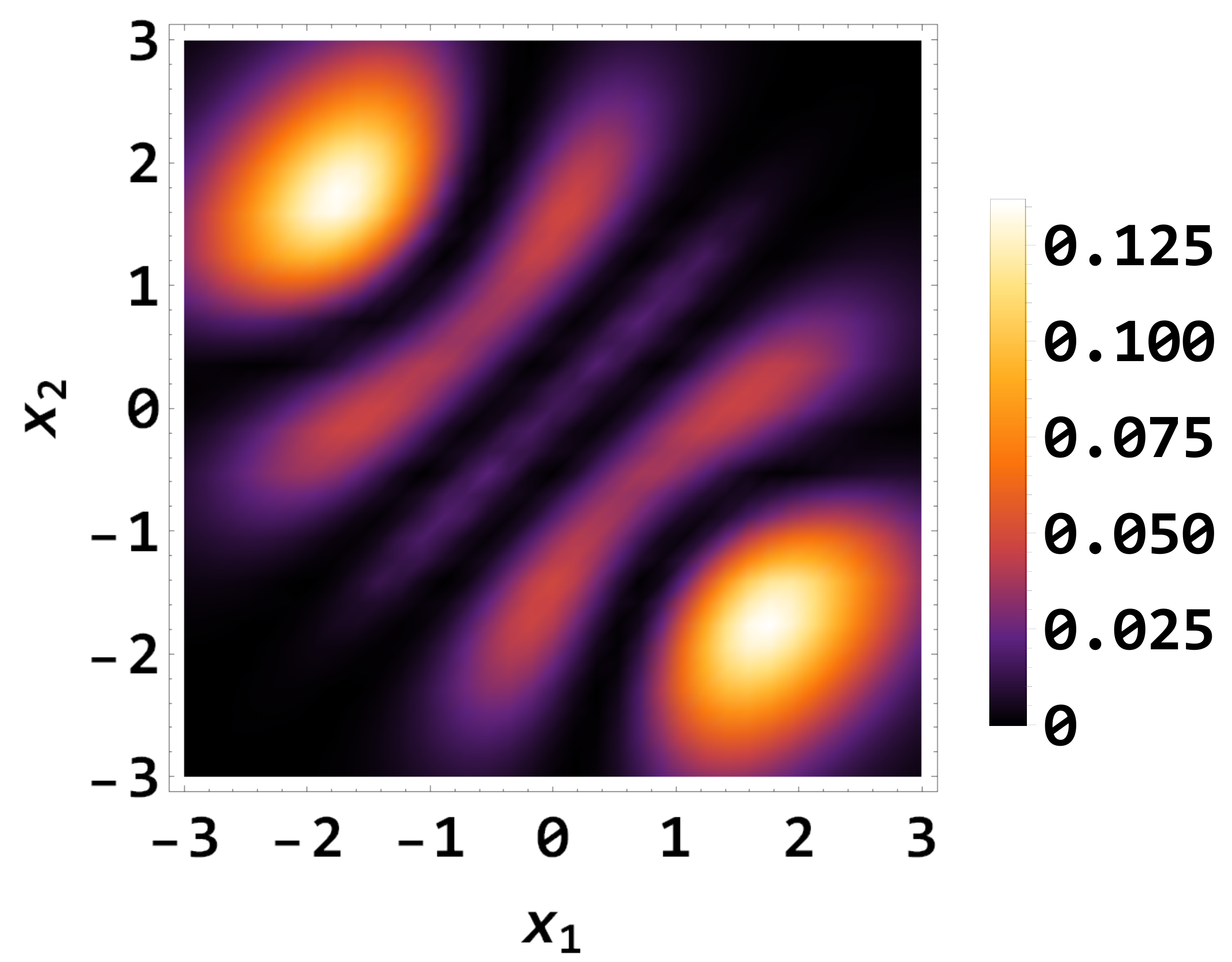} \\
	\end{tabular}
	\caption{(Color online) Probability density $|\Psi(x_1,x_2,t=0)|^2$ for different pre- and postquench stationary states.}\label{fwd_al_wf0}
\end{figure}

A distribution of probability densities $|\Psi(x_1,x_2,t=0)|^2$ of stationary states in Fig.~\ref{fwd_al_wf0} shows a quite complicated structure for the excited states. The anharmonic trap is more wider than the harmonic one, which leads to a more larger area for the wave function localization. Thus, humps of the wave function spread farther from the center.

\begin{figure}
	\centering
	\textbf{$\alpha=-0.03$}\par\medskip
	\begin{tabular}{ccc}
		$t=0.1$ & $t=\pi/8$ & $t=\pi/4$ \\
		\includegraphics[width=.15\textwidth]{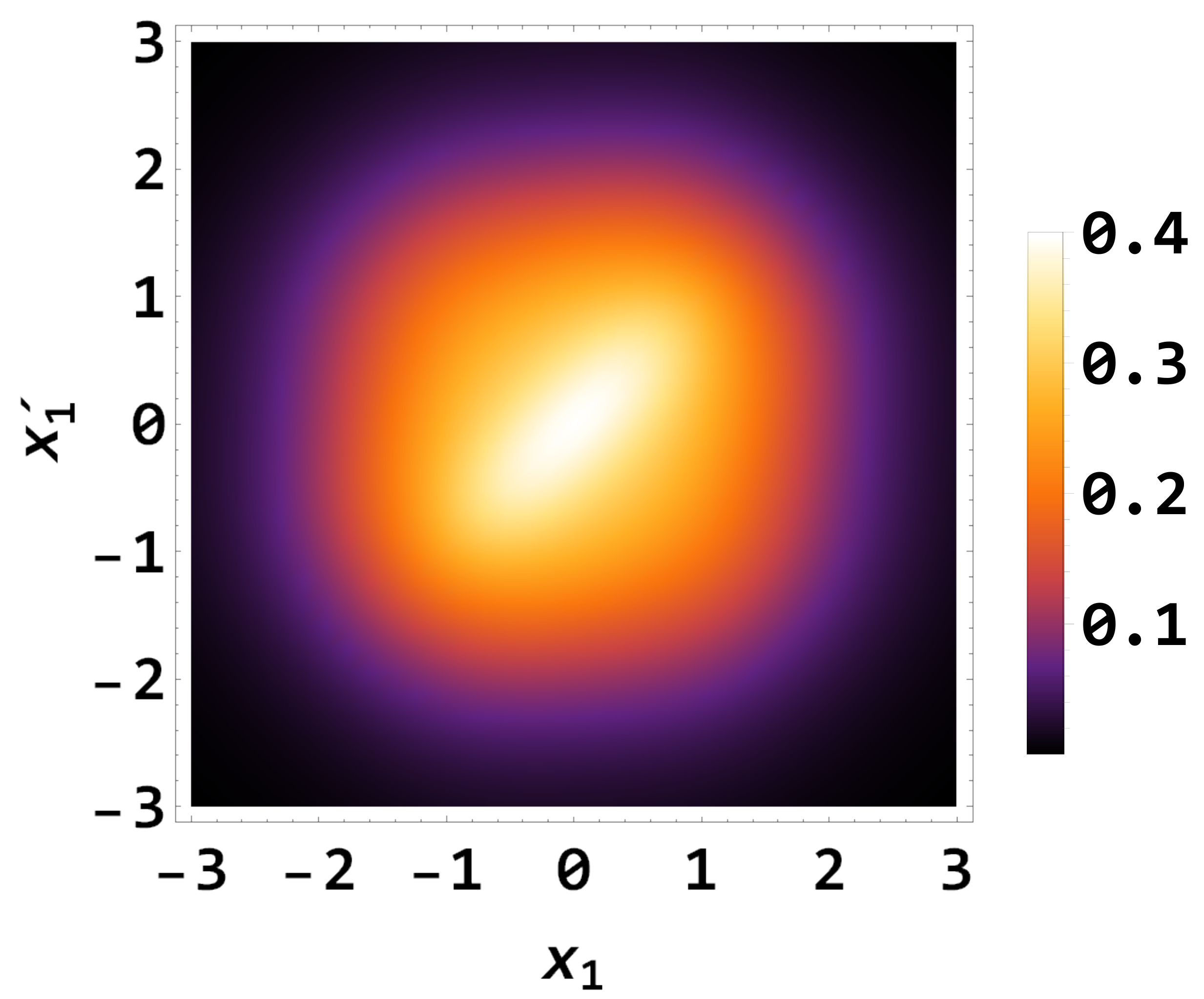} &
		\includegraphics[width=.15\textwidth]{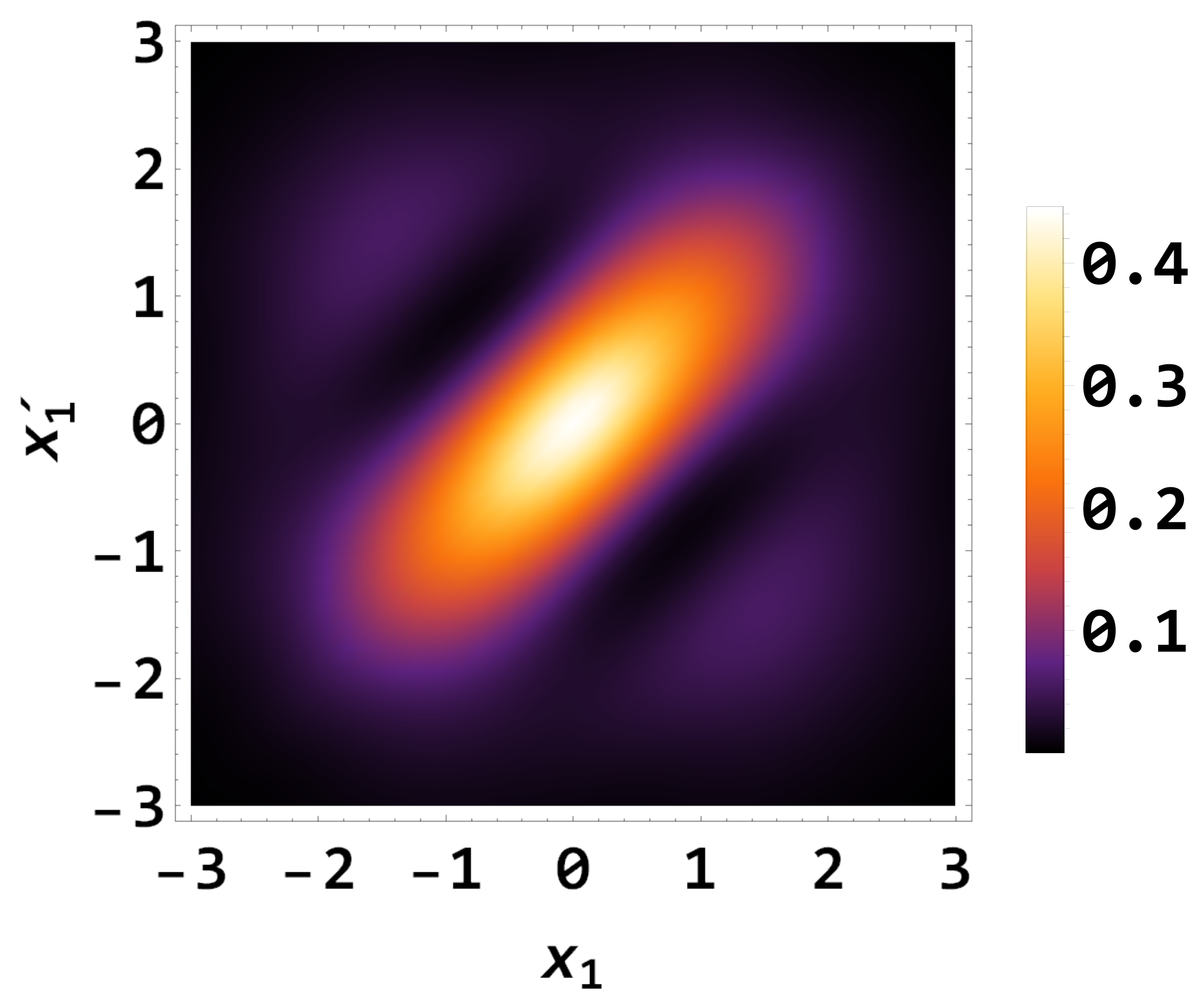} &
		\includegraphics[width=.15\textwidth]{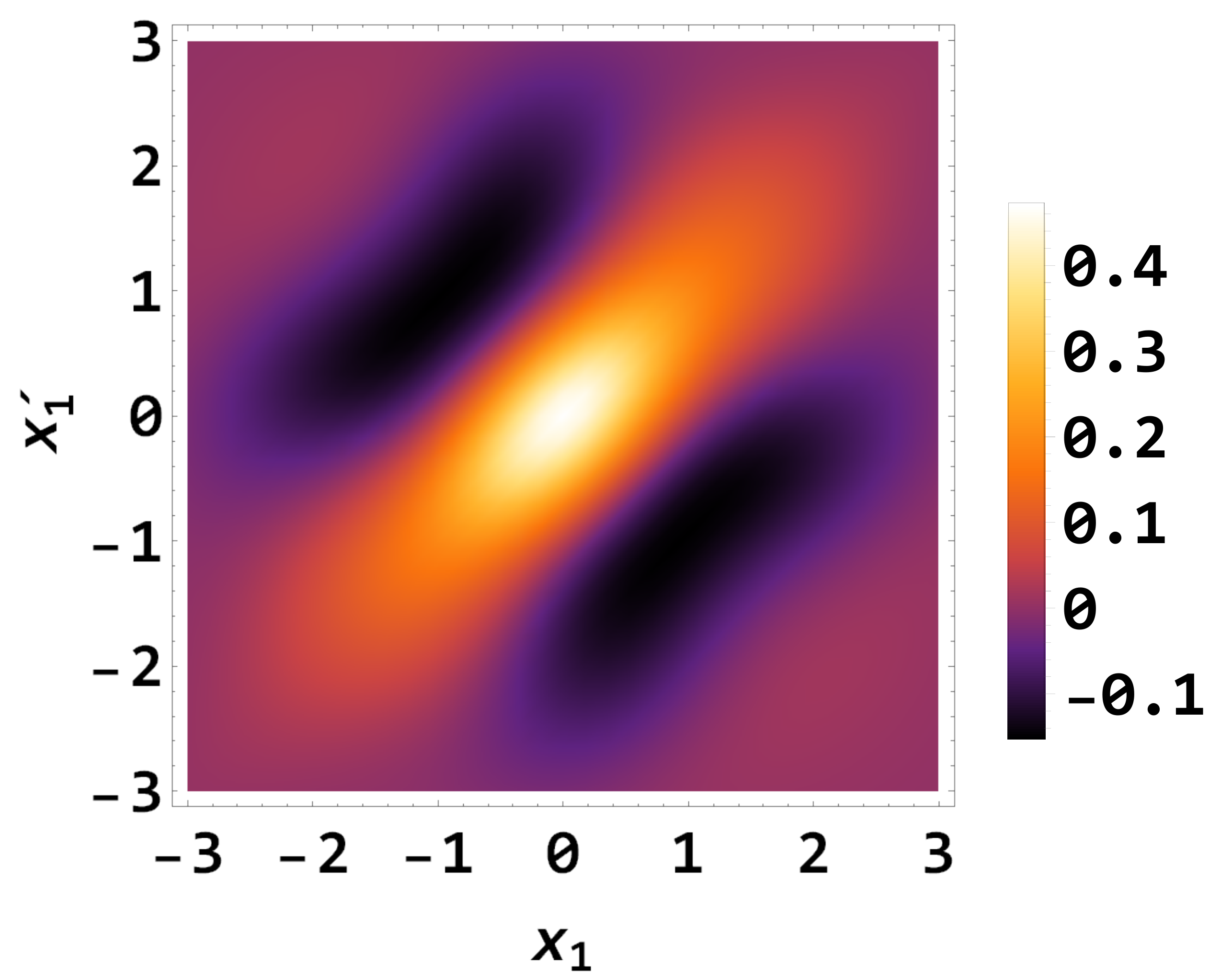} \\
		$t=\pi/2$ & $t=\pi$ \\
		\includegraphics[width=.15\textwidth]{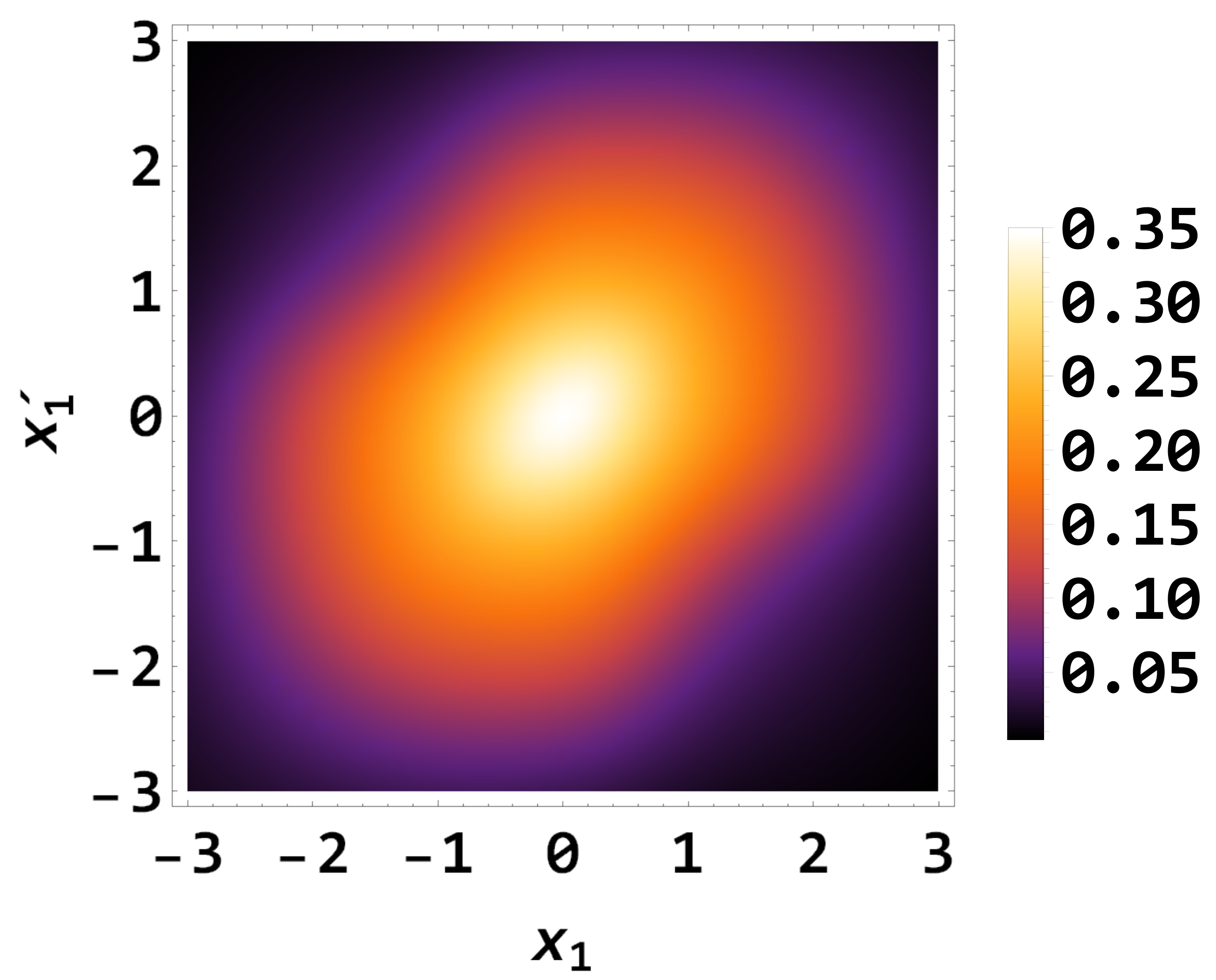} &
		\includegraphics[width=.15\textwidth]{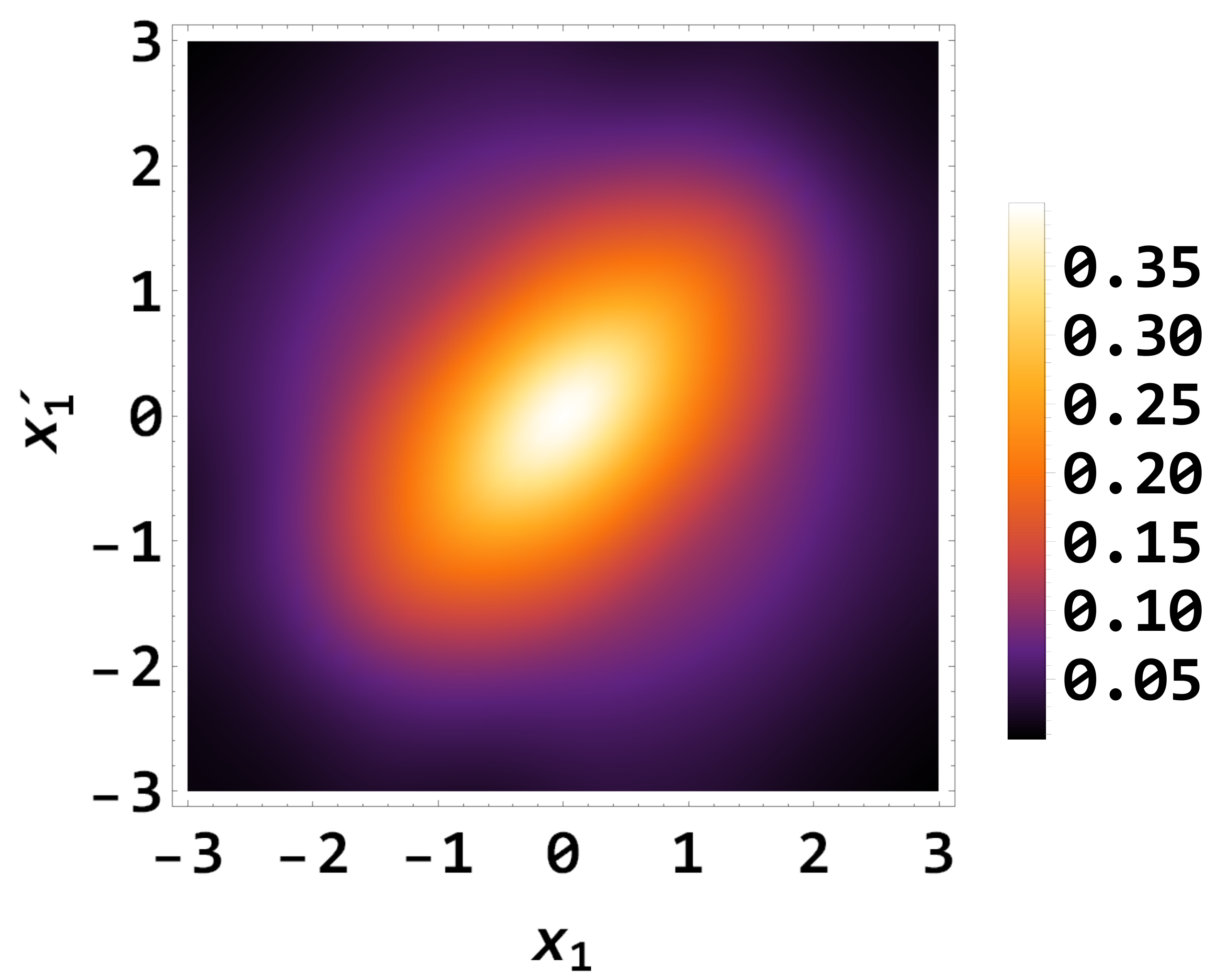} \\
	\end{tabular}
	\caption{(Color online) One-body reduced density matrix, $\rho^{(1)}(x_1,x_1';t)$, at different time-instants.}\label{fwd_al_rho}
\end{figure}

An analysis of the one-body reduced density matrix $\rho^{(1)}(x_1,x_1';t)$ in Fig.~\ref{fwd_al_rho} gives another glimpse of what happens during the quench dynamics. The initial spatial distribution of $\rho^{(1)}(x_1,x_1';t)$ at $t=0.1$ is quite large, which is a consequence of the repulsive interaction between the particles. By the time reaches $t=\pi/4$ the initial hump of $\rho^{(1)}(x_1,x_1';t)$ splits into three humps, which is an indication of a transition into some other state. Afterwards, the three-hump structure of $\rho^{(1)}(x_1,x_1';t)$ reverts to the one-hump one. At $t=\pi$, as opposed to the case of the harmonic trap, the spatial distribution of $\rho^{(1)}(x_1,x_1';t)$ is still different from the initial one. This is due to the size of the anharmonic trap, which requires larger times for the wave packet to take its initial shape.

\begin{figure}
	\centering
	\textbf{$\alpha=-0.03$}\par\medskip
	\begin{tabular}{ccc}
		$g=2,(0,0)$  & $g=2,(2,0)$ & $g=2,(4,0)$ \\
		\includegraphics[width=.15\textwidth]{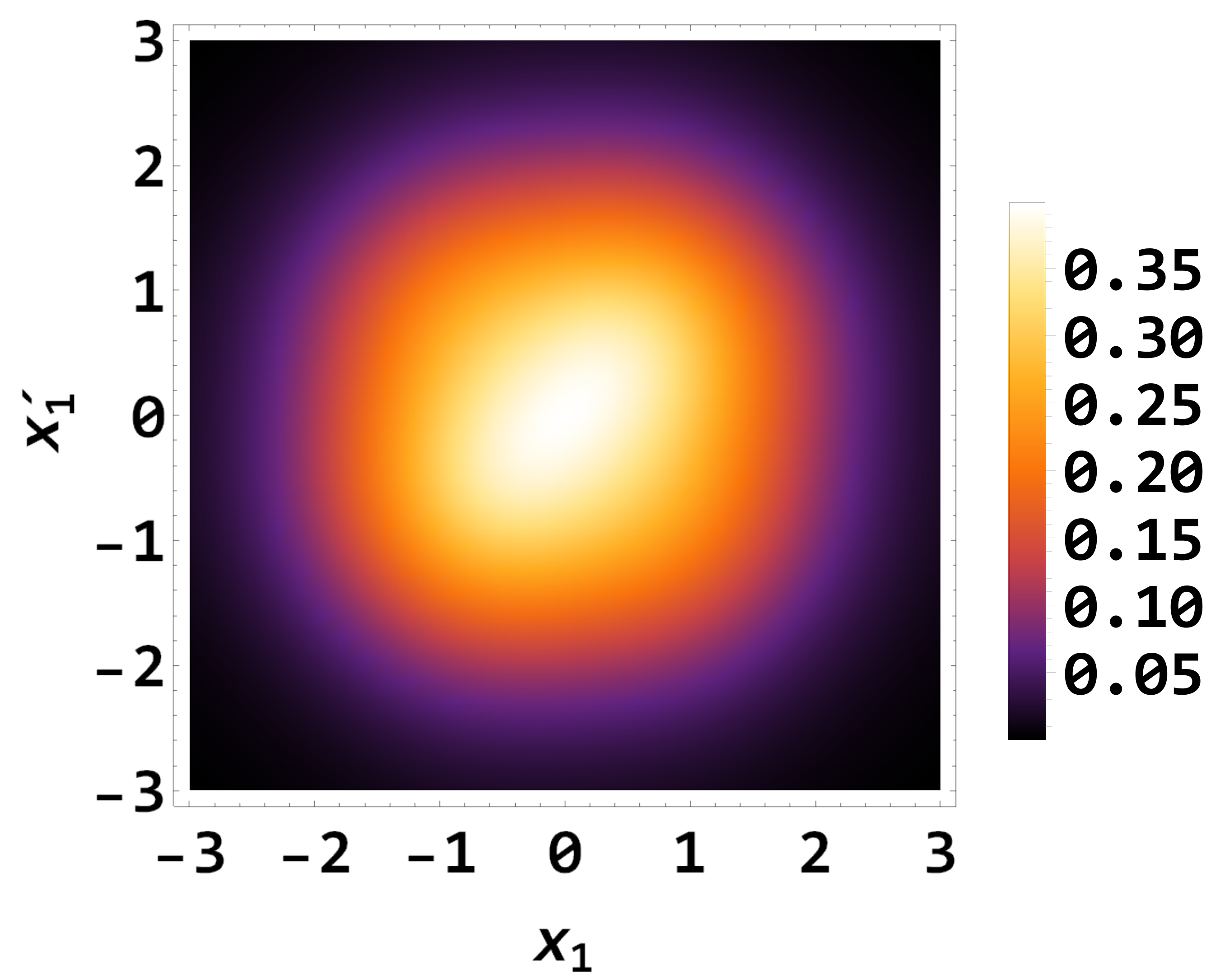} &
		\includegraphics[width=.15\textwidth]{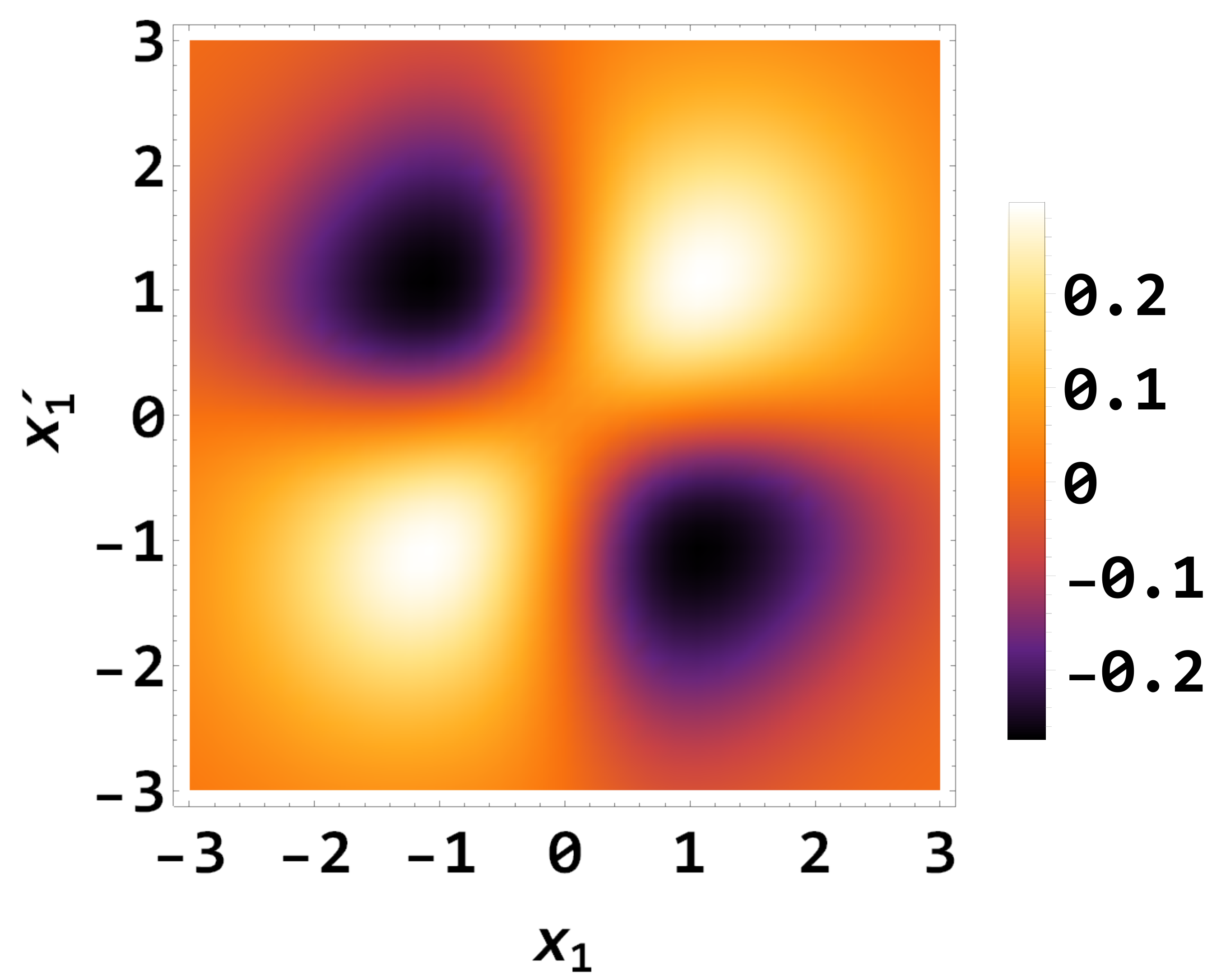} &
		\includegraphics[width=.15\textwidth]{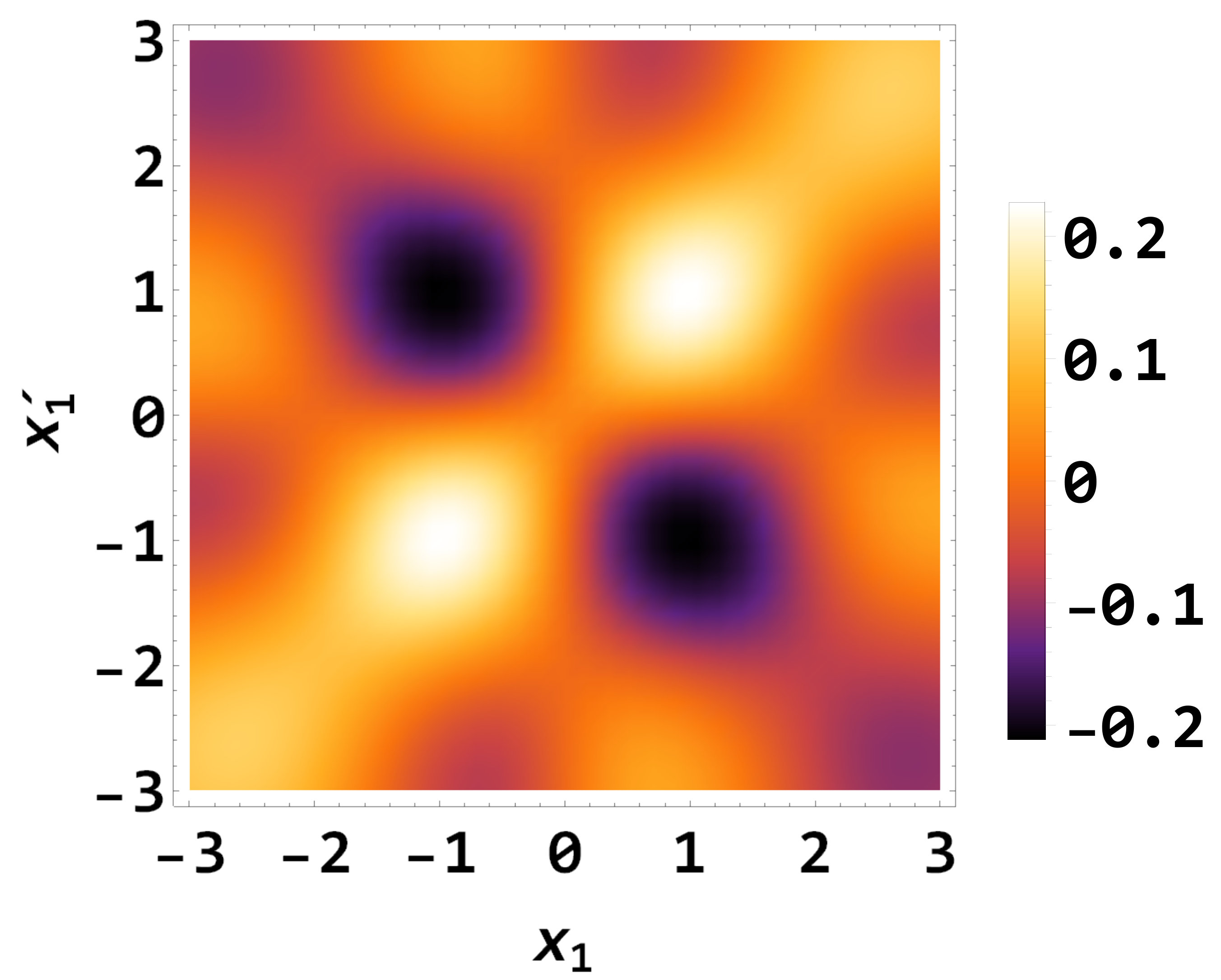} \\
		$g=-2,(0,0)$  & $g=-2,(2,0)$ & $g=-2,(4,0)$ \\
		\includegraphics[width=.15\textwidth]{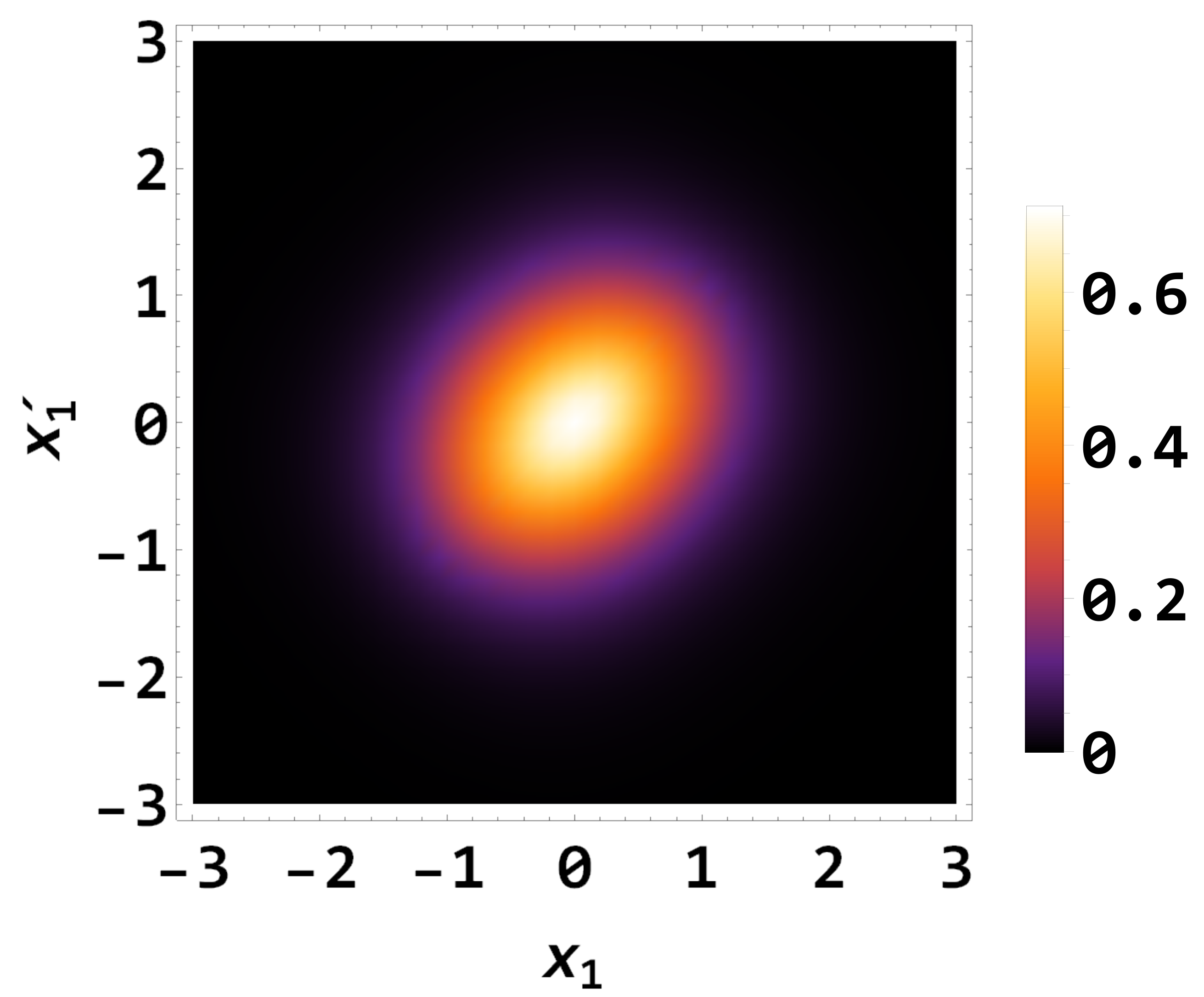} &
		\includegraphics[width=.15\textwidth]{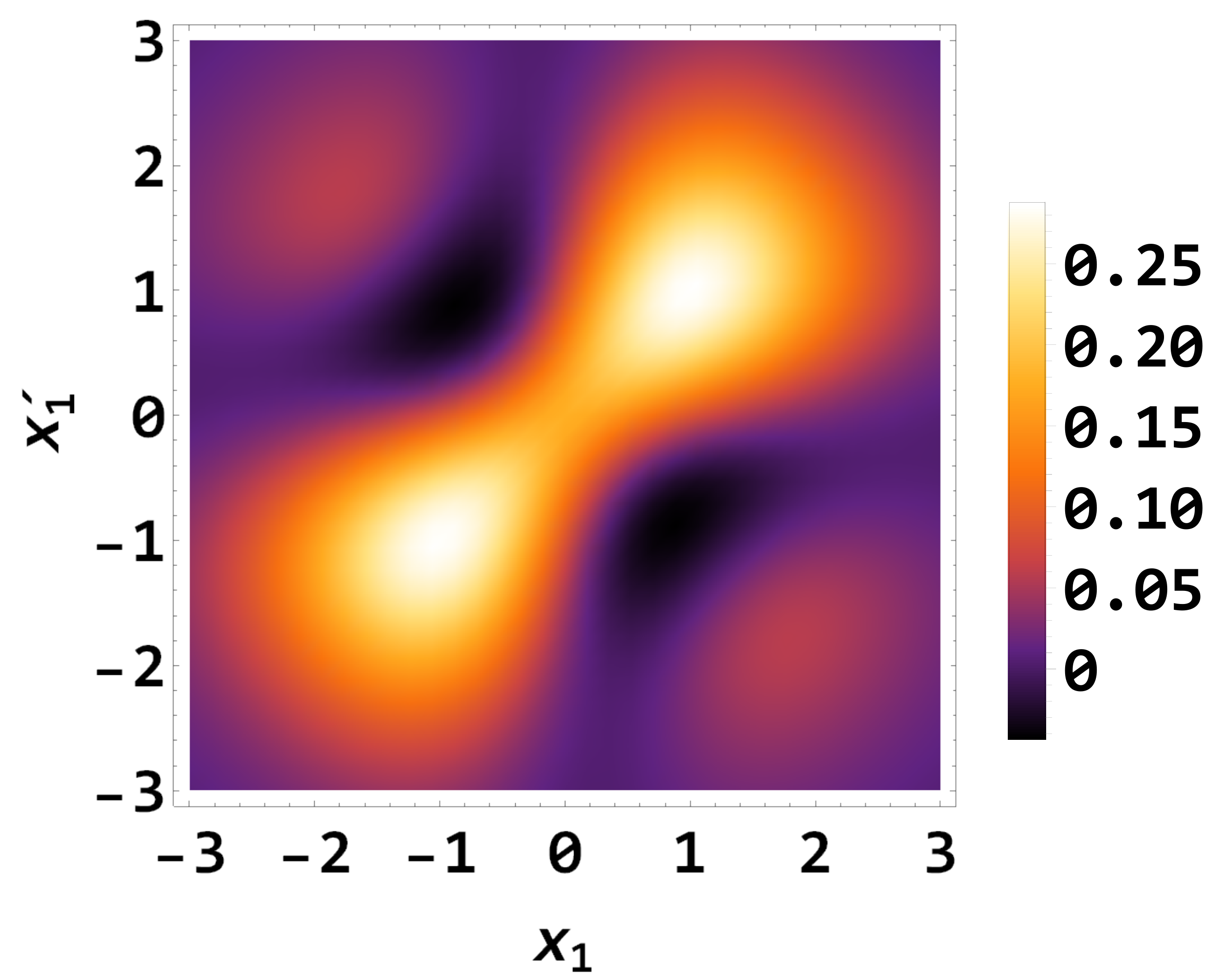} &
		\includegraphics[width=.15\textwidth]{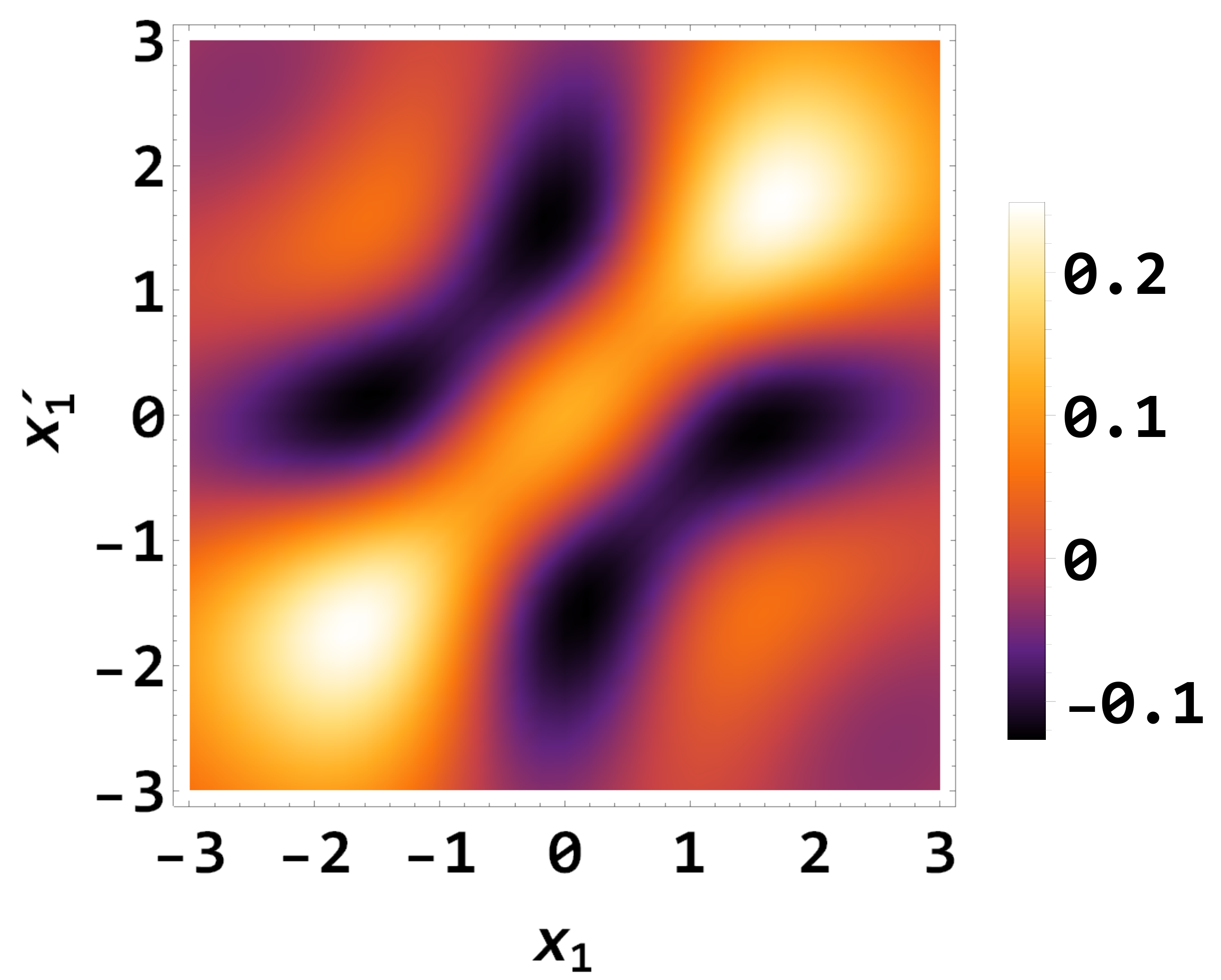} \\
	\end{tabular}
	\caption{(Color online) One-body reduced density matrix, $\rho^{(1)}(x_1,x_1';t=0)$, for different stationary states.}\label{fwd_al_irho}
\end{figure}

The shapes of the one-body reduced density matrix $\rho^{(1)}(x_1,x_1';t=0)$ for the pre- and postquench stationary states is shown in Fig.~\ref{fwd_al_irho}. The states of the pre- and postquench excited states differ significantly, yet both of them have a similarity in that there are humps on the sides of the diagonal axis $x_1=x_2$. This suggests the at least some part of them may overlap with the wave packet.

\begin{figure}
	\centering
	\textbf{$\alpha=-0.03$}\par\medskip
	\includegraphics[width=7cm,clip]{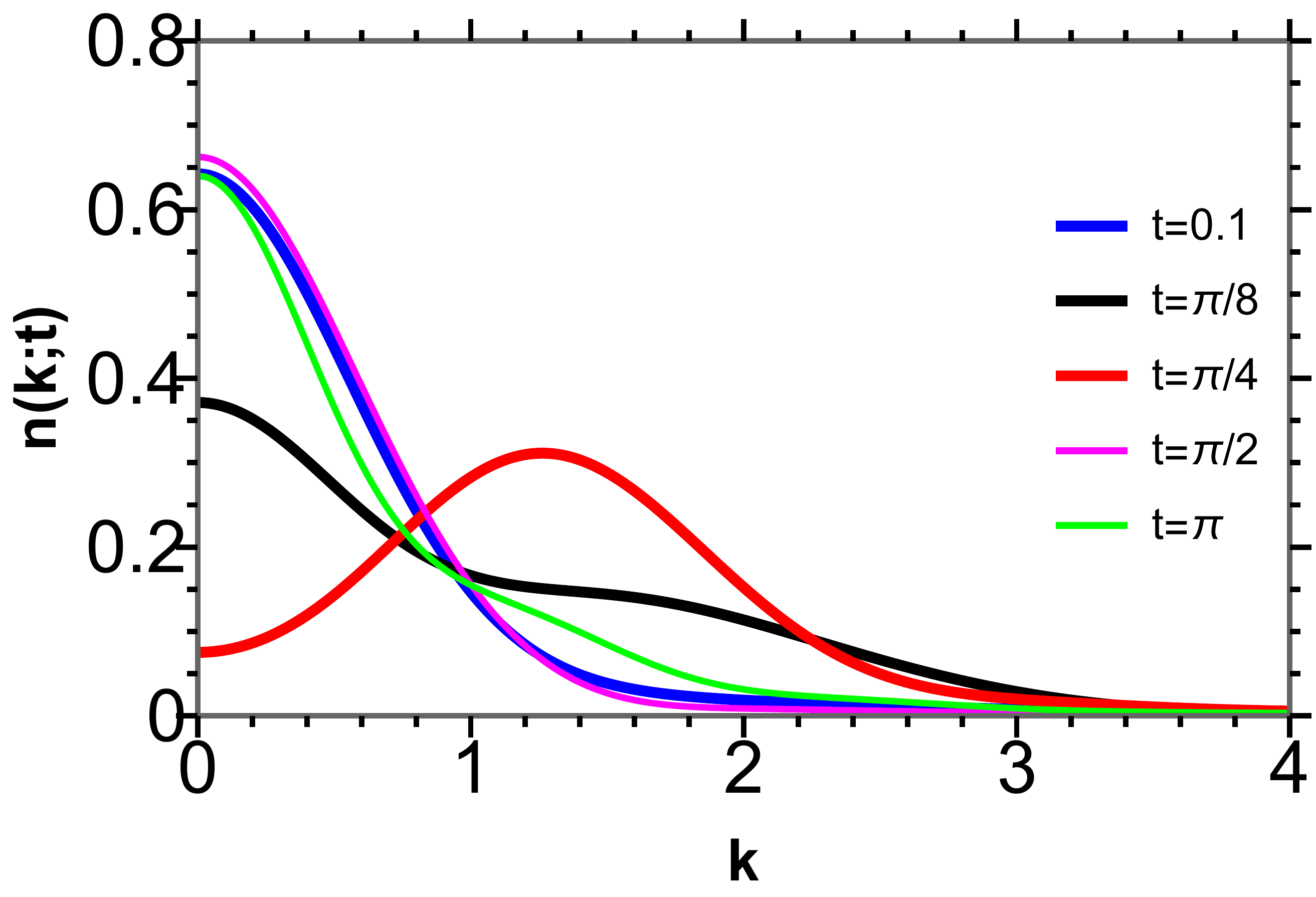}
	\caption{(Color online) Evolution of the momentum distribution $n(k;t)$ for the anharmonic trap.}\label{fwd_al_nk_all}
\end{figure}

We now consider the dynamics of the momentum distribution $n(k;t)$ in Fig.~\ref{fwd_al_nk_all} in the case of the anharmonic trap, $\alpha=-0.03$. The deformation of the shape of $n(k;t)$ with time is quite similar to that of the harmonic trap case. The initial shape, at $t=0.1$, possesses a pronounced zero-momentum peak. By the time $t=\pi/4$ the peak shifts to a higher momentum and the overall shape becomes more stretched, having a lesser maximum value. Then, at $t=\pi/2$, the shape again has a high zero-momentum peak, since, according to Fig.~\ref{fwd_al}, the fidelity $F(t)$ almost reaches quite a high value. At $t=\pi$, however, the shape is slightly different from the shape at $t=0.1$, meaning some deviation of the period from the period in the case of the harmonic trap.

\begin{figure}
	\centering
	\textbf{$\alpha=-0.03$,~~~$t=\pi/8$}\par\medskip
	\begin{tabular}{cc}
		Prequench states &  Postquench states \\
		\hspace{-.5cm}\includegraphics[width=.25\textwidth]{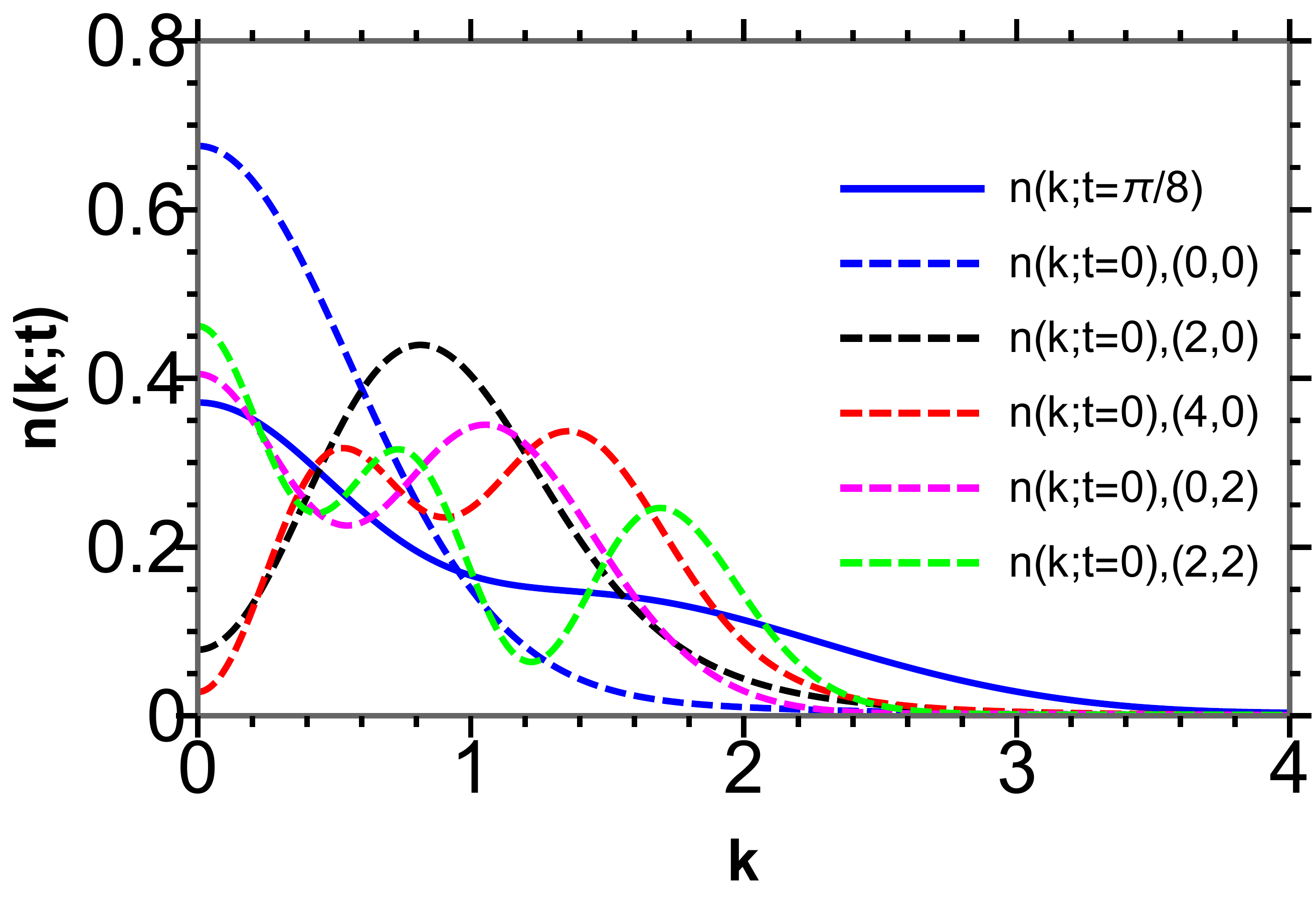} &
		\includegraphics[width=.25\textwidth]{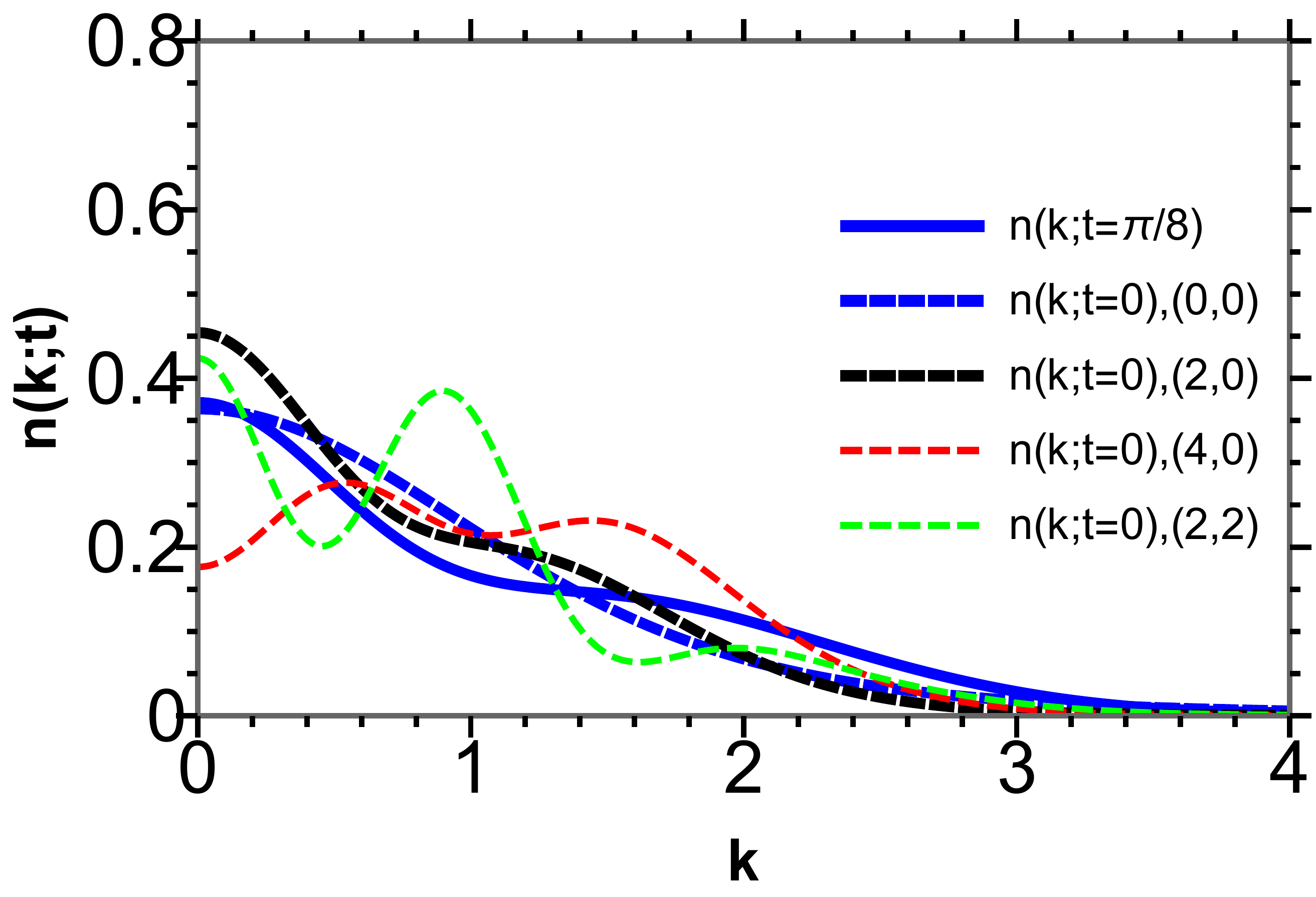} \\
	\end{tabular}
	\caption{(Color online) Comparison of the momentum distribution $n(k;t)$ (blue line) with the momentum distributions $n(k;t=0)$ for the pre- and postquench states (dashed lines).}\label{fwd_al_nk_12}
\end{figure}

\begin{figure}
	\centering
	\textbf{$\alpha=-0.03$,~~~$t=\pi/4$}\par\medskip
	\begin{tabular}{cc}
		Prequench states &  Postquench states \\
		\hspace{-.5cm}\includegraphics[width=.25\textwidth]{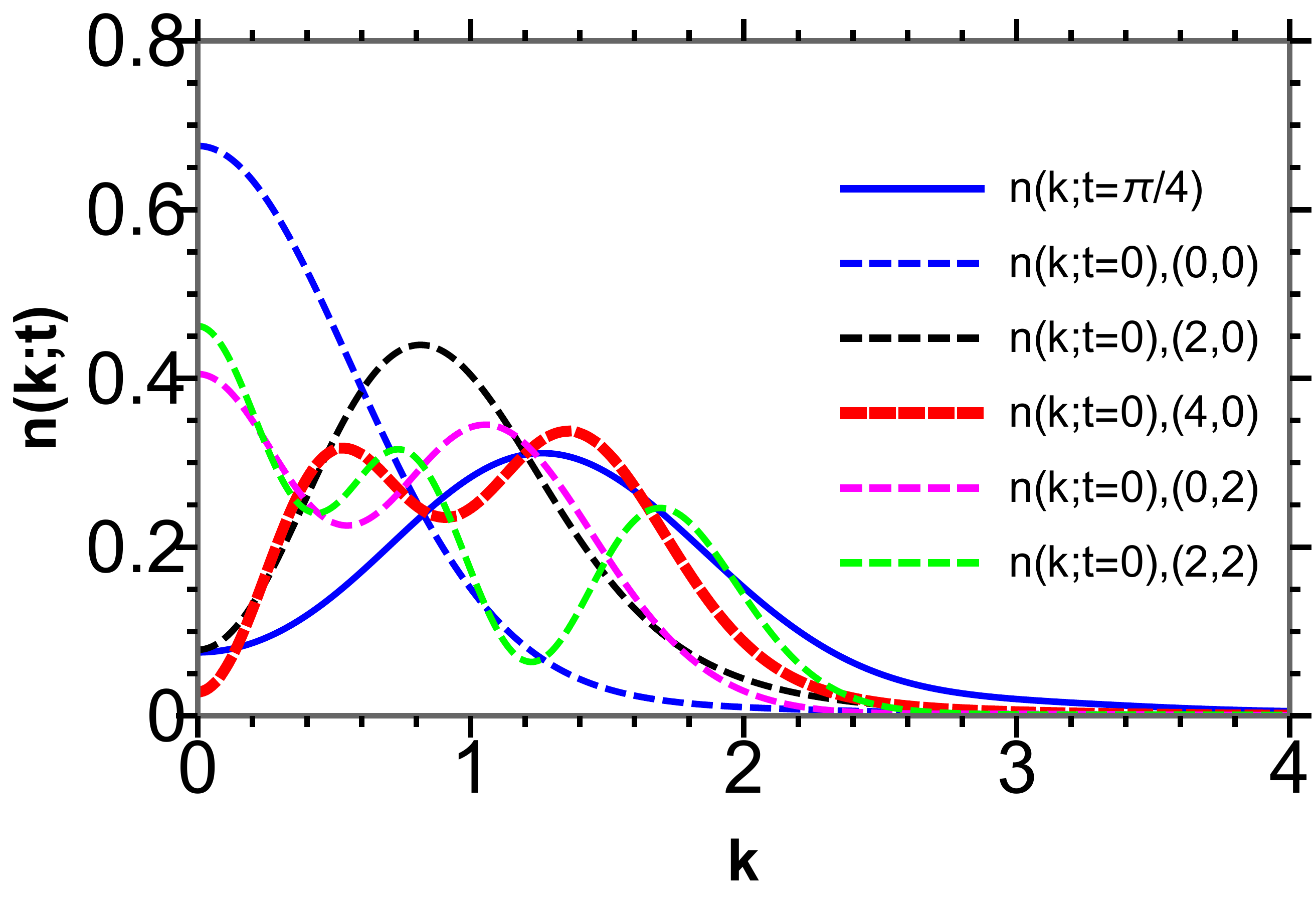} &
		\includegraphics[width=.25\textwidth]{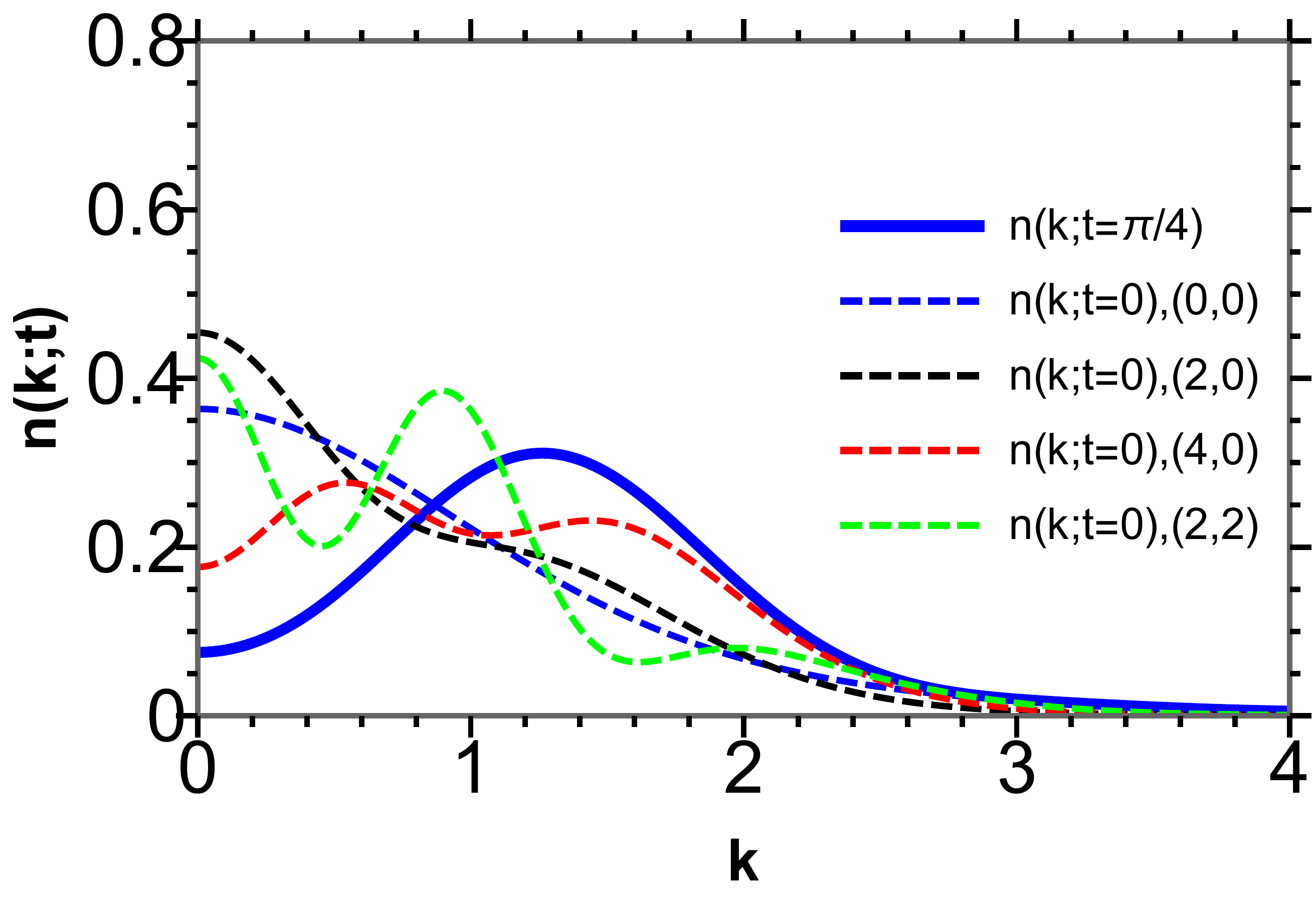} \\
	\end{tabular}
	\caption{(Color online) Comparison of the momentum distribution $n(k;t)$ (blue line) with the momentum distributions $n(k;t=0)$ for the pre- and postquench states (dashed lines).}\label{fwd_al_nk_34}
\end{figure}

Due to a larger number of states involved in the dynamics in the case of the anharmonic trap it is difficult to match shapes of the momentum distribution $n(k;t)$ with any particular shape of the momentum distribution $n(k,t=0)$ for the considered pre- and postquench states. However, at $t=\pi/8$ in Fig.~\ref{fwd_al_nk_12} the shape of $n(k;t)$ (blue line) is quite similar to the shapes of $n(k,t=0)$ for the postquench ground state (blue dashed line) and excited state $(2,0)$ (black dashed line). In Fig.~\ref{fwd_al} we see that the overlaps for these states are, indeed, quite large. At $t=\pi/4$ in Fig.~\ref{fwd_al_nk_34} there is a remote resemblance of the shape of $n(k;t)$ (blue line) with that of $n(k,t=0)$ for the prequench excited state $(4,0)$ (red dashed line). The overlap for this prequench excited state in Fig.~\ref{fwd_al} is not dominant, yet considerable.

\subsection{Harmonic trap, $\alpha=0$, the excited state, $(0,0)$}

\begin{figure}
\centering
\textbf{$\alpha=0$}\par\medskip
\includegraphics[width=8.5cm,clip]{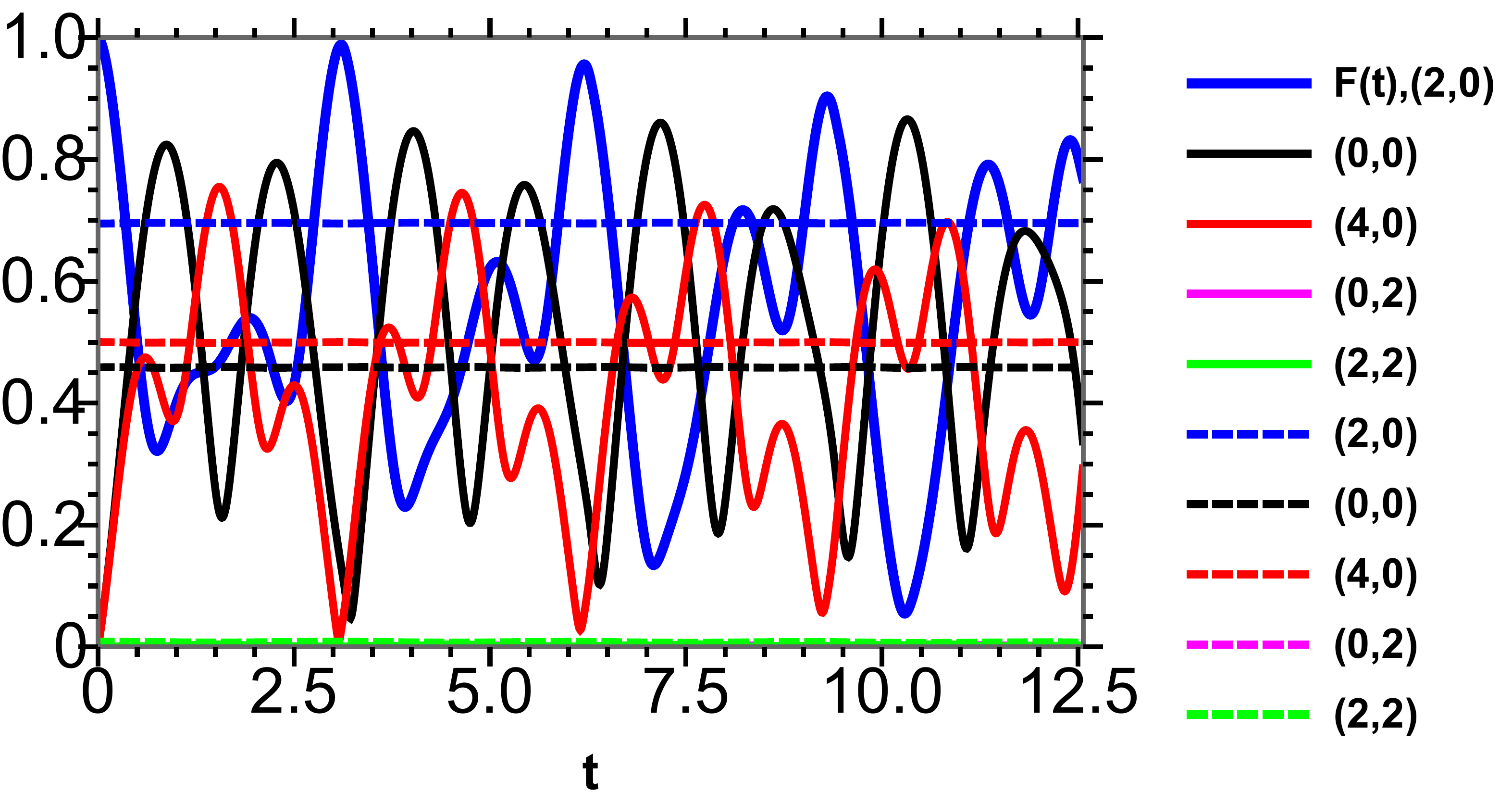}
\caption{(Color online) Fidelity $F(t)$ and the overlap integrals $\mathcal{Q}$ between the time-evolving state $\Psi(x_1,x_2,t)$ and different pre- (solid lines) and postquench (dashed lines) states in the case of the harmonic trap. The indices $(n,N)$ refer to the states with the quantum numbers of the relative and center-of-mass motions. The system of the two atoms is prepared in the excited state $(2,0)$ with $g=2$ and quenched to $g=-2$.}\label{fwd_exc}
\end{figure}

One other possible scenario to explore is to consider the quench dynamics from a higher state of the initial state, namely the excited state $(2,0)$. The time-evolution of the fidelity $F(t)$ and the overlaps $\mathcal{Q}$ for the pre- and postquench states in the case of the harmonic trap, $\alpha=0$, are plotted in Fig.~\ref{fwd_exc}.
The behavior of the fidelity $F(t)$ in Fig.~\ref{fwd_exc} is quite different from the case, shown in Fig.~\ref{fwd}. Here we can see that the amplitude of $F(t)$ is not as constant as in the case of the ground state dynamics and the peaks between the maximum peaks are now significantly distorted. 

It is interesting to note that the overlap for the prequench ground state shows the same pattern as the overlap for the prequench excited state $(2,0)$ in the case of Fig.~\ref{fwd}. This behavior can be explained by considering the wave packet expansion for the center-of-mass state $N=0$ \cite{mistakidis}:
\begin{equation}\label{expansion}
	\begin{aligned}
	&\Psi_{n_i,0}(x_1,x_2,t)\\&=\sum\limits_{0\le n_f \le \infty}e^{-it(E_f+1/2)}C_{n_f,n_i}\Psi_{n_f,0}(x_1,x_2,t=0),
	\end{aligned}
\end{equation}
where $C_{n_f,n_i}=\langle n_f,0|n_i,0\rangle$. In \eqref{expansion} $n_i$ and $n_f$ are the quantum numbers of the relative motion for the pre- and postquench states, respectively, $E_f+1/2$ is the energy level of the postquench state. The summation goes over the postquench states. From \eqref{expansion} we can see that
\begin{equation}\label{ovreq}
\langle\Phi_{2,0}|\Psi_{0,0}(t)\rangle=\langle\Phi_{0,0}|\Psi_{2,0}(t)\rangle,
\end{equation}
where $\Phi_{2,0}$ and $\Phi_{0,0}$ are the prequench excited state $(2,0)$ and the ground state, respectively. Eq. \eqref{ovreq} is the reason why we have the same dynamics for the both overlaps. As we will see below, this equation holds for the anharmonic case and the reverse quench scenario as well.

\begin{figure}
	\centering
	\textbf{$\alpha=0$}\par\medskip
	\includegraphics[width=7cm,clip]{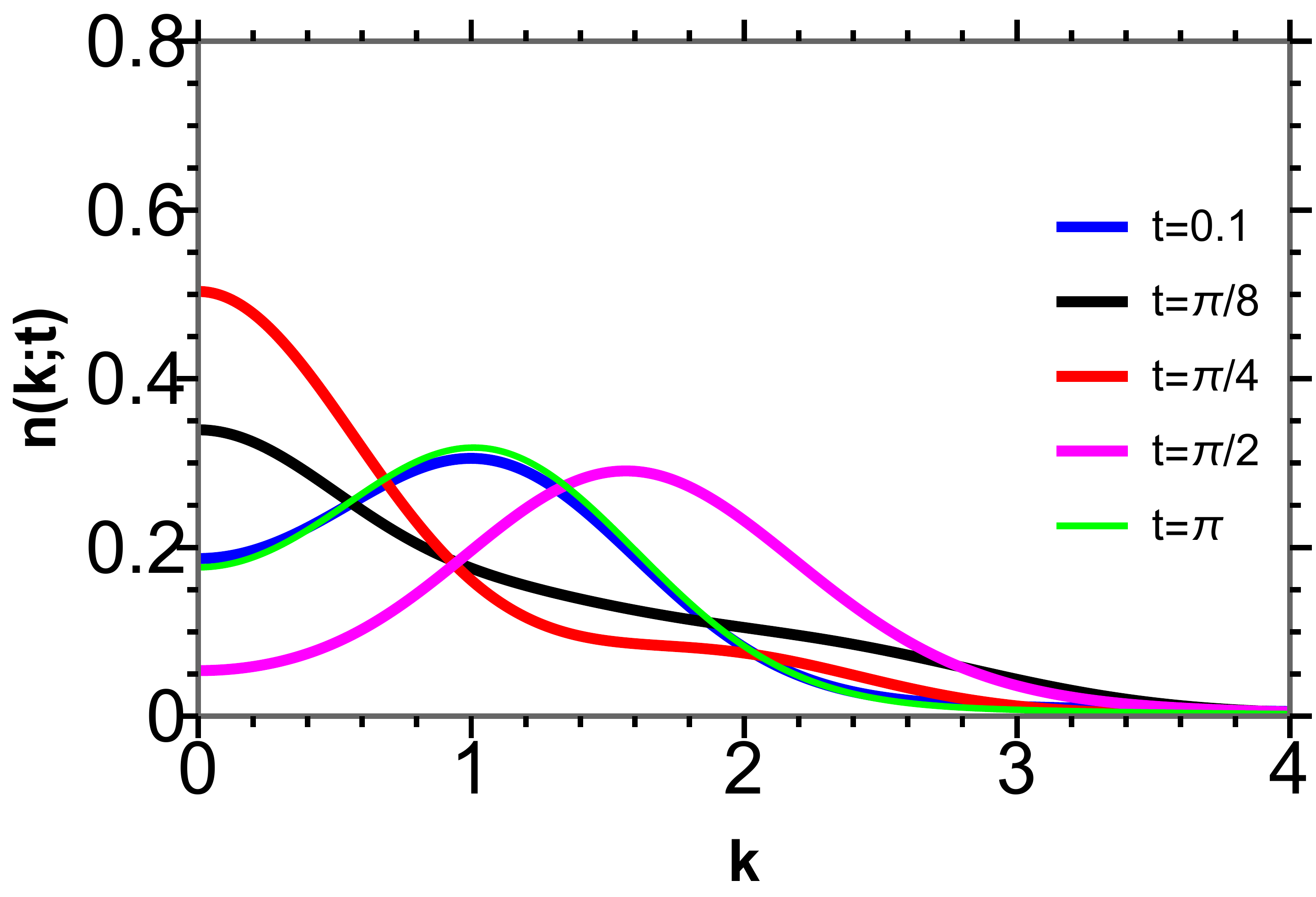}
	\caption{(Color online) Evolution of the momentum distribution $n(k;t)$ for the harmonic trap.}\label{fwd_exc_nk_all}
\end{figure}

The evolution of the momentum distribution $n(k;t)$ in the case of the excited initial state is shown in Fig.~\ref{fwd_exc_nk_all}. At $t=0.1$ the maximum value of the shape of $n(k;t)$ (blue line) is located farther from the origin.  which is due to the initial state being an excited one. Later on, at $t=\pi/4$, the shape (red line) has a quite pronounced zero-momentum peak. This indicates a transition of the system to the ground state. With increasing time the system again returns to the excited state and at $t=\pi$ it almost coincides with the initial excited state (green line).

\begin{figure}
	\centering
	\textbf{$\alpha=0$,~~~$t=\pi/8$}\par\medskip
	\begin{tabular}{cc}
		Prequench states &  Postquench states \\
		\hspace{-.5cm}\includegraphics[width=.25\textwidth]{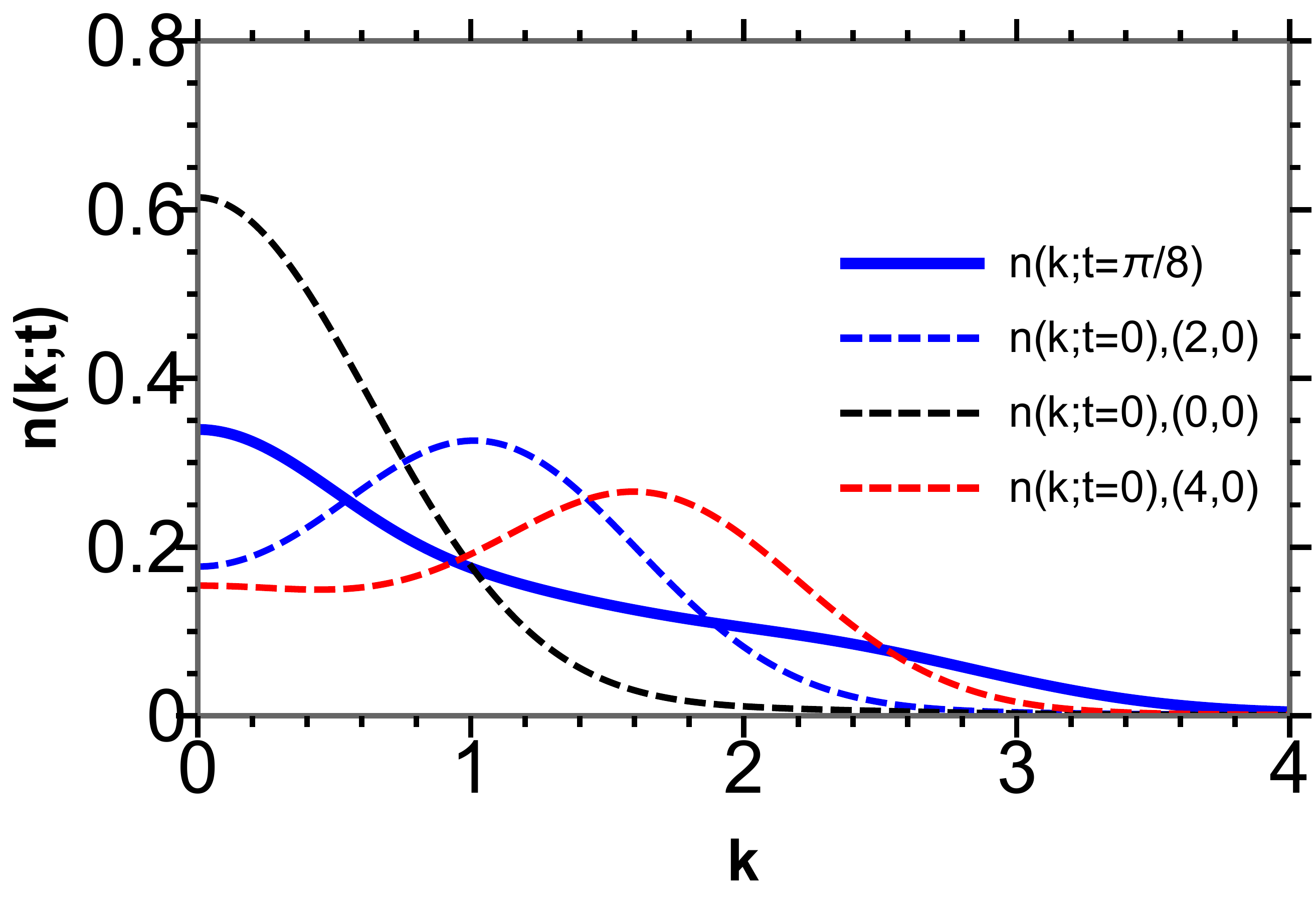} &
		\includegraphics[width=.25\textwidth]{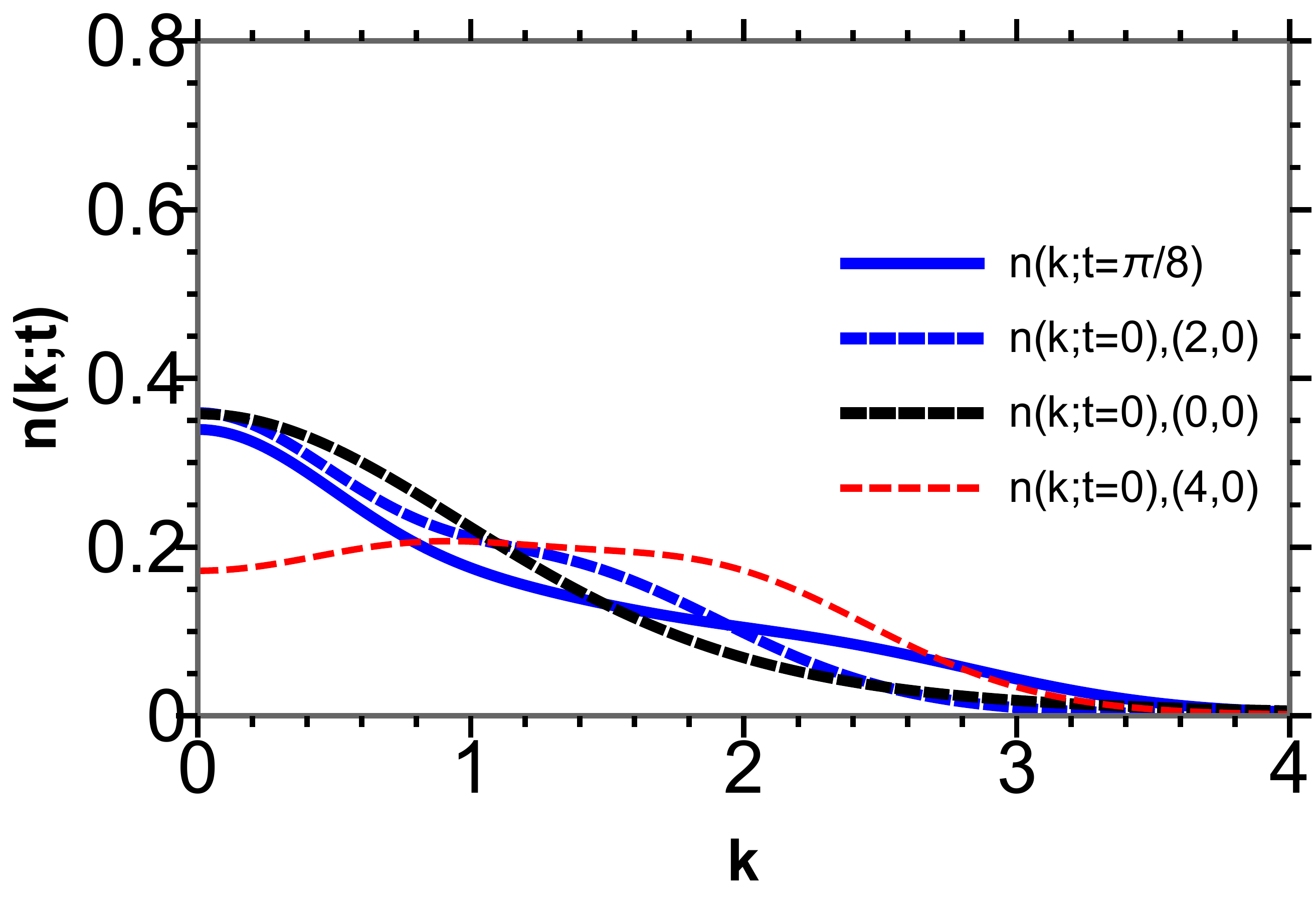} \\
	\end{tabular}
	\caption{(Color online) Comparison of the momentum distribution $n(k;t)$ (blue line) with the momentum distributions $n(k;t=0)$ for the pre- and postquench states (dashed lines).}\label{fwd_exc_nk_12}
\end{figure}

\begin{figure}
	\centering
	\textbf{$\alpha=0$,~~~$t=\pi/4$}\par\medskip
	\begin{tabular}{cc}
		Prequench states &  Postquench states \\
		\hspace{-.5cm}\includegraphics[width=.25\textwidth]{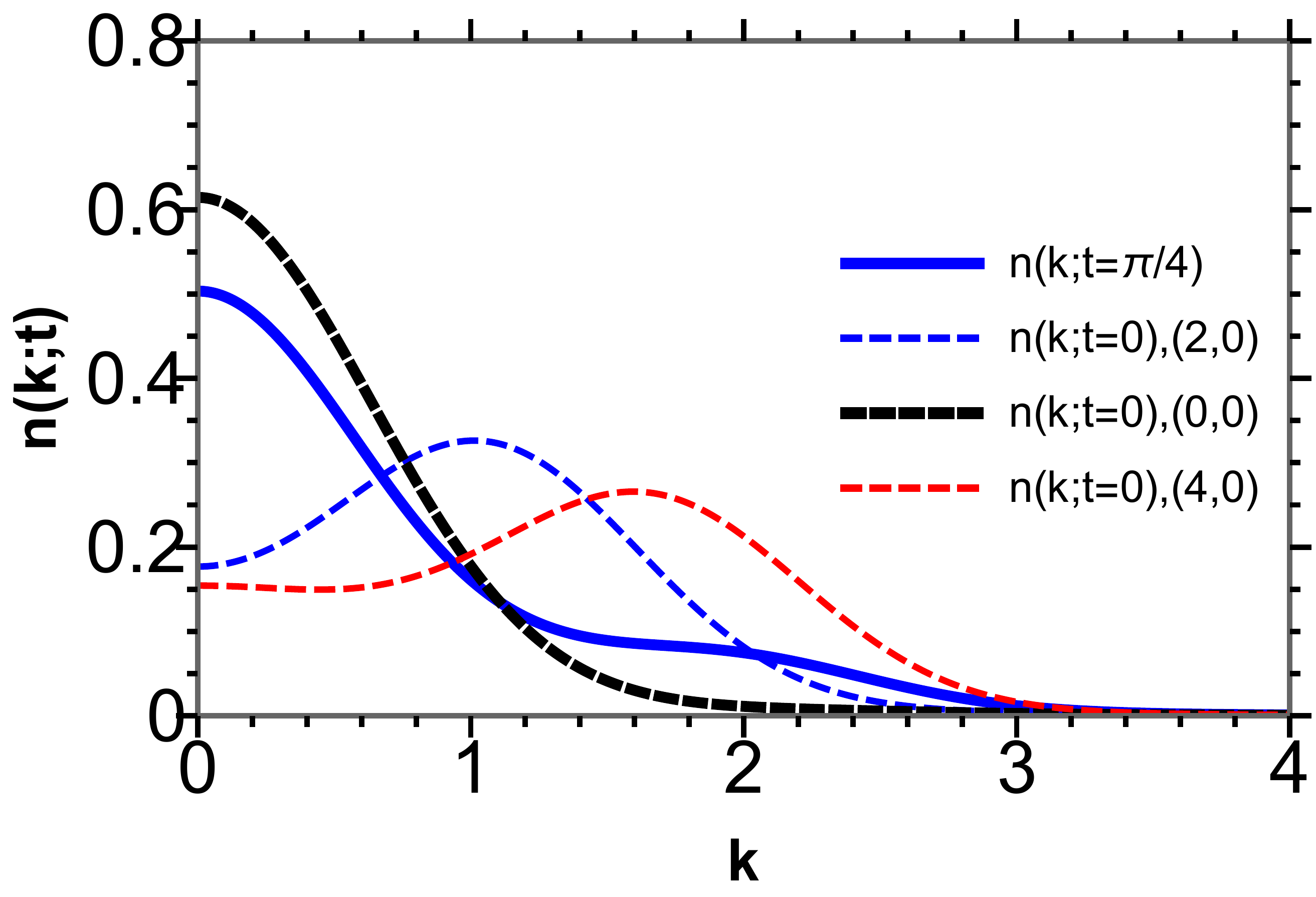} &
		\includegraphics[width=.25\textwidth]{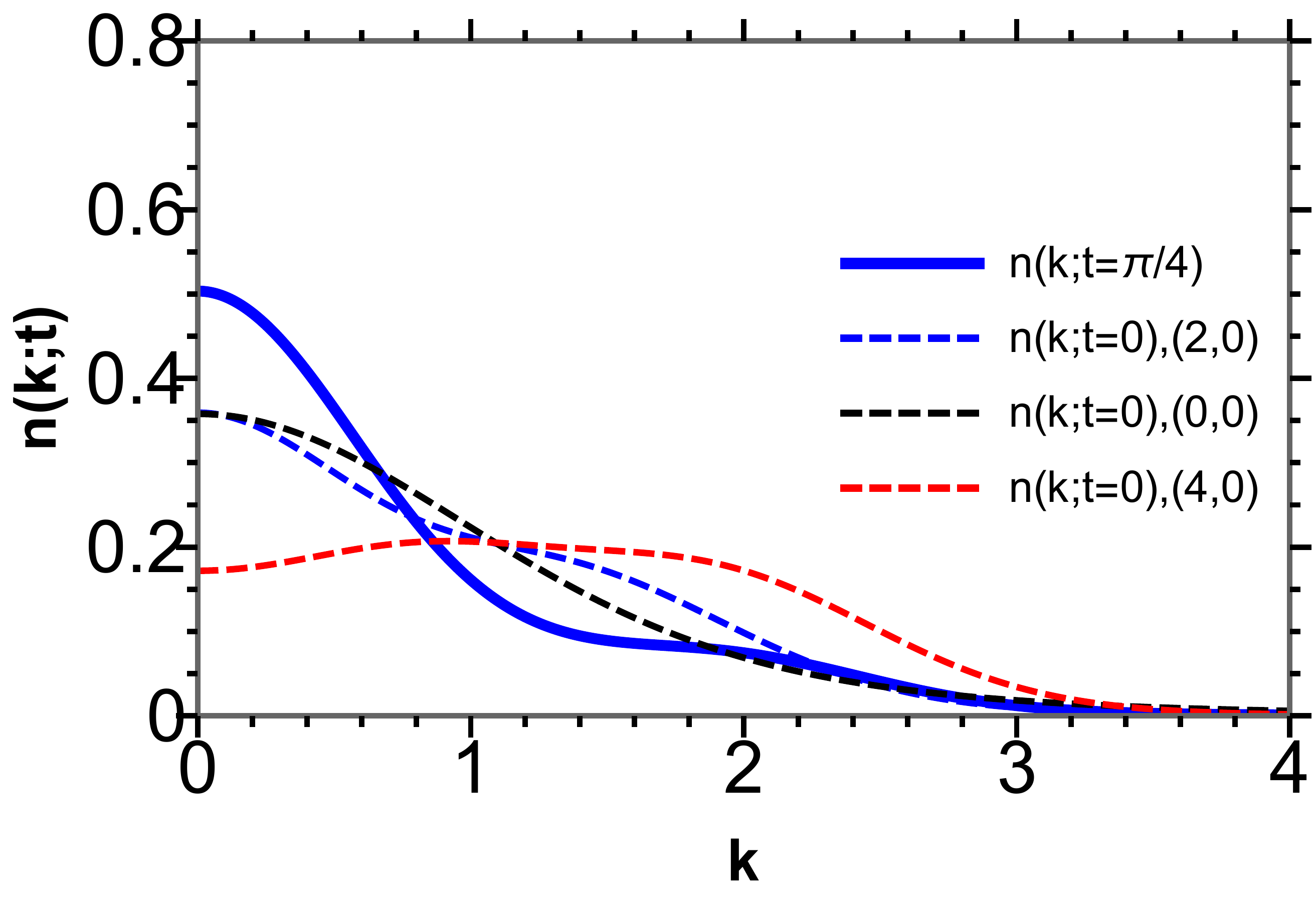} \\
	\end{tabular}
	\caption{(Color online) Comparison of the momentum distribution $n(k;t)$ (blue line) with the momentum distributions $n(k;t=0)$ for the pre- and postquench states (dashed lines).}\label{fwd_exc_nk_34}
\end{figure}

\begin{figure}
	\centering
	\textbf{$\alpha=0$,~~~$t=\pi/2$}\par\medskip
	\begin{tabular}{cc}
		Prequench states &  Postquench states \\
		\hspace{-.5cm}\includegraphics[width=.25\textwidth]{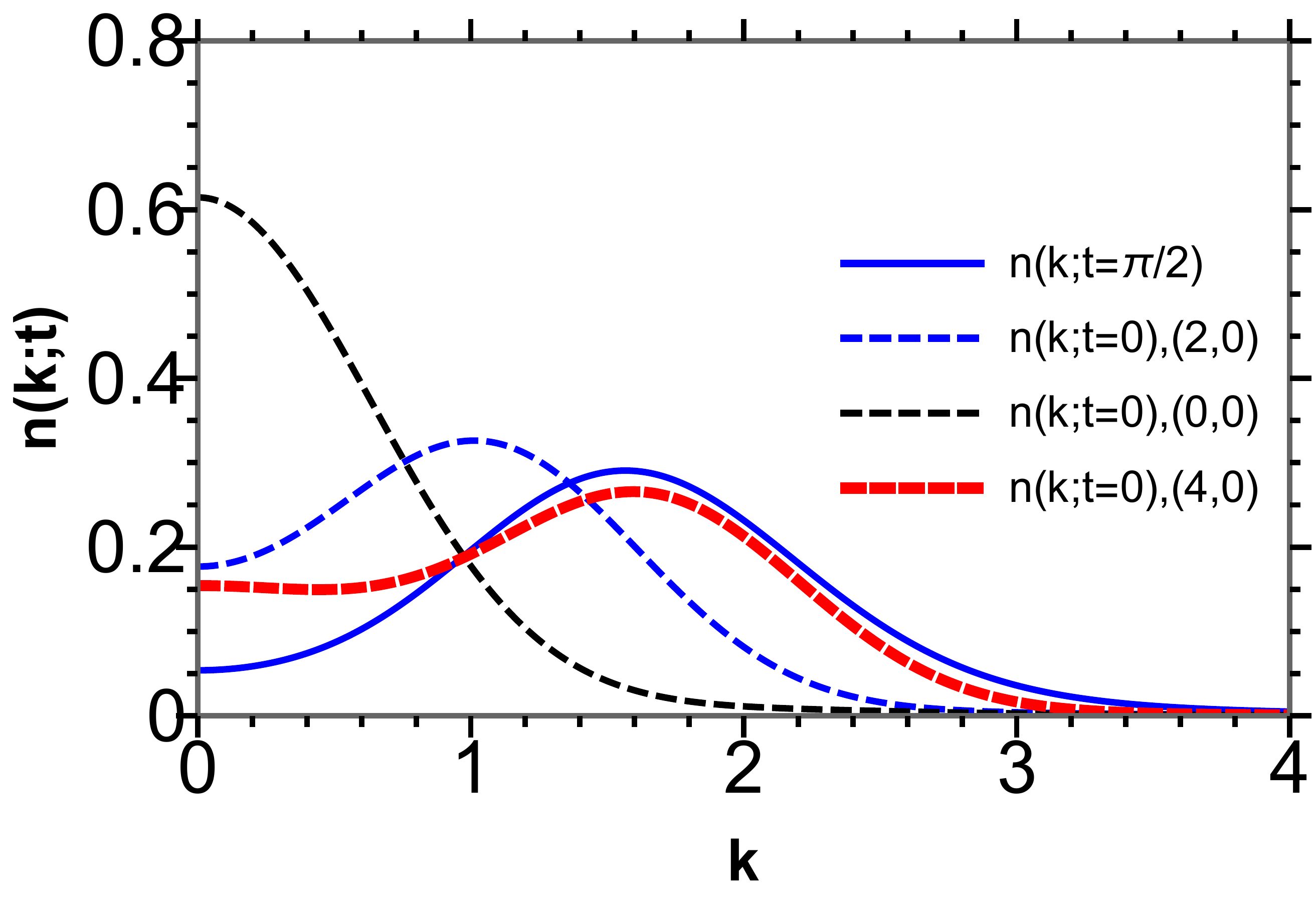} &
		\includegraphics[width=.25\textwidth]{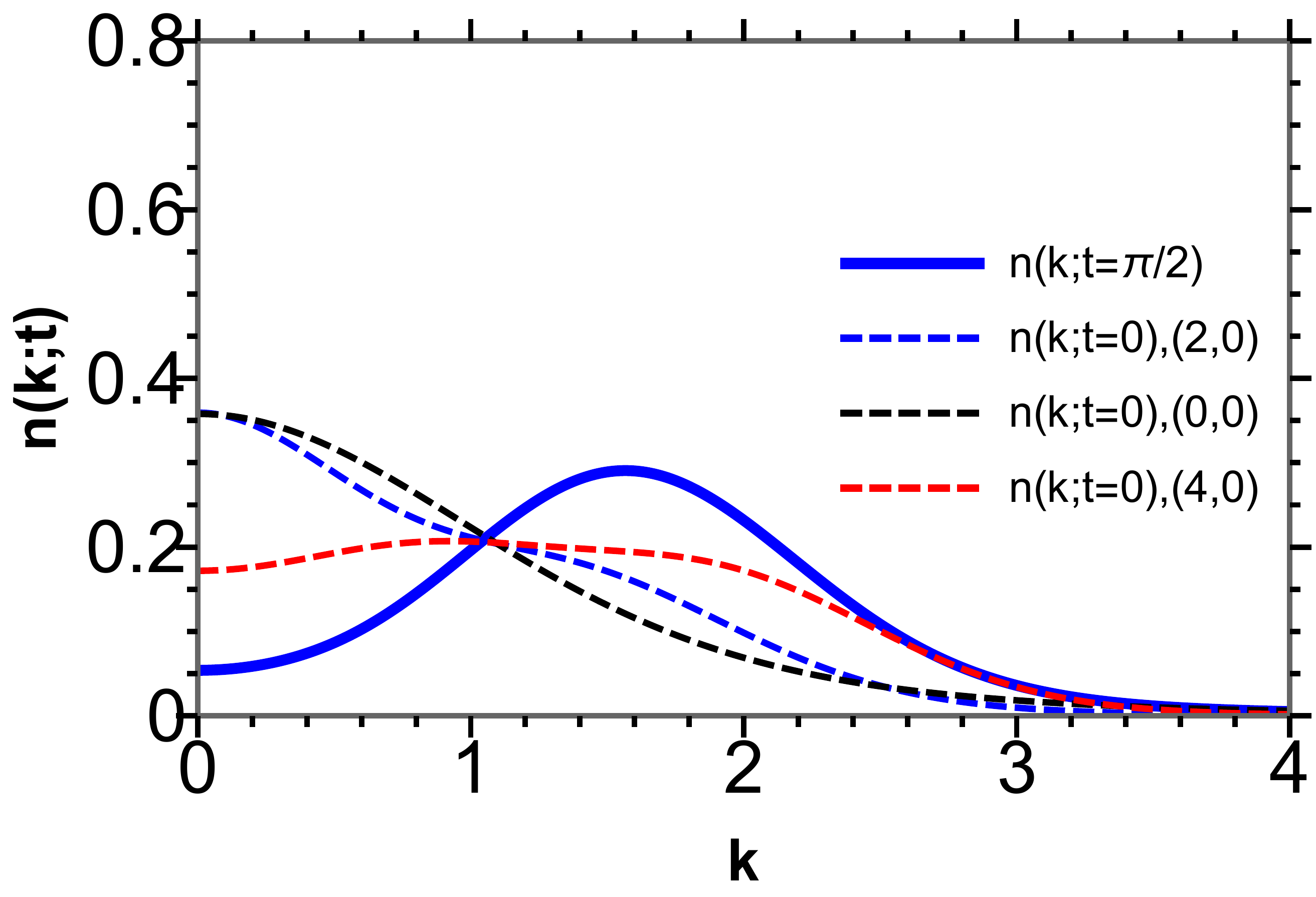} \\
	\end{tabular}
	\caption{(Color online) Comparison of the momentum distribution $n(k;t)$ (blue line) with the momentum distributions $n(k;t=0)$ for the pre- and postquench states (dashed lines).}\label{fwd_exc_nk_56}
\end{figure}

At $t=\pi/8$ in Fig.~\ref{fwd_exc_nk_12} the shape of $n(k,t)$ (blue line) is quite similar to those for the postquench ground state (black dashed line) and excited state $(2,0)$ (blue dashed line). This is because the overlaps for these states are large, as can be seen in Fig.~\ref{fwd_exc}. However, even though the overlap for the postquench excited state $(4,0)$ is large as well, the shape of its momentum distribution is less closer to $n(k,t)$. At $t=\pi/4$ in Fig.~\ref{fwd_exc_nk_34} there is an approximate match with the prequench ground state (black dashed line), since at this moment the overlap for this state takes one of its maximum value. Another match, this time with the prequench excited state $(4,0)$ (red dashed line), occurs at $t=\pi/2$ in Fig.~\ref{fwd_exc_nk_56}. This is also consistent with the overlap dynamics, where, at this moment of time, the overlap for this excited state is at its peak.

\subsection{Anharmonic trap, $\alpha=-0.03$, the excited state, $(2,0)$}

\begin{figure}
\centering
\textbf{$\alpha=-0.03$}\par\medskip
\includegraphics[width=8.5cm,clip]{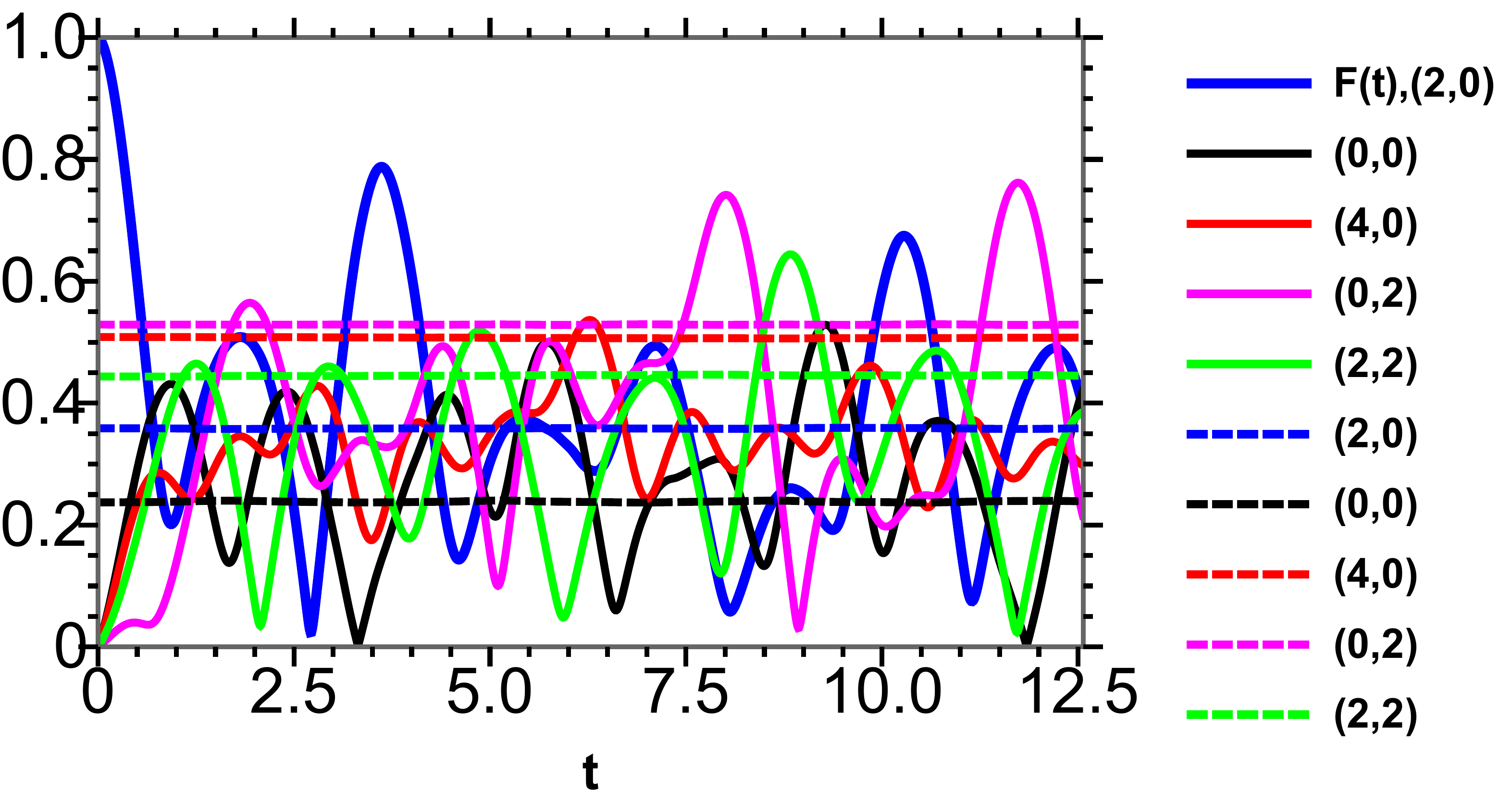}
\caption{(Color online) Fidelity $F(t)$ and the overlap integrals $\mathcal{Q}$ between the time-evolving state $\Psi(x_1,x_2,t)$ and different pre- (solid lines) and postquench (dashed lines) states in the case of the anharmonic trap. The indices $(n,N)$ refer to the states with the quantum numbers of the relative and center-of-mass motions. The system of the two atoms is prepared in the excited state $(2,0)$ with $g=2$ and quenched to $g=-2$.}\label{fwd_exc_al}
\end{figure}

In the anharmonic case, $\alpha=-0.03$, when the initial state is prepared in the excited state $(2,0)$, every considered state overlaps with the wave packet and for the end result we get a quite complicated dynamics for the all overlaps (Fig.~\ref{fwd_exc_al}). The center-of-mass excited states are now included into the dynamics. The oscillation of the fidelity $F(t)$ loses its uniformity and the fidelity no longer reaches a unity. The overlap between the wave packet and the prequench ground state (black line) has the same time dependence as the overlap in the case of the ground state dynamics in Fig.~\ref{fwd_al}. This is in agreement with the equation \eqref{ovreq}.

\begin{figure}
	\centering
	\textbf{$\alpha=-0.03$}\par\medskip	
	\includegraphics[width=7cm,clip]{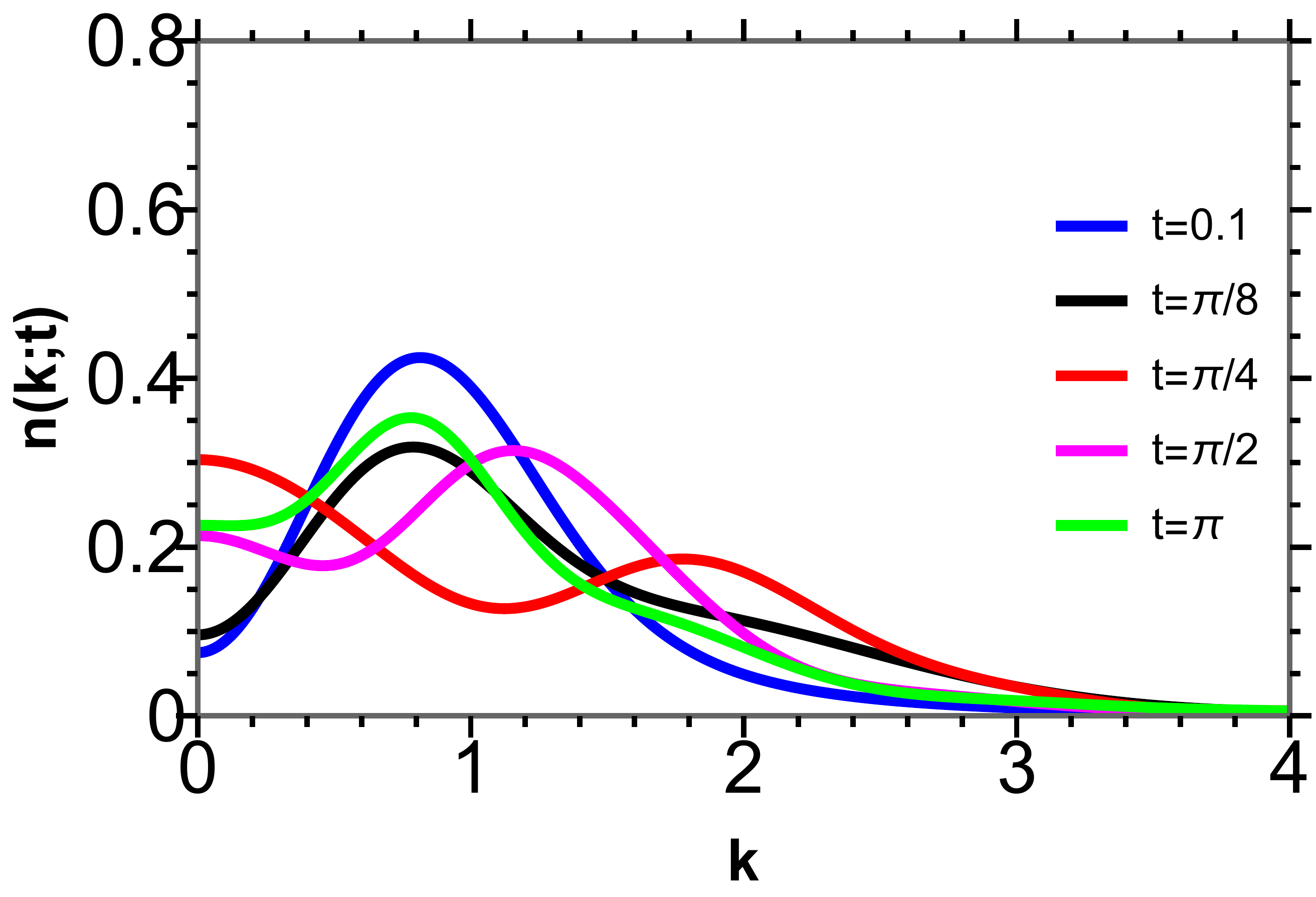}
	\caption{(Color online) Evolution of the momentum distribution $n(k;t)$ for the anharmonic trap.}\label{fwd_exc_al_nk_all}
\end{figure}

As in the case of the harmonic trap the shape of the momentum distribution in Fig.~\ref{fwd_exc_al_nk_all} at $t=0.1$ (blue line) has a maximum for momentum away from zero, since the initial state is an excited one. With increasing time up to $t=\pi/4$ (red line) the zero-momentum peak increases and the overall shape broadens. After that the zero-momentum peak decreases again, moving towards a higher momentum value, so that the whole dynamics of the shape deformation is approximately repeated over (magenta and green lines).

\begin{figure}
	\centering
	\textbf{$\alpha=-0.03$,~~~$t=\pi/4$}\par\medskip
	\begin{tabular}{cc}
		Prequench states &  Postquench states \\
		\hspace{-.5cm}\includegraphics[width=.25\textwidth]{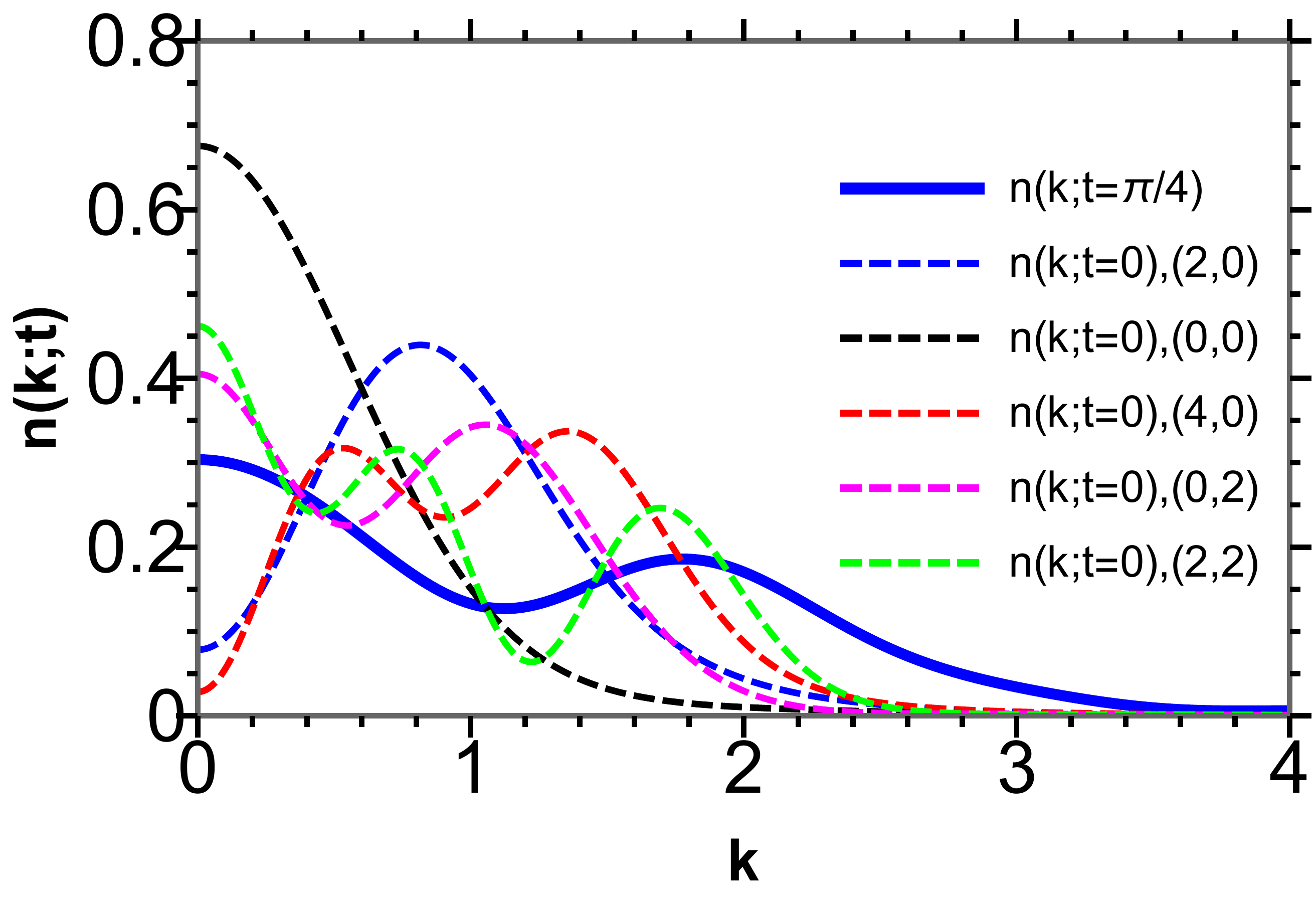} &
		\includegraphics[width=.25\textwidth]{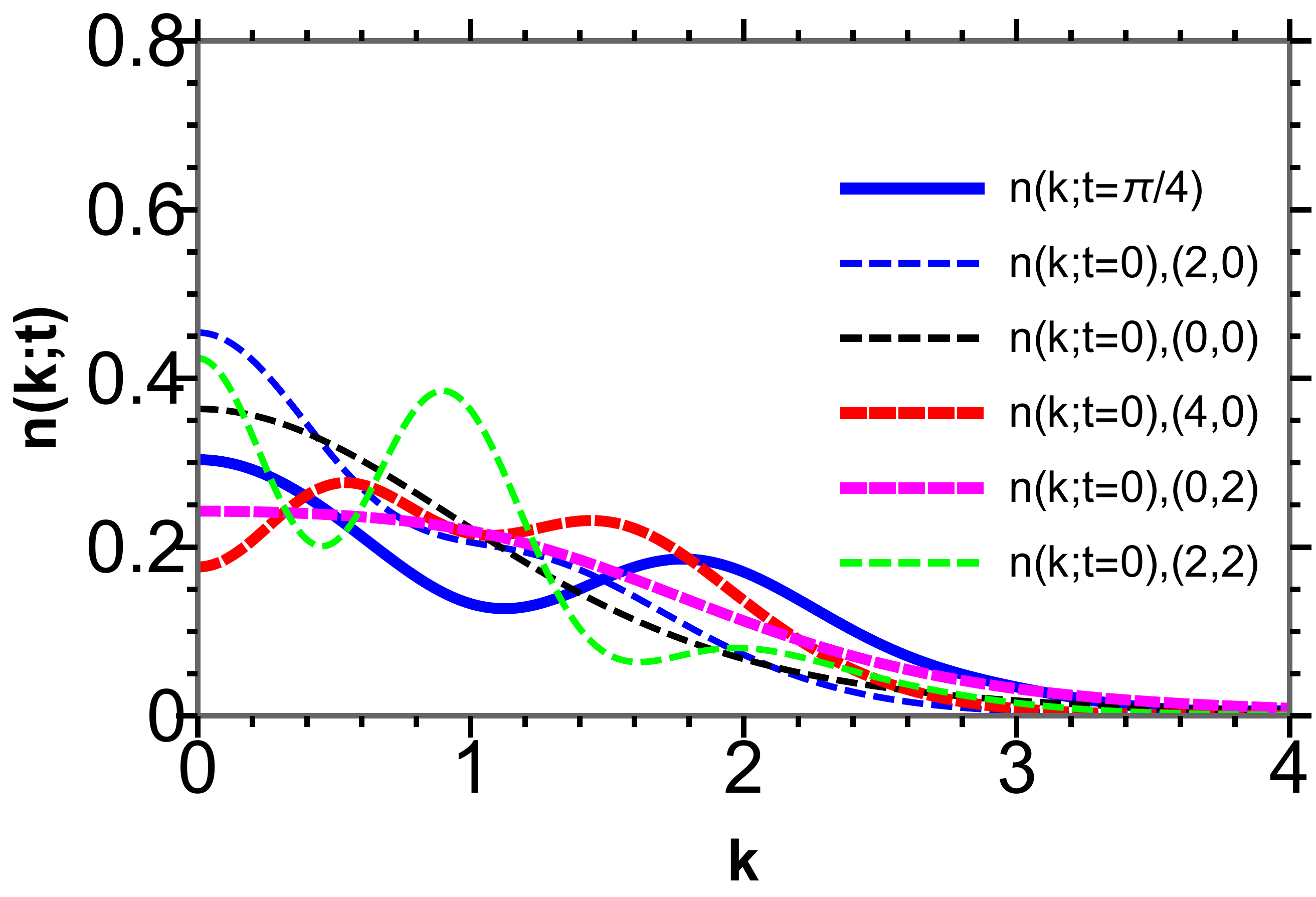} \\
	\end{tabular}
	\caption{(Color online) Comparison of the momentum distribution $n(k;t)$ (blue line) with the momentum distributions $n(k;t=0)$ for the pre- and postquench states (dashed lines).}\label{fwd_exc_al_nk_12}
\end{figure}

\begin{figure}
	\centering
	\textbf{$\alpha=-0.03$,~~~$t=\pi/2$}\par\medskip
	\begin{tabular}{cc}
		Prequench states &  Postquench states \\
		\hspace{-.5cm}\includegraphics[width=.25\textwidth]{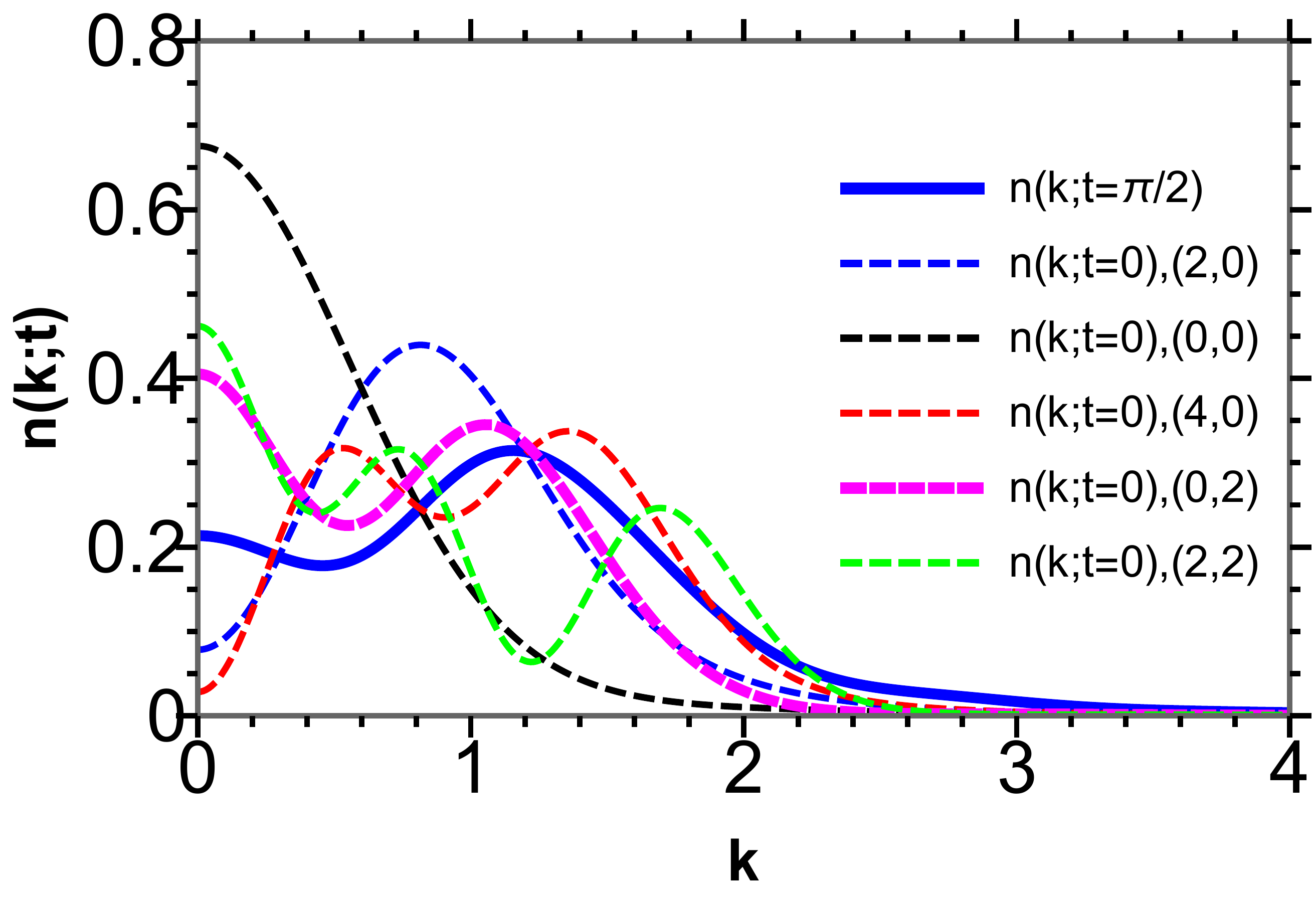} &
		\includegraphics[width=.25\textwidth]{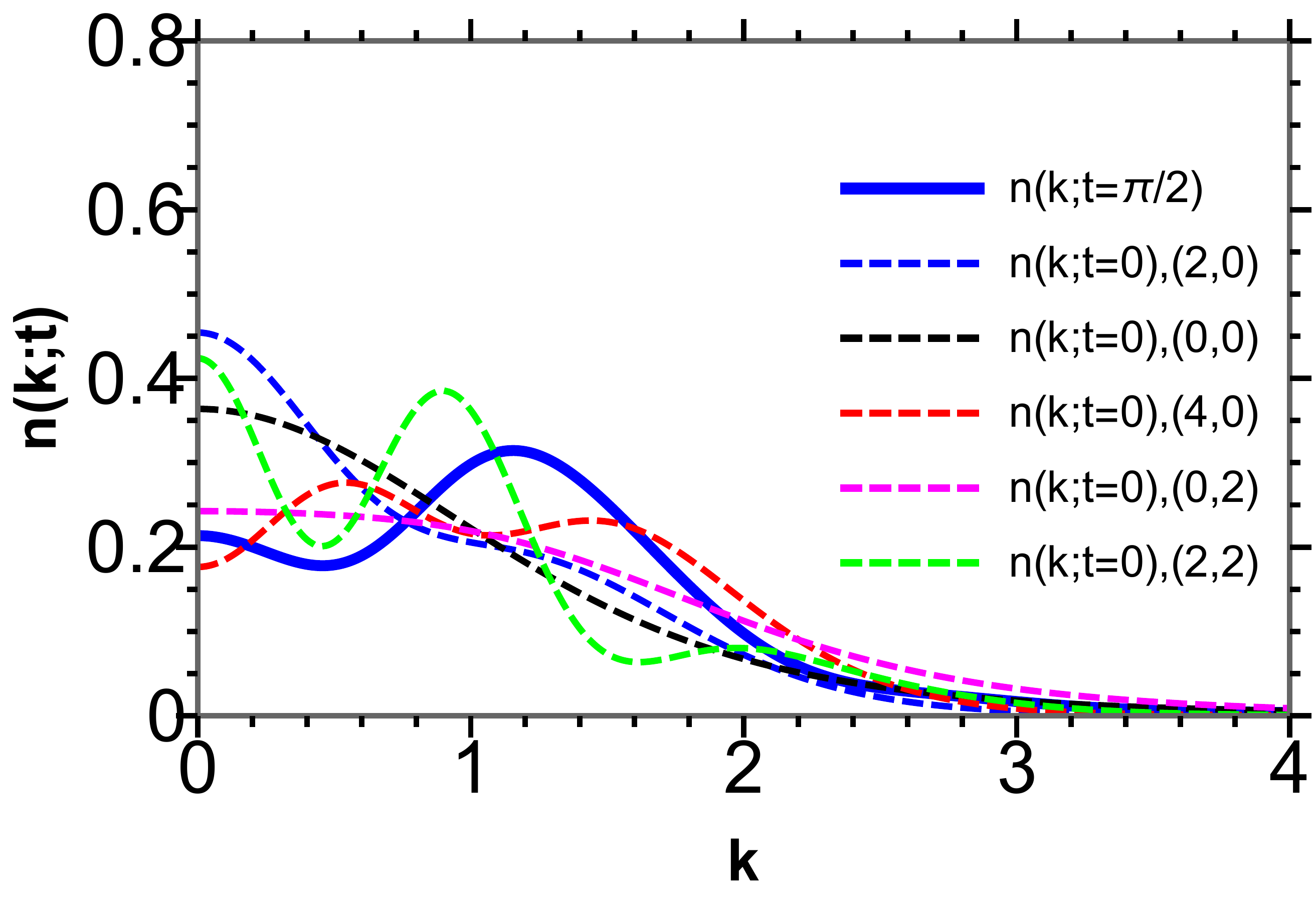} \\
	\end{tabular}
	\caption{(Color online) Comparison of the momentum distribution $n(k;t)$ (blue line) with the momentum distributions $n(k;t=0)$ for the pre- and postquench states (dashed lines).}\label{fwd_exc_al_nk_34}
\end{figure}

In Fig.~\ref{fwd_exc_al_nk_12} and Fig.~\ref{fwd_exc_al_nk_34} we compare snapshots of the momentum distribution $n(k;t)$ at several moments of time with the momentum distributions $n(k;t=0)$ for the different pre- and postquench states, respectively, in the case of the anharmonic trap, $\alpha=-0.03$. In this case, there is hardly any complete match of shapes of $n(k;t)$ with any shapes of $n(k;t=0)$. At $t=\pi/4$ in Fig.~\ref{fwd_exc_al_nk_12} we can find some slight resemblance of the shape of $n(k;t)$ (blue line) with the shapes of $n(k;t=0)$ for the postquench excited states $(4,0)$ (red dashed line) and $(0,2)$ (magenta dashed line). This is in agreement with the overlap dynamics in Fig.~\ref{fwd_exc_al}, where approximately at this moment of time the fidelity reaches one of its minimums and the overlaps for these postquench excited states prevail over it. The shape of $n(k;t)$ at $t=\pi/2$ (blue line) in Fig.~\ref{fwd_exc_al_nk_34} has some similarities with the shape of $n(k;t=0)$ for the prequench excited state $(0,2)$ (magenta dashed line), which is also quite consistent with Fig.~\ref{fwd_exc_al}, since at this moment of time the overlap for this excited state almost reaches one of its maximum values. 

\section{Quench dynamics from attractive to repulsive interactions}\label{quench2}

In this section we consider the reverse scenario - the change of the attractive interaction $g=-2$ to the repulsive one $g=2$. This case is different due to the presence of a repulsive barrier which alters the dynamics significantly. The barrier serves as an obstacle for the particles to travel easily from one side of the trap to another. As will be seen below, the oscillation frequency is approximately decreased twice in comparison with the case of the previous scenario. As before, we analyze the quench dynamics for the harmonic and anharmonic traps starting from the ground and excited states. 

\subsection{Harmonic trap, $\alpha=0$, the ground state, $(0,0)$}

\begin{figure}
	\centering
	\textbf{$\alpha=0$}\par\medskip
	\includegraphics[width=8.5cm,clip]{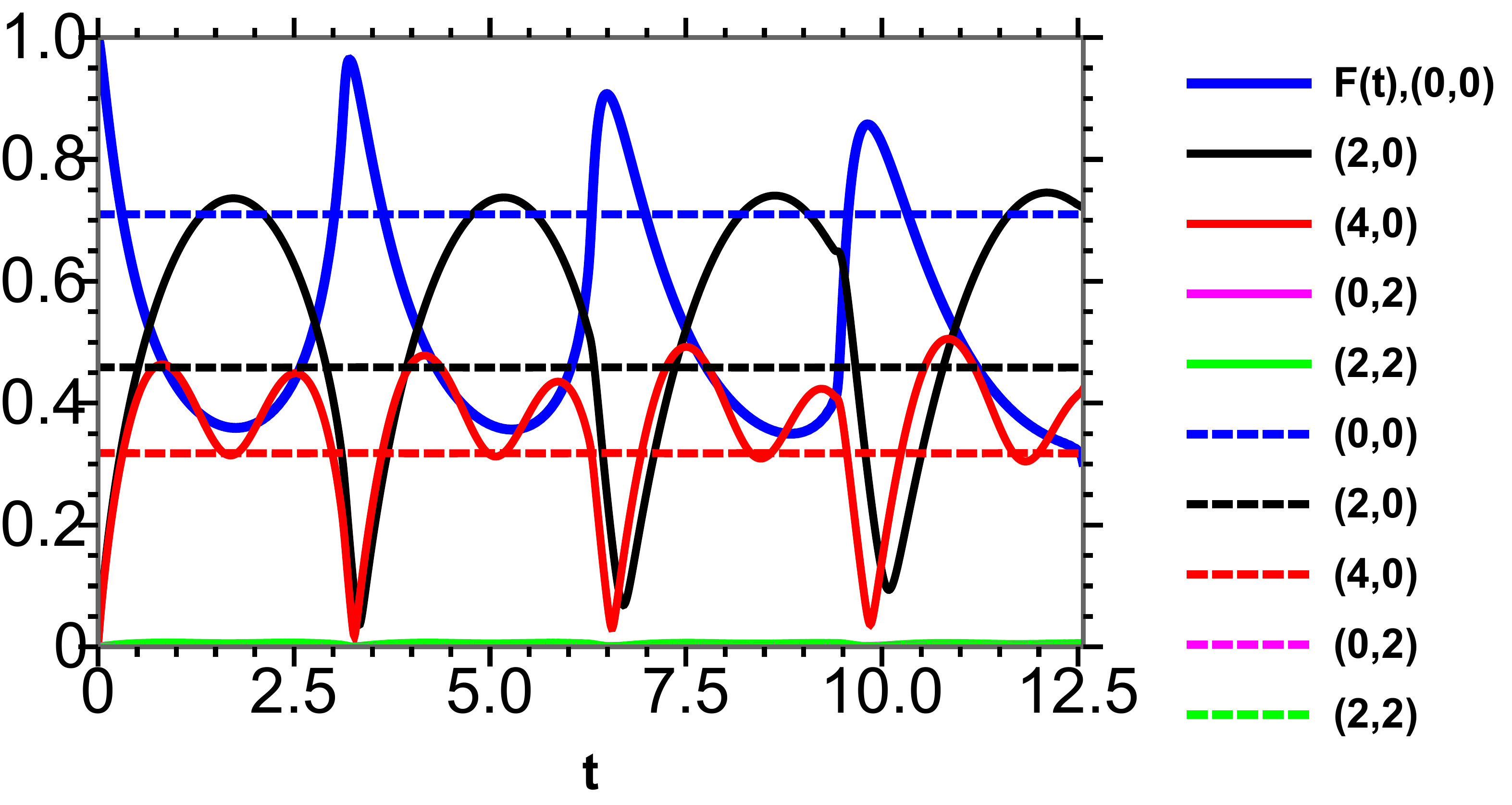}
	\caption{(Color online) Fidelity $F(t)$ and the overlap integrals $\mathcal{Q}$ between the time-evolving state $\Psi(x_1,x_2,t)$ and different pre- (solid lines) and postquench (dashed lines) states in the case of the harmonic trap. The indices $(n,N)$ refer to the states with the quantum numbers of the relative and center-of-mass motions. The system of the two atoms is prepared in the ground state with $g=-2$ and quenched to $g=2$.}\label{bwd}
\end{figure}

First, let us consider the overlaps between the wave packet and the pre- and postquench stationary states. It can be noticed in Fig.~\ref{bwd} that the oscillation amplitude of the fidelity $F(t)$ is damping over time. This is because the original positions of the particles are tightly localized around the center and the repulsive barrier prevents the particles to return to that position. As previously, the fidelity $F(t)$ and the overlaps between prequench states oscillate approximately in antiphase. Due to the separability of the coordinates in the case of the harmonic trap, the only contributing states are the relative motion states.

\begin{figure}
	\centering
	\textbf{$\alpha=0$}\par\medskip
	\begin{tabular}{ccc}
		$t=0.001$ & $t=0.1$ & $t=\pi/8$ \\
		\includegraphics[width=.15\textwidth]{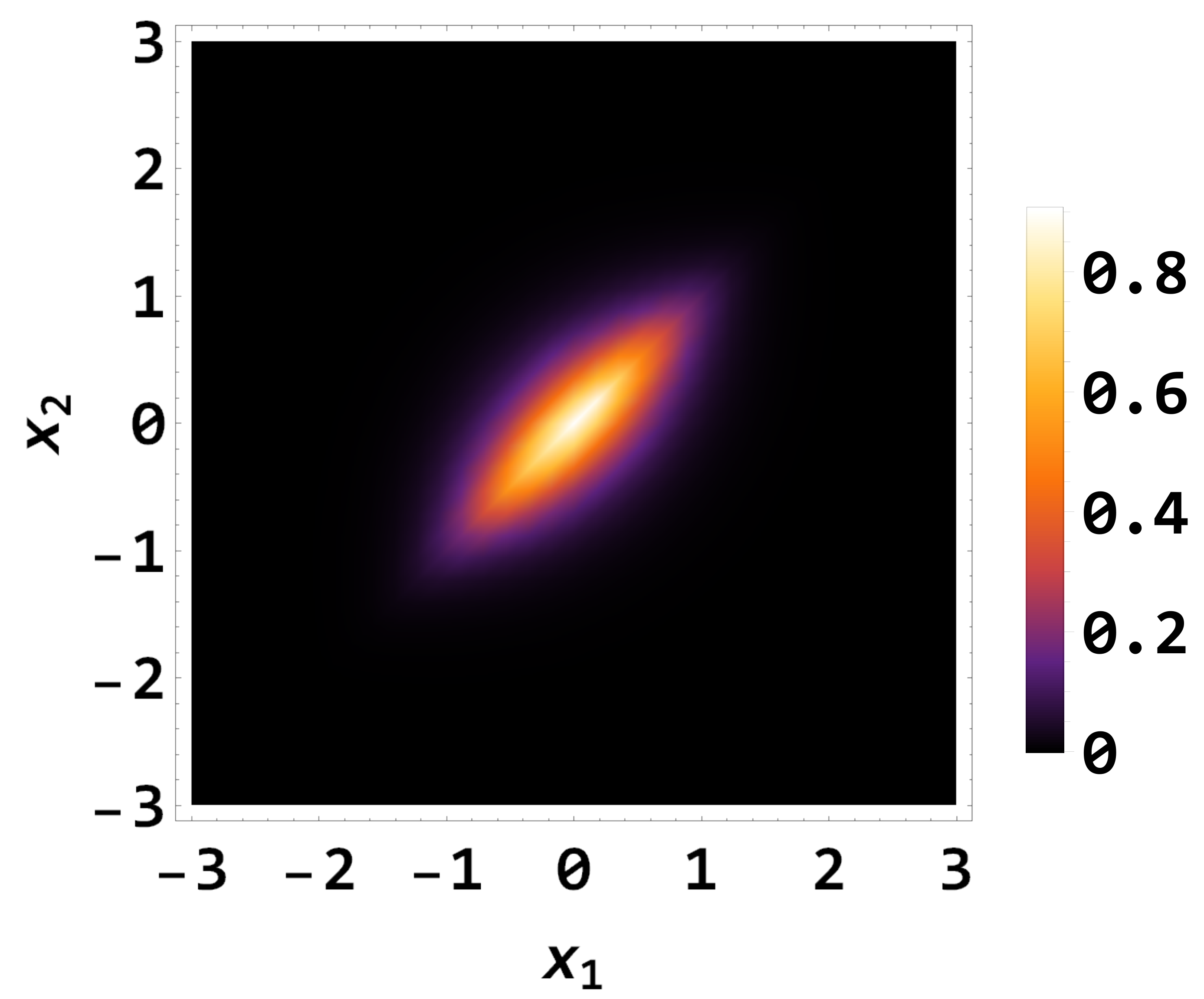} &
		\includegraphics[width=.15\textwidth]{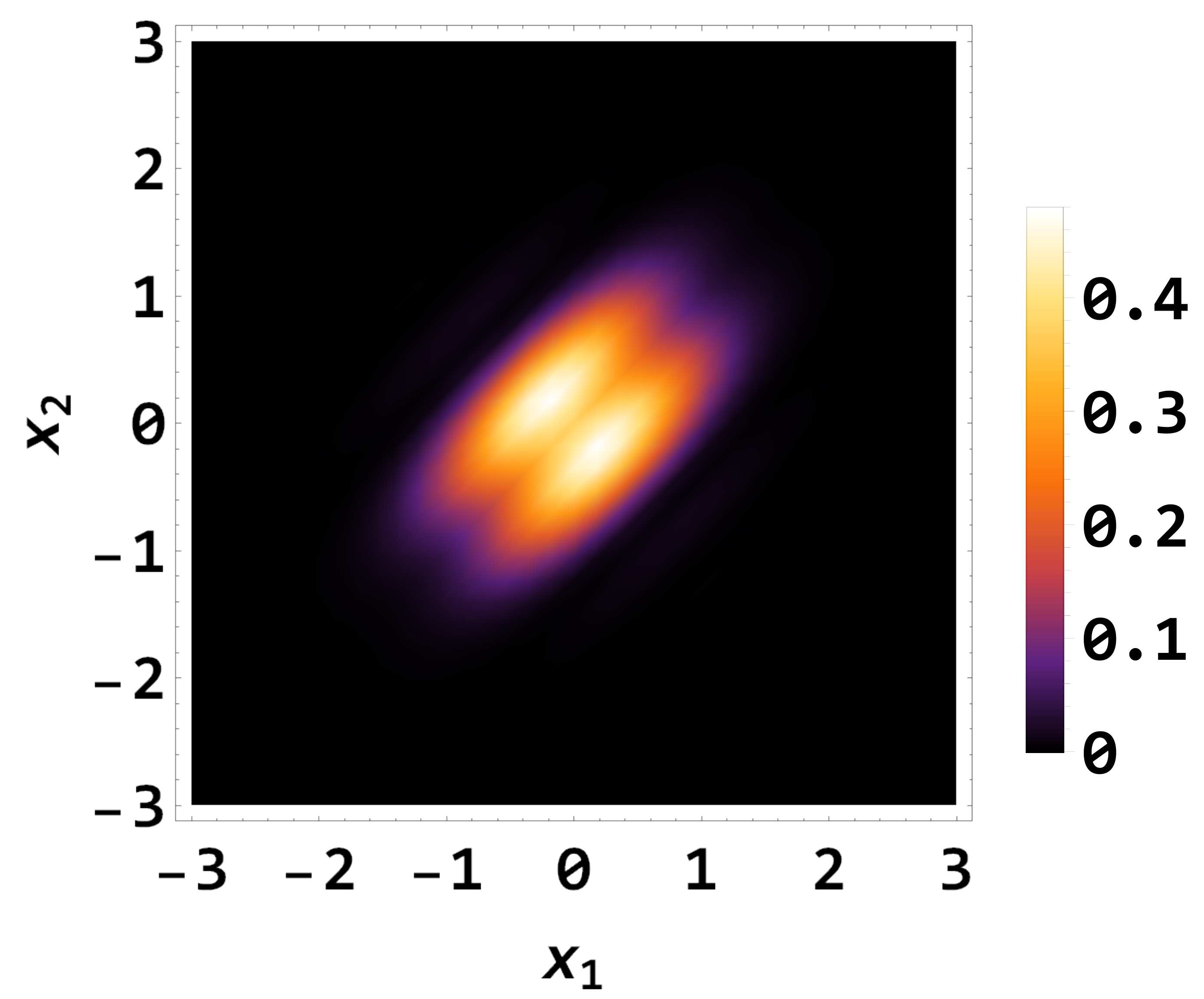} &
		\includegraphics[width=.15\textwidth]{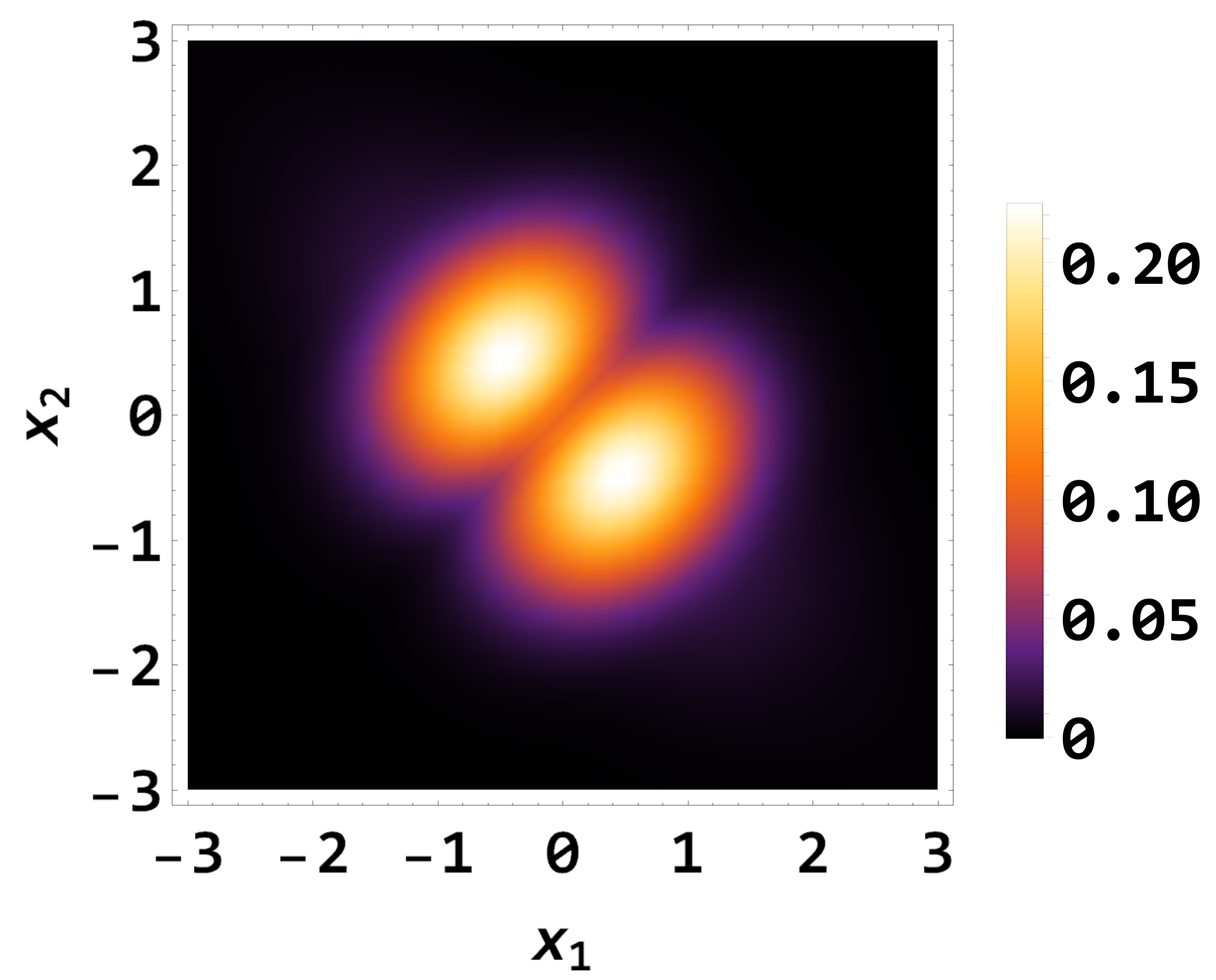} \\
		$t=\pi/4$ & $t=\pi/2$ & $t=\pi$\\
		\includegraphics[width=.15\textwidth]{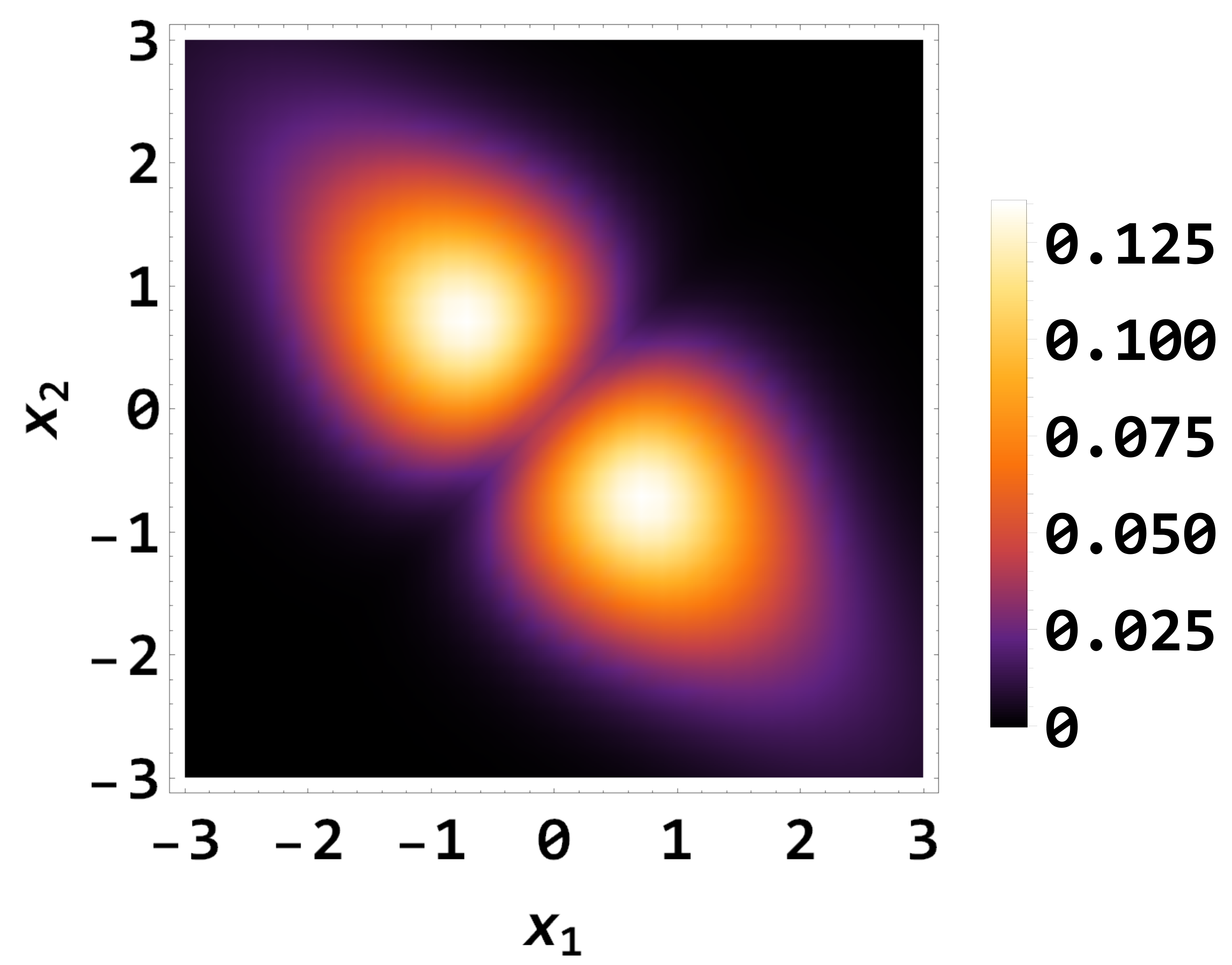} &
		\includegraphics[width=.15\textwidth]{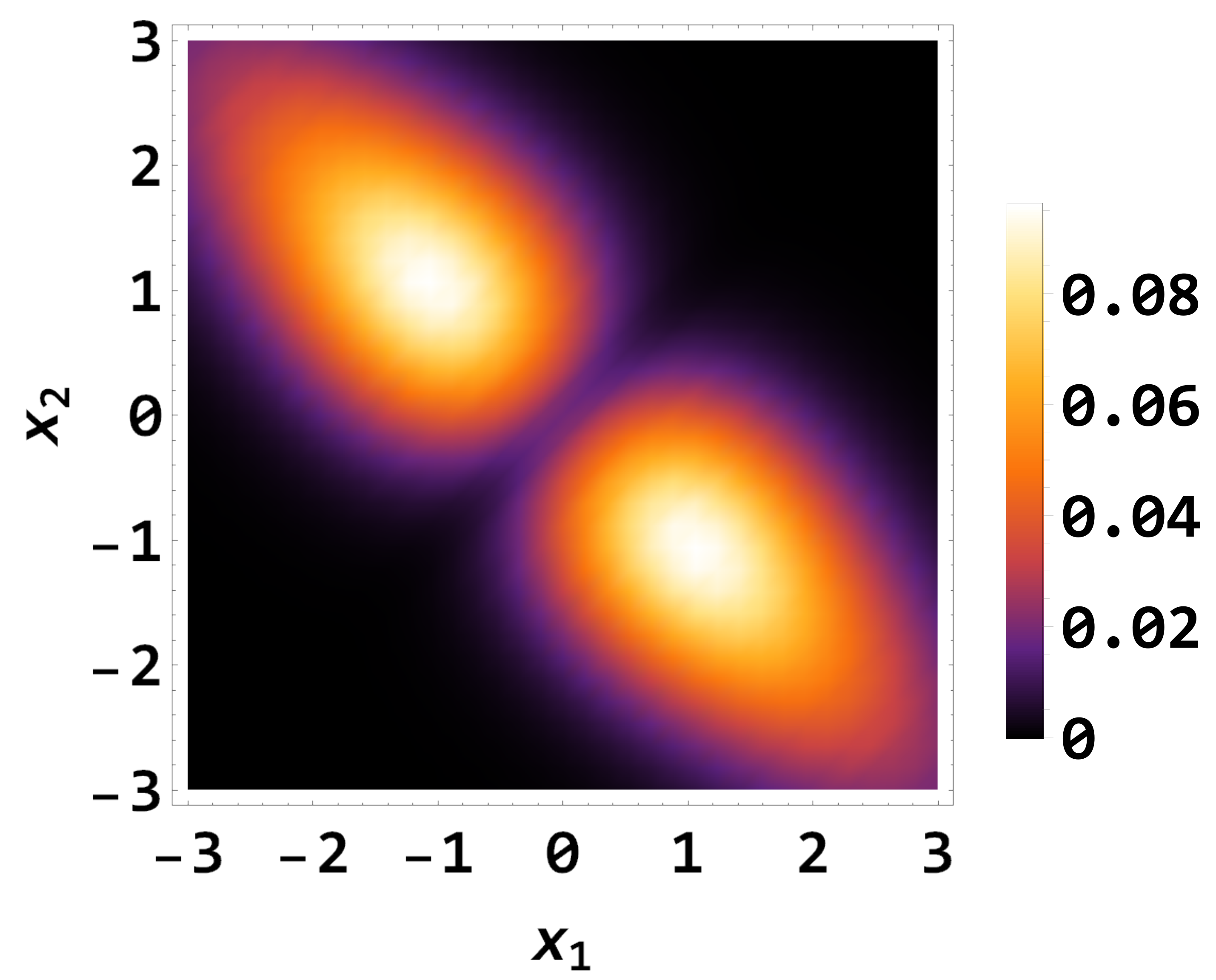} &
		\includegraphics[width=.15\textwidth]{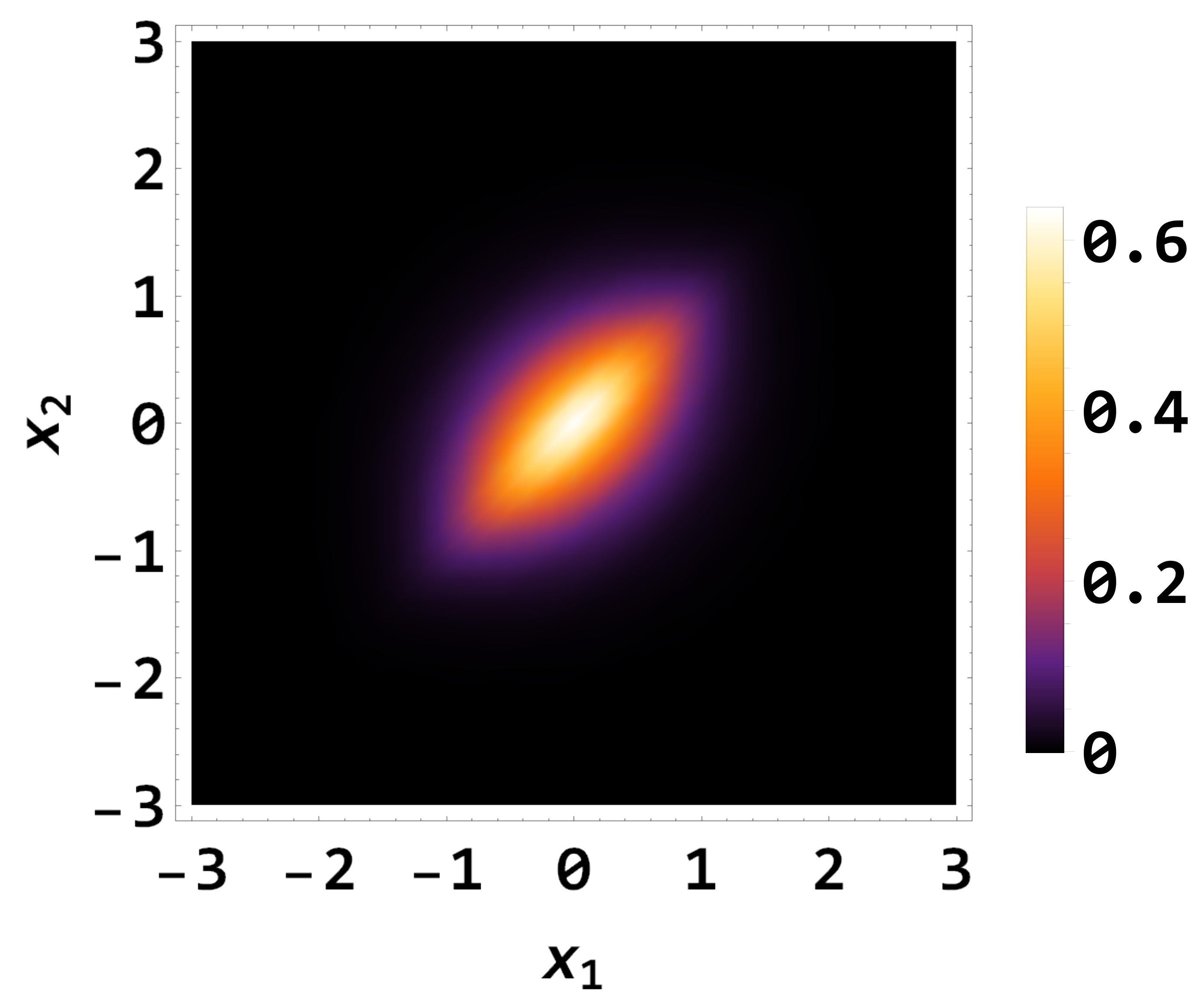} \\
	\end{tabular}
	\caption{(Color online) Evolution of the probability density $|\Psi(x_1,x_2,t)|^2$.}\label{bwd_wf}
\end{figure}

The evolution of the probability density $|\Psi(x_1,x_2,t)|^2$ is shown in Fig.~\ref{bwd_wf}. Due to the initial attractive interaction the spatial distribution of $|\Psi(x_1,x_2,t)|^2$ is quite narrow. With increasing time $|\Psi(x_1,x_2,t)|^2$ possesses a two-hump structure. The barrier prevents the probability density to localize around the center, which is seen over the course of the evolution. However, at the moment $t=\pi$, the system almost returns to its initial state, which tells us about an almost periodic oscillation of the system.

\begin{figure}
	\centering
	\textbf{$\alpha=0$}\par\medskip
	\begin{tabular}{ccc}
		$g=-2,(0,0)$  & $g=-2,(2,0)$ & $g=-2,(4,0)$ \\
		\includegraphics[width=.15\textwidth]{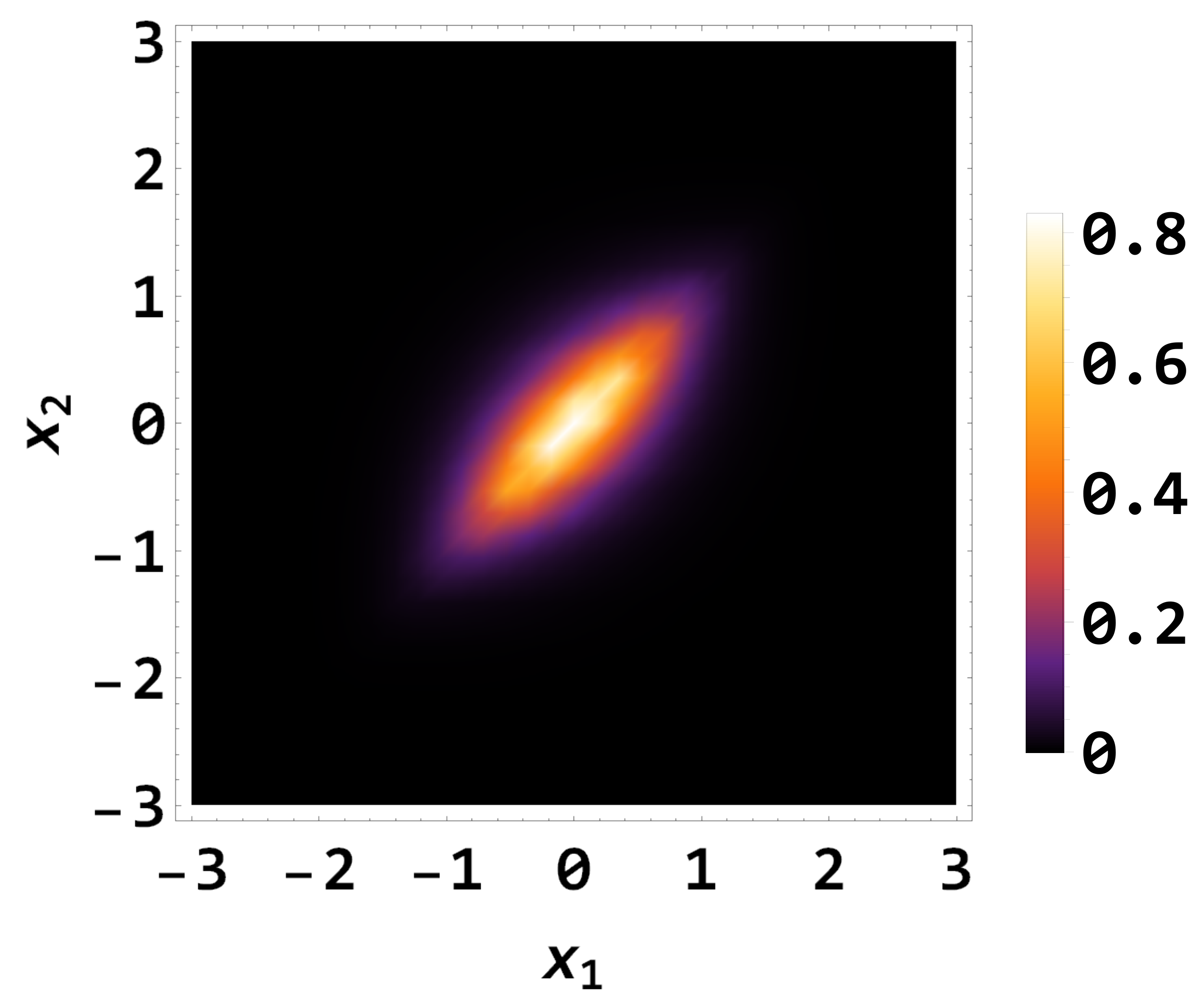} &
		\includegraphics[width=.15\textwidth]{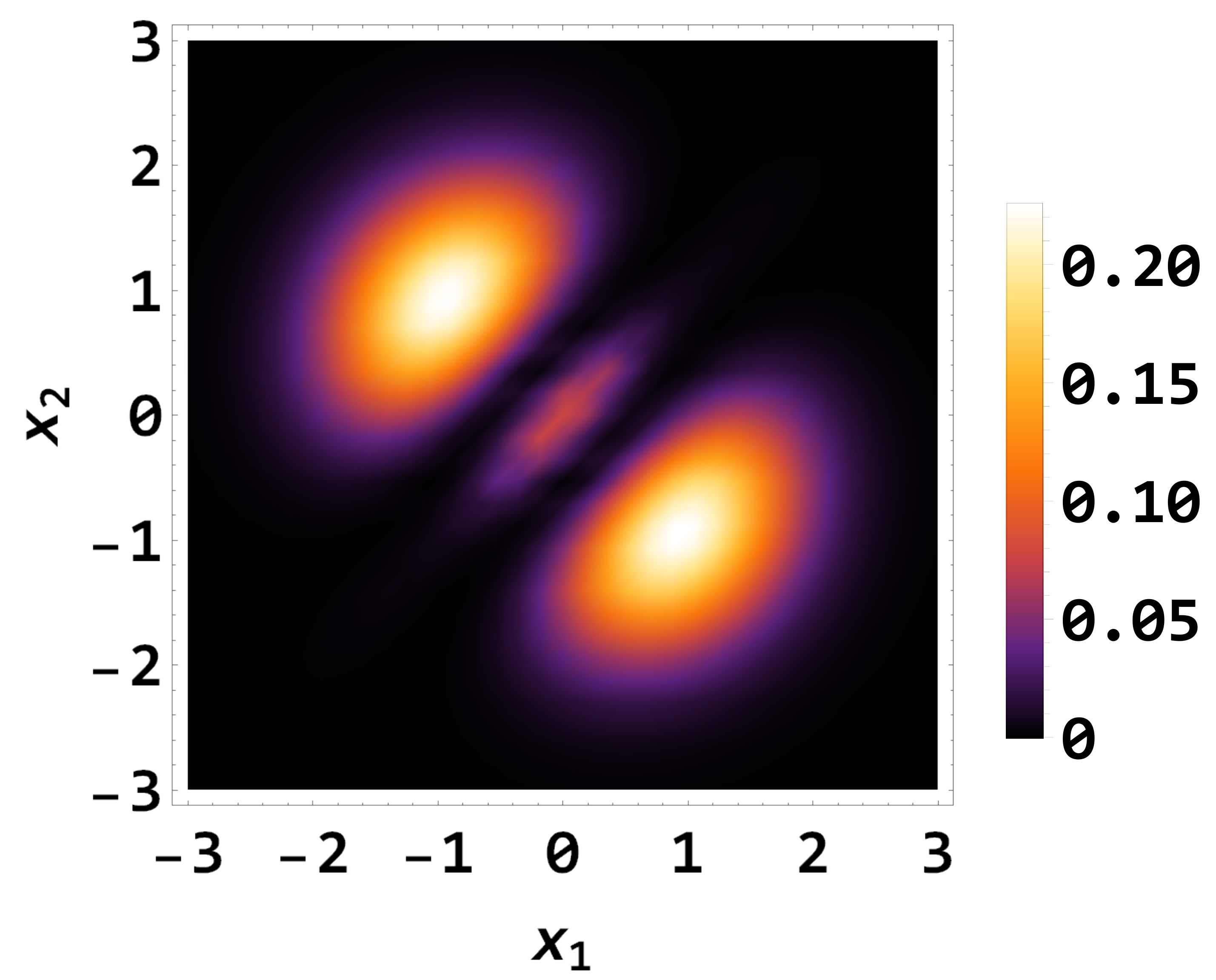} &
		\includegraphics[width=.15\textwidth]{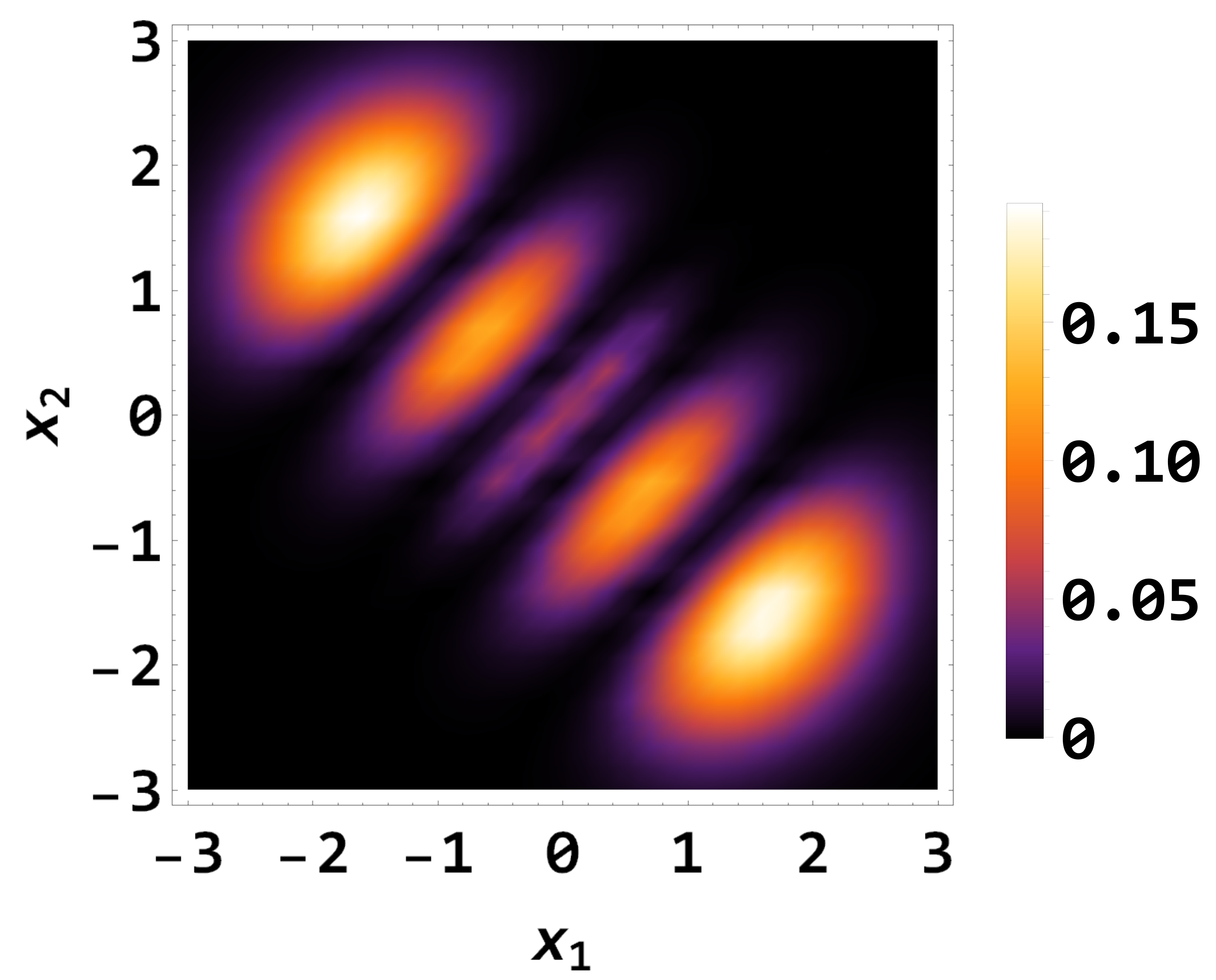} \\
		$g=2,(0,0)$  & $g=2,(2,0)$ & $g=2,(4,0)$ \\
		\includegraphics[width=.15\textwidth]{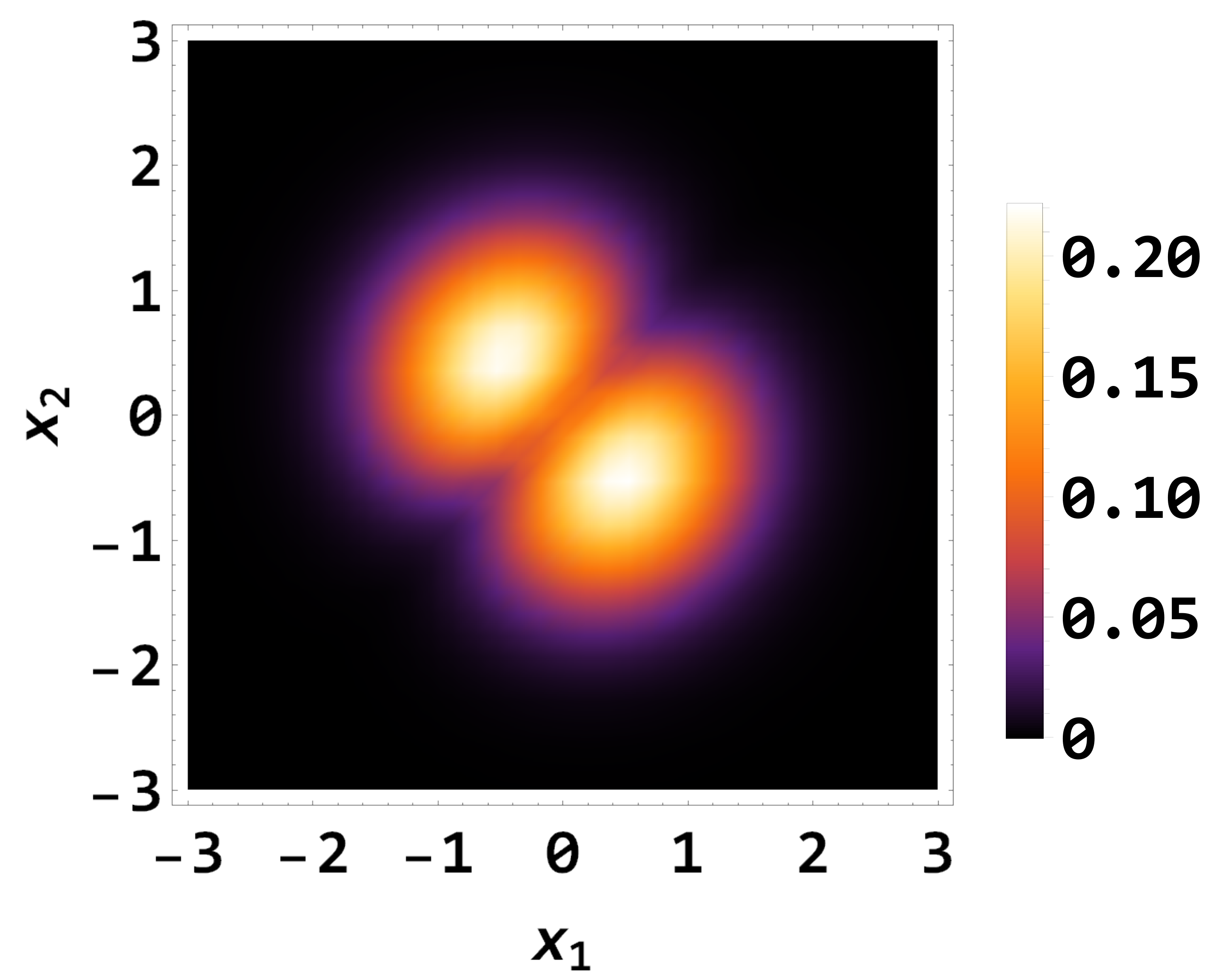} &
		\includegraphics[width=.15\textwidth]{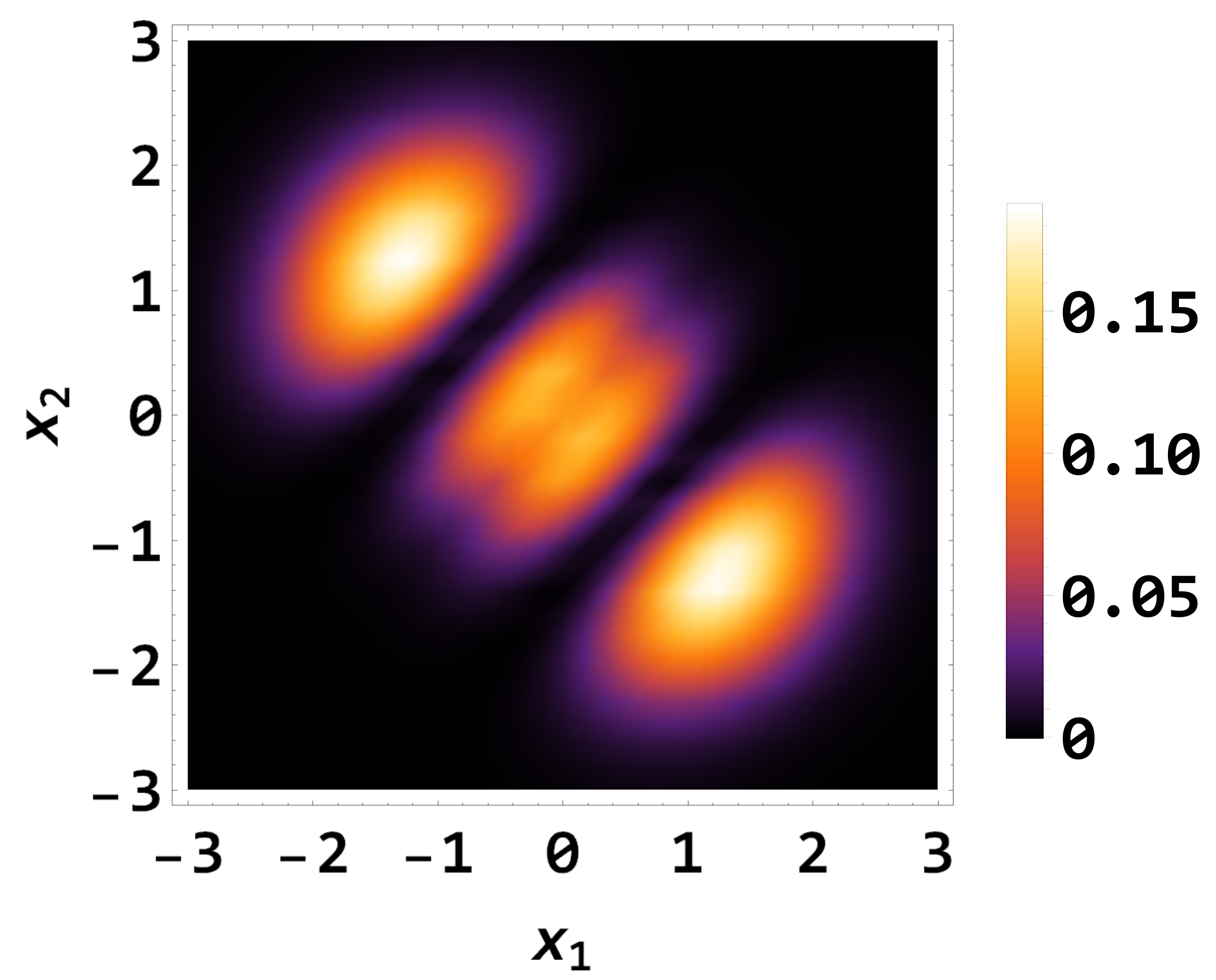} &
		\includegraphics[width=.15\textwidth]{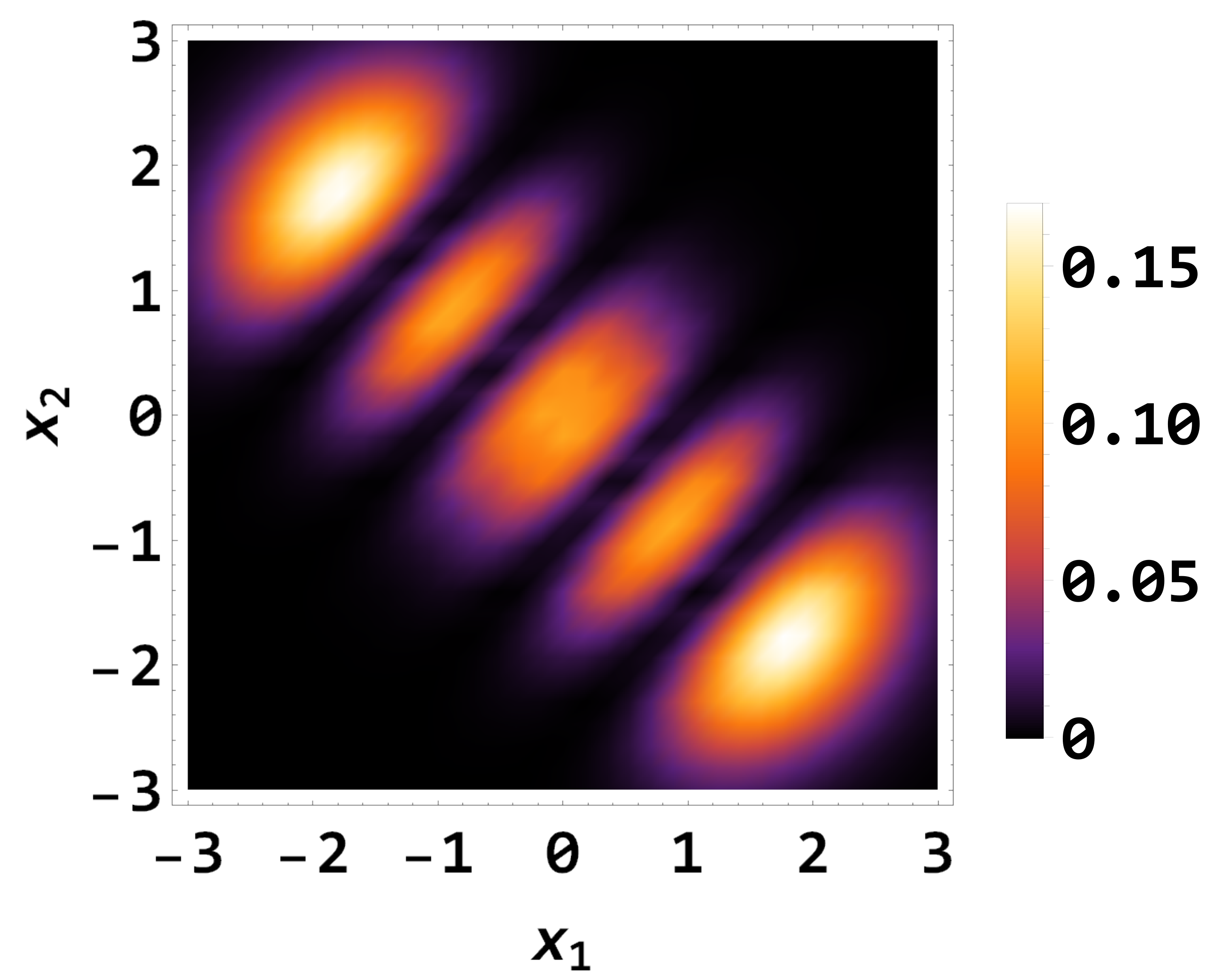} \\
	\end{tabular}
	\caption{(Color online) Probability density $|\Psi(x_1,x_2,t=0)|^2$ for different pre- and postquench stationary states.}\label{bwd_wf0}
\end{figure}

Let us plot the probability densities $|\Psi(x_1,x_2,t)|^2$ of pre- and postquench states in Fig.~\ref{bwd_wf0}. The probability densities of the excited states consists of several humps which localize along the axis $x_1=-x_2$ and stretch even farther along this axis and get multiplied when higher excited states are considered. The initial hump in Fig.~\ref{bwd_wf} splits into two humps, which then get stretched along $x_1=-x_2$ axis, as time passes. This two hump pattern resembles the similar distributions of the stationary states in Fig.~\ref{bwd_wf0}, which suggests that during the quench the system undergoes transitions to states depicted in Fig.~\ref{bwd_wf0}.

\begin{figure}
	\centering
	\textbf{$\alpha=0$}\par\medskip
	\begin{tabular}{ccc}
		$t=0.1$ & $t=\pi/8$ & $t=\pi/4$ \\
		\includegraphics[width=.15\textwidth]{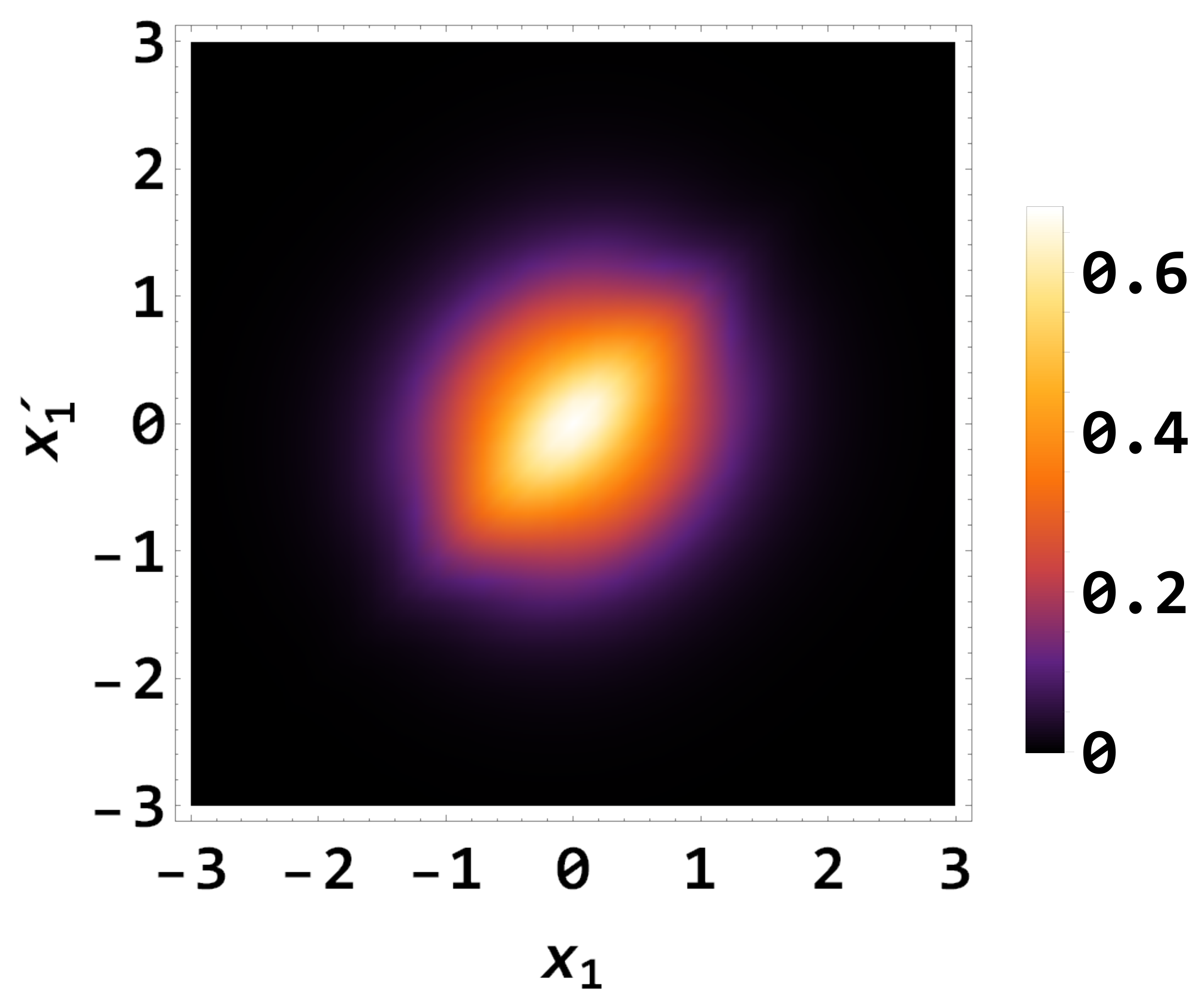} &
		\includegraphics[width=.15\textwidth]{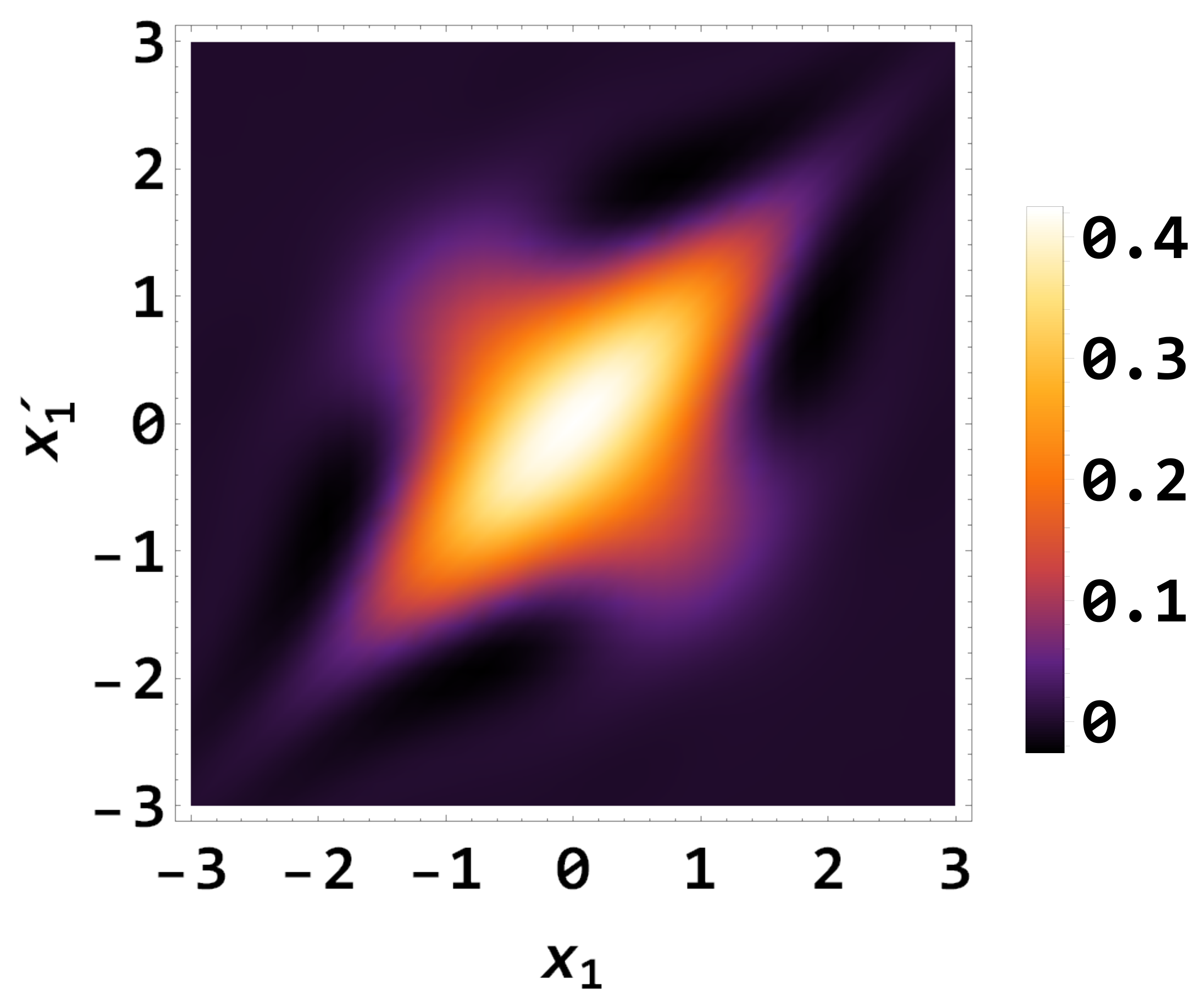} &
		\includegraphics[width=.15\textwidth]{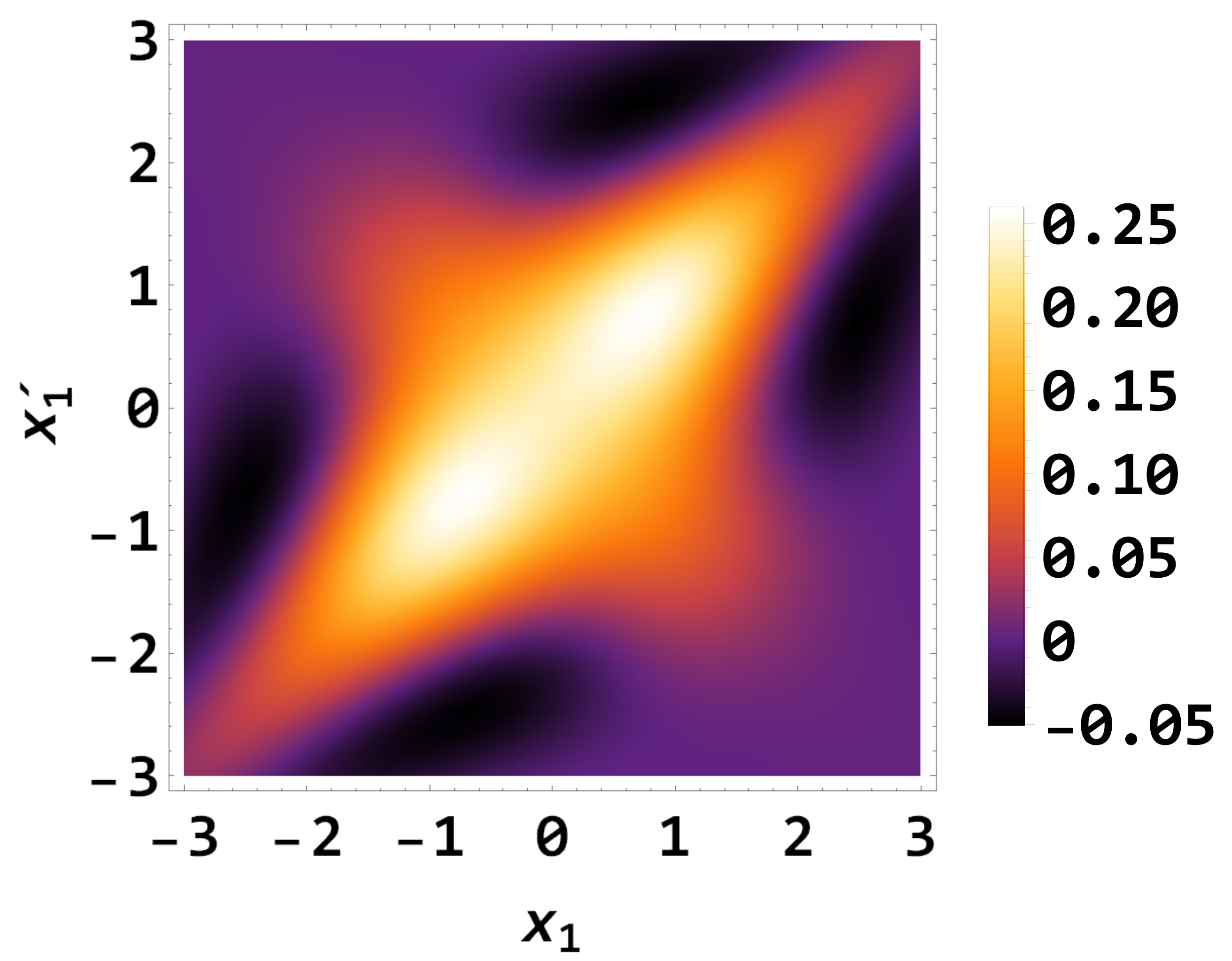} \\
		$t=\pi/2$ & $t=\pi$ \\
		\includegraphics[width=.15\textwidth]{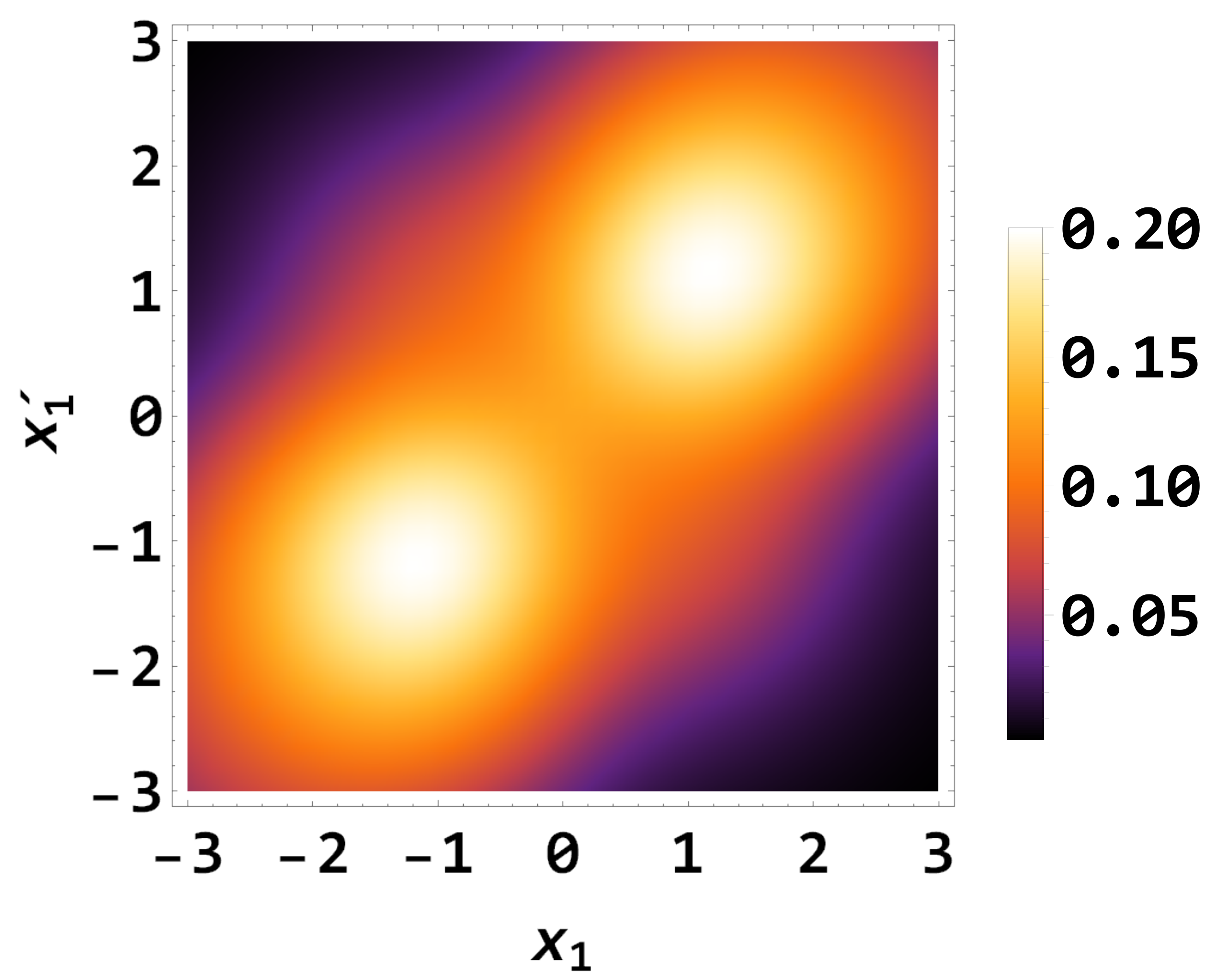} &
		\includegraphics[width=.15\textwidth]{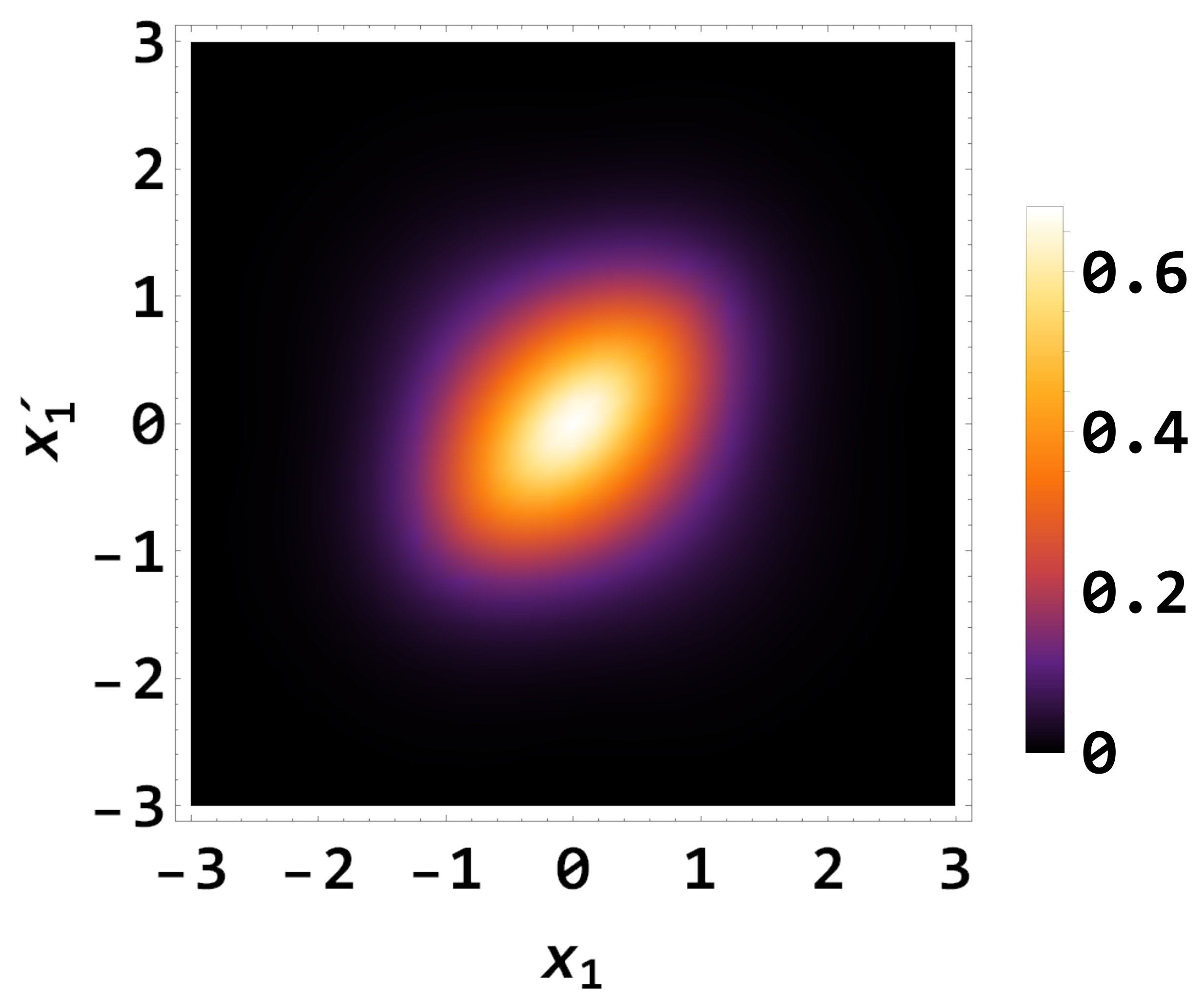} \\
	\end{tabular}
	\caption{(Color online) One-body reduced density matrix, $\rho^{(1)}(x_1,x_1';t)$, at different time-instants.}\label{bwd_rho}
\end{figure}

The one-body reduced density matrix $\rho^{(1)}(x_1,x_1';t)$, shown in Fig.~\ref{bwd_rho}, during the evolution splits into a two-hump structure. This indicates spatial delocalization of an atom. This is the result of the repulsive potential barrier, which tends to separate the particles from each other. At $t=\pi$ the system almost returns to its initial state.

\begin{figure}
	\centering
	\textbf{$\alpha=0$}\par\medskip
	\begin{tabular}{ccc}
		$g=-2,(0,0)$  & $g=-2,(2,0)$ & $g=-2,(4,0)$ \\
		\includegraphics[width=.15\textwidth]{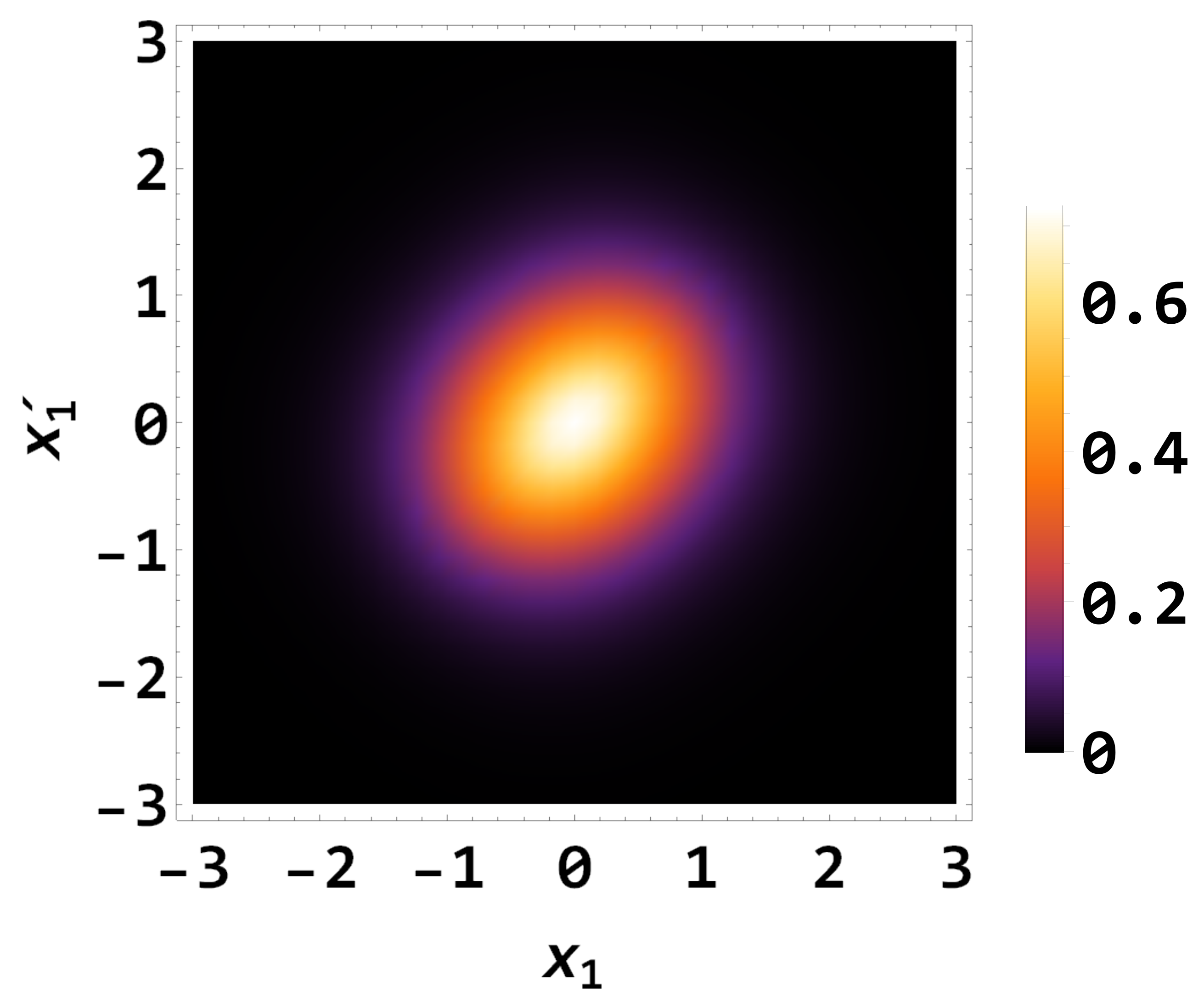} &
		\includegraphics[width=.15\textwidth]{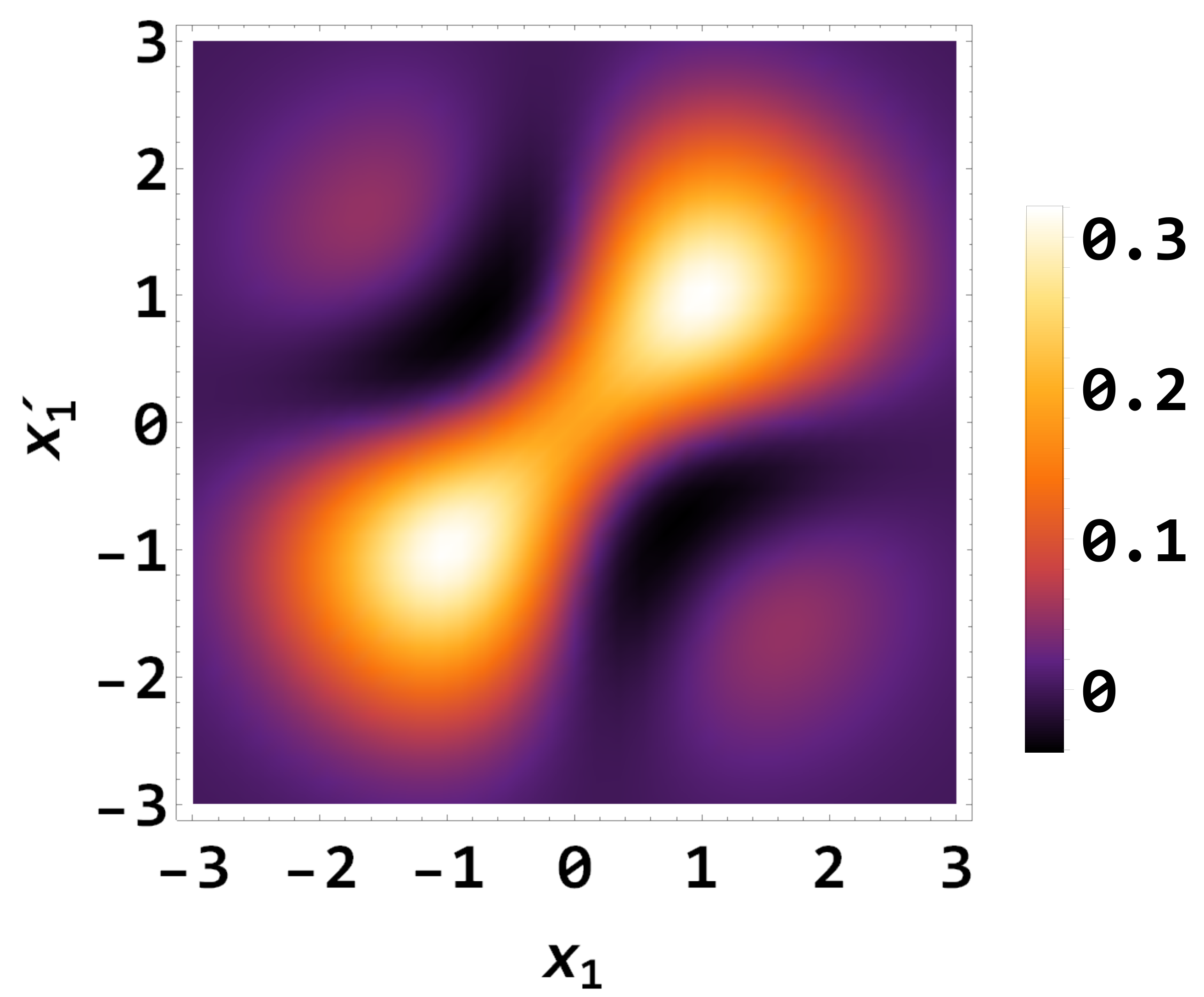} &
		\includegraphics[width=.15\textwidth]{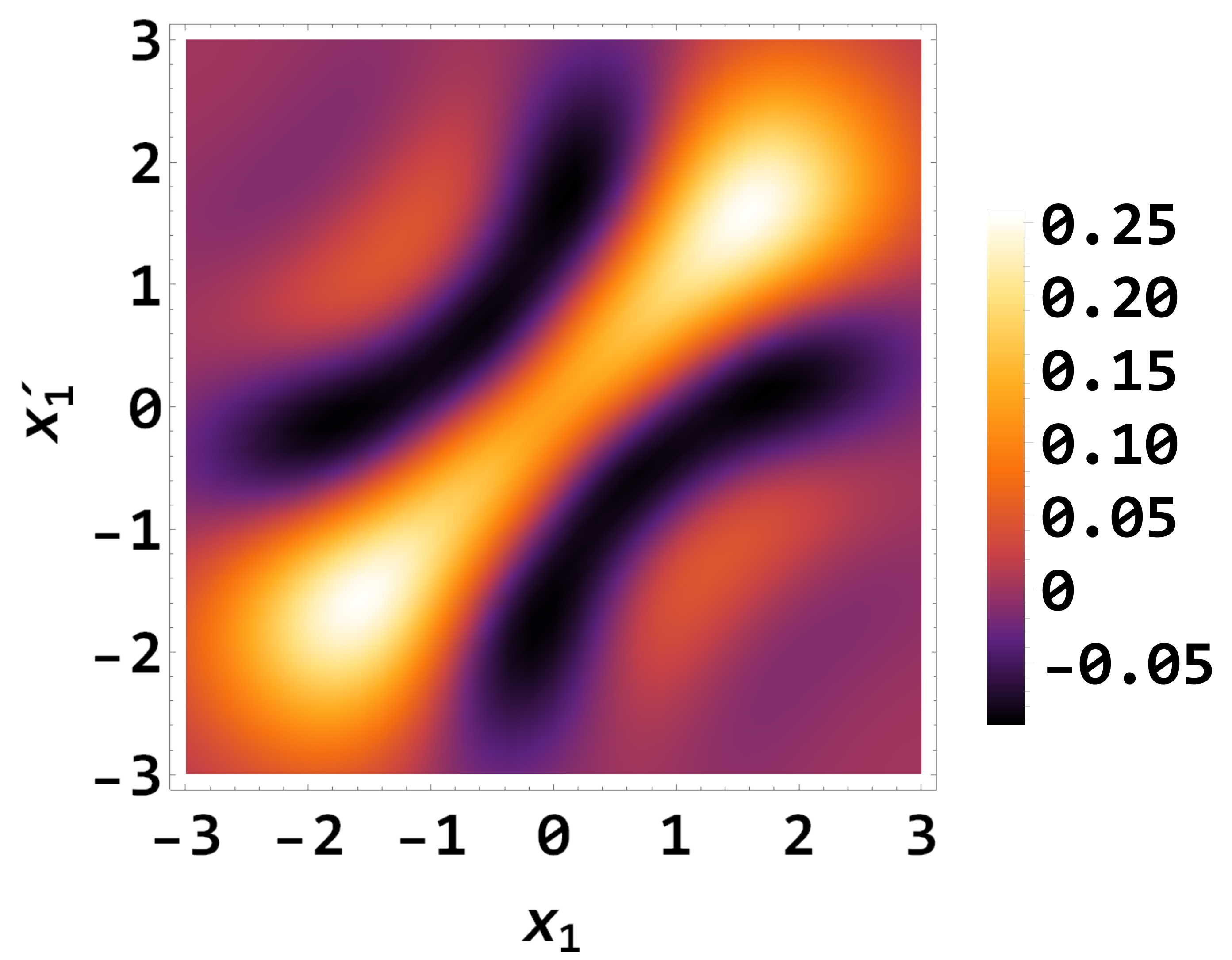} \\
		$g=2,(0,0)$  & $g=2,(2,0)$ & $g=2,(4,0)$ \\
		\includegraphics[width=.15\textwidth]{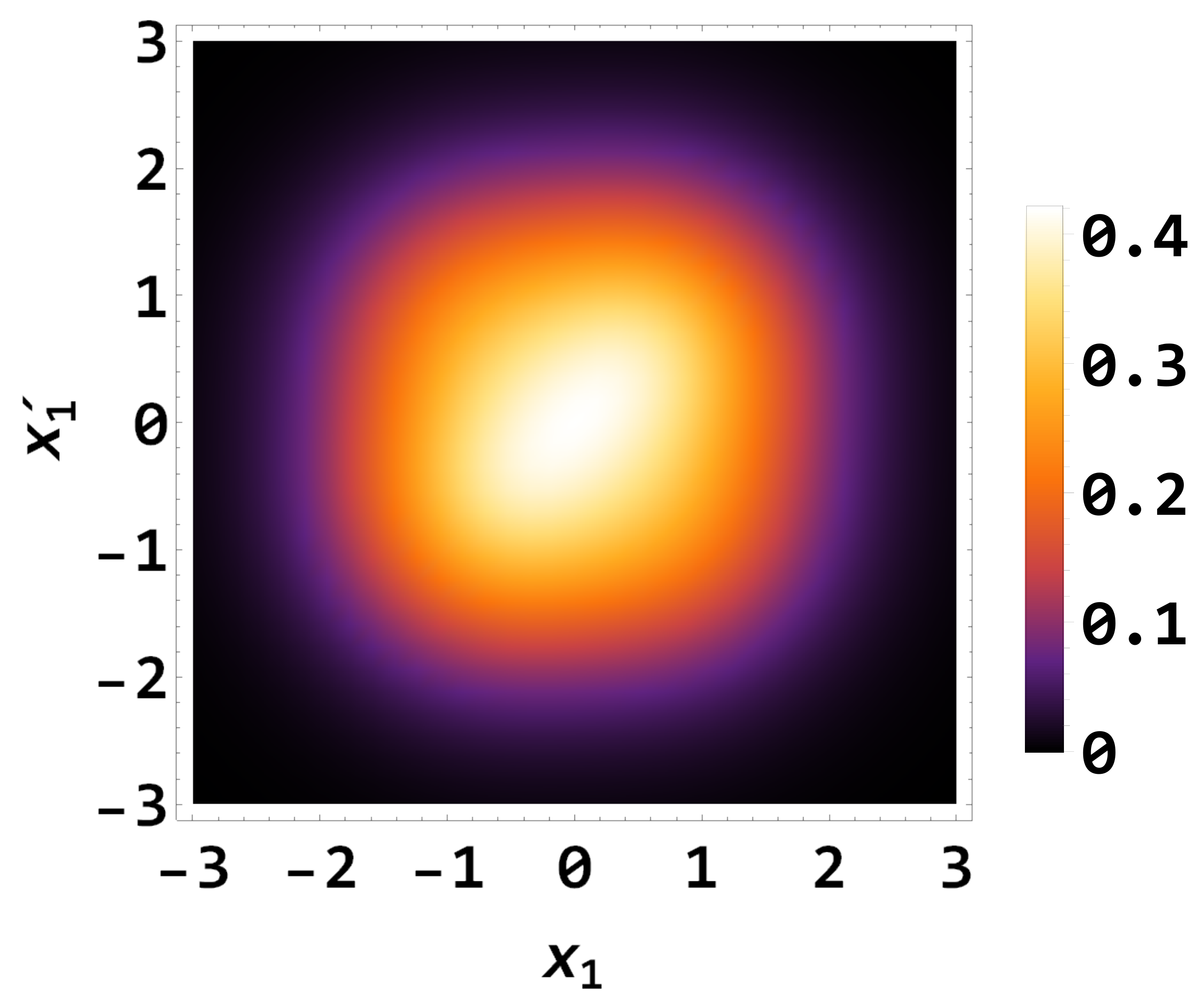} &
		\includegraphics[width=.15\textwidth]{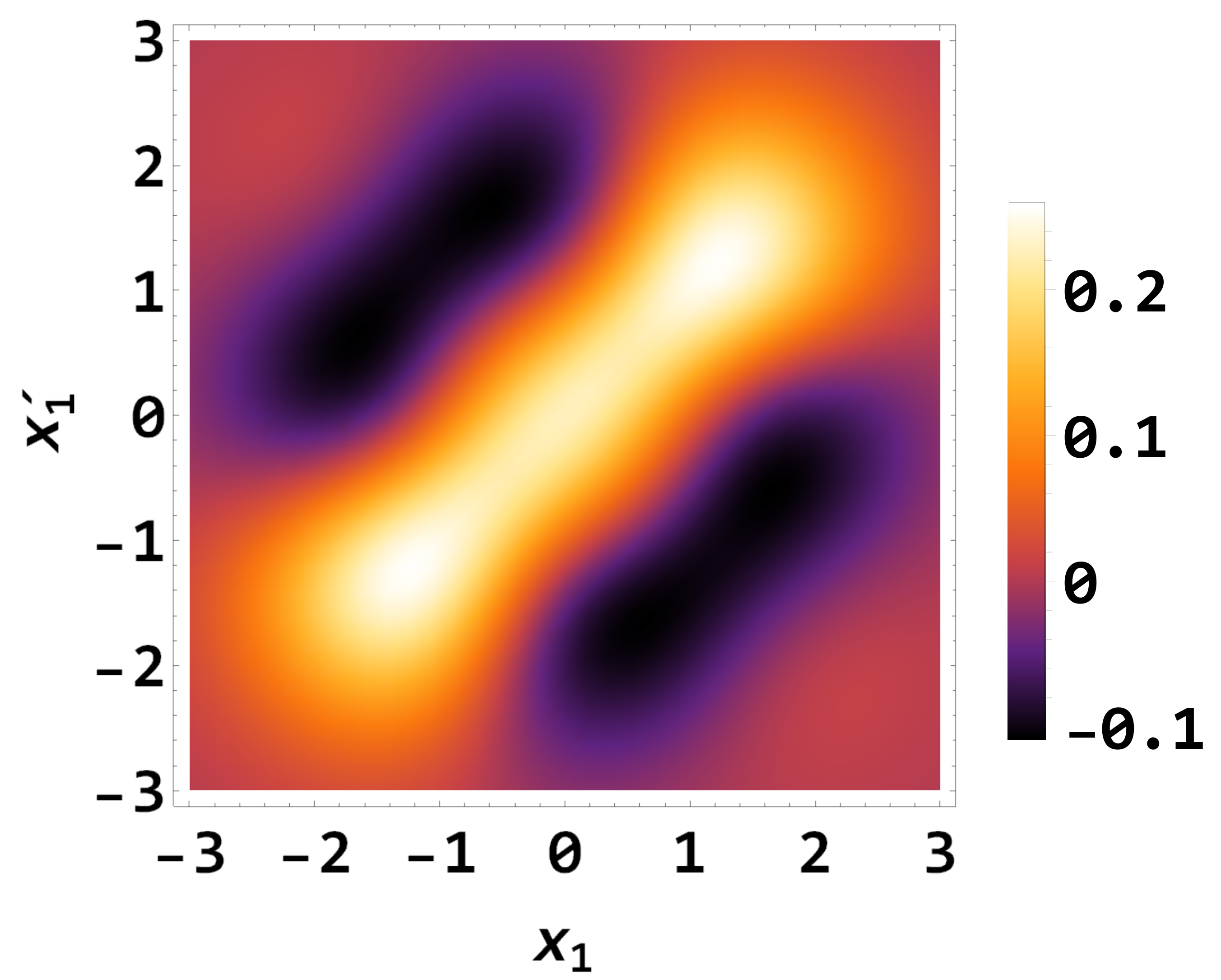} &
		\includegraphics[width=.15\textwidth]{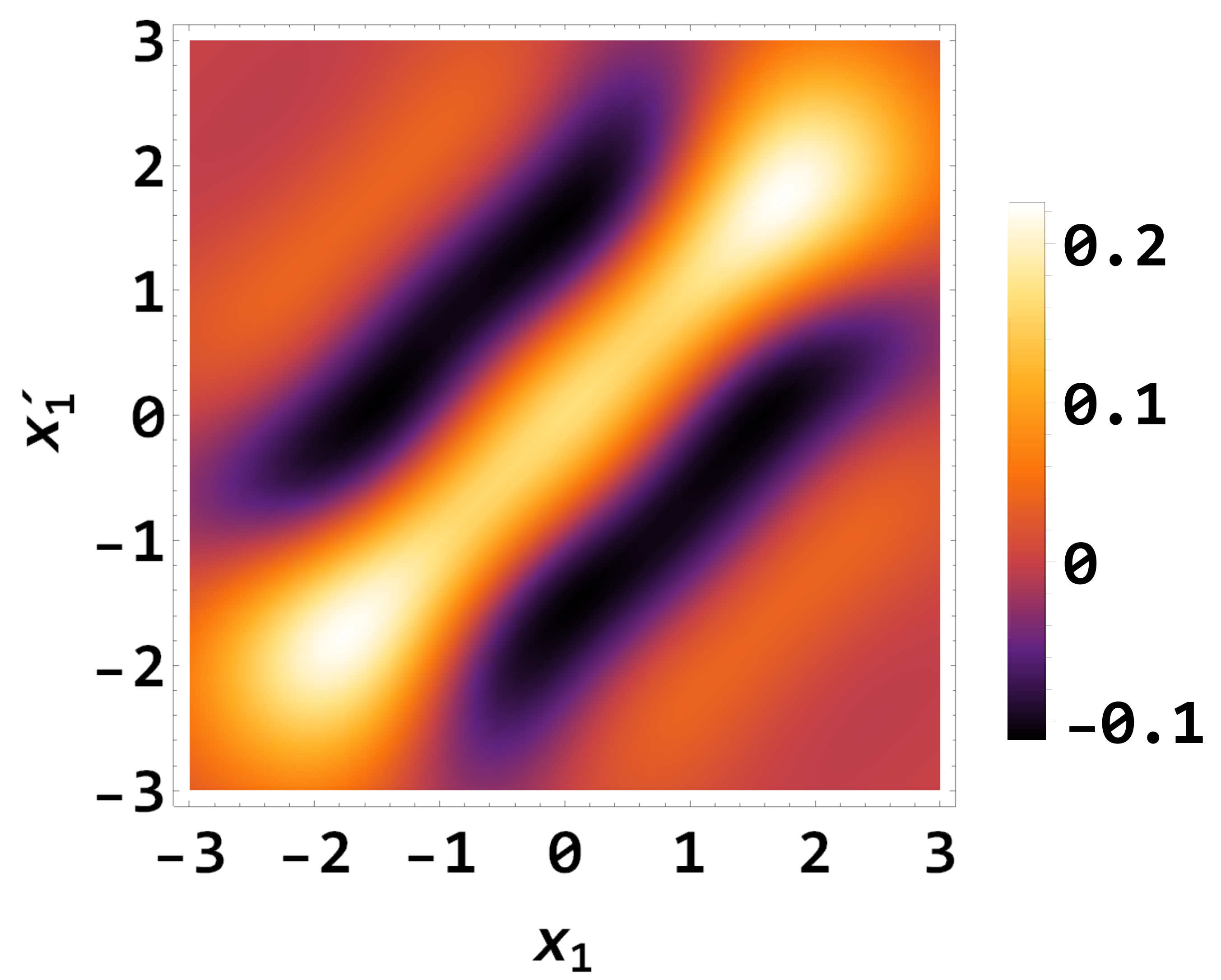} \\
	\end{tabular}
	\caption{(Color online) One-body reduced density matrix, $\rho^{(1)}(x_1,x_1';t=0)$, for different stationary states.}\label{bwd_irho}
\end{figure}

In Fig.~\ref{bwd_irho} we plot the one-body reduced density matrces $\rho^{(1)}(x_1,x_1';t=0)$ for different pre- and postquench states. It can be noticed that for the prequench excited states $\rho^{(1)}(x_1,x_1';t=0)$ shows two-hump structures, which are localized along the $x_1=x_2$ axis, while for the postquench excited states there is lesser distinction between these humps and their spatial distribution is rather elongated along this axis. These kind of distributions are similar to the two-hump distributions of $\rho^{(1)}(x_1,x_1';t)$ during its evolution course. This situation demonstrates which possible states the system occupies when quenched.

\begin{figure}
	\centering
	\textbf{$\alpha=0$}\par\medskip
	\includegraphics[width=7cm,clip]{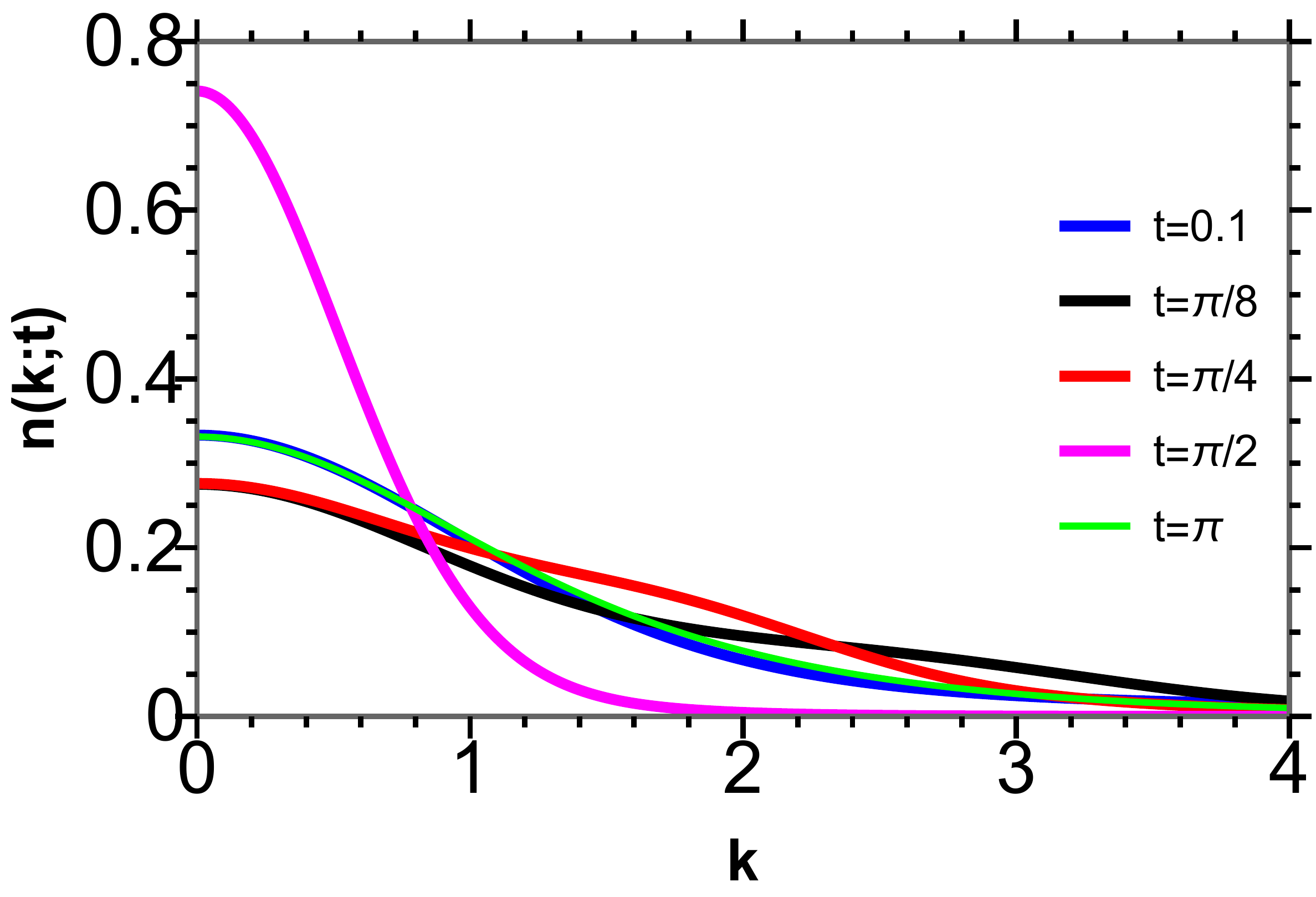}
	\caption{(Color online) Evolution of the momentum distribution $n(k;t)$ for the harmonic trap.}\label{bwd_nk_all}
\end{figure}

The evolution of the momentum distribution $n(k;t)$ is shown in Fig.~\ref{bwd_nk_all}. The blue line, corresponding to the shape of $n(k;t)$ at $t=0.1$, does not have quite a pronounced peak around $k=0$. This is because of the initial attractive interaction between the particles, which leads to a quite narrow spatial distribution of the one-body reduced density matrix, $\rho^{(1)}(x_1,x_1';t)$ (cf. Fig~\ref{bwd_rho}). From $t=0.1$ up to $t=\pi/4$ we observe that there is only a little change of the shape of $n(k;t)$. Approximately around $t=\pi/2$ the spatial distribution of $\rho^{(1)}(x_1,x_1';t)$ reaches quite a large size and hence the corresponding highly pronounced zero-momentum peak of $n(k;t)$. The green line, which corresponds to the moment $t=\pi$, is almost indistinguishable from the blue line which tells us that the system returns to its initial state.

\begin{figure}
	\centering
	\textbf{$\alpha=0$,~~~$t=\pi/4$}\par\medskip
	\begin{tabular}{cc}
		Prequench states &  Postquench states \\
		\hspace{-.5cm}\includegraphics[width=.25\textwidth]{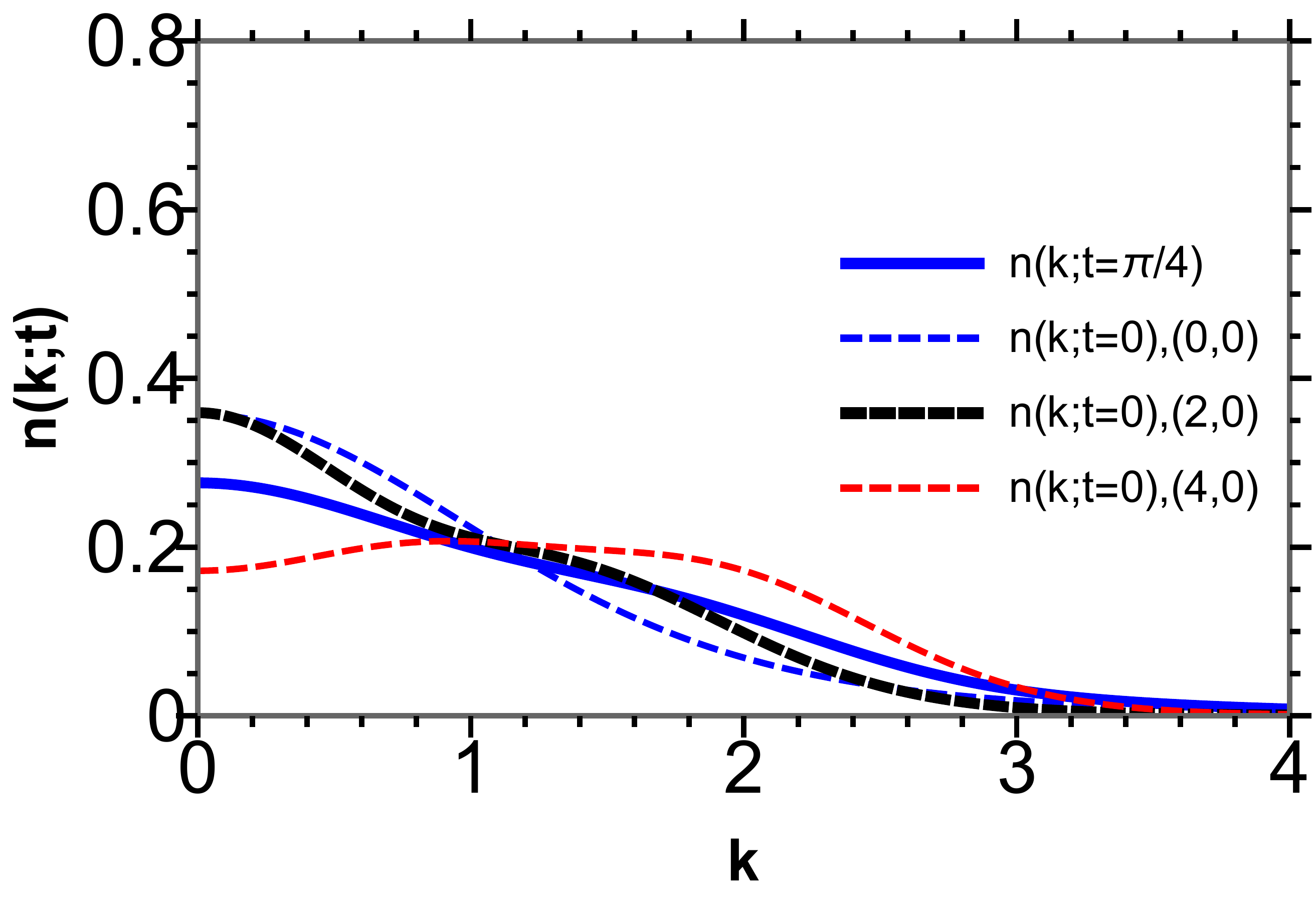} &
		\includegraphics[width=.25\textwidth]{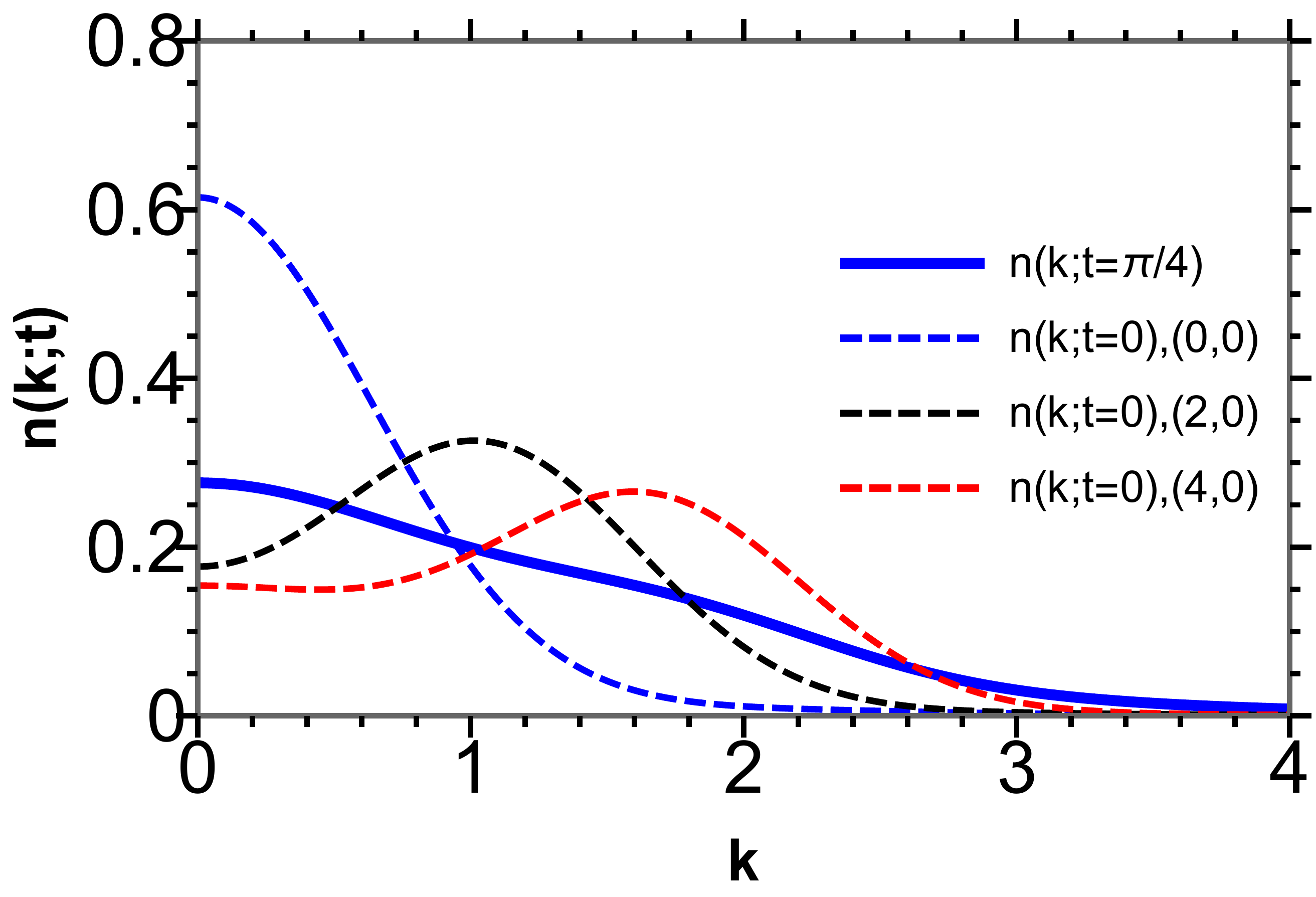} \\
	\end{tabular}
	\caption{(Color online) Comparison of the momentum distribution $n(k;t)$ (blue line) with the momentum distributions $n(k;t=0)$ for the pre- and postquench states (dashed lines).}\label{bwd_nk_12}
\end{figure}

\begin{figure}
	\centering
	\textbf{$\alpha=0$,~~~$t=\pi/2$}\par\medskip
	\begin{tabular}{cc}
		Prequench states &  Postquench states \\
		\hspace{-.5cm}\includegraphics[width=.25\textwidth]{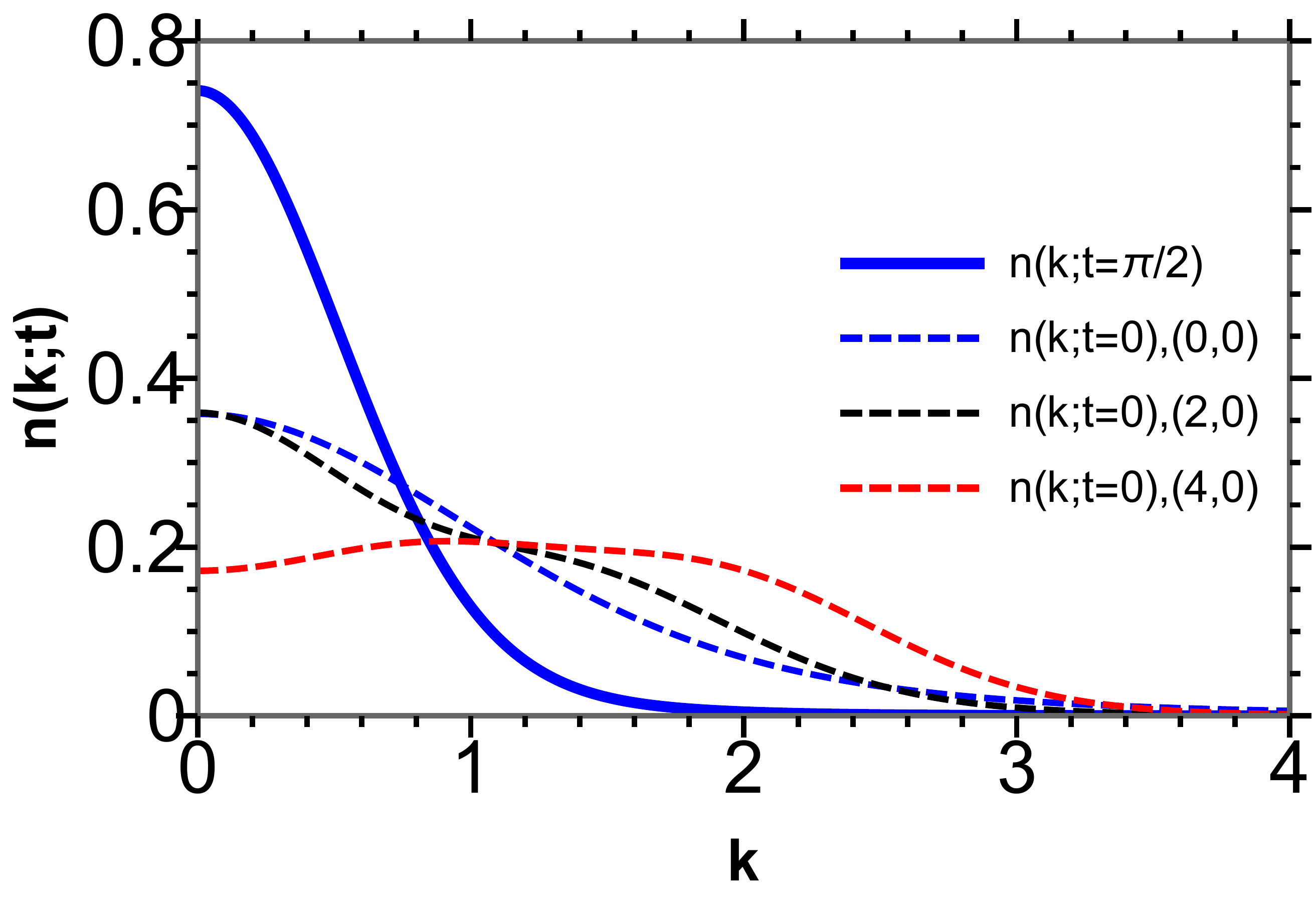} &
		\includegraphics[width=.25\textwidth]{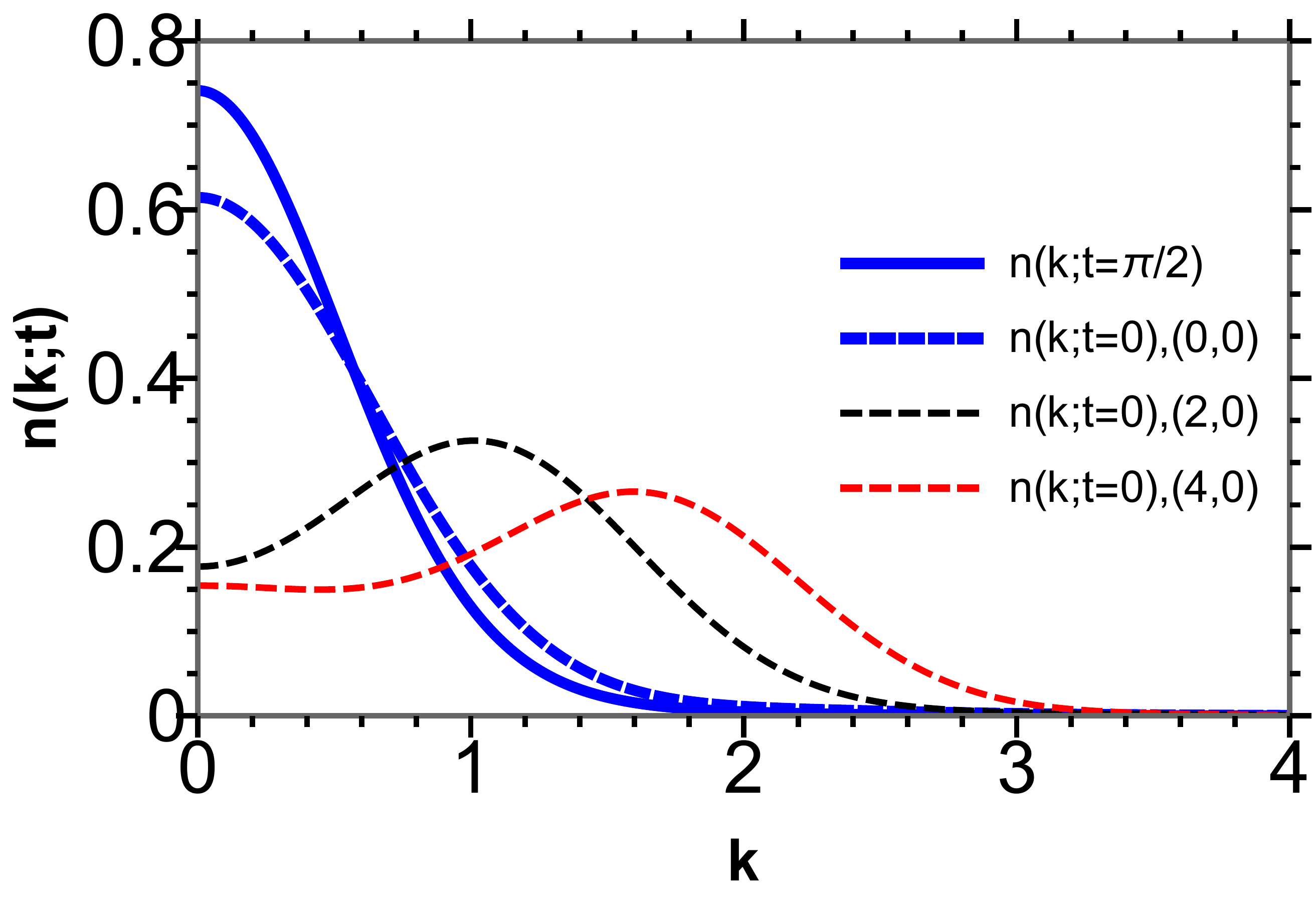} \\
	\end{tabular}
	\caption{(Color online) Comparison of the momentum distribution $n(k;t)$ (blue line) with the momentum distributions $n(k;t=0)$ for the pre- and postquench states (dashed lines).}\label{bwd_nk_34}
\end{figure}

In Fig.~\ref{bwd_nk_12} the shape of the momentum distribution $n(k;t)$ at $t=\pi/4$ (blue line) is quite similar to the shape of $n(k;t=0)$ for the prequench excited state $(2,0)$ (black dashed line). It also shares some similarities with the prequench ground state (blue dashed line). According to Fig.~\ref{bwd} these prequench states, indeed, have quite comparable overlaps at that moment of time. In Fig.~\ref{bwd_nk_34}, at $t=\pi/2$, the shape of $n(k;t)$ (blue line) resembles the shape of $n(k;t=0)$ for the postquench ground state (blue dashed line). This correspondence is also confirmed by Fig.~\ref{bwd}, where the overlap between the wave packet and this postquench ground state is large as well.

\subsection{Anharmonic trap, $\alpha=-0.03$, the ground state, $(0,0)$}\label{section_bwd_al}

\begin{figure}
	\centering
	\textbf{$\alpha=-0.03$}\par\medskip
	\includegraphics[width=8.5cm,clip]{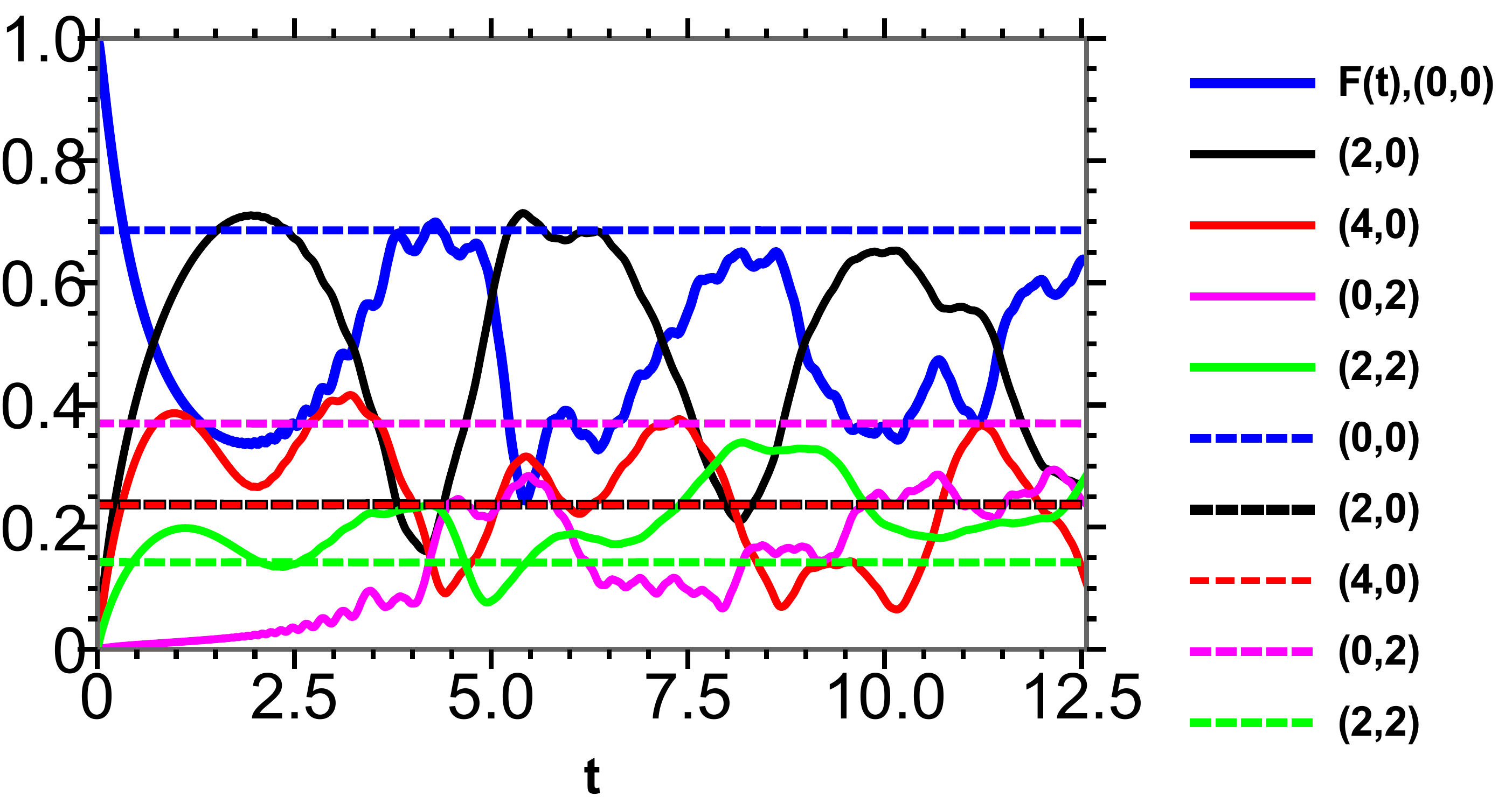}
	\caption{(Color online) Fidelity $F(t)$ and the overlap integrals $\mathcal{Q}$ between the time-evolving state $\Psi(x_1,x_2,t)$ and different pre- (solid lines) and postquench (dashed lines) states in the case of the anharmonic trap. The indices $(n,N)$ refer to the states with the quantum numbers of the relative and center-of-mass motions. The system of the two atoms is prepared in the ground state with $g=-2$ and quenched to $g=2$.}\label{bwd_al}
\end{figure}

The dynamics of the overlaps in the anharmonic case, $\alpha=-0.03$, show quite a complicated time-dependent behavior (Fig.~\ref{bwd_al}). There is no longer such a stable uniform periodic pattern as before. The fidelity does not reach unity after the quench. The coupling between the relative and center-of-mass motions causes additional center-of-mass excited states to contribute to the dynamics. 

\begin{figure}
	\centering
	\textbf{$\alpha=-0.03$}\par\medskip
	\begin{tabular}{ccc}
		$t=0.001$ & $t=0.1$ & $t=\pi/8$ \\
		\includegraphics[width=.15\textwidth]{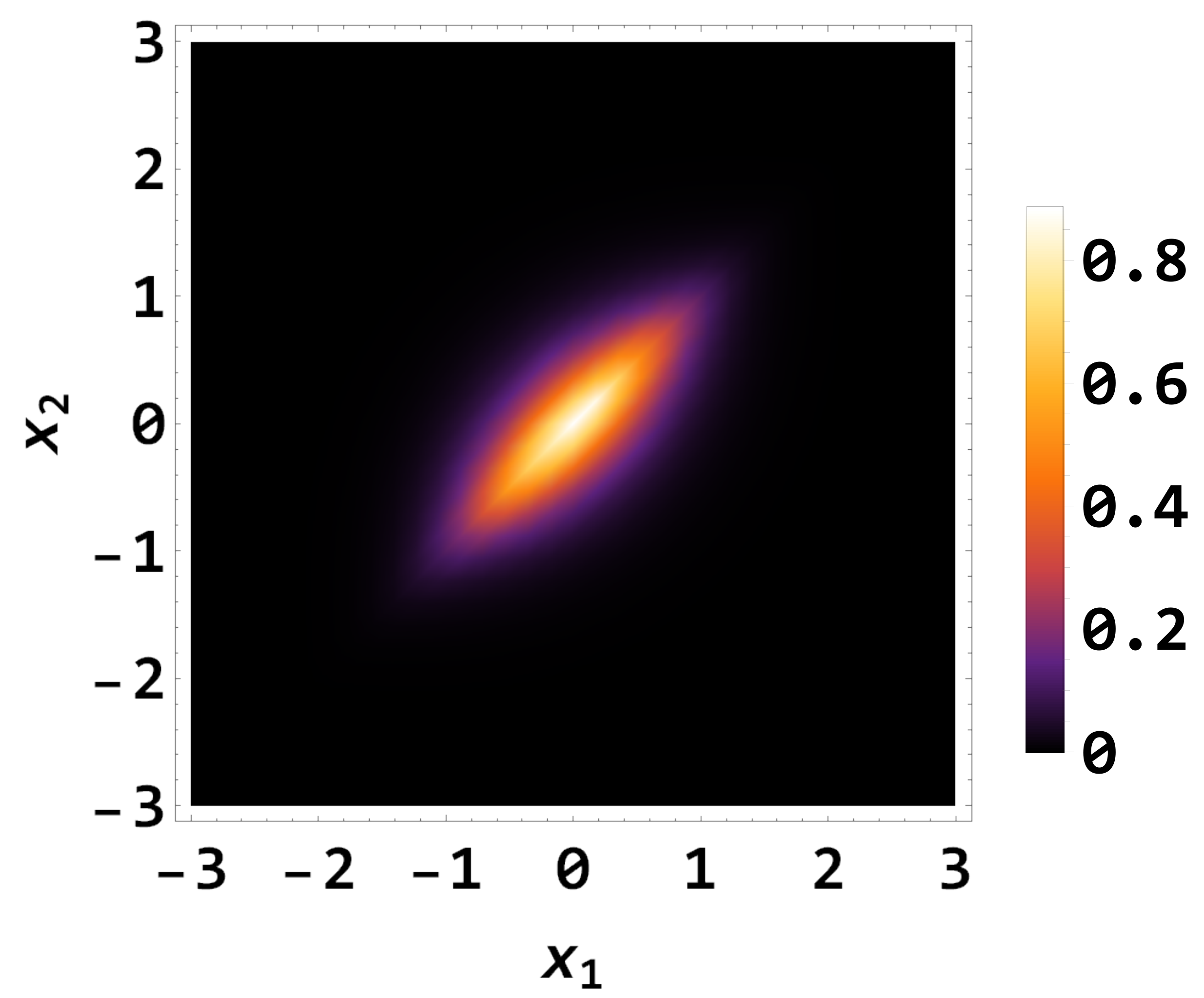} &
		\includegraphics[width=.15\textwidth]{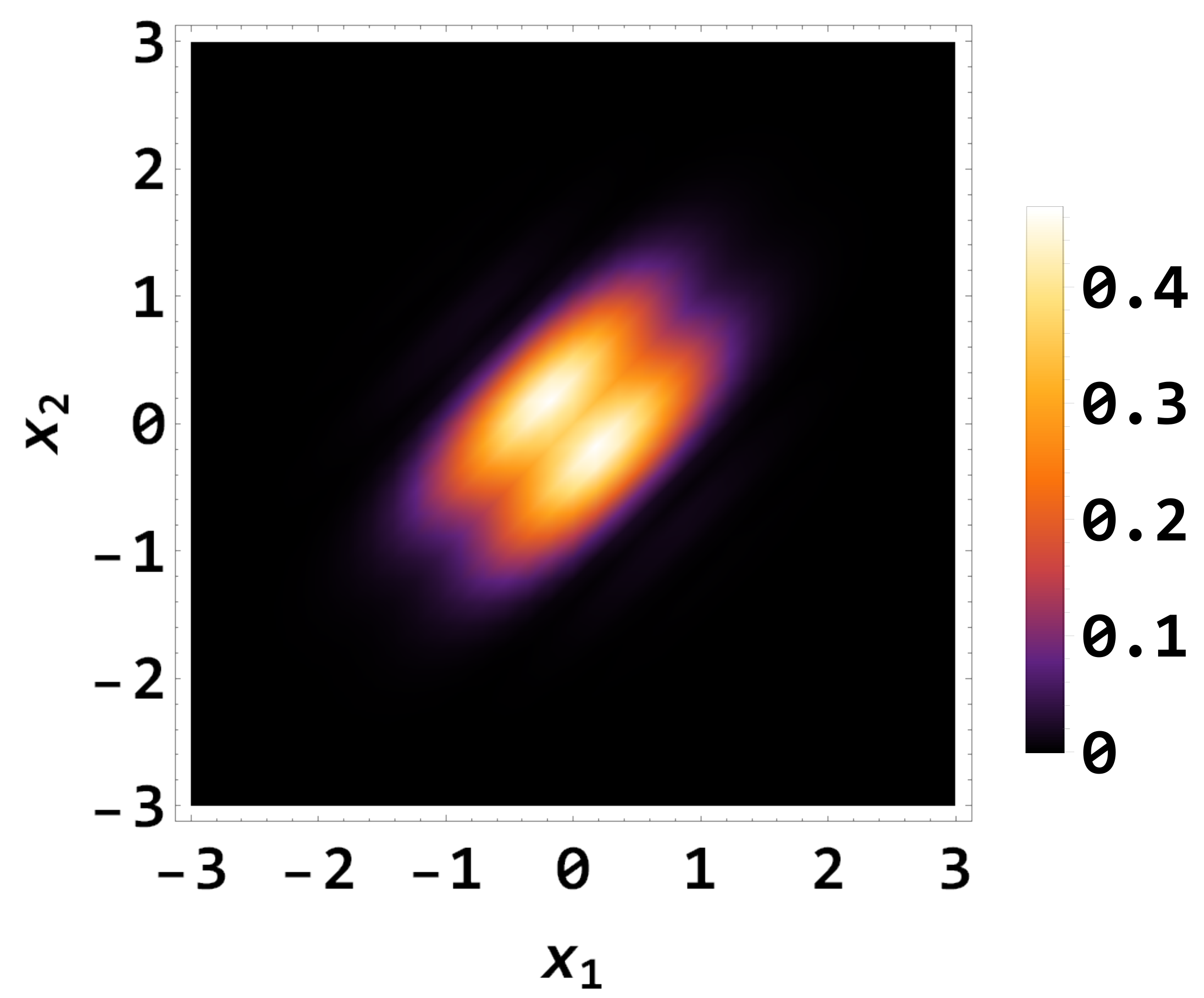} &
		\includegraphics[width=.15\textwidth]{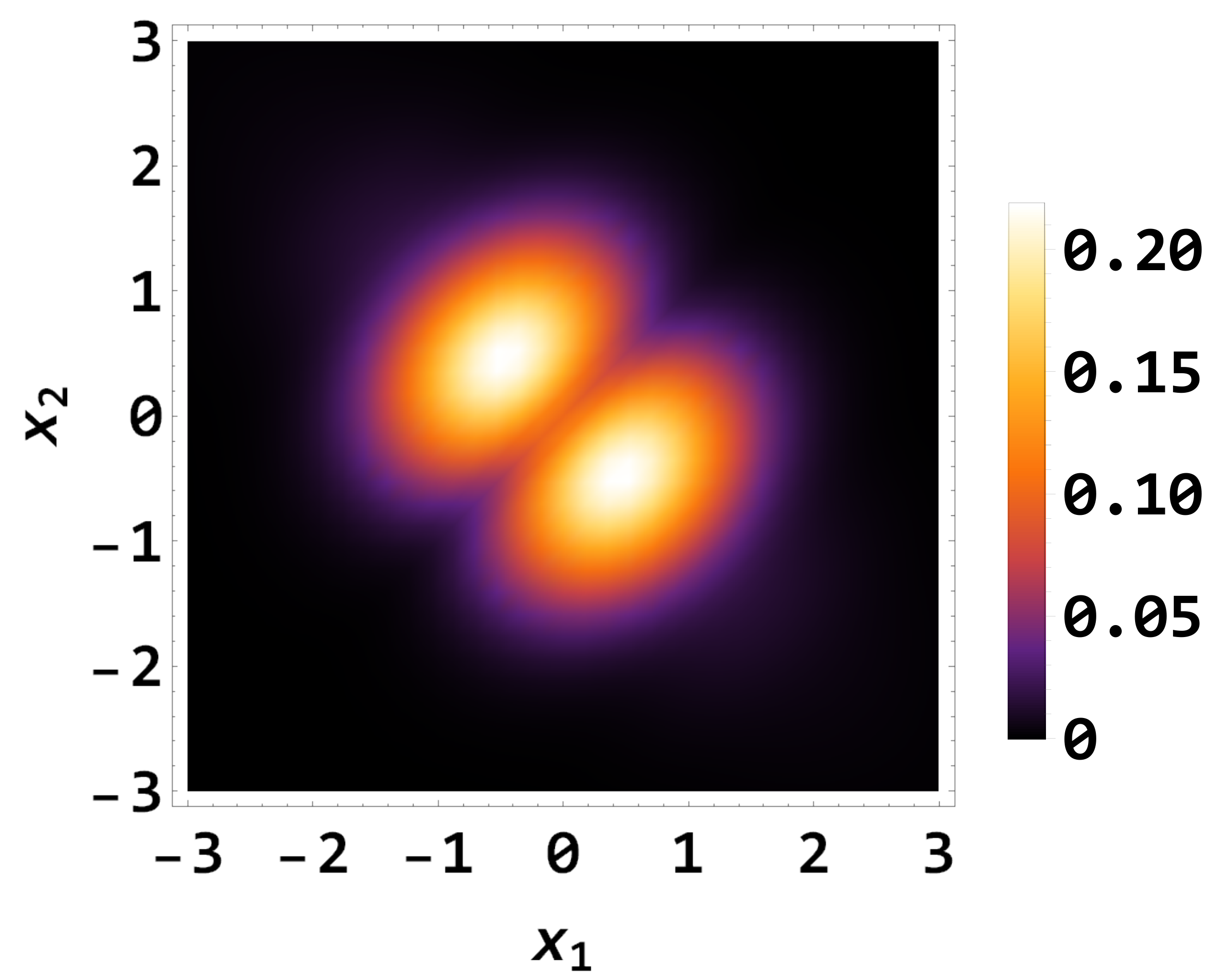} \\
		$t=\pi/4$ & $t=\pi/2$ & $t=\pi$\\
		\includegraphics[width=.15\textwidth]{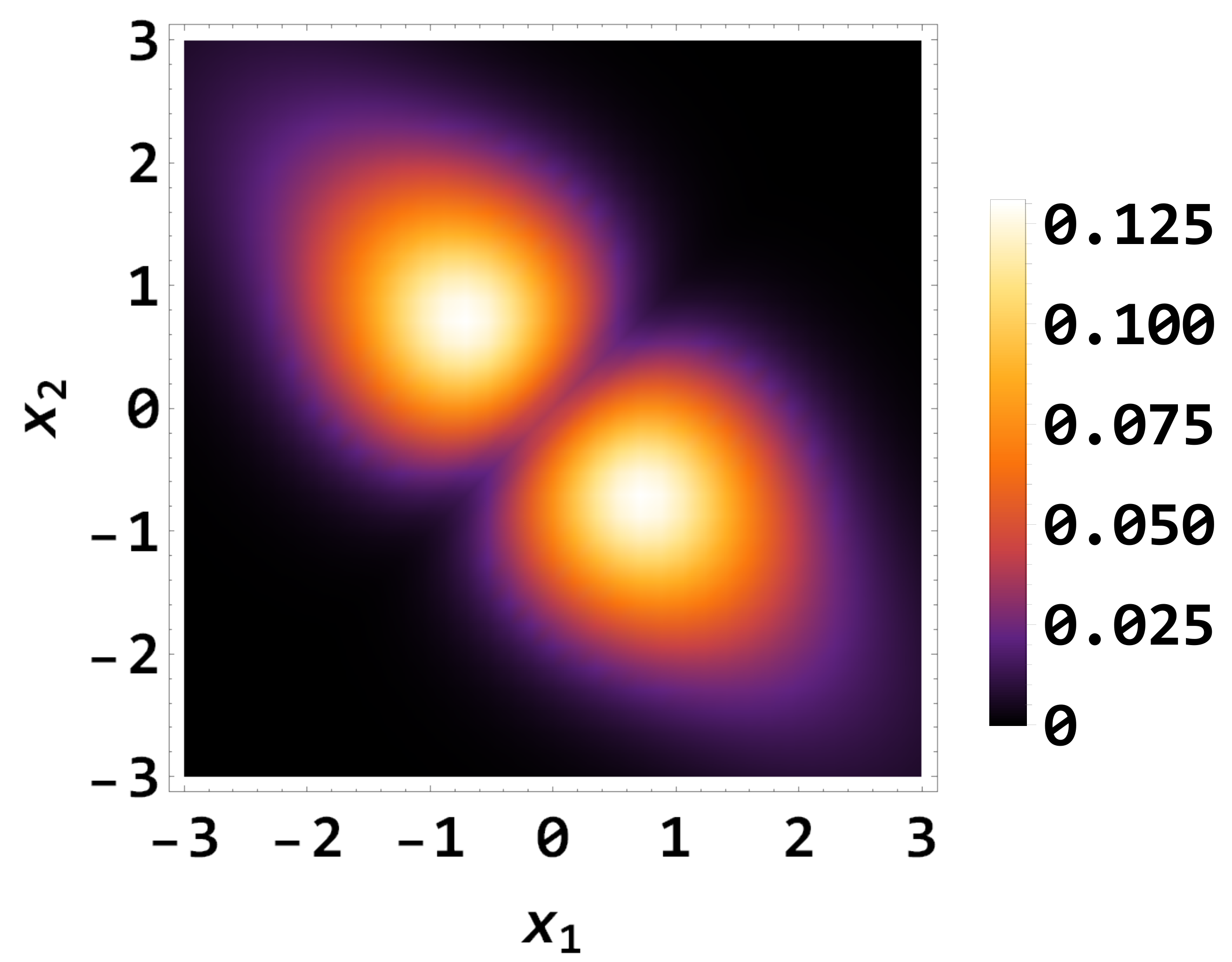} &
		\includegraphics[width=.15\textwidth]{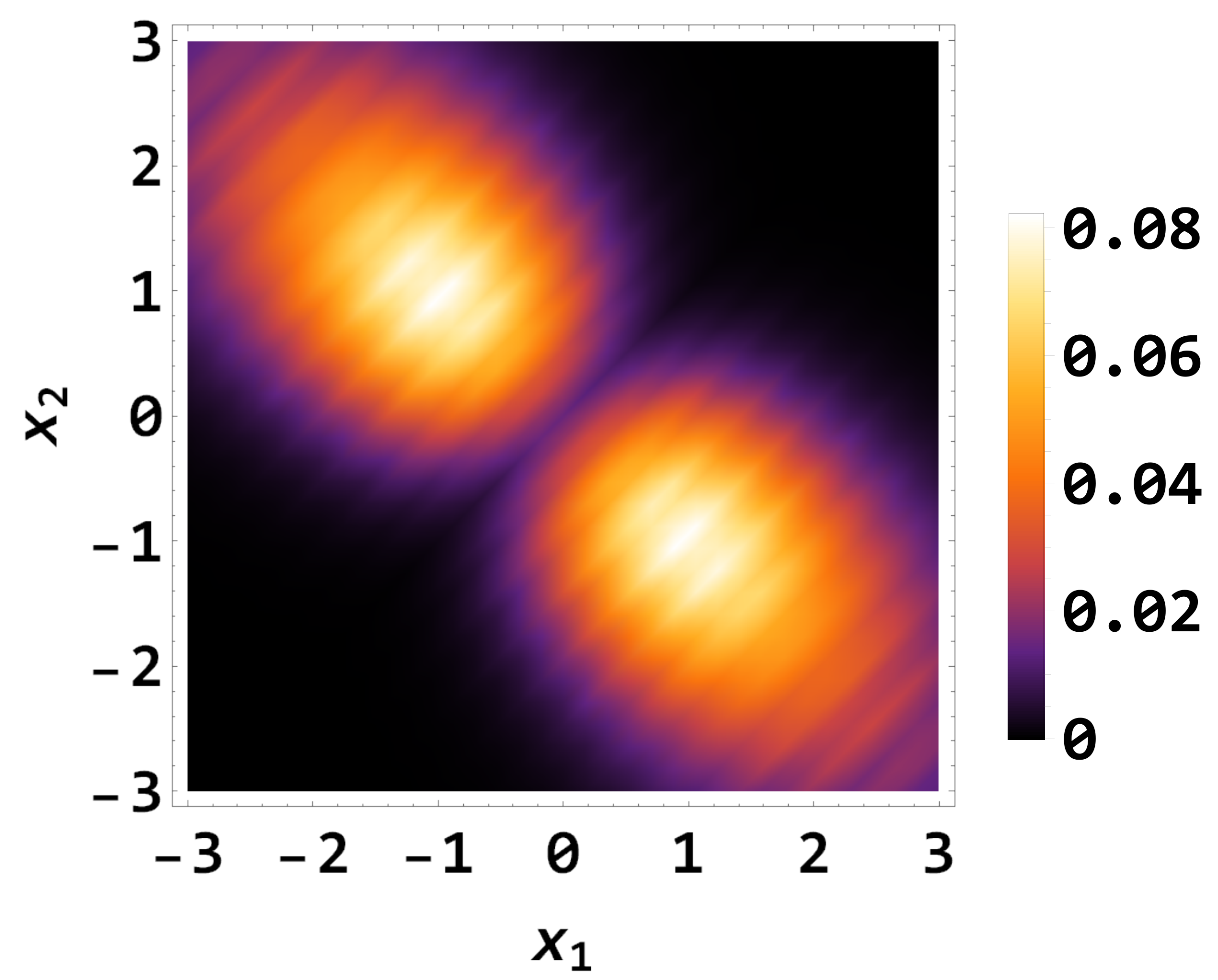} &
		\includegraphics[width=.15\textwidth]{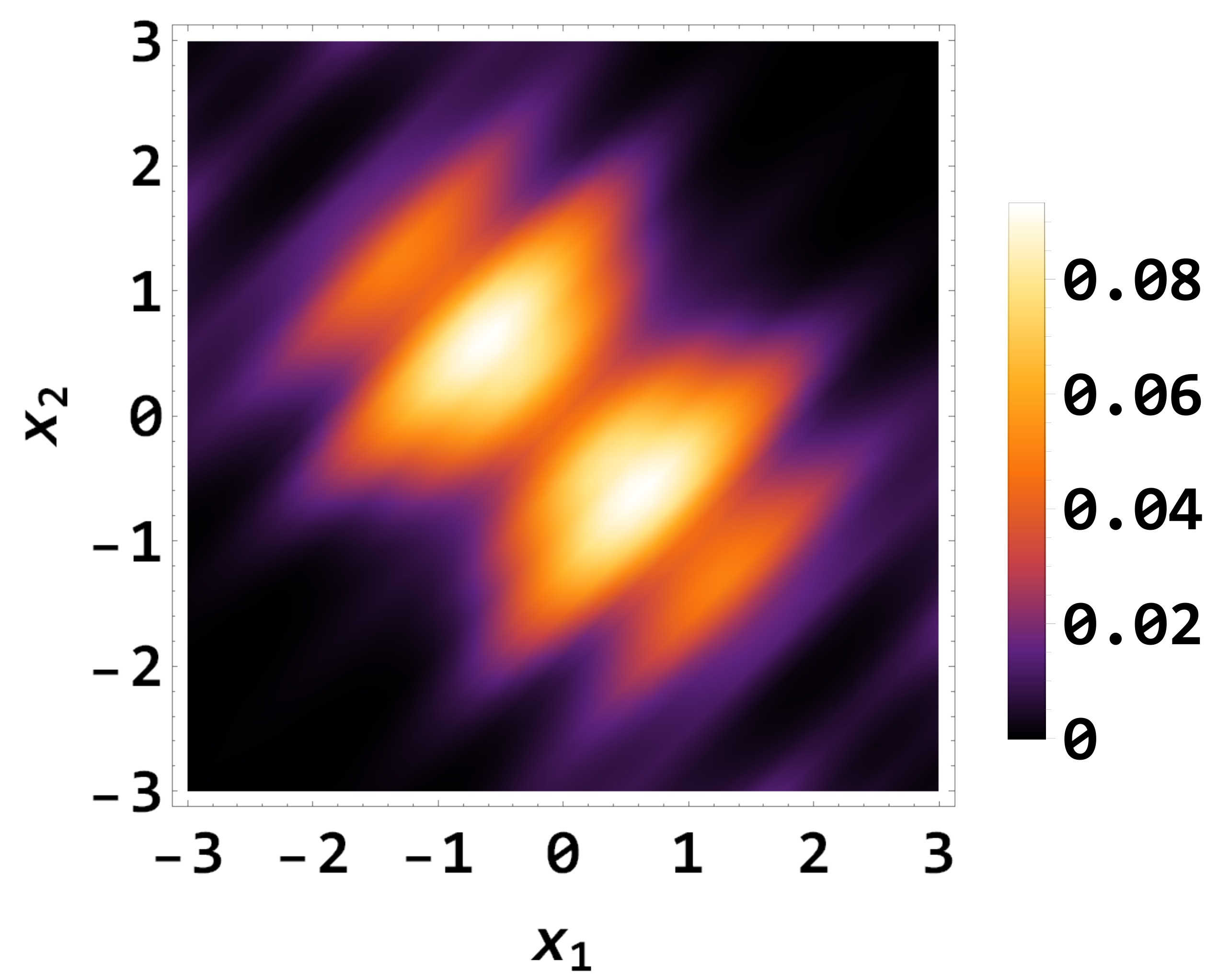} \\
	\end{tabular}
	\caption{(Color online) Evolution of the probability density $|\Psi(x_1,x_2,t)|^2$.}\label{bwd_al_wf}
\end{figure}

The evolution of the wave packet in the anharmonic case is shown in Fig.~\ref{bwd_al_wf}. As in the harmonic case, the wave packet of the initial attractive state has quite a narrow localization around the center. As time passes, the initial peak splits into two peaks, which spread and expand along the anti-diagonal axis, $x_1=-x_2$. Starting from $t=\pi/2$ one can notice some sort of ripples in the spatial distribution of the wave packet. This can be attributed to the additional reflections of the wave packet from both the interaction potential barrier at the center and anharmonic boundaries of the trap potential. At $t=\pi$ the shape of the time-evolving state is different from its initial spatial distribution, which has been already indicated by the fidelity dynamics above.

\begin{figure}
	\centering
	\textbf{$\alpha=-0.03$}\par\medskip
	\begin{tabular}{ccc}
		$g=-2,(0,0)$  & $g=-2,(2,0)$ & $g=-2,(4,0)$ \\
		\includegraphics[width=.15\textwidth]{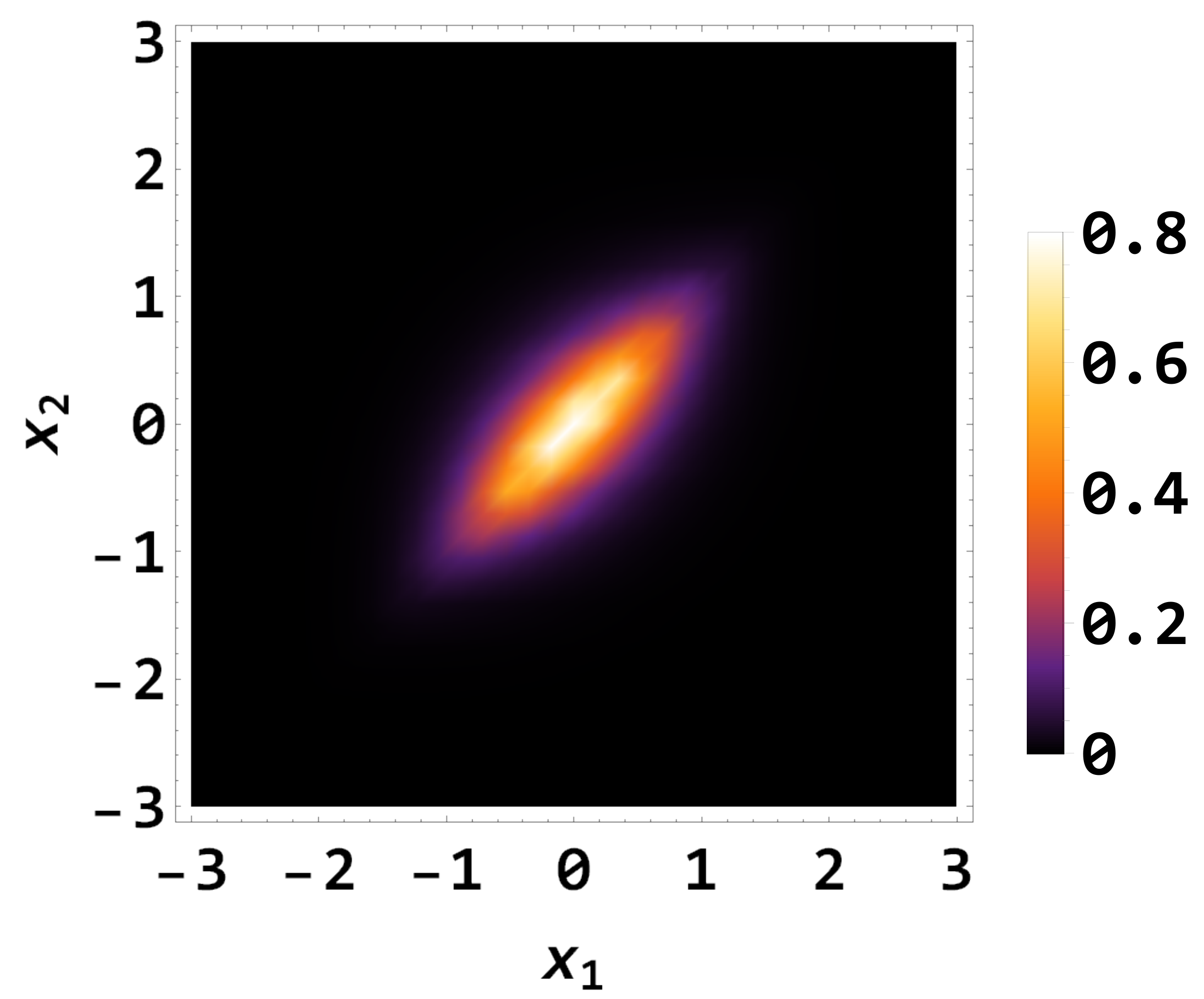} &
		\includegraphics[width=.15\textwidth]{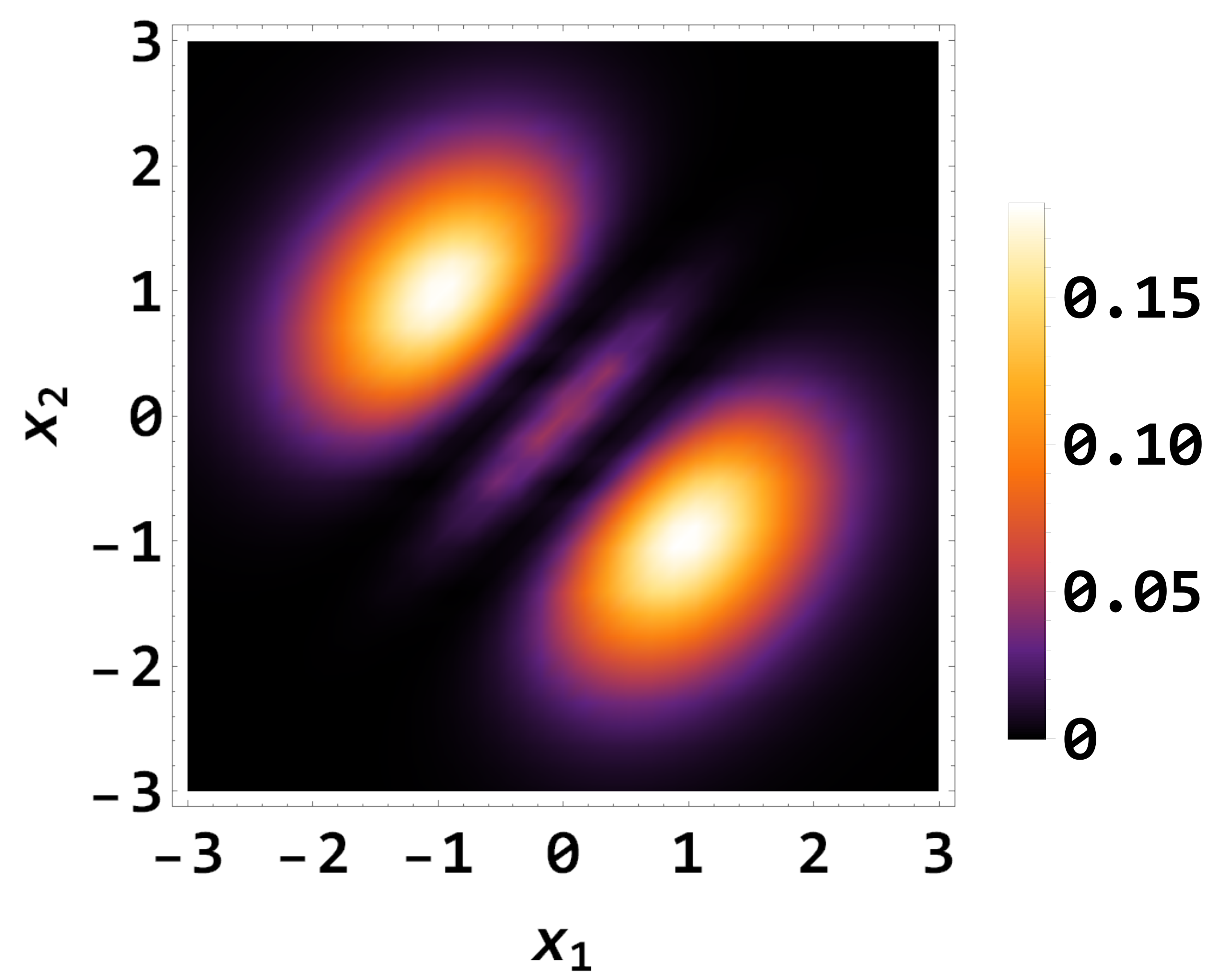} &
		\includegraphics[width=.15\textwidth]{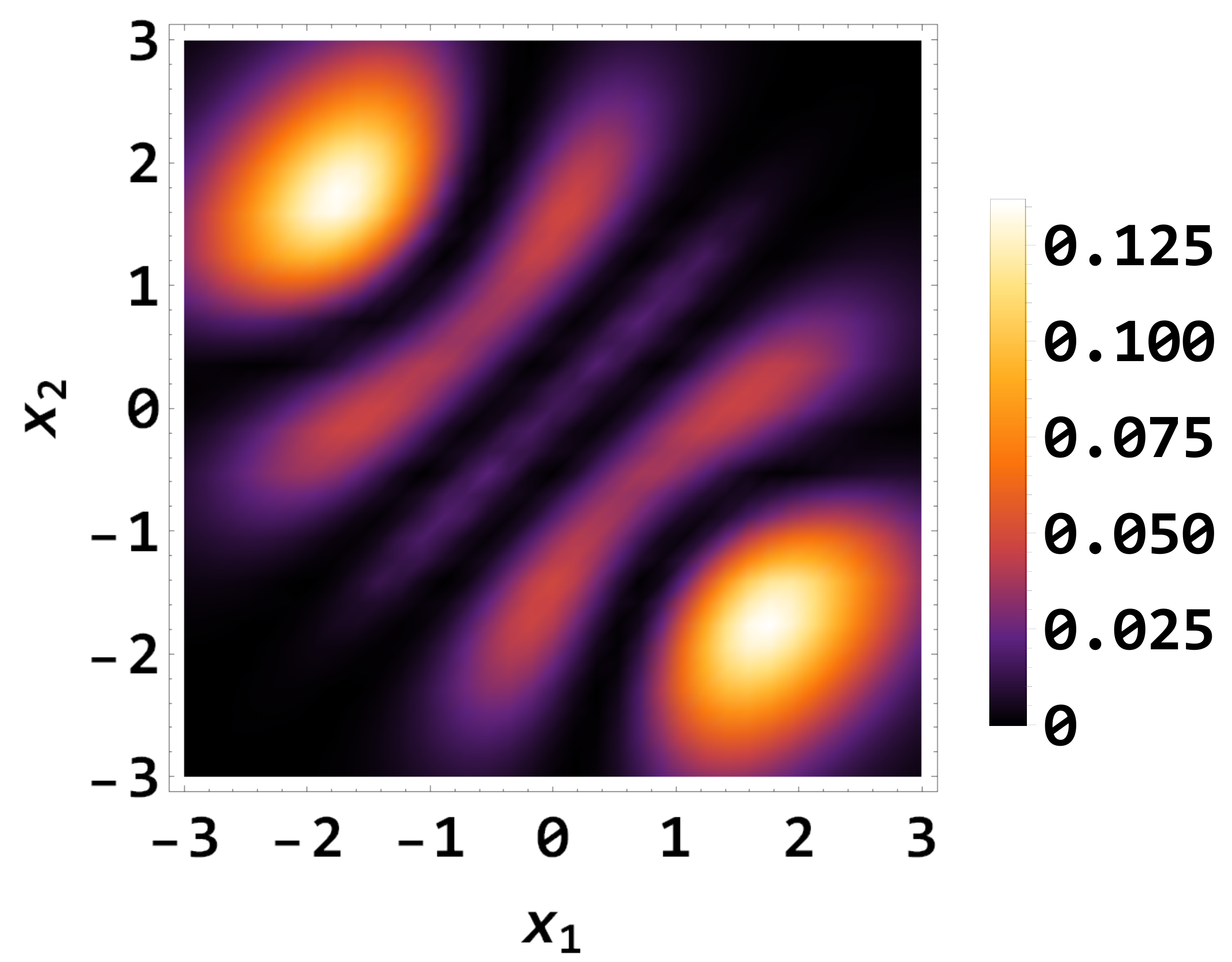} \\
		$g=2,(0,0)$  & $g=2,(2,0)$ & $g=2,(4,0)$ \\
		\includegraphics[width=.15\textwidth]{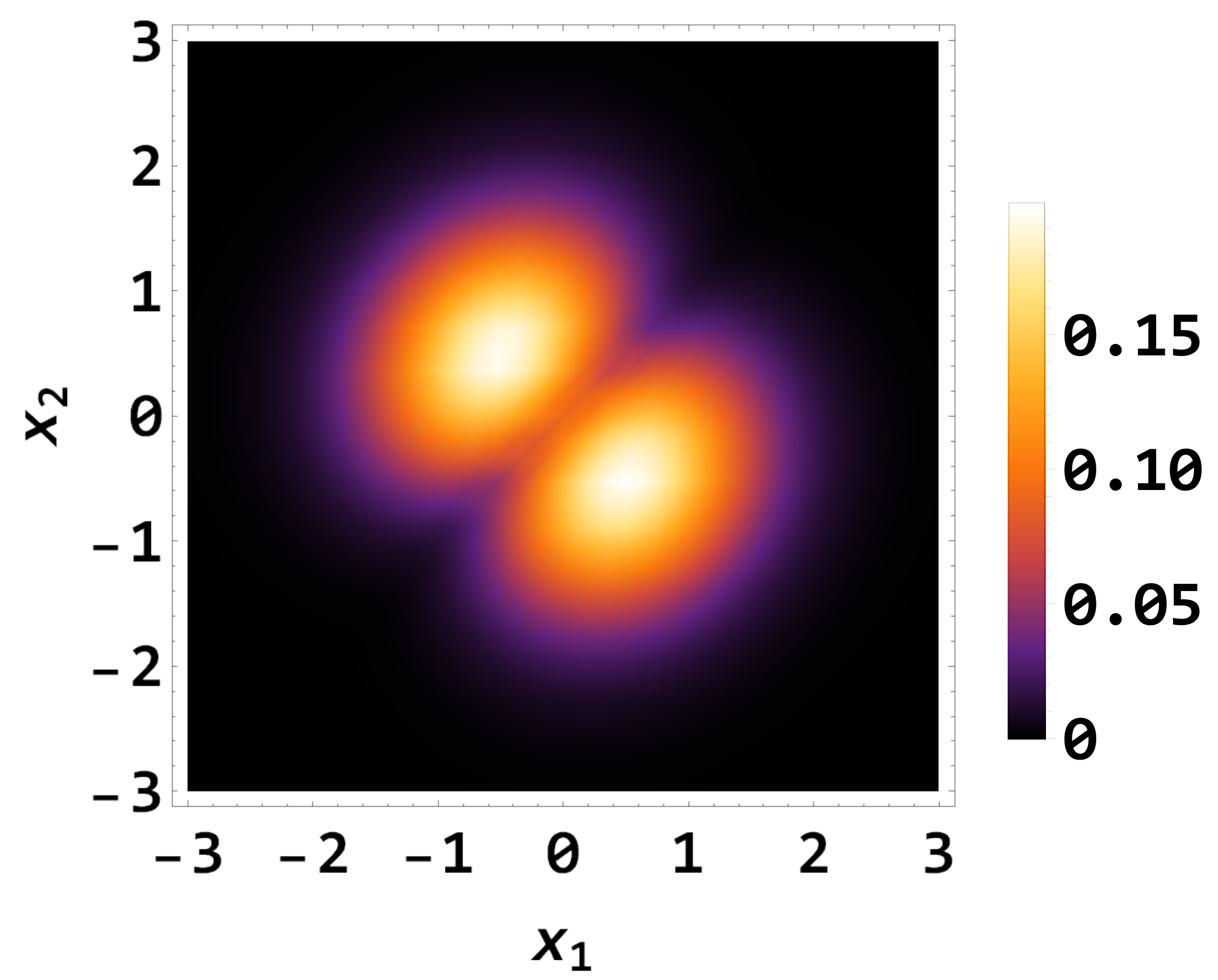} &
		\includegraphics[width=.15\textwidth]{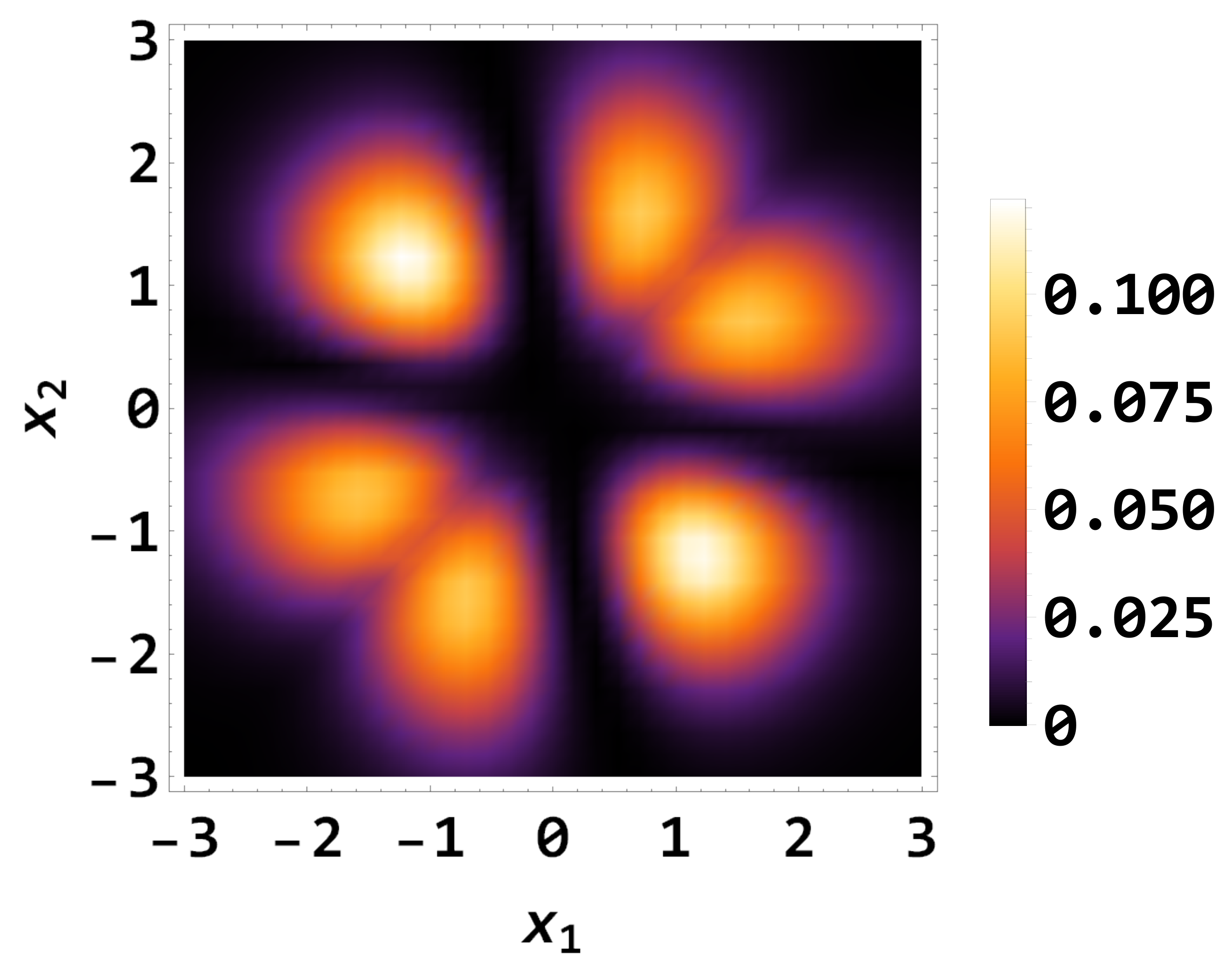} &
		\includegraphics[width=.15\textwidth]{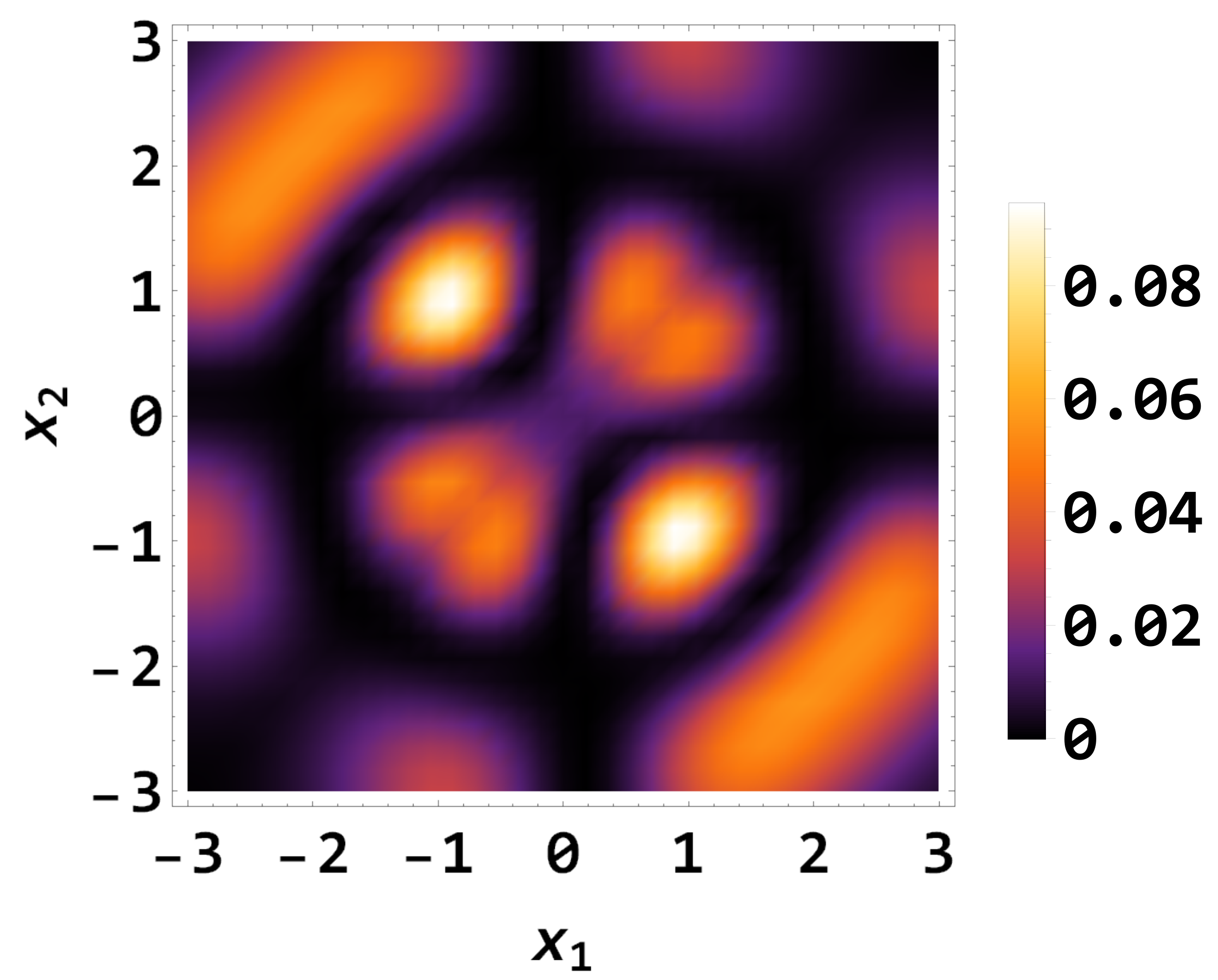} \\
	\end{tabular}
	\caption{(Color online) Probability density $|\Psi(x_1,x_2,t=0)|^2$ for different pre- and postquench stationary states.}\label{bwd_al_wf0}
\end{figure}

For the next thing we consider the probability densities $|\Psi(x_1,x_2,t=0)|^2$ of the pre- and postquench stationary ground and excited states in the case of the anharmonic trap (Fig.~\ref{bwd_al_wf0}). The prequench excited states extent along the anti-diagonal axis, having two pronounced humps on its edges. This behavior has some similarities with the distribution of the wave packet at $t\ge\pi/4$ (cf. Fig.~\ref{bwd_al_wf}). The postquench excited states spread in the both directions and have quite a complicated distribution of the humps, so it is not that easy to attribute these spatial distributions to the distributions in Fig.~\ref{bwd_al_wf}.

\begin{figure}
	\centering
	\textbf{$\alpha=-0.03$}\par\medskip
	\begin{tabular}{ccc}
		$t=0.1$ & $t=\pi/8$ & $t=\pi/4$ \\
		\includegraphics[width=.15\textwidth]{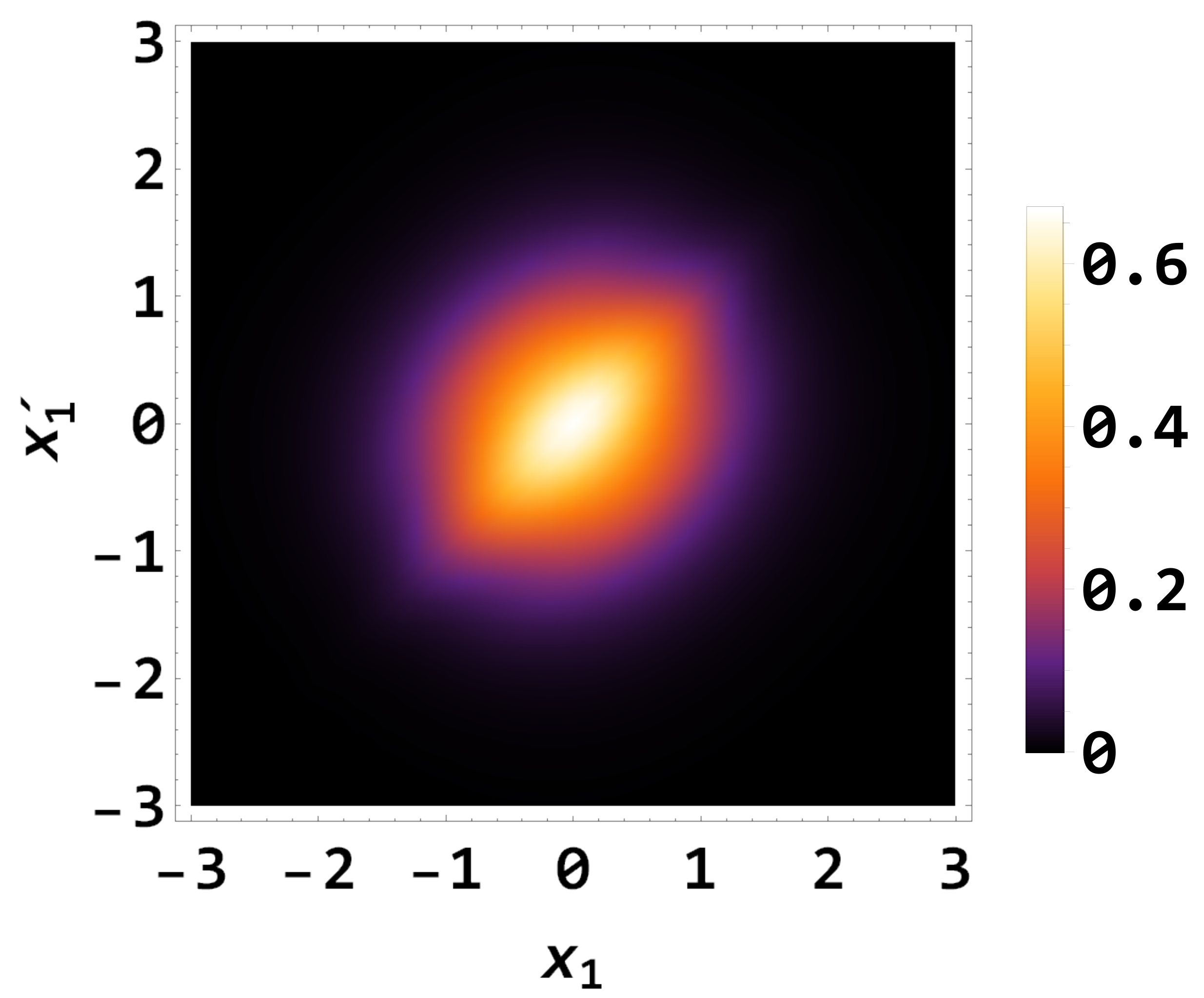} &
		\includegraphics[width=.15\textwidth]{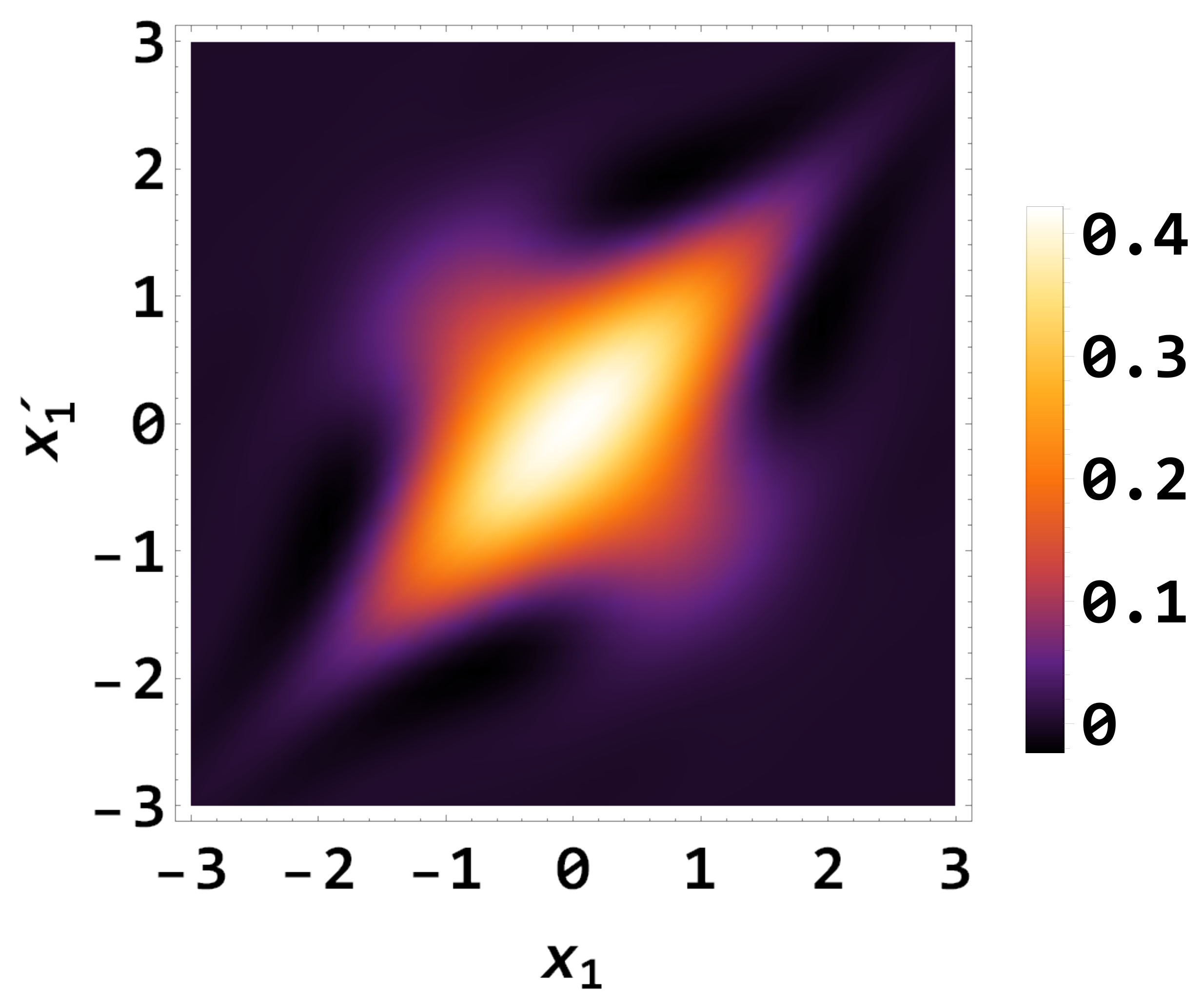} &
		\includegraphics[width=.15\textwidth]{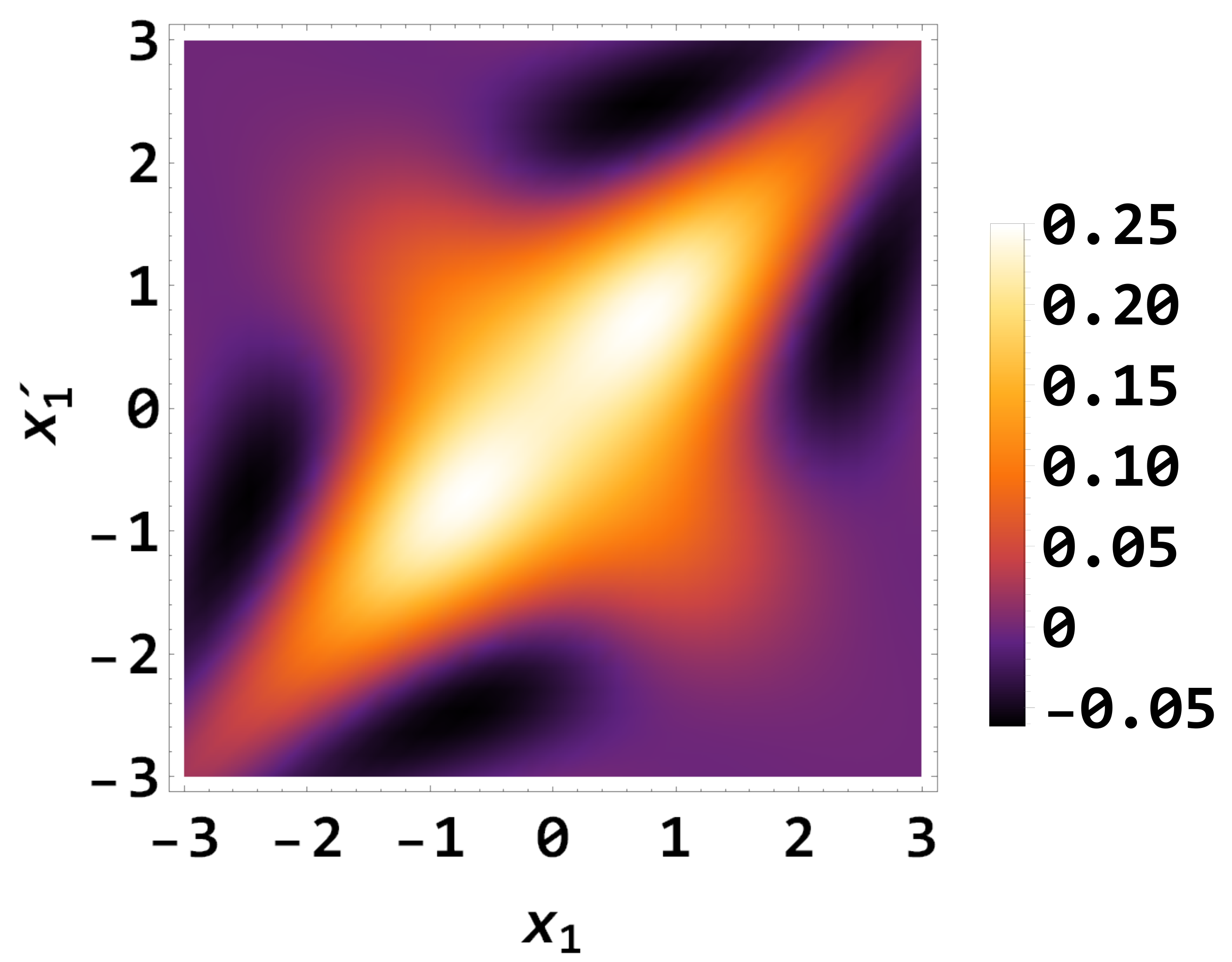} \\
		$t=\pi/2$ & $t=2$ &$t=\pi$ \\
		\includegraphics[width=.15\textwidth]{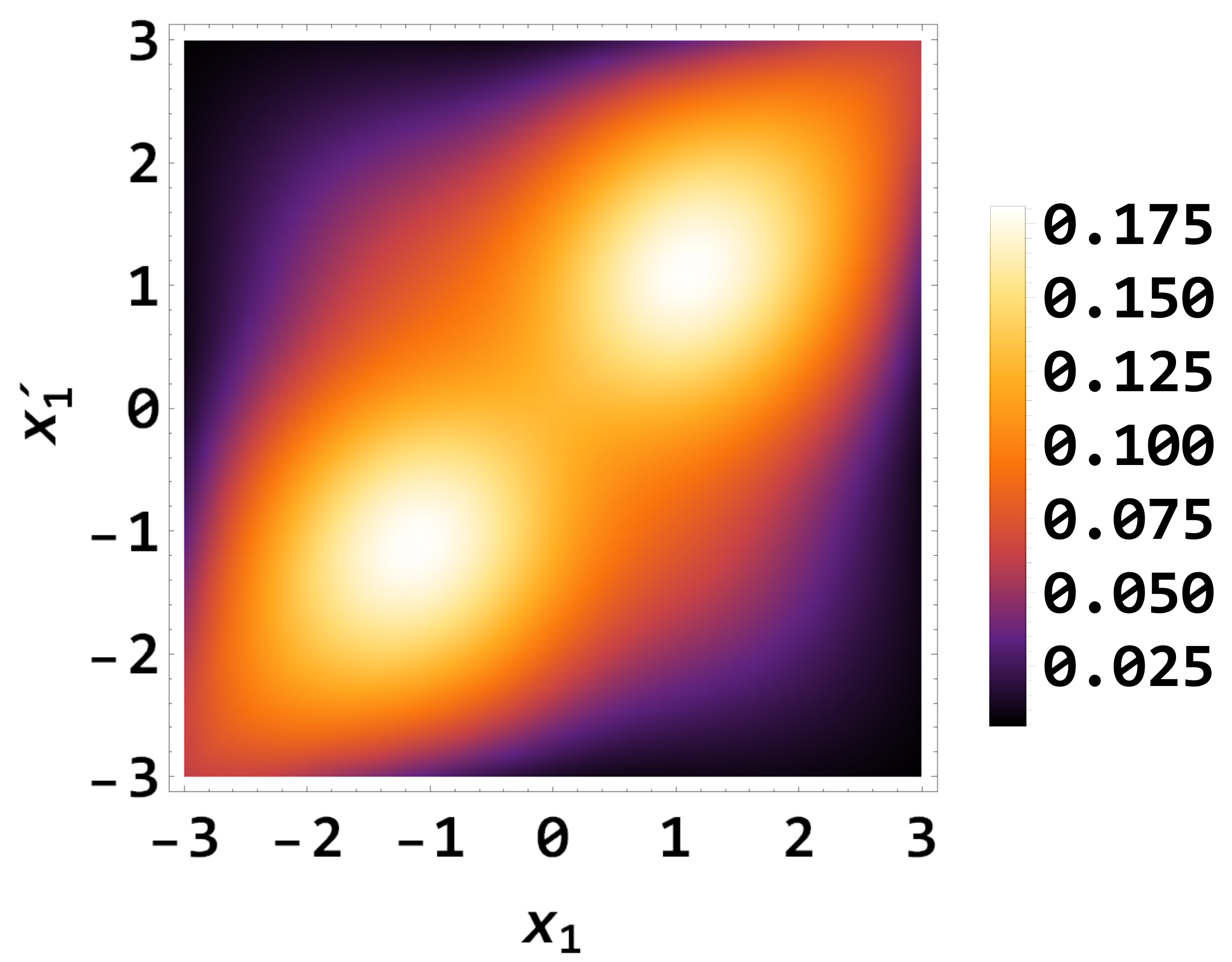} &
		\includegraphics[width=.15\textwidth]{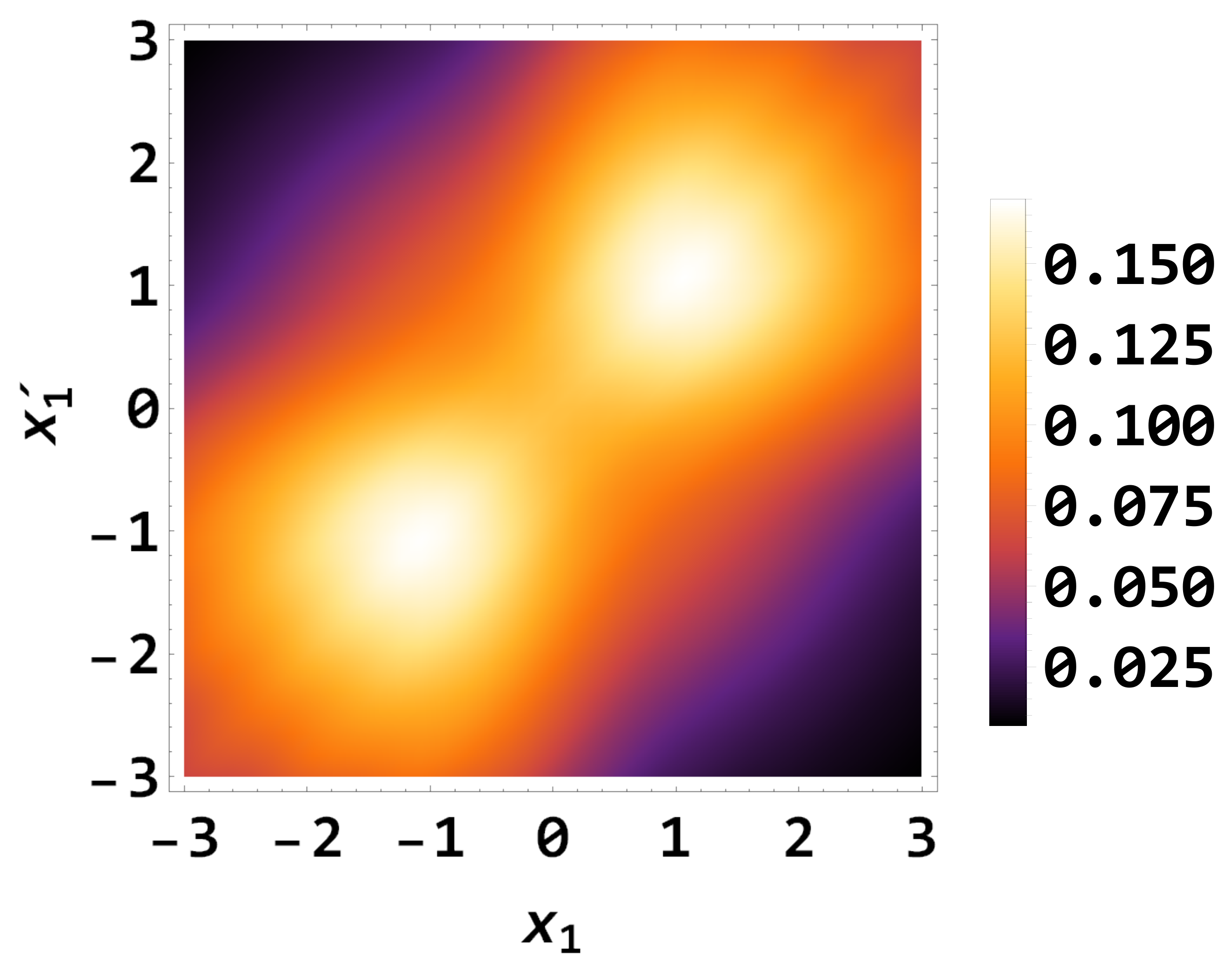} &
		\includegraphics[width=.15\textwidth]{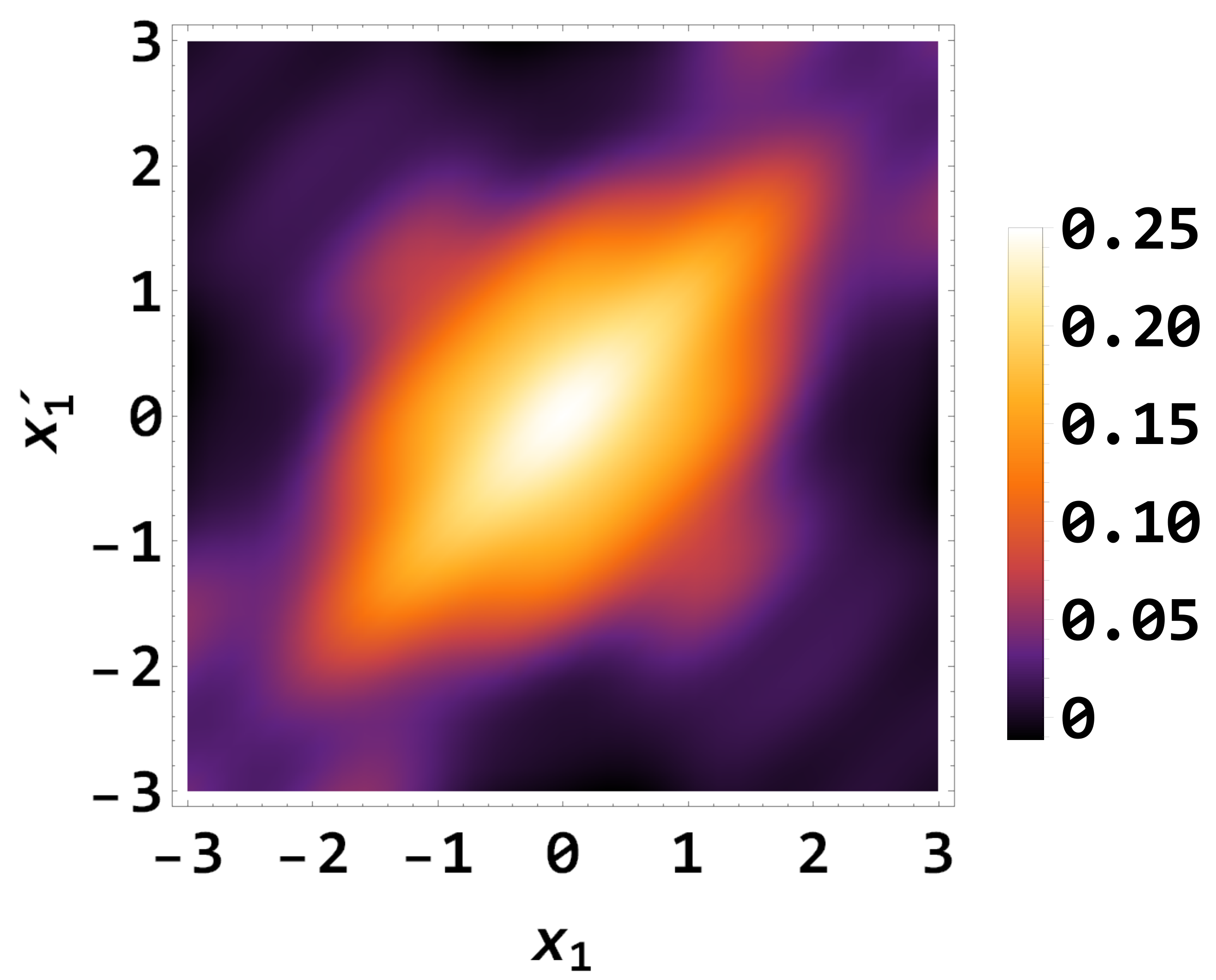} \\
	\end{tabular}
	\caption{(Color online) One-body reduced density matrix, $\rho^{(1)}(x_1,x_1';t)$, at different time-instants.}\label{bwd_al_rho}
\end{figure}

The evolution of the one-body reduced density matrix $\rho^{(1)}(x_1,x_1';t)$ (Fig.~\ref{bwd_al_rho}) shows a splitting of the initial peak into two humps along the diagonal axis. This indicates spatial delocalization of a particle, which is due to the repulsive interaction between the particles. By the time $t=\pi$ the spatial distribution of $\rho^{(1)}(x_1,x_1';t)$ contracts into a single hump, yet not having the same shape of the initial peak.

\begin{figure}
	\centering
	\textbf{$\alpha=-0.03$}\par\medskip
	\begin{tabular}{ccc}
		$g=-2,(0,0)$  & $g=-2,(2,0)$ & $g=-2,(4,0)$ \\
		\includegraphics[width=.15\textwidth]{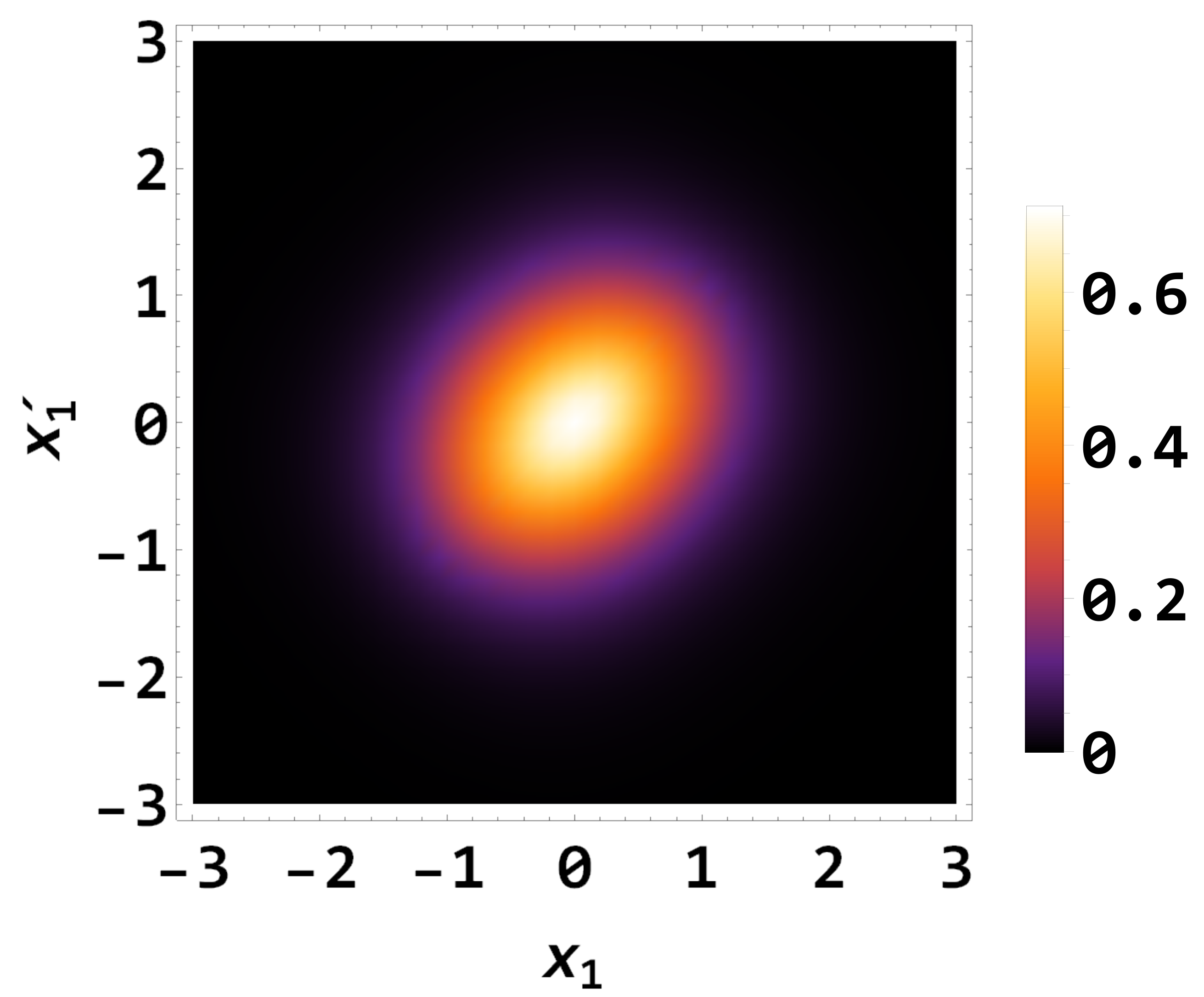} &
		\includegraphics[width=.15\textwidth]{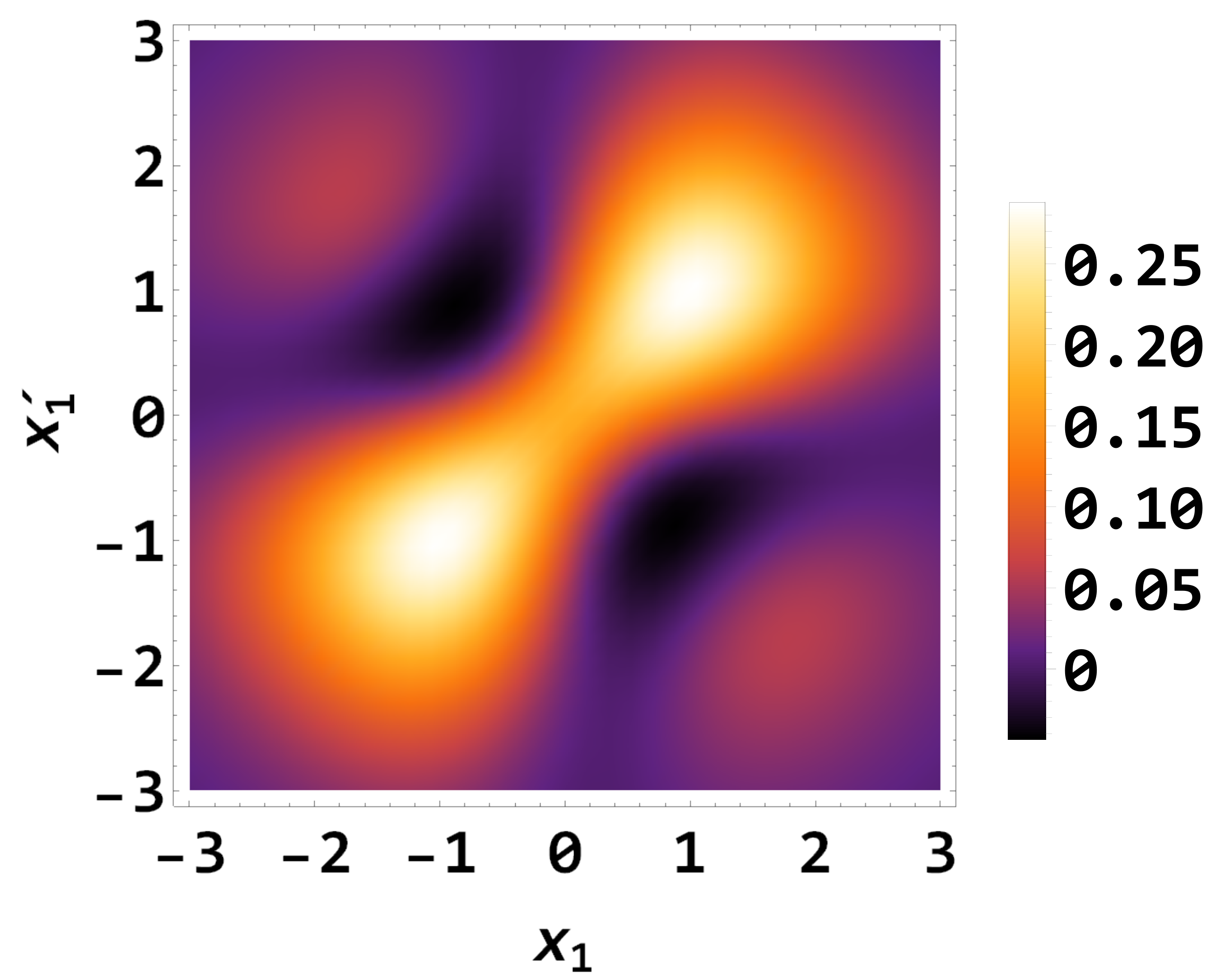} &
		\includegraphics[width=.15\textwidth]{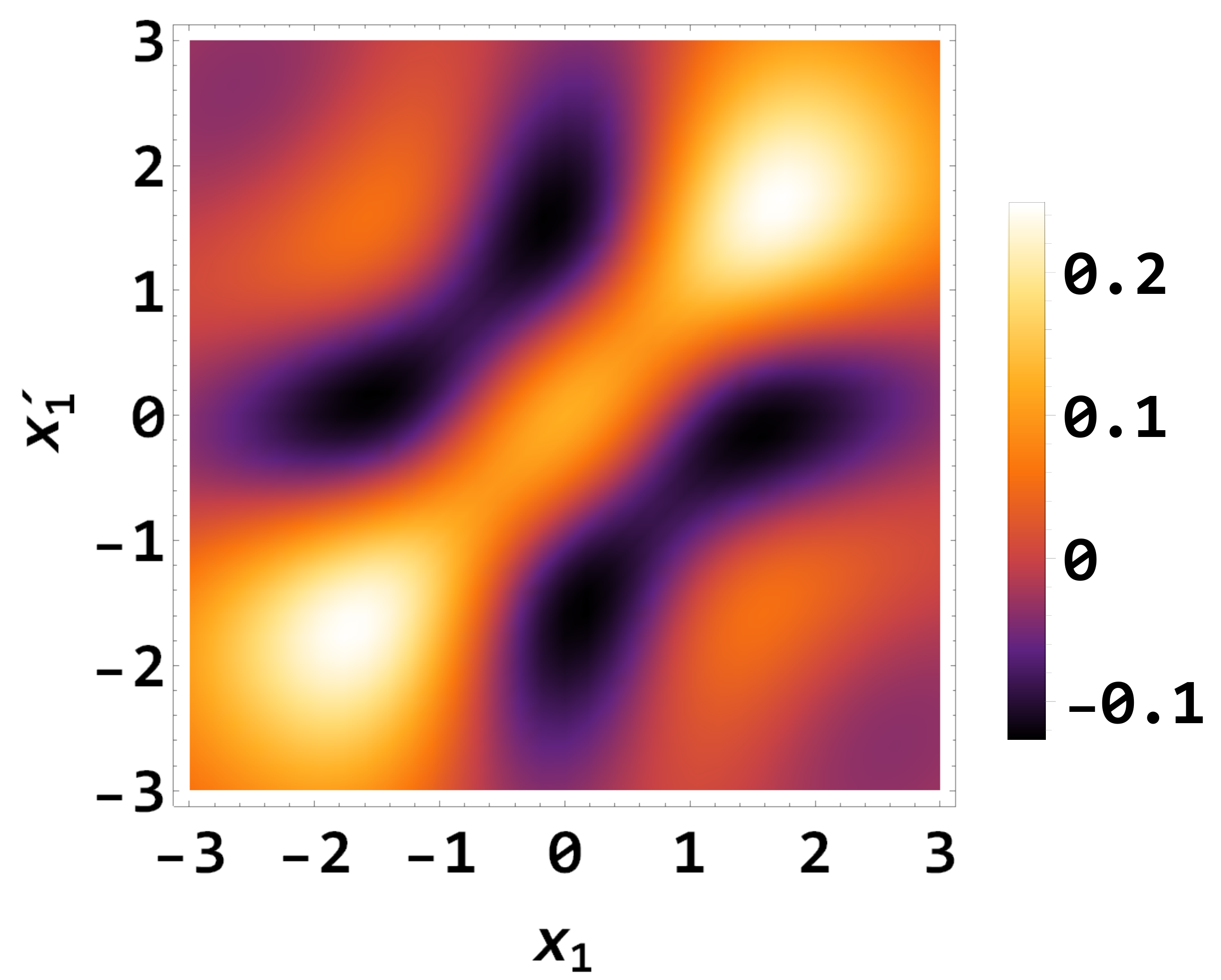} \\
		$g=2,(0,0)$  & $g=2,(2,0)$ & $g=2,(4,0)$ \\
		\includegraphics[width=.15\textwidth]{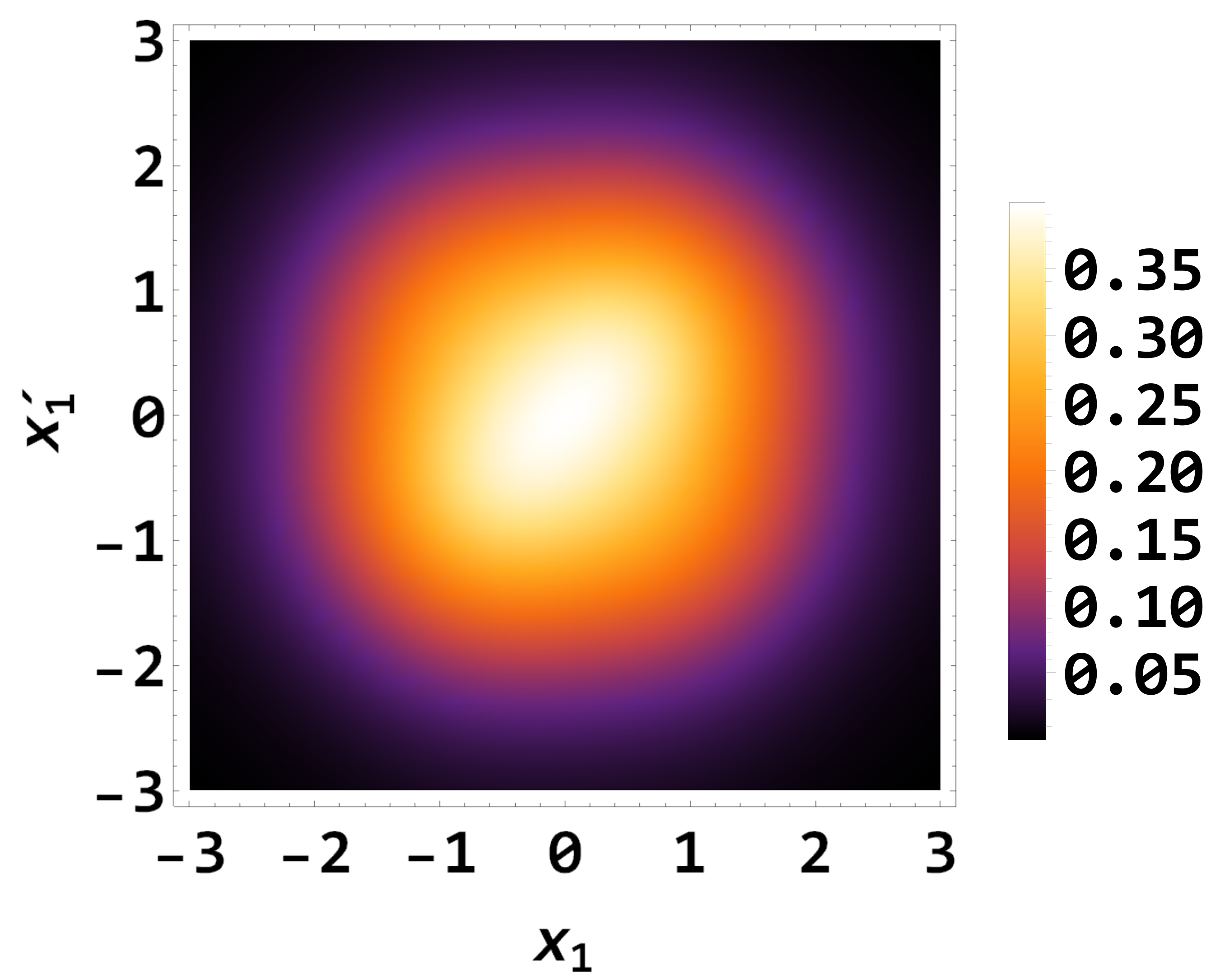} &
		\includegraphics[width=.15\textwidth]{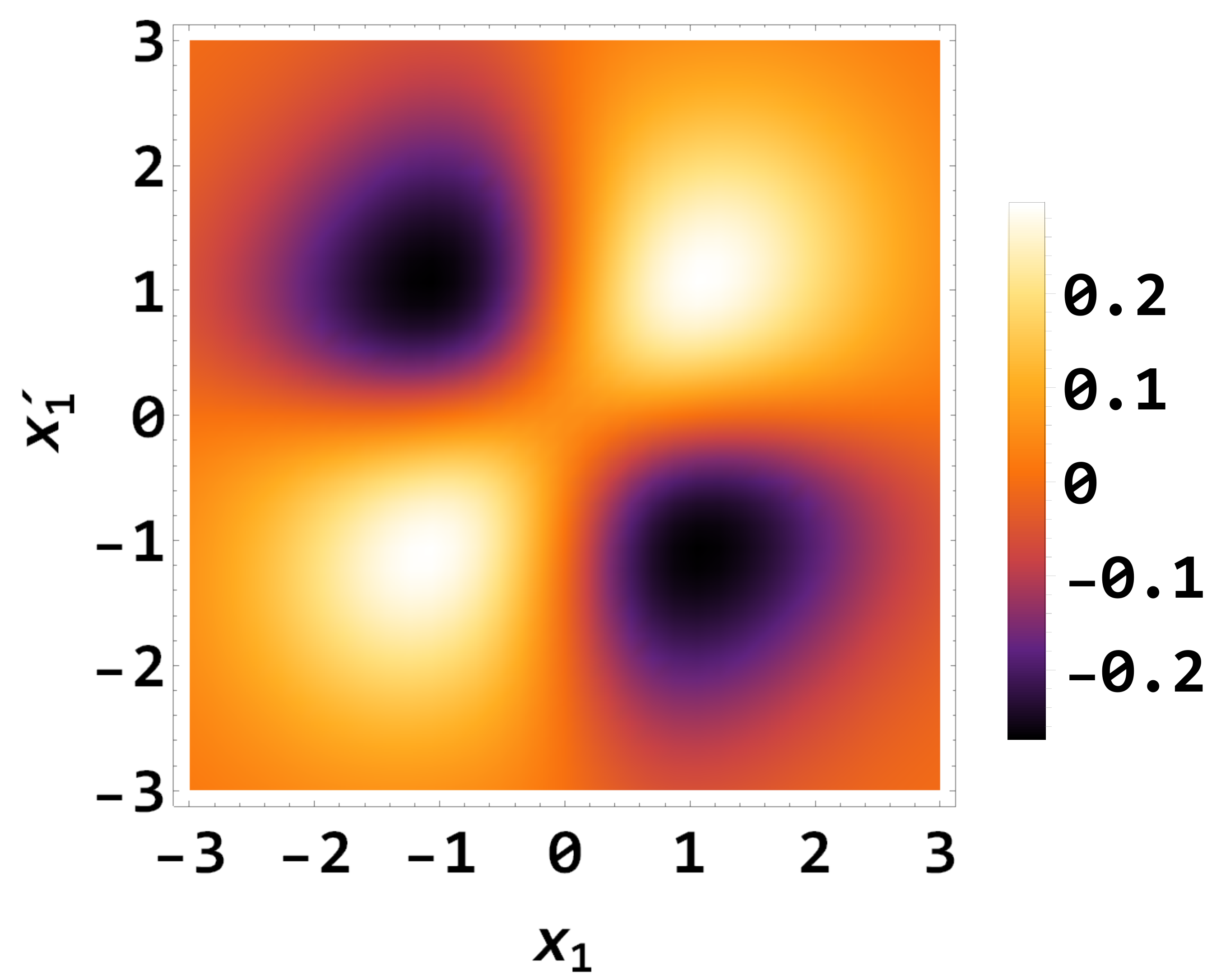} &
		\includegraphics[width=.15\textwidth]{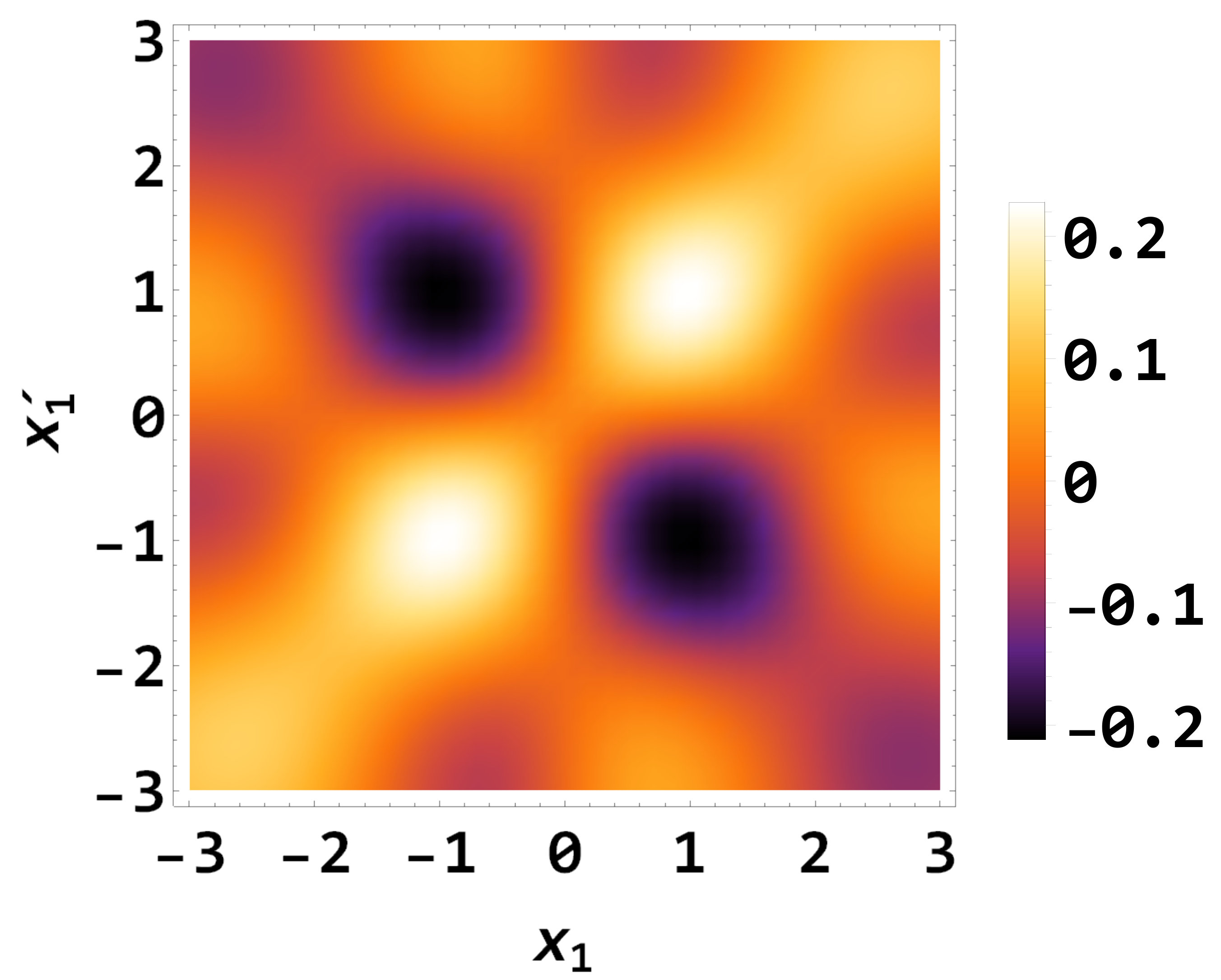} \\
	\end{tabular}
	\caption{(Color online) One-body reduced density matrix, $\rho^{(1)}(x_1,x_1';t=0)$, for different stationary states.}\label{bwd_al_irho}
\end{figure}

Figure~\ref{bwd_al_irho} shows the one-body reduced density matrix $\rho^{(1)}(x_1,x_1';t=0)$ for the pre- and postquench stationary states in the case of the anharmonic trap. The difference of spatial distributions of $\rho^{(1)}(x_1,x_1';t=0)$ from the distributions in the case of the harmonic trap are mainly noticeable for the postquench excited states. The once elongated shapes of $\rho^{(1)}(x_1,x_1';t=0)$ for these states are now break into several pronounced humps. The two-hump structure of $\rho^{(1)}(x_1,x_1';t=0)$ for the prequench excited state $(2,0)$ is quite similar to that of $\rho^{(1)}(x_1,x_1';t)$ at $t=\pi/2$ (cf. Fig.~\ref{bwd_al_rho}). 

\begin{figure}
	\centering
	\textbf{$\alpha=-0.03$}\par\medskip
	\includegraphics[width=7cm,clip]{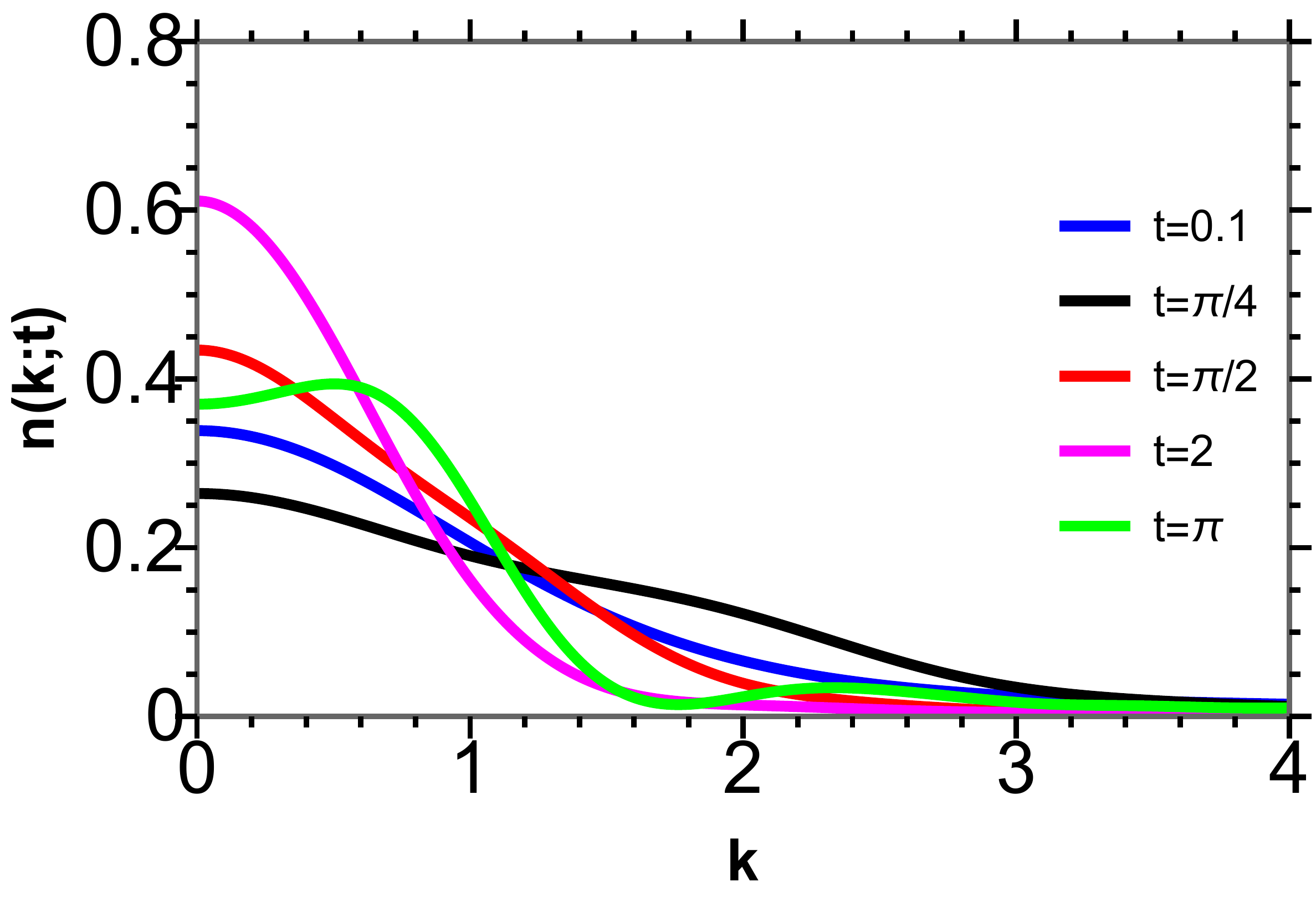}
	\caption{(Color online) Evolution of the momentum distribution $n(k;t)$ for the anharmonic trap.}\label{bwd_al_nk_all}
\end{figure}

The time dynamics of the momentum distribution $n(k;t)$ in the case of the anharmonic trap is shown in Fig.~\ref{bwd_al_nk_all}. The initial shape, at $t=0.1$ (blue line), has a small zero-momentum peak, which, as mentioned above in the case of the harmonic trap, is the result of the initial attractive interaction between the particles. By the time $t=2$ the peak (magenta line) reaches a large value due to the larger spatial expansion of the one-body reduced density matrix $\rho^{(1)}(x_1,x_1';t)$ (cf. Fig.~\ref{bwd_al_rho}). Then $\rho^{(1)}(x_1,x_1';t)$ reduces in size which leads to a reduction of the zero-momentum peak of $n(k;t)$ by the time $t=\pi$ (green line).

\begin{figure}
	\centering
	\textbf{$\alpha=-0.03$,~~~$t=\pi/2$}\par\medskip
	\begin{tabular}{cc}
		Prequench states &  Postquench states \\
		\hspace{-.5cm}\includegraphics[width=.25\textwidth]{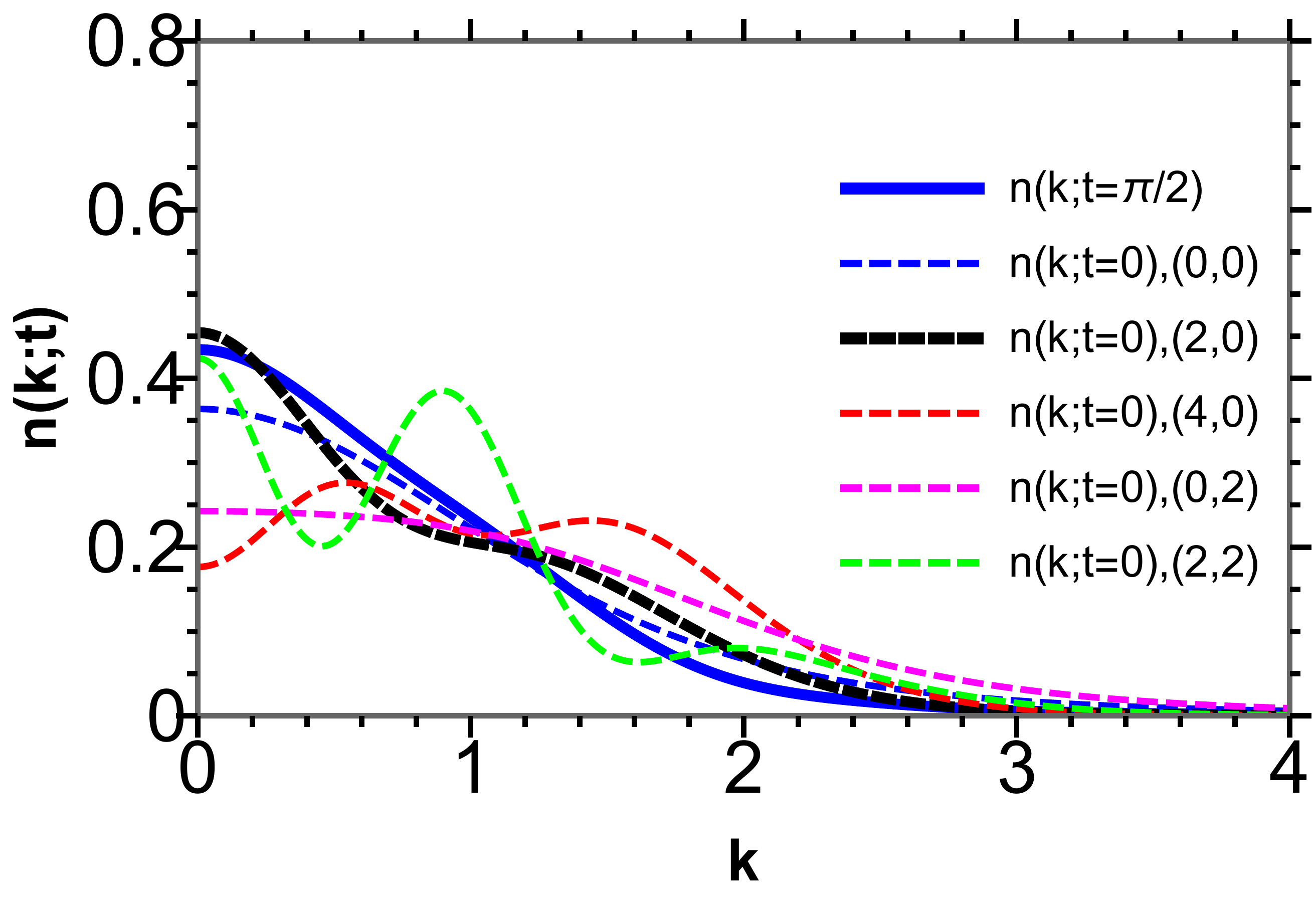} &
		\includegraphics[width=.25\textwidth]{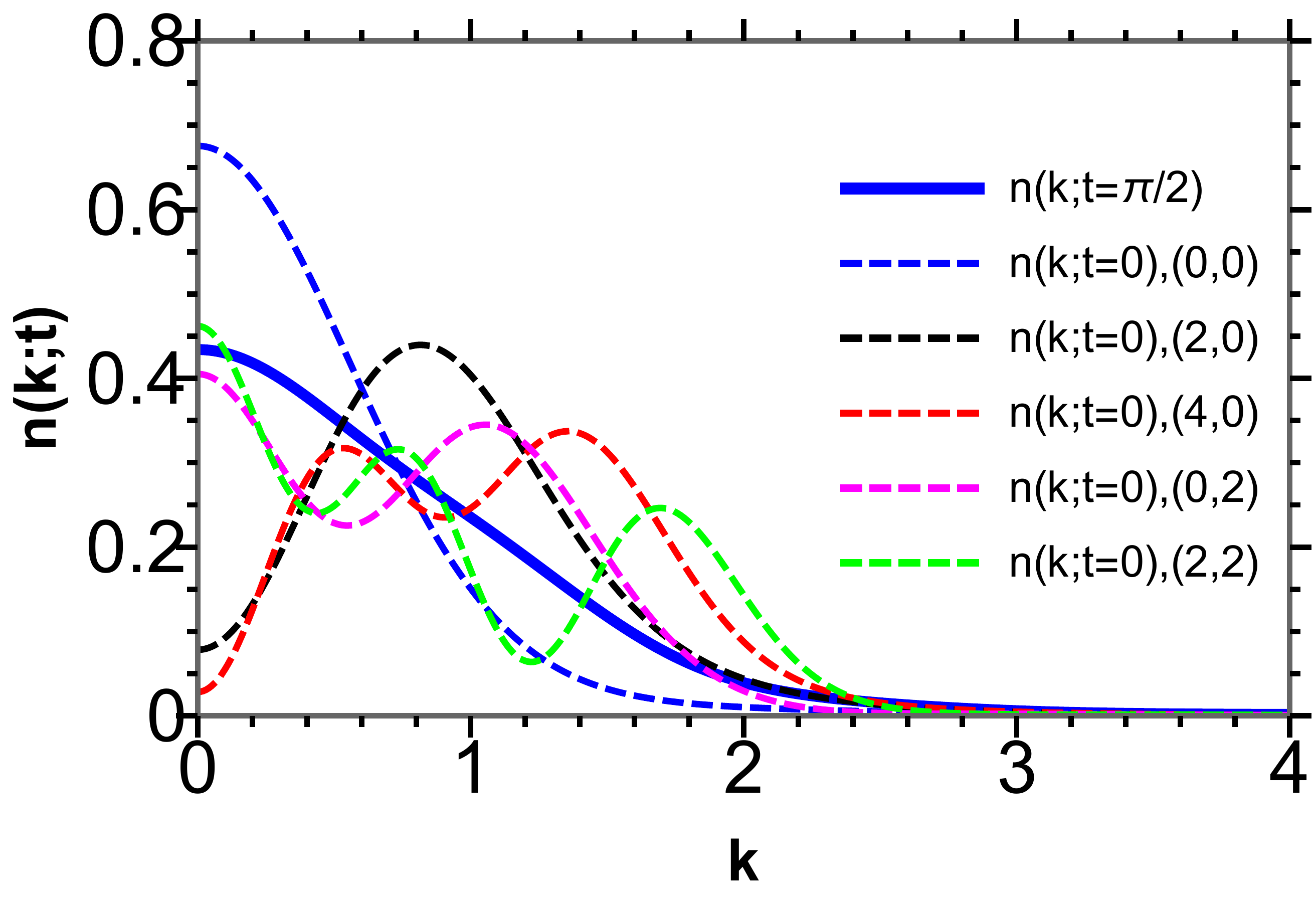} \\
	\end{tabular}
	\caption{(Color online) Comparison of the momentum distribution $n(k;t)$ (blue line) with the momentum distributions $n(k;t=0)$ for the pre- and postquench states (dashed lines).}\label{bwd_al_nk_12}
\end{figure}

\begin{figure}
	\centering
	\textbf{$\alpha=-0.03$,~~~$t=2$}\par\medskip
	\begin{tabular}{cc}
		Prequench states &  Postquench states \\
		\hspace{-.5cm}\includegraphics[width=.25\textwidth]{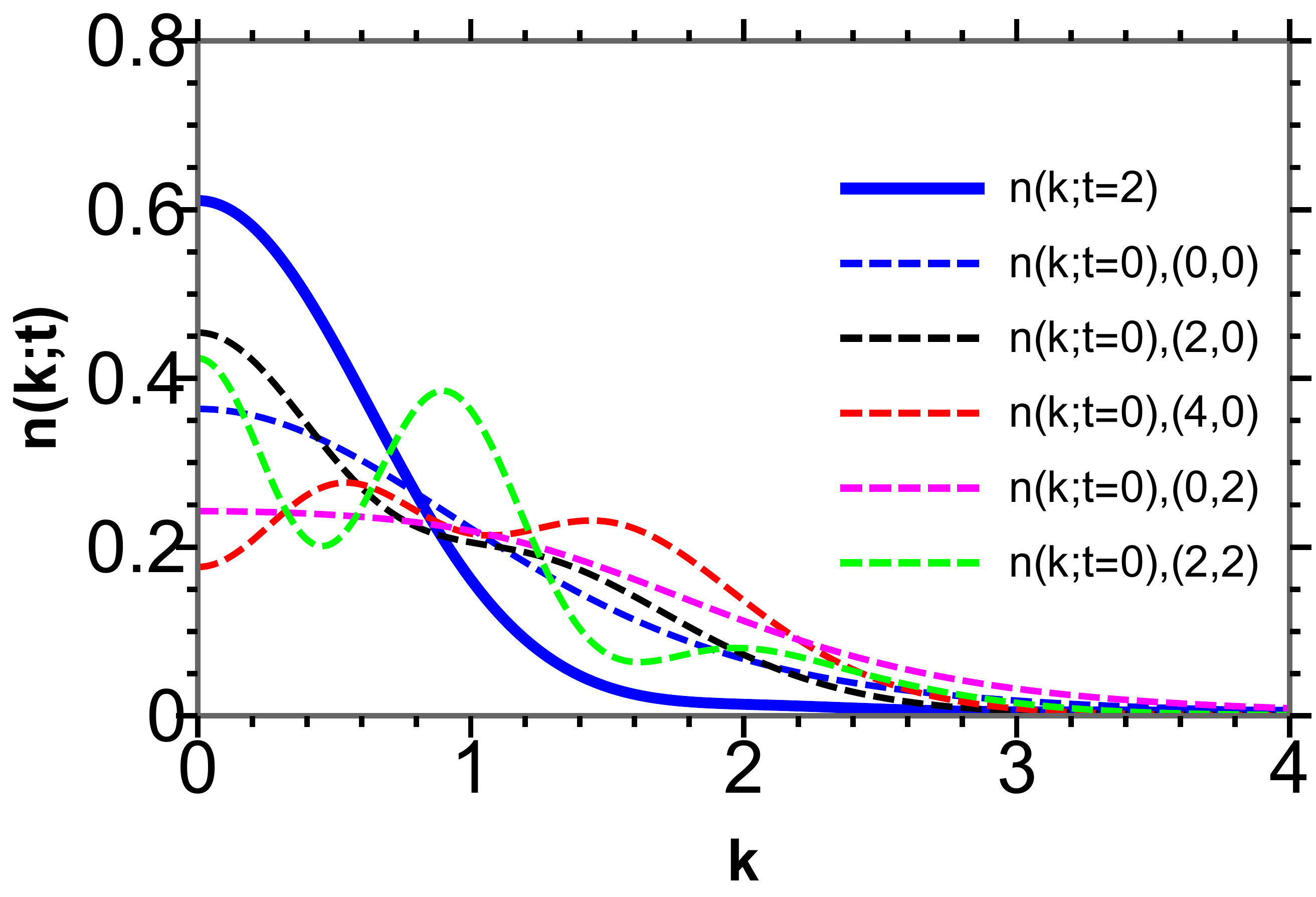} &
		\includegraphics[width=.25\textwidth]{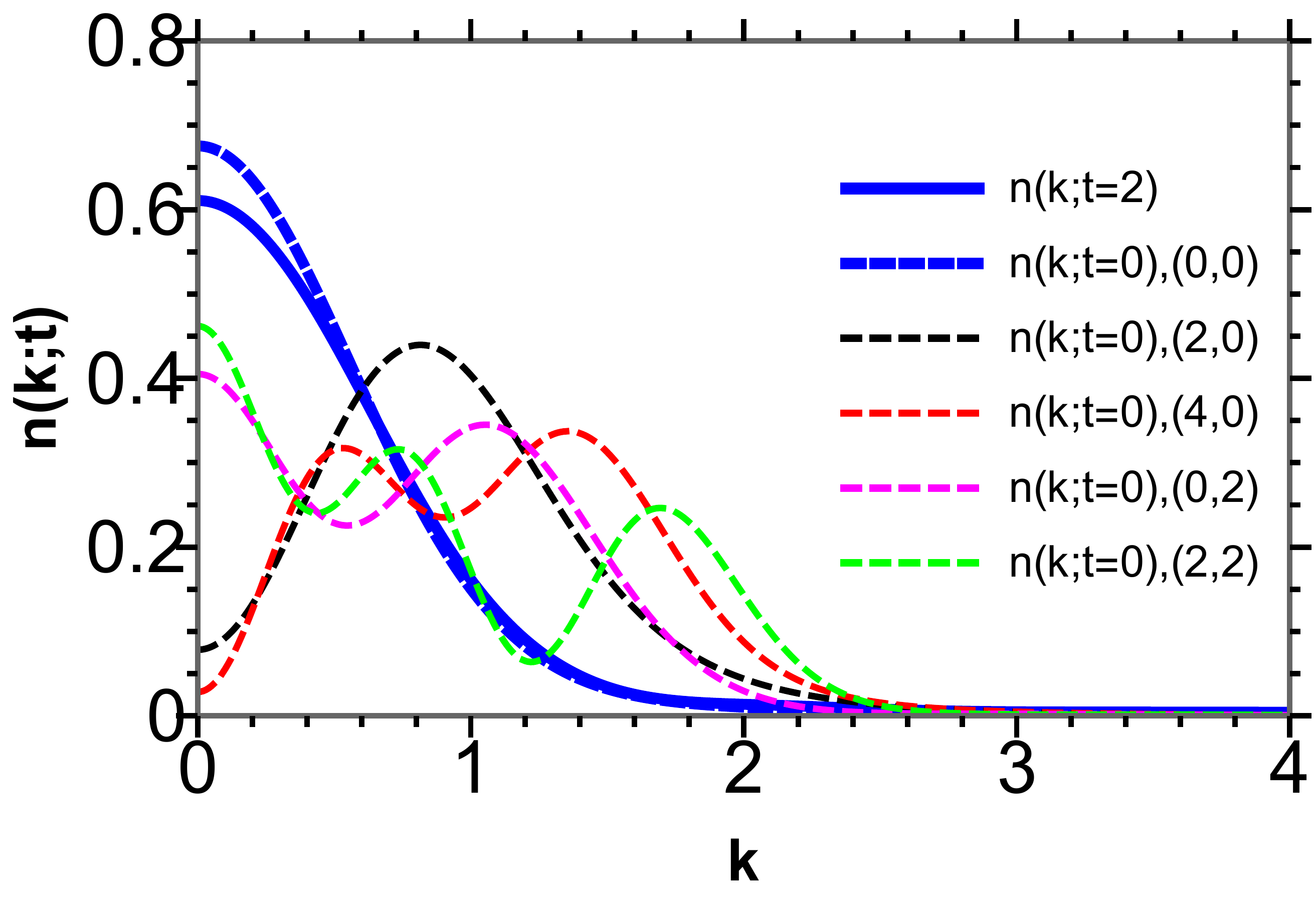} \\
	\end{tabular}
	\caption{(Color online) Comparison of the momentum distribution $n(k;t)$ (blue line) with the momentum distributions $n(k;t=0)$ for the pre- and postquench states (dashed lines).}\label{bwd_al_nk_34}
\end{figure}

We pick two moments of time $t=\pi/2$ and $t=2$ on which it could be more clearly to match the shapes of $n(k;t)$ with the shapes of $n(k;t=0)$ for the pre- and postquench states. In Fig.~\ref{bwd_al_nk_12} we can see that from all of the considered shapes of $n(k;t=0)$ we can approximately match the shape of $n(k;t)$ at $t=\pi/2$ (blue line) with that of $n(k;t=0)$ for the prequench excited state $(2,0)$ (black dashed line). In the next moment $t=2$ in Fig.~\ref{bwd_al_nk_34} the shape of $n(k;t)$ (blue line) is very close to the shape of $n(k;t=0)$ for the postquench ground state (blue dashed line). Both these cases are quite in a agreement with the overlap dynamics in Fig.~\ref{bwd_al}, where at these instants the overlaps between the wave packet and the considered states are large.

\subsection{Harmonic trap, $\alpha=0$, the excited state, $(2,0)$}

\begin{figure}
	\centering
	\textbf{$\alpha=0$}\par\medskip
	\includegraphics[width=8.5cm,clip]{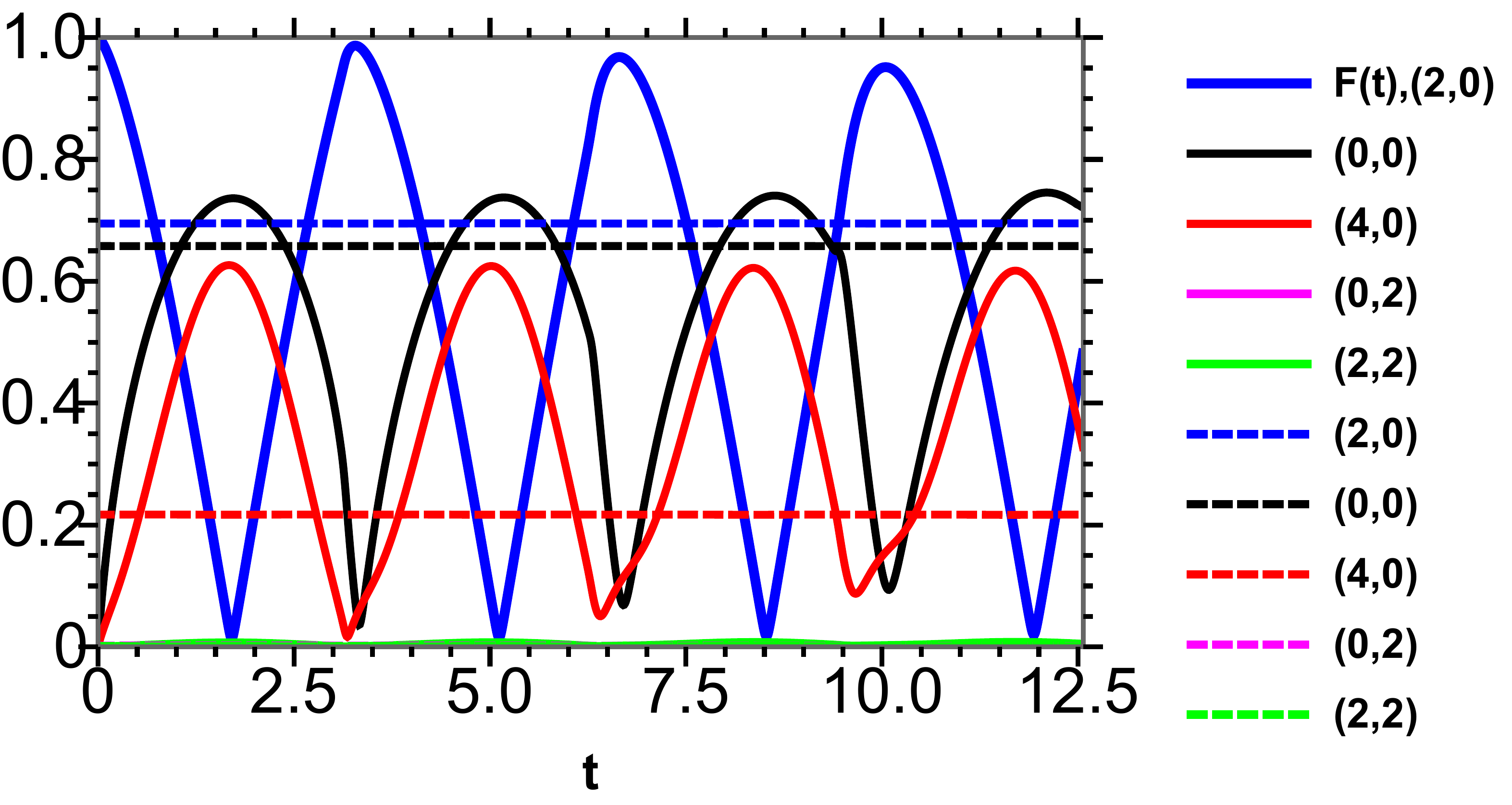}
	\caption{(Color online) Fidelity $F(t)$ and the overlap integrals $\mathcal{Q}$ between the time-evolving state $\Psi(x_1,x_2,t)$ and different pre- (solid lines) and postquench (dashed lines) states in the case of the harmonic trap. The indices $(n,N)$ refer to the states with the quantum numbers of the relative and center-of-mass motions. The system of the two atoms is prepared in the excited state $(2,0)$ with $g=-2$ and quenched to $g=2$.}\label{bwd_exc}
\end{figure}

Let us turn to the quench dynamics starting from the excited state $(2,0)$ in the case of the harmonic trap (Fig.~\ref{bwd_exc}). The fidelity $F(t)$ oscillates in an almost periodic manner with a period of $\pi$. In comparison with the Fig.~\ref{bwd} damping here is less pronounced. The reason for this is that the original positions of the particles here are mainly located farther from the center (cf. Fig.~\ref{bwd_wf0}), where the influence of the repulsive barrier is high. Thus, the particles experience less trouble travelling back to their initial positions. The fidelity and the overlaps between the wave packet and the prequench ground state and excited state $(4,0)$ oscillates approximately in antiphase. The overlap for the prequench ground state is the same as the overlap for the prequench excited state $(2,0)$ in the case of the quench dynamics from the ground state (cf. Fig.~\ref{bwd}). This is due to the Eq.~\eqref{ovreq}. The overlap for the postquench ground state and excited state $(2,0)$ are close to each other and quite large.

\begin{figure}
	\centering
	\textbf{$\alpha=0$}\par\medskip
	\includegraphics[width=7cm,clip]{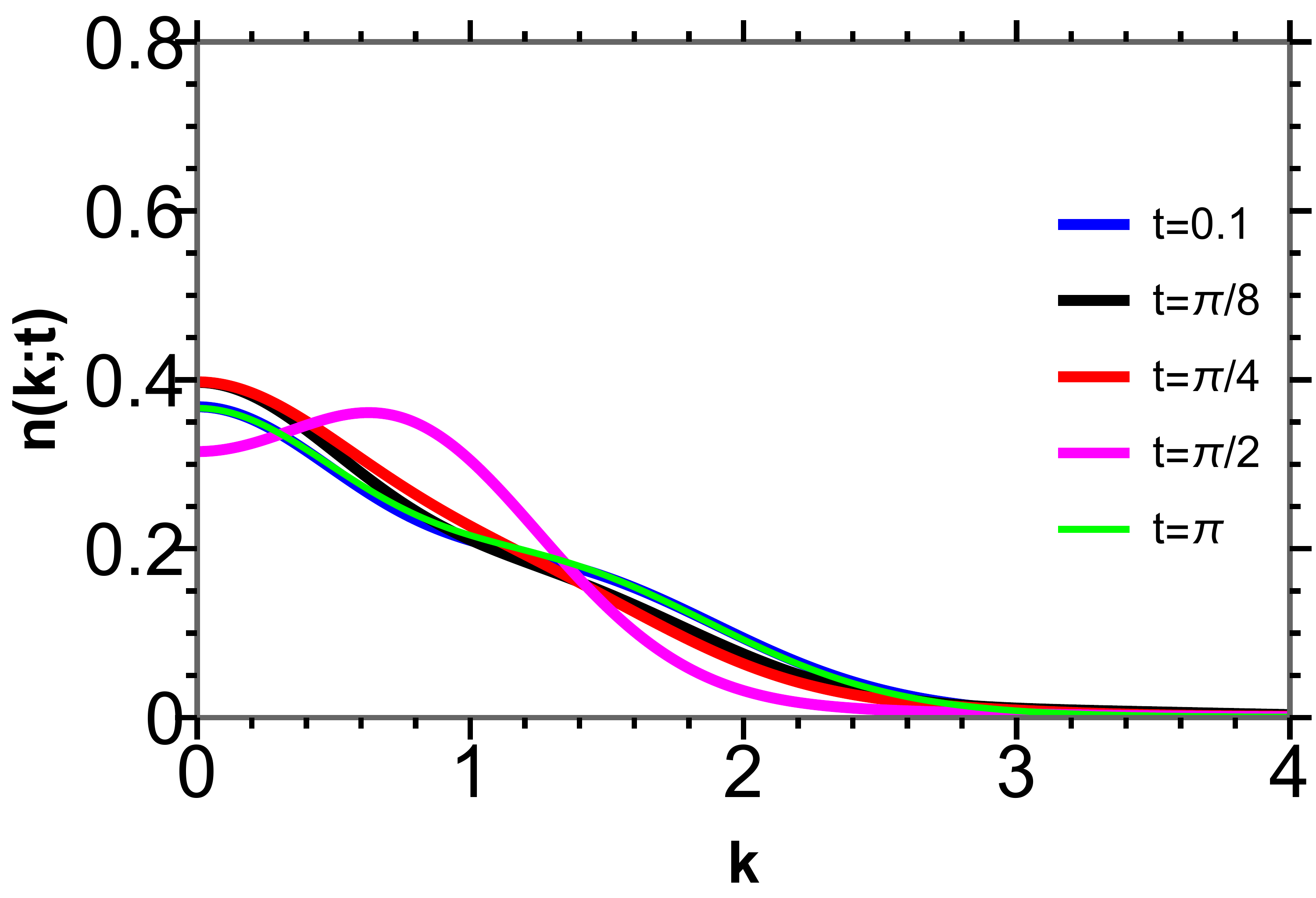}
	\caption{(Color online) Evolution of the momentum distribution $n(k;t)$ for the harmonic trap.}\label{bwd_exc_nk_all}
\end{figure}

The evolution of the momentum distribution $n(k;t)$ is shown in Fig.~\ref{bwd_exc_nk_all}. The shape of $n(k;t)$ up to $t=\pi/4$ changes a little and at $t=\pi/2$ (magenta line) there is a shift of the maximum of $n(k;t)$ to higher momentum. This shifts clearly suggests an impact of an another state on the dynamics. Then, by the time $t=\pi$, the shape of $n(k;t)$ (green line) returns to its initial form, as can be expected from the fidelity dynamics in Fig.~\ref{bwd_exc}.

\begin{figure}
	\centering
	\textbf{$\alpha=0$,~~~$t=\pi/4$}\par\medskip
	\begin{tabular}{cc}
		Prequench states &  Postquench states \\
		\hspace{-.5cm}\includegraphics[width=.25\textwidth]{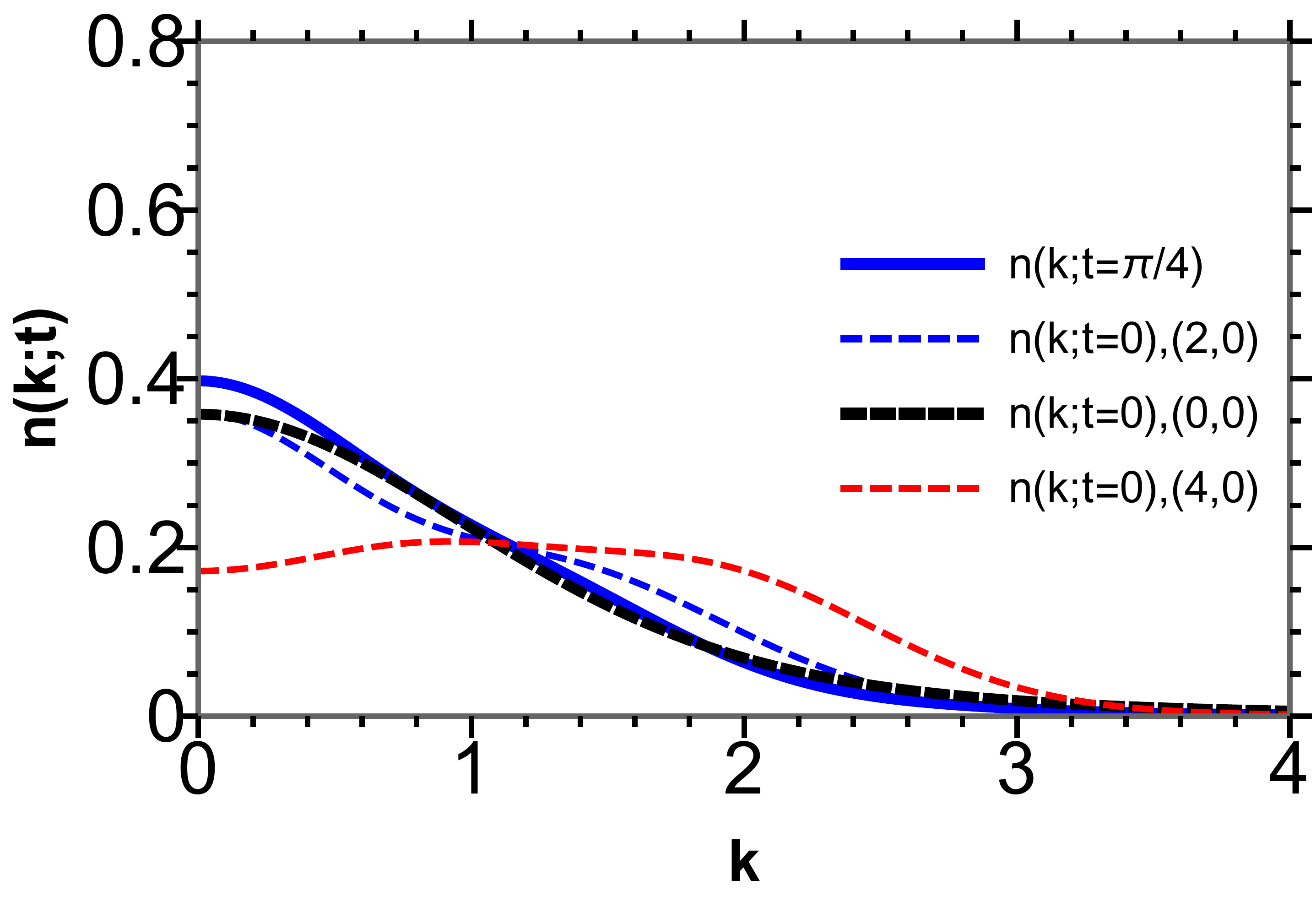} &
		\includegraphics[width=.25\textwidth]{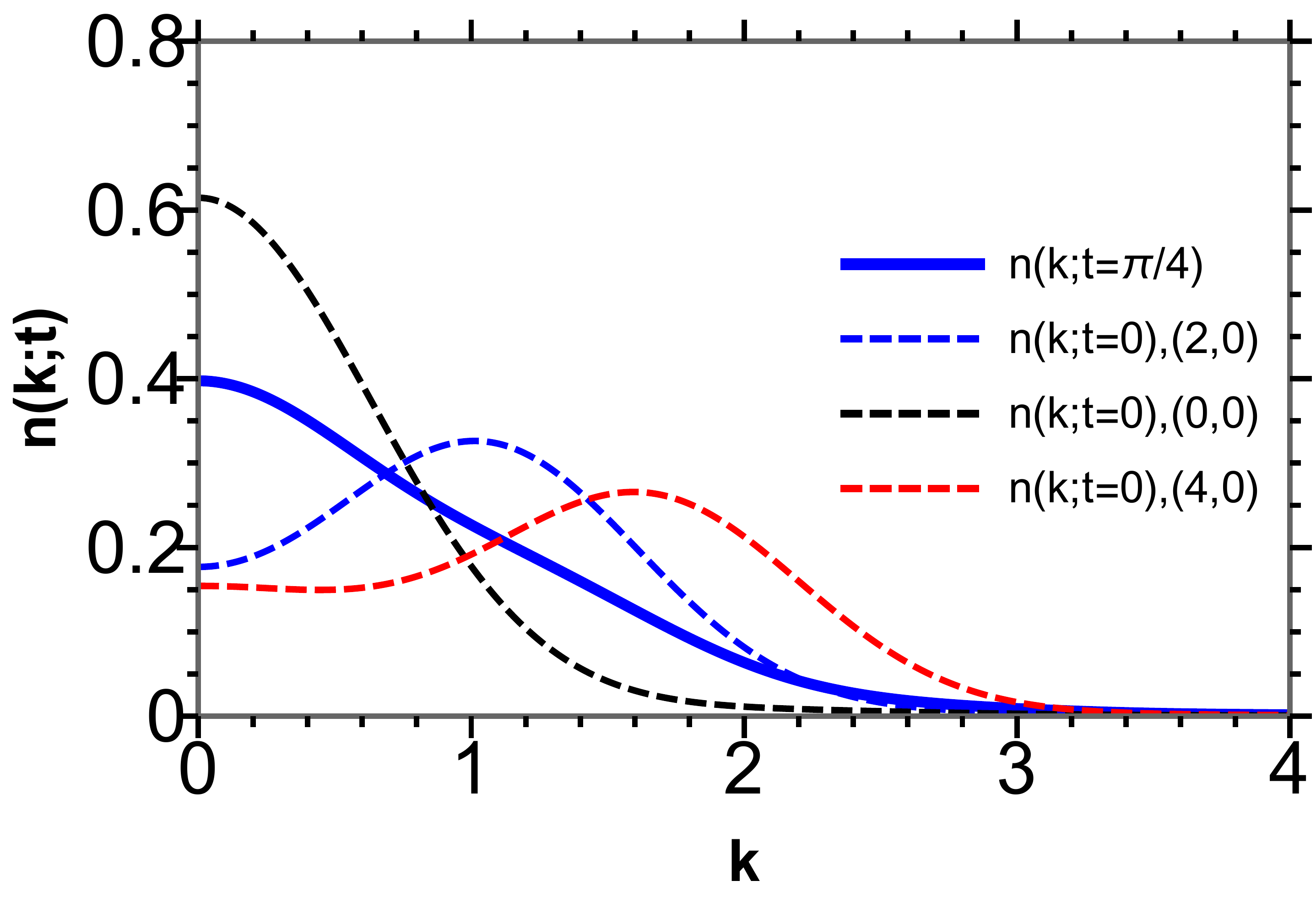} \\
	\end{tabular}
	\caption{(Color online) Comparison of the momentum distribution $n(k;t)$ (blue line) with the momentum distributions $n(k;t=0)$ for the pre- and postquench states (dashed lines).}\label{bwd_exc_nk_12}
\end{figure}

\begin{figure}
	\centering
	\textbf{$\alpha=0$,~~~$t=\pi/2$}\par\medskip
	\begin{tabular}{cc}
		Prequench states &  Postquench states \\
		\hspace{-.5cm}\includegraphics[width=.25\textwidth]{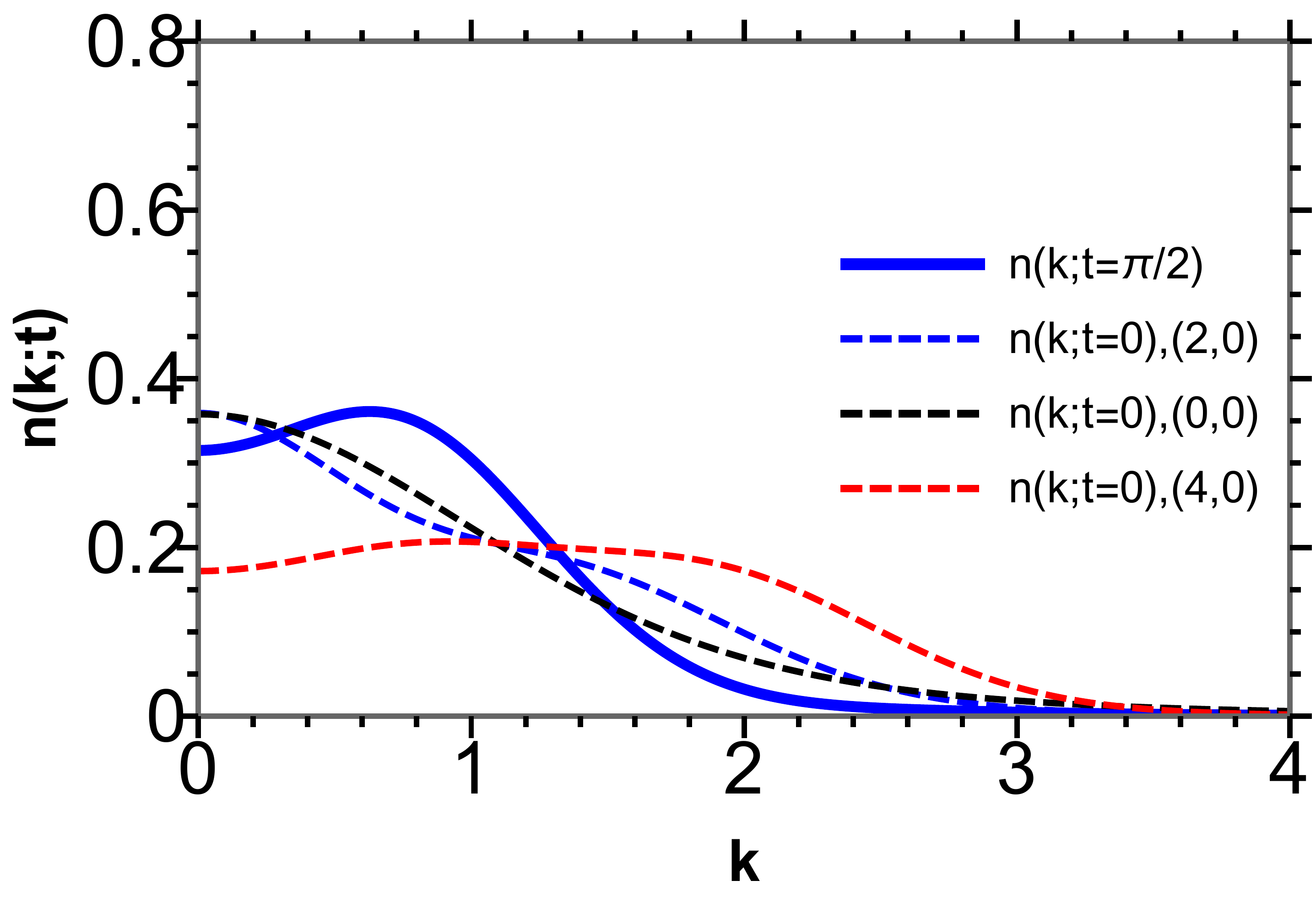} &
		\includegraphics[width=.25\textwidth]{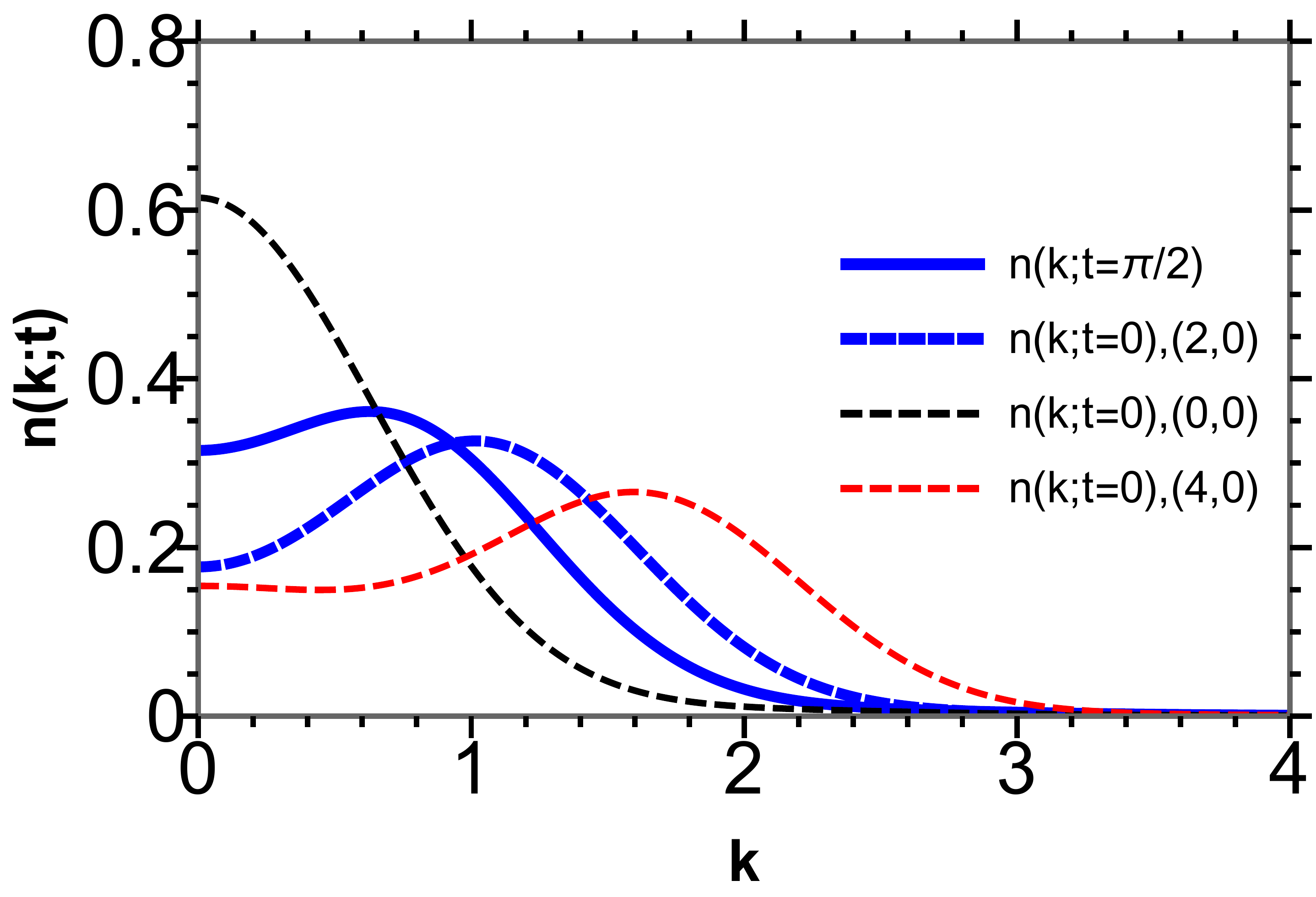} \\
	\end{tabular}
	\caption{(Color online) Comparison of the momentum distribution $n(k;t)$ (blue line) with the momentum distributions $n(k;t=0)$ for the pre- and postquench states (dashed lines).}\label{bwd_exc_nk_34}
\end{figure}

In Fig.~\ref{bwd_exc_nk_12} the shape of the momentum distribution $n(k;t)$ at $t=\pi/4$ (blue line) almost matches the shape of the momentum distribution for the prequench ground state (black dashed line). This prequench state surely should have quite an impact due to a large overlap for it around this moment of time (cf. Fig.~\ref{bwd_exc}). The shape of $n(k;t)$ at $t=\pi/2$ (blue line) in Fig.~\ref{bwd_exc_nk_34} resembles the shape of $n(k;t=0)$ for the postquench excited state $(2,0)$ (blue dashed line), which is also quite consistent with Fig.~\ref{bwd_exc}, since the overlap for this state is almost dominant.

\subsection{Anharmonic trap, $\alpha=-0.03$, the excited state, $(2,0)$}

\begin{figure}
	\centering
	\textbf{$\alpha=-0.03$}\par\medskip
	\includegraphics[width=8.5cm,clip]{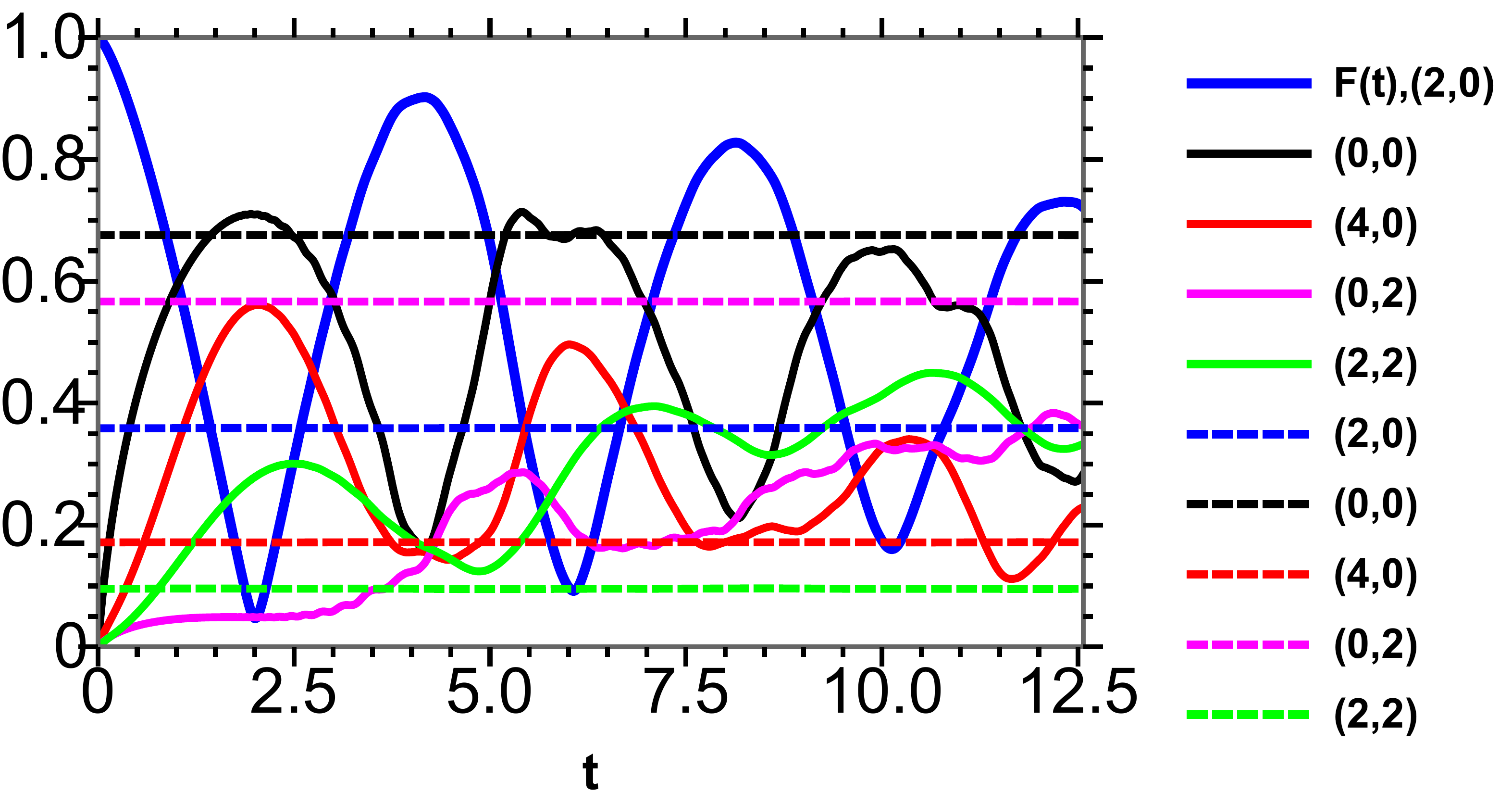}
	\caption{(Color online) Fidelity $F(t)$ and the overlap integrals $\mathcal{Q}$ between the time-evolving state $\Psi(x_1,x_2,t)$ and different pre- (solid lines) and postquench (dashed lines) states in the case of the anharmonic trap. The indices $(n,N)$ refer to the states with the quantum numbers of the relative and center-of-mass motions. The system of the two atoms is prepared in the excited state $(2,0)$ with $g=-2$ and quenched to $g=2$.}\label{bwd_exc_al}
\end{figure}

Now we consider the quench dynamics for the excited state $(2,0)$ in the case of the anharmonic trap potential. The anharmonic terms in the trap potential lead all of the considered states to impact on the quench dynamics (Fig.~\ref{bwd_exc_al}) and the whole periodic behavior of all the overlaps becomes significantly distorted. The oscillation of $F(t)$, within the considered time window, is damping, so are the oscillations of the $(0,0)$ and $(4,0)$ prequench states. The overlaps for the center-of-mass prequench excited states are, on the other hand, increasing. The overlap for the postquench ground state becomes dominant over all other overlaps for the postquench states.

\begin{figure}
	\centering
	\textbf{$\alpha=-0.03$}\par\medskip
	\includegraphics[width=7cm,clip]{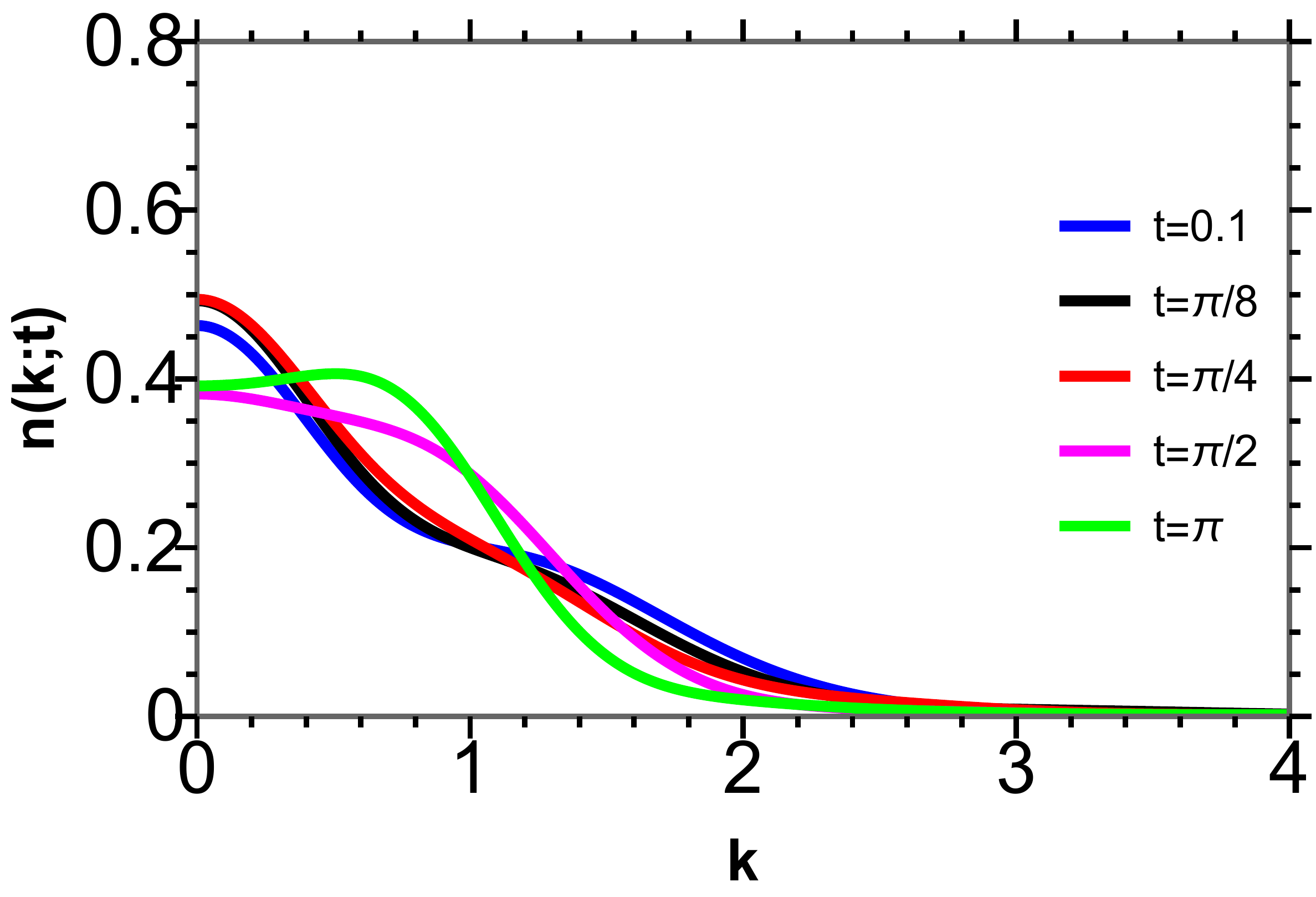}
	\caption{(Color online) Evolution of the momentum distribution $n(k;t)$ for the anharmonic trap.}\label{bwd_exc_al_nk_all}
\end{figure}

The evolution of the momentum distribution $n(k;t)$ in the case of the anharmonic trap in Fig.~\ref{bwd_exc_al_nk_all} is quite similar to that of the harmonic trap case up to $t=\pi/4$ such that there is a similar shape undergoing similar small deviations. At $t=\pi/2$ the shape (magenta line) changes noticeably with its peak being reduced and the width broadened. The shape at $t=\pi$ (green line), however, does not revert to the initial form in this case and its peak is shifted towards higher momentum.

\begin{figure}
	\centering
	\textbf{$\alpha=-0.03$,~~~$t=\pi/2$}\par\medskip
	\begin{tabular}{cc}
		Prequench states &  Postquench states \\
		\hspace{-.5cm}\includegraphics[width=.25\textwidth]{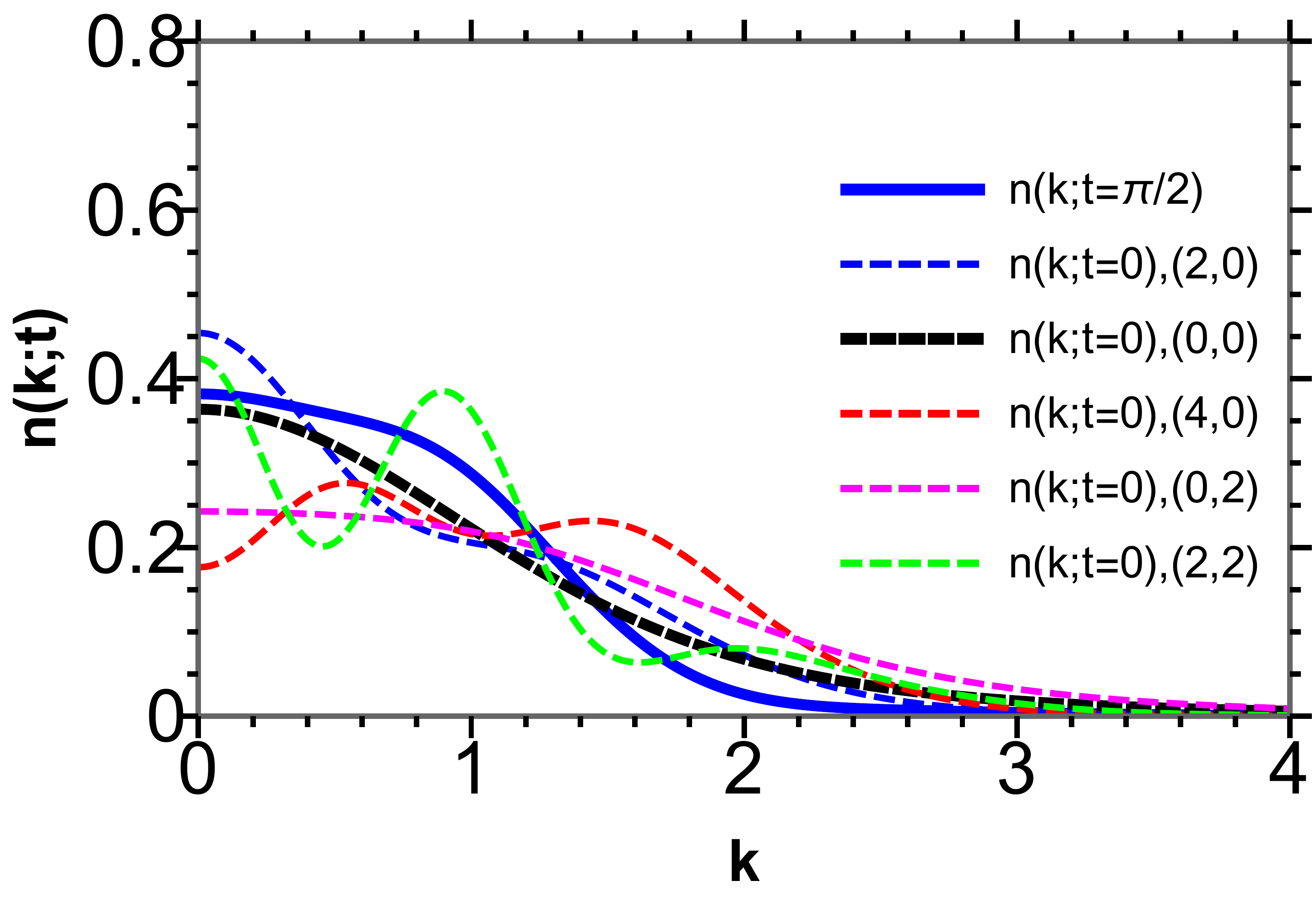} &
		\includegraphics[width=.25\textwidth]{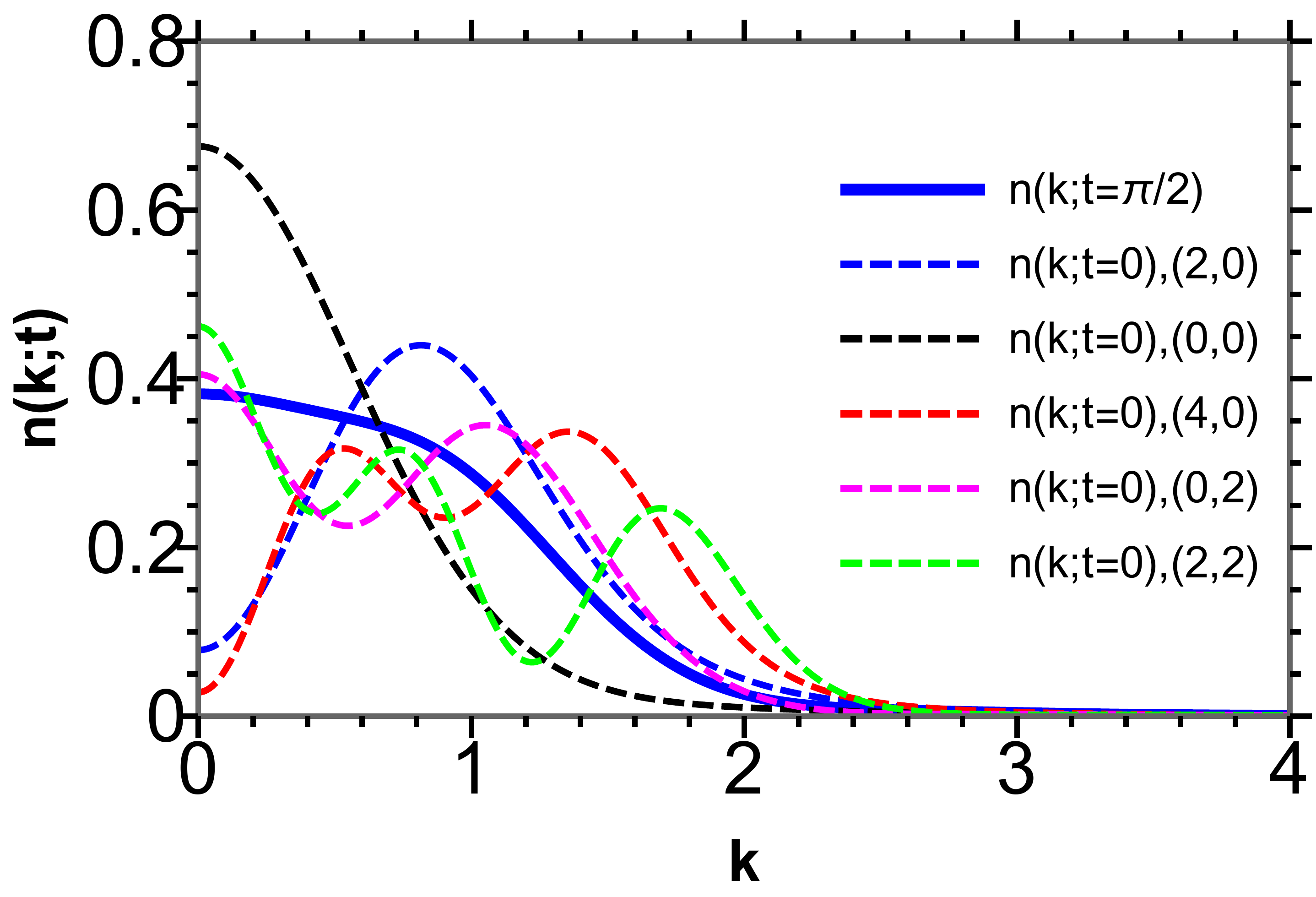} \\
	\end{tabular}
	\caption{(Color online) Comparison of the momentum distribution $n(k;t)$ (blue line) with the momentum distributions $n(k;t=0)$ for the pre- and postquench states (dashed lines).}\label{bwd_exc_al_nk_12}
\end{figure}

The shape of the momentum distribution $n(k;t)$ in Fig.~\ref{bwd_exc_al_nk_12} at $t=\pi/2$ (blue line) is quite similar to the shape of $n(k;t=0)$ for the prequench ground state (black dashed line) among all the other shapes. This should be, indeed, the case according to Fig.~\ref{bwd_exc_al}, where around this moment of time the overlap between the wave packet and this state is large.

\section{Summary}
\label{summary}

The quench dynamics of the two-particle system in the one-dimensional harmonic and anharmonic traps was analyzed in details. The initial coupling strength of the particles is quenched from the repulsive to attractive interaction and vice versa. The terms pre- and postquench states are used for a clarity purpose. The both cases showed quite different and yet approximately periodic pattern in the harmonic trap potential. The analysis was performed in terms of the evolution of the fidelity, overlap integrals between the wave-packet and the pre- and postquench states, reduced one-body density matrix and momentum distribution. The evolution of all this quantities was compared to the corresponding quantities of the pre- and postquench states. In particular, matching the momentum distribution with the prequench states in most cases confirmed their high contribution to the dynamics. The presence of the repulsive interaction during the evolution reduces the frequency of the fidelty and damps the amplitude of the fidelity when the system is quenched from the ground state. The anharmonic corrections to the trap potential induces the excited states connected with the center-of-mass motion to impact on the system dynamics, which leads to a more distorted oscillation pattern of the all considered quantities. Matching the momentum distribution with the pre- and postquench states in this case becomes less clear.

Such an analysis can be extended to a more complex systems with a larger particle number and higher dimensions. A more realistic interaction potential, such as the Morse potential \cite{vinitsky2}, which is not symmetric, can also be considered. Including the spin interaction is also one of the steps towards matching a theory with an experiment \cite{bergschneider}.

\end{document}